\documentclass[acmtog,screen]{acmart}

\usepackage{xcolor}
\usepackage{float}
\usepackage{verbatim}
\usepackage{algorithm}  
\usepackage{algorithmicx}  
\usepackage{algpseudocode}
\usepackage{tabularx}
\usepackage{booktabs}
\usepackage{enumitem}
\usepackage{dsfont}
\usepackage{listofitems}
\usepackage{graphicx}
\usepackage[labelfont=bf,format=plain,singlelinecheck=false]{caption}
\usepackage{subcaption}

\definecolor{mypurple}{HTML}{8E7DE4}
\definecolor{mygreen}{HTML}{00A444}
\definecolor{myorange}{HTML}{FF9F36}
\definecolor{myblue}{HTML}{4271D6}
\definecolor{mypink}{HTML}{F97BA2}
\definecolor{myRed}{HTML}{fcc0f6}
\definecolor{myCyan}{HTML}{00b2cd}

\AtBeginDocument{%
  \providecommand\BibTeX{{%
    \normalfont B\kern-0.5em{\scshape i\kern-0.25em b}\kern-0.8em\TeX}}}

\newcommand{\rr}{\mathbf{r}}

\newcommand{\pp}{\mathbf{p}}

\renewcommand{\tt}{\mathbf{t}}

\newcommand\restr[2]{{\left.\kern-\nulldelimiterspace{}#1\right|_{#2}}}

\newcommand{\FLIP}{\textnormal{\reflectbox{F}LIP }}

\usepackage{tikz}

\newcommand{\flamewithbox}[3]{%
\begin{tikzpicture}%
\node[anchor=south west,inner sep=0](image) at (0,0){\includegraphics[trim=200 200 200 200,clip,width=#1]{#3}};%
\begin{scope}[x={(image.south east)},y={(image.north west)}]%
\definecolor{boxcolor1}{RGB}{#2}%
\draw[rounded corners=0pt,color=boxcolor1,thick] (0.160, 0.481) rectangle (0.320, 0.641);%
\end{scope}
\end{tikzpicture}
}

\setcopyright{acmcopyright}
\copyrightyear{2018}
\acmYear{2018}
\acmDOI{XXXXXXX.XXXXXXX}

\acmConference[ ]{ }{ }{ }
\acmPrice{15.00}
\acmISBN{978-1-4503-XXXX-X/18/06}

\renewcommand\footnotetextcopyrightpermission[1]{} 

\begin{document}
\newcommand{\yang}[1]{{\color{blue}[\textbf{Yang:} {\em #1}]}}

\newcommand{\junqiu}[1]{{\color{green}[\textbf{Junqiu:} {\em #1}]}}

\newcommand{\daqi}[1]{{\color{purple}[\textbf{Daqi:} {\em #1}]}}

\newcommand{\tao}[1]{{\color{red}[\textbf{Tao:} {\em #1}]}}

\def\imgtrimleft{0}
\def\imgtrimbottom{0}
\def\imgtrimright{0}
\def\imgtrimtop{0}

\def\topimgtrimleft{0}
\def\topimgtrimbottom{0}
\def\topimgtrimright{0}
\def\topimgtrimtop{0}

\def\midimgtrimleft{0}
\def\midimgtrimbottom{0}
\def\midimgtrimright{0}
\def\midimgtrimtop{0}

\def\extractimgdim#1{\twowords#1\relax}
\def\extractimgtrim#1{\fourwords#1\relax}
\def\twowords#1 #2\relax{%
    \def\imgtrimwidth{#1}%
    \def\imgtrimheight{#2}%
}
\def\fourwords#1 #2 #3 #4\relax{
    \def\imgtrimleft{#1}%
    \def\imgtrimbottom{#2}%
    \def\imgtrimright{#3}%
    \def\imgtrimtop{#4}%
}

\newcommand{\testdrawer}[3]{%
\def\tempbase{#1}%
\def\tempw{#2}%
\def\temph{#3}%
\testdrawercont}

\newcommand{\testdrawercont}[9]{%
\definecolor{topcolor}{RGB}{255,255,255}
\definecolor{botcolor}{RGB}{255,255,255}
\begin{minipage}{#9}%
        \begin{tikzpicture}%
        \node[anchor=south west,inner sep=0](image) at (0,0) {\includegraphics[trim=0 0 0
        0,clip,width=\columnwidth]{\tempbase}};%
        \begin{scope}[x={(image.south east)},y={(image.north west)}]%
        \pgfmathsetmacro{\ttlx}{(#1)/\tempw};%
        \pgfmathsetmacro{\ttly}{(#2)/\temph};%
        \pgfmathsetmacro{\tbrx}{(\tempw-#3)/\tempw};%
        \pgfmathsetmacro{\tbry}{(\temph-#4)/\temph};%
        \draw[color=orange,thick] (\ttlx, \ttly) rectangle (\tbrx, \tbry);%
        \pgfmathsetmacro{\btlx}{(#5)/\tempw};%
        \pgfmathsetmacro{\btly}{(#6)/\temph};%
        \pgfmathsetmacro{\bbrx}{(\tempw-#7)/\tempw};%
        \pgfmathsetmacro{\bbry}{(\temph-#8)/\temph};%
        \draw[color=myCyan,thick] (\btlx, \btly) rectangle (\bbrx, \bbry);%
        \end{scope}%
        \end{tikzpicture}%
\end{minipage}%
}

\newcommand{\testdriver}[5]{%
\extractimgdim{#2}%
\extractimgtrim{#3}%
\let\topimgtrimleft\imgtrimleft%
\let\topimgtrimbottom\imgtrimbottom%
\let\topimgtrimright\imgtrimright%
\let\topimgtrimtop\imgtrimtop%
\extractimgtrim{#4}%
\testdrawer{#1}{\imgtrimwidth}{\imgtrimheight}
{\topimgtrimleft}{\topimgtrimbottom}{\topimgtrimright}{\topimgtrimtop}
{\imgtrimleft}{\imgtrimbottom}{\imgtrimright}{\imgtrimtop}{#5}%
}

\setlength{\fboxsep}{0pt}%
\setlength{\fboxrule}{1pt}%

\newcommand{\incgraphics}[3][0 0 0 0]{\includegraphics[trim=#1,clip,width=#2]{#3}}

\newcommand{\imgWithZoom}[4][1450 620]{%
\begin{minipage}{0.33\linewidth}\centering%
\testdriver{#4}{#1}{#2}{#3}{\linewidth}\vspace{0.005\linewidth}\\%
\fcolorbox{orange}{white}{\incgraphics[#2]{0.485\linewidth}{#4}}\hfill
\fcolorbox{myCyan}{white}{\incgraphics[#3]{0.485\linewidth}{#4}}\\
\end{minipage}%
}

\newcommand{\imgWithZoomThree}[5]{%
\imgWithZoom{#1}{#2}{#3}\hfill%
\imgWithZoom{#1}{#2}{#4}\hfill%
\imgWithZoom{#1}{#2}{#5}%
}

%

\title{Real-time Level-of-detail Strand-based Hair Rendering}



\author{Tao Huang}
\authornote{Part of this work was done when Tao Huang was an intern at LightSpeed Studios.}
\orcid{}
\email{tao_huang@ucsb.edu}
\affiliation{%
  \institution{University of California, Santa Barbara \& LightSpeed Studios}
  \city{Santa Barbara}
  \state{CA}
  \country{USA}
}

\author{Yang Zhou}
\orcid{0009-0006-3469-8058}
\email{yzhou426@cs.ucsb.edu}
\affiliation{%
  \institution{University of California, Santa Barbara}
  \city{Santa Barbara}
  \state{CA}
  \country{USA}
}

\author{Daqi Lin}
\orcid{0000-0002-5139-6418}
\email{daqil@nvidia.com}
\affiliation{%
  \institution{NVIDIA}
  \city{Redmond}
  \state{WA}
  \country{USA}
}

\author{Junqiu Zhu}
\orcid{}
\email{junqiuzhu@ucsb.edu}
\affiliation{%
  \institution{University of California, Santa Barbara}
  \city{Santa Barbara}
  \state{CA}
  \country{USA}
}

\author{Ling-Qi Yan}
\orcid{}
\email{lingqi@cs.ucsb.edu}
\affiliation{%
  \institution{University of California, Santa Barbara}
  \city{Santa Barbara}
  \state{CA}
  \country{USA}
}

\author{Kui Wu}
\orcid{0000-0003-3326-7943}
\email{kwwu@global.tencent.com}
\affiliation{%
  \institution{LightSpeed Studios}
    \city{Los Angeles}
  \state{CA}
  \country{USA}
}

\begin{abstract}
We present a real-time strand-based hair rendering framework that ensures seamless transitions between different levels of detail (LoD) while maintaining a consistent appearance. Our LoD framework dynamically, adaptively, and independently replaces clusters of individual hair strands with \emph{thick hairs} based on their projected screen widths. We introduce a Bidirectional Curve Scattering Distribution Function (BCSDF) model for the thick hairs, which accurately captures both single and multiple scattering within the hair cluster. Through tests on diverse hairstyles, various hair colors, and animations, our framework closely replicates the appearance of multiple-scattered full hair geometries at various viewing distances, achieving up to a $13\times$ speedup.
\end{abstract}

%

%



\maketitle
\section{Introduction}

Hair contributes significantly to believable characters and immersive environments in films, video games, or virtual reality. However, an average human scalp has about $100K$ - $150K$ hair follicles. So, rendering and simulating realistic hair with complex geometries at such a scale poses significant challenges to computational efficiency for real-time applications. In the gaming industry, a common technique is to use hair cards~\cite{YIbing2016}, flat, textured planes that simulate the visual appearance of hair. While this technique provides computational efficiency, the inherent coarse approximation limits the realism achievable in physics simulations and rendering.

Strand-based hair simulation and rendering have gained prominence in recent production practices for their capacity to offer a more authentic portrayal of hair behavior and appearance~\cite{tafuri2019strand,unrealengine,Huang2023,Hsu2023}. Given the vast number of hairs, a level-of-detail (LoD) strategy that simplifies the model at different viewing distances is essential to optimize computational resources for real-time applications. However, existing LoD solutions for strand-based hair struggle with smooth transitions in dynamics and appearance between LoD levels. One approach is to use hair cards when the hair is at a distance. However, this introduces a notable discontinuity in dynamics and appearance during the transition from strands to cards due to their fundamentally different representations. An alternative strategy reduces the number of hair strands while increasing each remaining strand's thickness. Regrettably, this method leads to an excessively bold and dry hair appearance compared to the realistic brightness and hue, as the thick hair meant to represent a cluster of hair strands should have distinct optical properties from a single hair strand. This distinction becomes particularly pronounced during multiple light bounces, exacerbating the visual disparity. Recent research proposes aggregated scattering models for a collection of furs~\cite{ZhuZWXY22} and fibers~\cite{ZhuZJYA23} that capture the aggregated appearance, incorporating the effect of multiple scattering. However, neither of their methods can be used in real-time hair rendering due to the slow neural network inference \cite{ZhuZWXY22} and the inaccurate handling of multiple scattering within an aggregated ply, which is exaggerated at the model level~\cite{ZhuZJYA23}, respectively.

This work presents a novel real-time technique for rendering strand-based hairs with seamless transitions between different LoD levels while ensuring consistent hair appearance. Our approach starts with a hair hierarchy construction method, where we build hair clusters at each LoD level and fit each cluster with a thick hair to better conform to the overall shape. To shade these thick hairs, we consider the sparsity of hairs in a cluster and reframe existing aggregation scattering models~\cite{ZhuZWXY22, ZhuZJYA23} to derive a Bidirectional Curve Scattering Distribution Functions (BCSDF) model, capturing both single and multiple scattering interactions between individual hairs within the hair cluster. Our multiple scattering component modifies the dual scattering \cite{ZinkeYWK08} approximation by redistributing the lobe energies and introducing an extra light path series. We present a real-time strand-based hair rendering framework that dynamically selects the appropriate hair LoD level, along with an on-the-fly hair width calculation method to tightly cover represented fine hair strands. We implement our real-time rendering pipeline in modern GPU rasterization pipeline and conduct tests on various hairstyles with dynamics. In our experiments, our LoD framework, combined with our new BCSDF model, facilitates smooth transitions with a consistent appearance, resulting in neglectable computational overhead (<1\%) for close-up views, a speedup of 2$\times$ for middle views, and up to 13$\times$ for far views.

\section{RELATED WORK} \label{sec:related}

\paragraph{Hair BCSDF model}
\citet{KajiyaK89} introduced the first physically-based analytic fiber shading model, considering the variation of reflectance in azimuthal directions and accounting for internal absorption and caustics. This model is later extended by \citet{MarschnerJCWH03} into the well-known Bidirectional Curve Scattering Distribution Functions (BCSDF) model for hair, which remains actively used in the games and film industries. The analysis involved ray paths within dielectric cylinders, and the scattering is split into $R$, $TT$, and $TRT$ modes. These modes are represented as separable products of azimuthal and longitudinal functions, where $R$ and $T$ signify one reflection and transmission inside the fiber, respectively. This model has undergone improvements by various researchers. Notable enhancements include adding a non-physical diffuse term for accurate fitting of measured data~\cite{ZinkeRLWAK09}, decoupling into artist-friendly lobes~\cite{Sadeghi2010}, adding a $TRRT$ lobe for energy conservation~\cite{dEonFHLA11}, and adding lobe for high-order internal reflections~\cite{ChiangBTB16}. Additionally, it has been extended for elliptical hairs~\cite{Khungurn2017}, animal hairs with interior medullas~\cite{Yan2015}, and microfacet-based hair scattering model ~\cite{Huang2022hair}. In recent work, \citet{Xia2020, Xia2023} propose a wave optics scattering model to capture the colorful glints of hair and fur.

\paragraph{Multiple scattering among hairs}
Multiple scattering is essential for realistic hair rendering but is computationally expensive.
\citet{MoonM06} use photon mapping to track indirect light bounces, storing incident radiance in a 3D grid of spherical harmonic (SH) coefficients~\cite{moon08}. \citet{kt2023} use MLP to learn the high-order scattered radiance between hairs. \citet{ZinkeYWK08} introduce the dual scattering technique, dividing computations into global multiple scattering and local multiple scattering. Using a deep opacity map~\cite{Yuksel2008} to accelerate global scattering, they achieved real-time strand-based hair rendering. Our work also builds upon the dual scattering by introducing an additional local scattering component to efficiently approximate the complex scattering effects between aggregated hairs in real-time scenarios.

\paragraph{Level-of-detail construction}
The level-of-detail (LoD) strategy is commonly employed in real-time applications to optimize performance by using models with varying complexities based on viewer distance~\cite{karis2021deep}. Various simplification techniques have been developed to create LoD meshes at different complexity levels, including element removal~\cite{garland1997surface, hoppe1996progressive, luebke2002LoD}, re-triangulation~\cite{VSA2004, Li2021}, and primitives fitting~\cite{PolyFit2017, ksr2020}. However, traditional LoD methods do not suit filament-like shapes such as hair due to their unique geometry. \citet{Weier2023} present an adaptive neural representation for correlation-aware LoD voxels designed for general purposes. 
In simulation~\cite{Bertails2003} and hair editing~\cite{Ward2003}, a bottom-up LoD strategy was employed to group hair strands into clusters by their scalp locations. Their approach, however, introduces noticeable appearance differences between LoD levels due to applying a single hair appearance model for a hair cluster, ignoring light interaction within the cluster. We use a top-down strategy to construct the hair strand hierarchy, which is more precise.

\paragraph{Surface-based hair simplification -- hair cards}  
\citet{Koh2001} pioneer the use of a polygon strip to represent a cluster of hair strands with alpha maps for real-time applications, a technique still prevalent in the game industry~\cite{YIbing2016}. However, creating hair cards for different LoD levels is time-consuming for artists and can cause discontinuities in dynamics and appearance during transitions.  In contrast, the LoD solution proposed in this paper only requires a model with full hair strands, eliminating the need for additional manual authoring in LoD asset creation.

\paragraph{Strand-based hair simplification -- aggregation model}
Based on the hierarchical structure of yarn geometries, the aggregation model has recently been employed to depict the optical characteristics of a cluster of fibers in woven ~\cite{MontazeriGZJ20} and knitted structures~\cite{Montazeri2021}. Their model integrates normal mapping with a specialized BSDF that combines a specular component for specular reflection and transmission and a body component for multiple scattering within the fibers bundle. \citet{Cook2007} introduce a stochastic technique by randomly selecting a subset of the geometric elements and altering those elements statistically to preserve the overall appearance. \citet{ZhuZWXY22} proposed an aggregation BCSDF model for a bunch of fibers using neural networks. In contrast, \citet{ZhuZJYA23} derived an analytical single and multiple light scattering model for yarns based on dual-scattering theory, eliminating the need for data collection for parameter fitting. 

Note that the analytical aggregated scattering model~\cite{ZhuZJYA23} is based on a single fiber BCSDF with three lobes, $R$, $TT$, and $D$ (instead of $TRT$). Our work builds heavily on this approach with two significant improvements, and our framework is generalizable to also support the Marschner hair BCSDF~\cite{MarschnerJCWH03} with $R$, $TT$, and $TRT$ lobes. Firstly, we redistribute the lobe energies and introduce an extra light path series in the aggregation scattering model for more accurate light scattering approximation within the hair cluster. Secondly, our model is generalized to support an elliptical cross-section, creating a tighter fitting bound for hair clusters. 
\section{Background} \label{sec:background}

This section briefly overviews the single hair shading model, the dual scattering technique used for multiple scattering approximation, and the previous aggregated shading model that serves as the foundation for our proposed method.

The general rendering equation for outgoing radiance $L_o$ towards a direction parameterized by longitudinal-azimuthal angles ($\theta$, $\phi$) is defined as: 
\begin{align*} 
L_o(\theta_o, \phi_o) = \int_{-\pi}^\pi \int_{-\frac{\pi}{2}}^{\frac{\pi}{2}}  L_i(\theta_i, \phi_i) f(\theta_i, \phi_i, \theta_o, \phi_o) \cos^2(\theta_i) d \theta_i d\phi_i ,
\end{align*}
where $L_i$ and $L_o$ denote incoming radiance from direction ($\theta_i$, $\phi_i$) and outgoing radiance in direction ($\theta_o$, $\phi_o$), respectively.

\subsection{Single Hair Shading Model} \label{sec:singlehair}

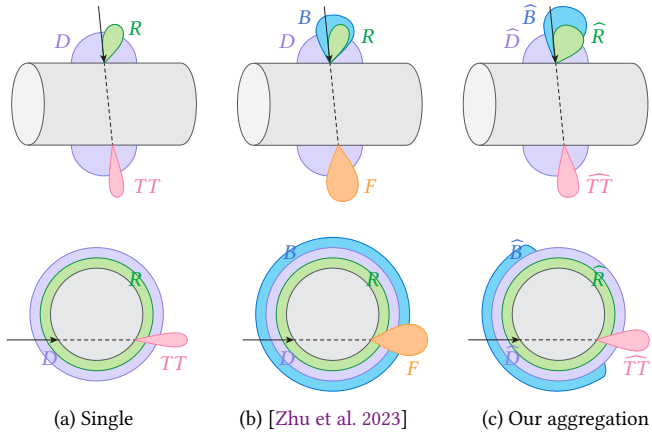
\begin{figure}[ht!]
\includegraphics[width=\linewidth]{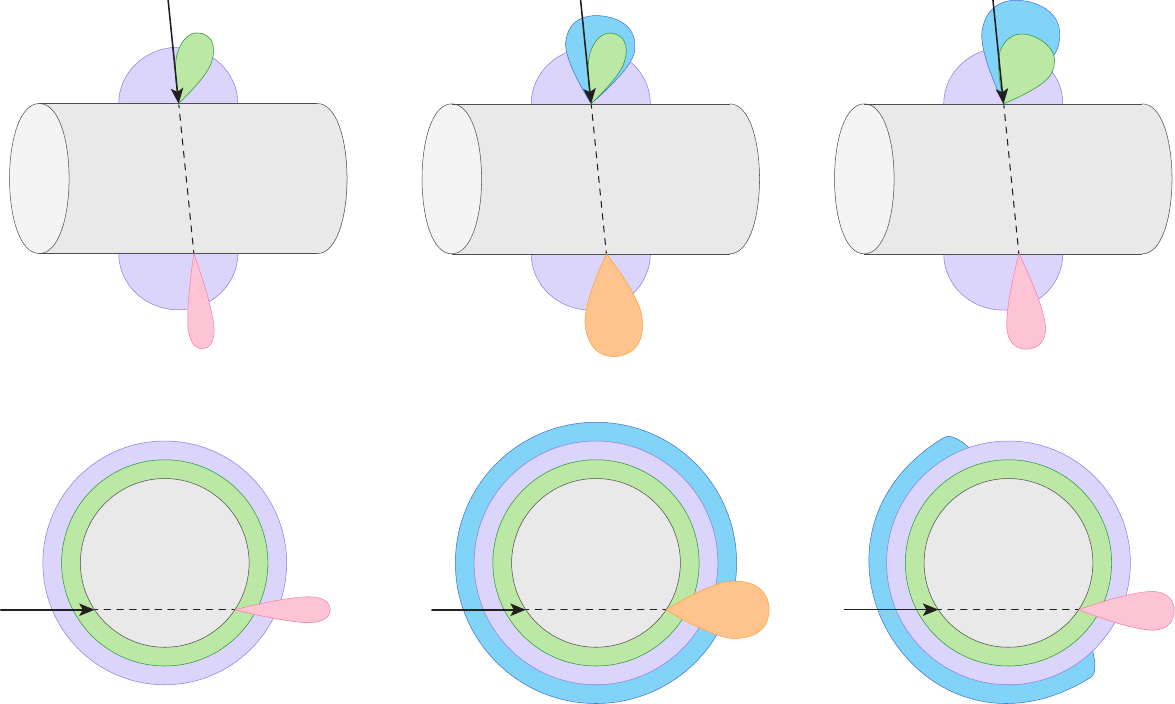}
\put(-225,130){\small\textcolor{mypurple}{$D$}}
\put(-140,130){\small\textcolor{mypurple}{$D$}}
\put(-55,130){\small\textcolor{mypurple}{$\widehat{D}$}}
\put(-230,10){\small\textcolor{mypurple}{$D$}}
\put(-140,10){\small\textcolor{mypurple}{$D$}}
\put(-55, 10){\small\textcolor{mypurple}{$\widehat{D}$}}
\put(-197,135){\small\textcolor{mygreen}{$R$}}
\put(-109,132){\small\textcolor{mygreen}{$R$}}
\put(-22, 132){\small\textcolor{mygreen}{$\widehat{R}$}}
\put(-197,40){\small\textcolor{mygreen}{$R$}}
\put(-107,40){\small\textcolor{mygreen}{$R$}}
\put(-22, 40){\small\textcolor{mygreen}{$\widehat{R}$}}
\put(-133,139){\small\textcolor{myblue}{$B$}}
\put(-48,139){\small\textcolor{myblue}{$\widehat{B}$}}
\put(-138.5,50){\small\textcolor{myblue}{$B$}}
\put(-53,50){\small\textcolor{myblue}{$\widehat{B}$}}
\put(-195,75){\small\textcolor{mypink}{$TT$}}
\put(-24,75){\small\textcolor{mypink}{$\widehat{TT}$}}
\put(-185,8){\small\textcolor{mypink}{$TT$}}
\put(-10, 6){\small\textcolor{mypink}{$\widehat{TT}$}}
\put(-108,75){\small\textcolor{myorange}{$F$}}
\put(-92,6){\small\textcolor{myorange}{$F$}}
\begin{flushleft}{\small \hspace{2em} (a) Single \hspace{4.5em} (b) ~\cite{ZhuZJYA23} \hspace{3.em} (c) Our aggregation}\end{flushleft}
\caption{Illustration of the single, aggregated~\cite{ZhuZJYA23}, and our BCSDF for longitudinal (top) and azimuthal (bottom) components. Our model replaces $F$ lobe with $\widehat{TT}$ lobe and distributes part of its contribution to $\widehat{R}$, $\widehat{TT}$, and $\widehat{D}$ with larger variances. Additionally, our $\widehat{B}$ lobe is designed to only account for the outgoing paths toward the back hemisphere.}
\label{fig:illustration}
\Description{}
\end{figure}

As suggested by~\citet{ZhuZJYA23}, the interaction between light and a single hair can be decomposed into reflection $R$ and transmission $TT$ lobes~\cite{MarschnerJCWH03}, along with an additional diffuse lobe $D$ to approximate multiple scattering inside the hair (see~\autoref{fig:illustration}a). The single hair BCSDF, denoted as $f$, is defined as:
\begin{align} \label{eq:bcsdf}
    f^\text{single}(\theta_i, \phi_i, \theta_o, \phi_o) = \sum_{p \in \{R,TT,D\}}A_p ~ M_p(\theta_h) ~ N_p(\phi),
\end{align}
where $p$ denotes different types of lobes, $\phi = \phi_o - \phi_i$ is the relative azimuthal angle, $\theta_h = (\theta_i+\theta_o)/2 $ is the longitudinal half angle, and $A_p$ is the attenuation function associated with each lobe. $M_p$ and $N_p$ are longitudinal and azimuthal scattering functions, respectively. In this paper, all results are using this BCSDF for a single fiber.

\subsection{Dual Scattering} \label{sec:ds}

\citet{ZinkeYWK08} proposed \underline{d}ual \underline{s}cattering (DS) to approximate multiple scattering between millions of hair segments efficiently. The key idea is to split the multiple scattering into two additional lobes: global multiple scattering and local multiple scattering. The global scattering approximates the amount of light remaining after \underline{f}orward scattering through a number of hairs, while local scattering estimates the total contribution through \underline{b}ackward scattering after multiple bounces in the neighborhood. Thus, two scattering lobes are added to the local shading (\autoref{eq:bcsdf}) as an approximation:
\begin{align}
    f^{ds}_F(\theta_i,\theta_o,\phi) &= \frac{I_F(\phi)}{\cos^2(\theta_i)}d_F A_F \sum_{p \in \{R,TT,D\}} G(\bar{\sigma}_F^2, \theta_o+\theta_i)N_p^F, \\
    f^{ds}_B(\theta_i,\theta_o,\phi) &= \frac{I_B(\phi)}{\pi \cos^2(\theta_i)} d_B A_B G(\bar{\sigma}_B^2, \theta_o+\theta_i),
\end{align}
where $A_{F|B}$, $d_{F|B}$, $\bar{\sigma}_{F|B}$, and $I_{F|B}$ represent the forward/backward attenuations, scattering density constants, averaged spread variance, and binary backward hemisphere indicators, respectively. $N_p^F$ is the averaged azimuthal lobe $p$ within the front hemisphere. Forward scattering attenuation $A_F$ can be computed by the sequential product of averaged forward attenuations that the light passes through, while $A_B = A_1 + A_3$ includes all light paths with single backward scattering $A_1$ and with three backward scattering $A_3$.  $G(\beta, \theta)=\frac{1}{\sqrt{2\pi}\beta} e^{-\theta^2/2\beta^2}$ is the traditional Gaussian distribution. We refer readers to the previous work~\cite{ZinkeYWK08} for more details. However, dual scattering only considers $R$ and $TT$ and assumes infinite hair strands behind the shading point.

\begin{figure}[ht]
\newcommand{\figcap}[1]{\begin{minipage}{0.24\linewidth}\centering#1\end{minipage}}
{$\quad$~}
\rotatebox[origin=c]{90}{\includegraphics[trim={400 340 350 400},clip,height=0.222\linewidth]{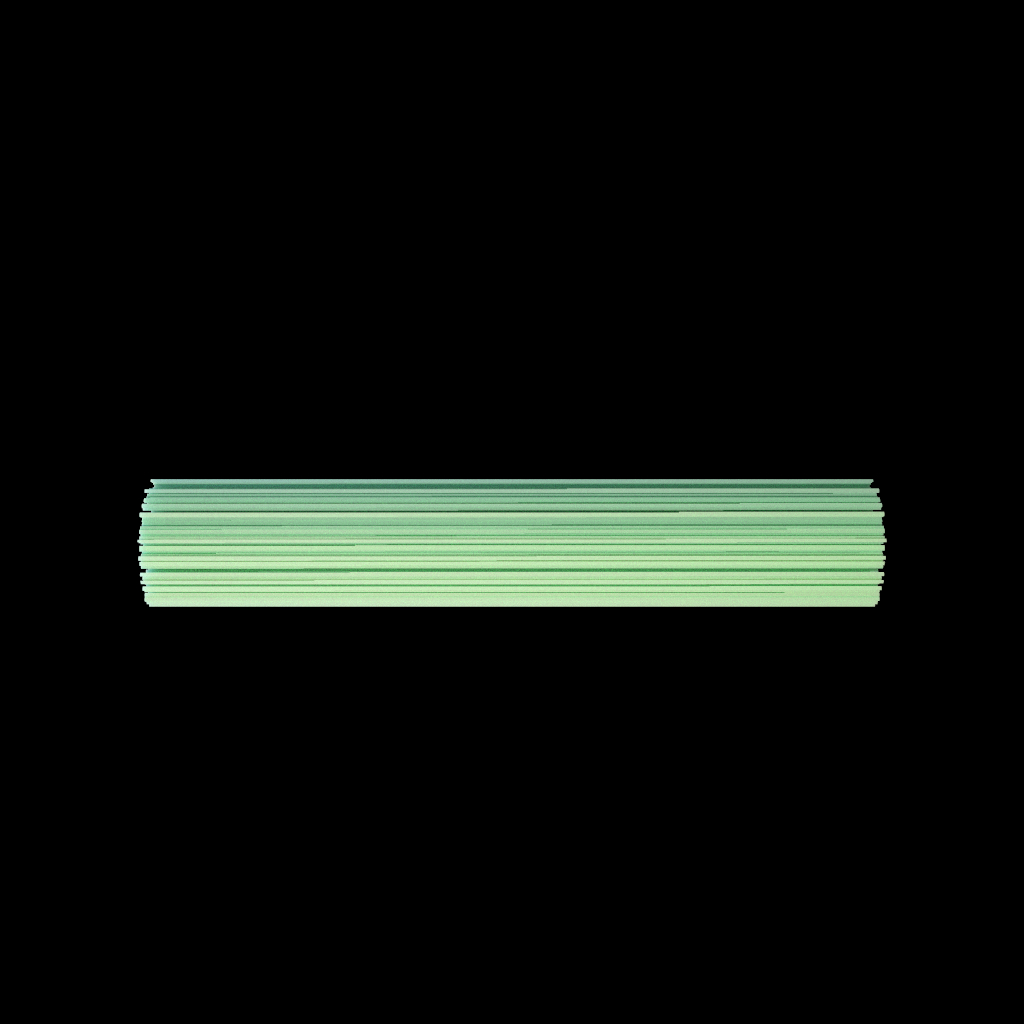}}\hfill%
\rotatebox[origin=c]{90}{\includegraphics[trim={400 340 350 400},clip,height=0.222\linewidth]{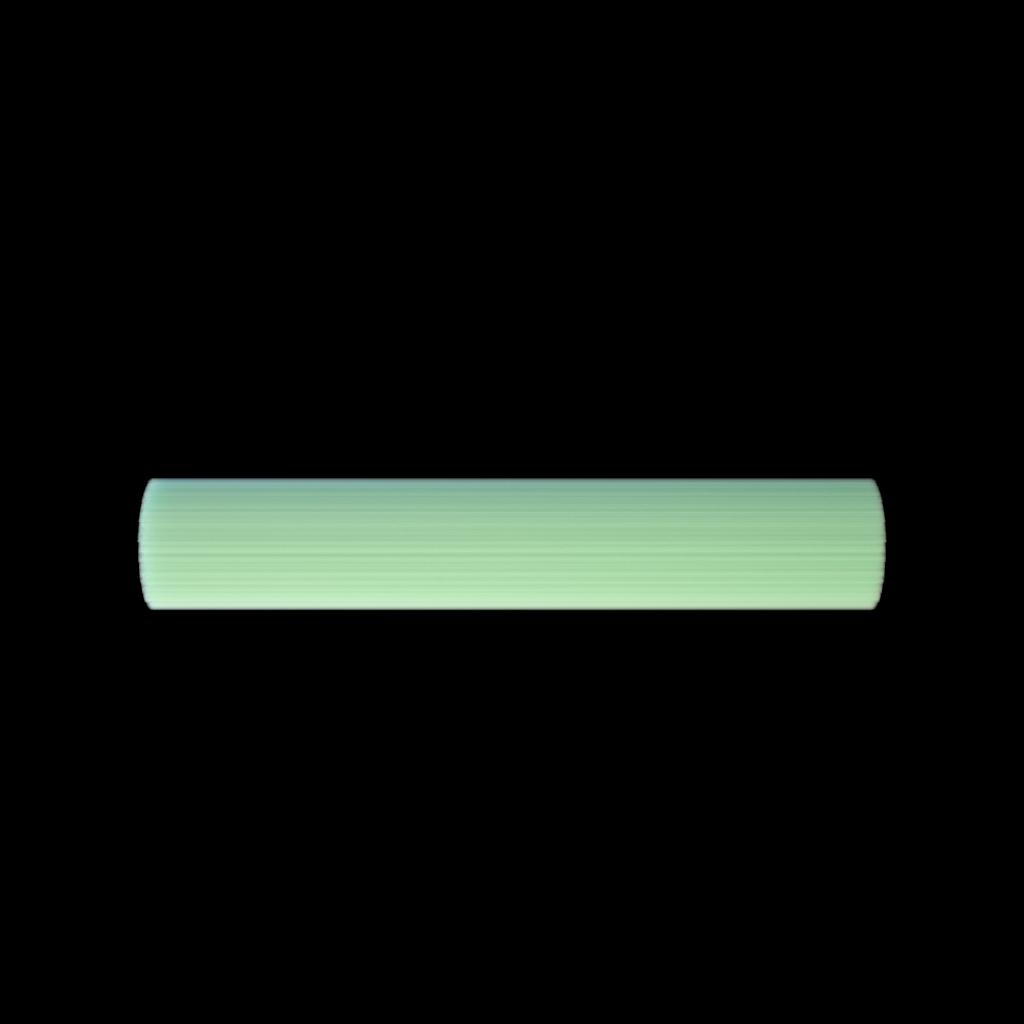}}\hfill%
\rotatebox[origin=c]{90}{\includegraphics[trim={400 340 350 400},clip,height=0.222\linewidth]{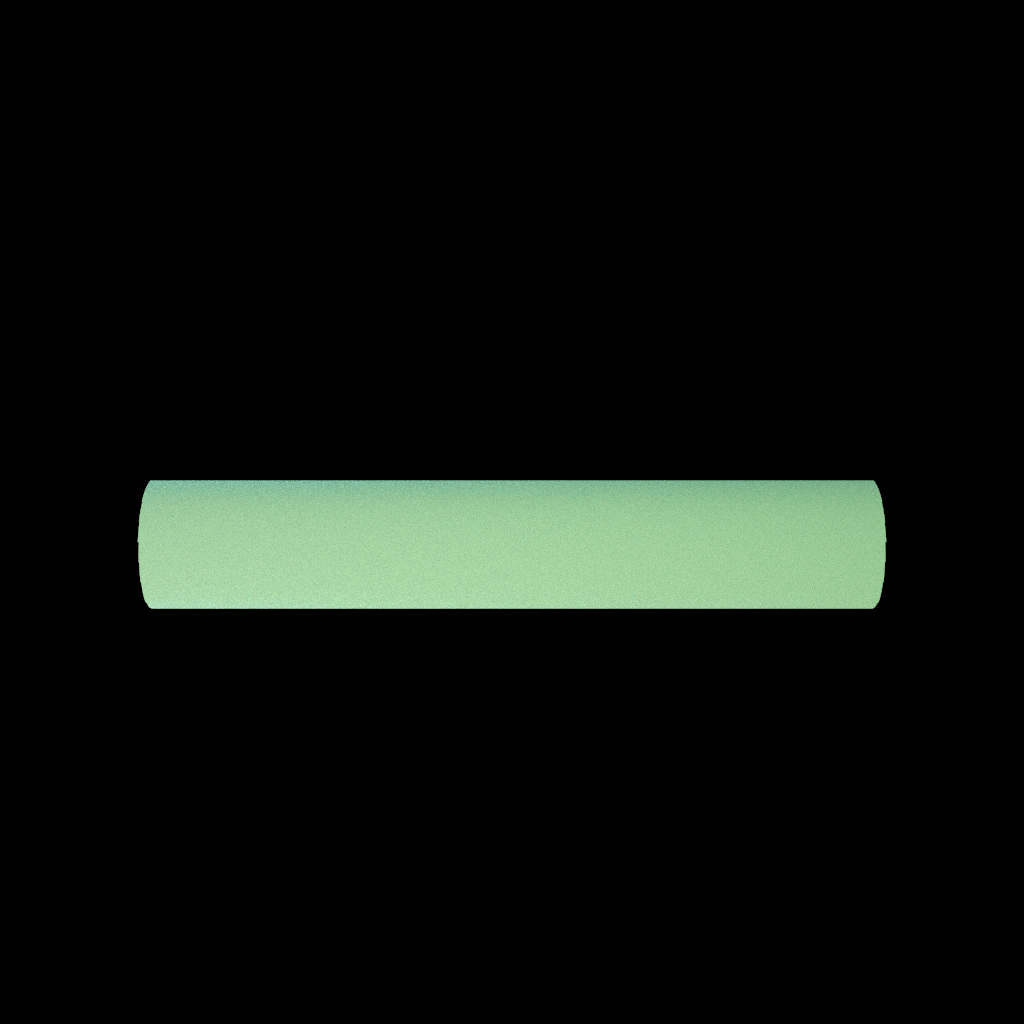}}\hfill%
\rotatebox[origin=c]{90}{\includegraphics[trim={400 340 350 400},clip,height=0.222\linewidth]{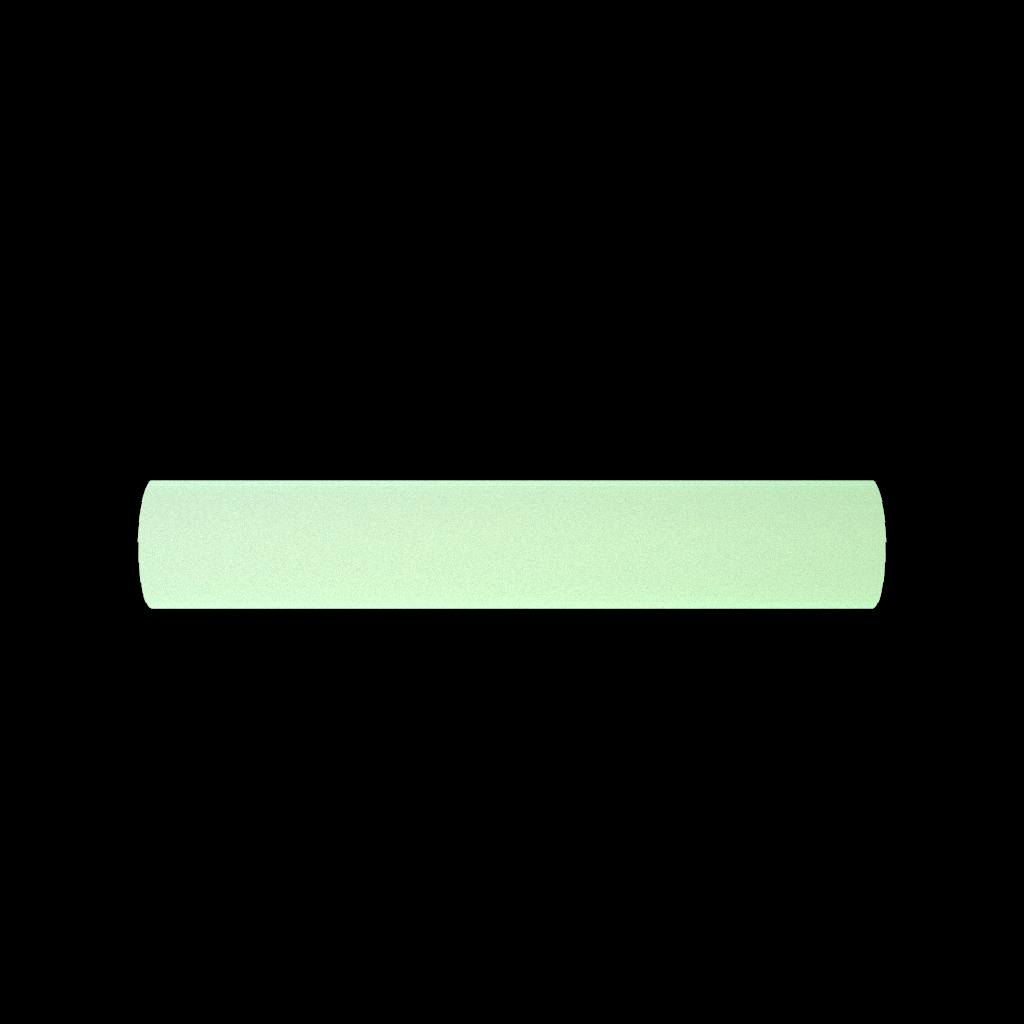}}
{$\quad$~}\\
{$\quad$~}
\rotatebox[origin=c]{90}{\includegraphics[trim={400 340 350 400},clip,height=0.222\linewidth]{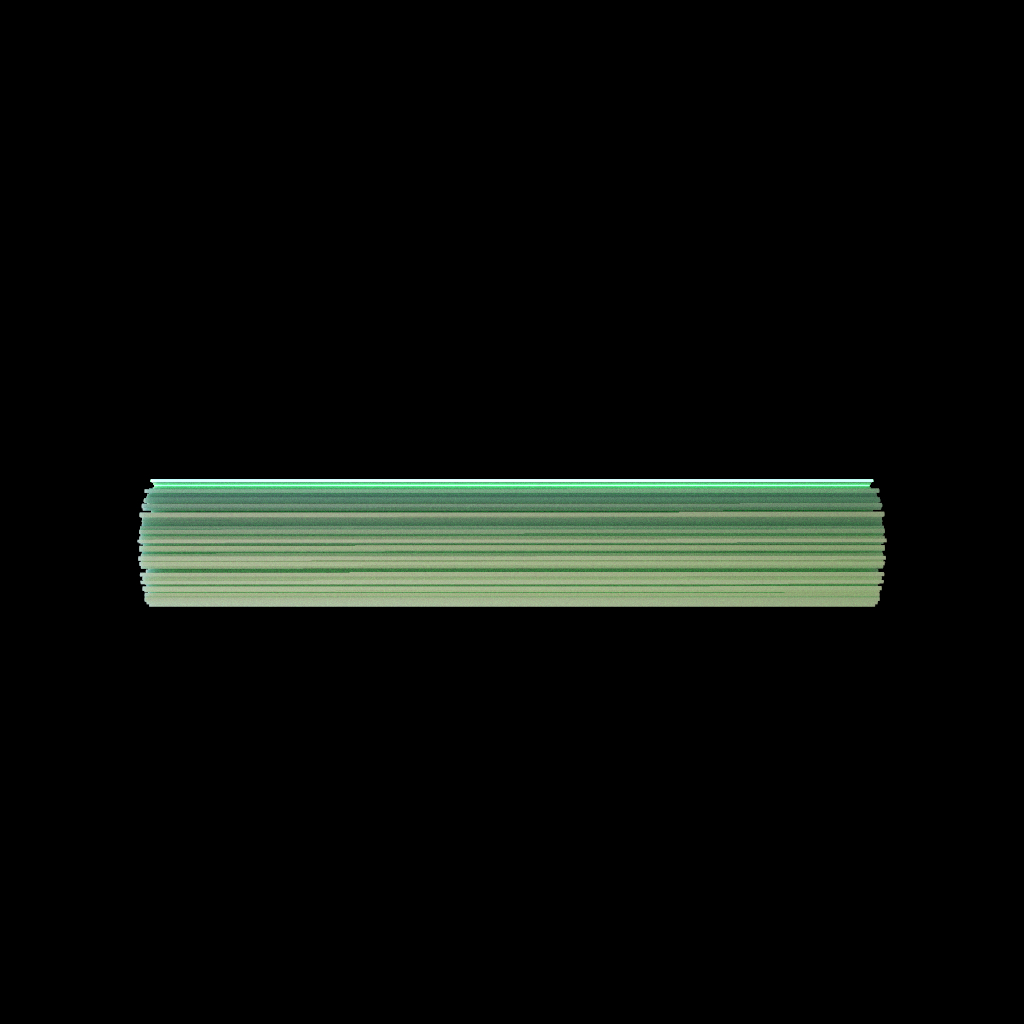}}\hfill%
\rotatebox[origin=c]{90}{\includegraphics[trim={400 340 350 400},clip,height=0.222\linewidth]{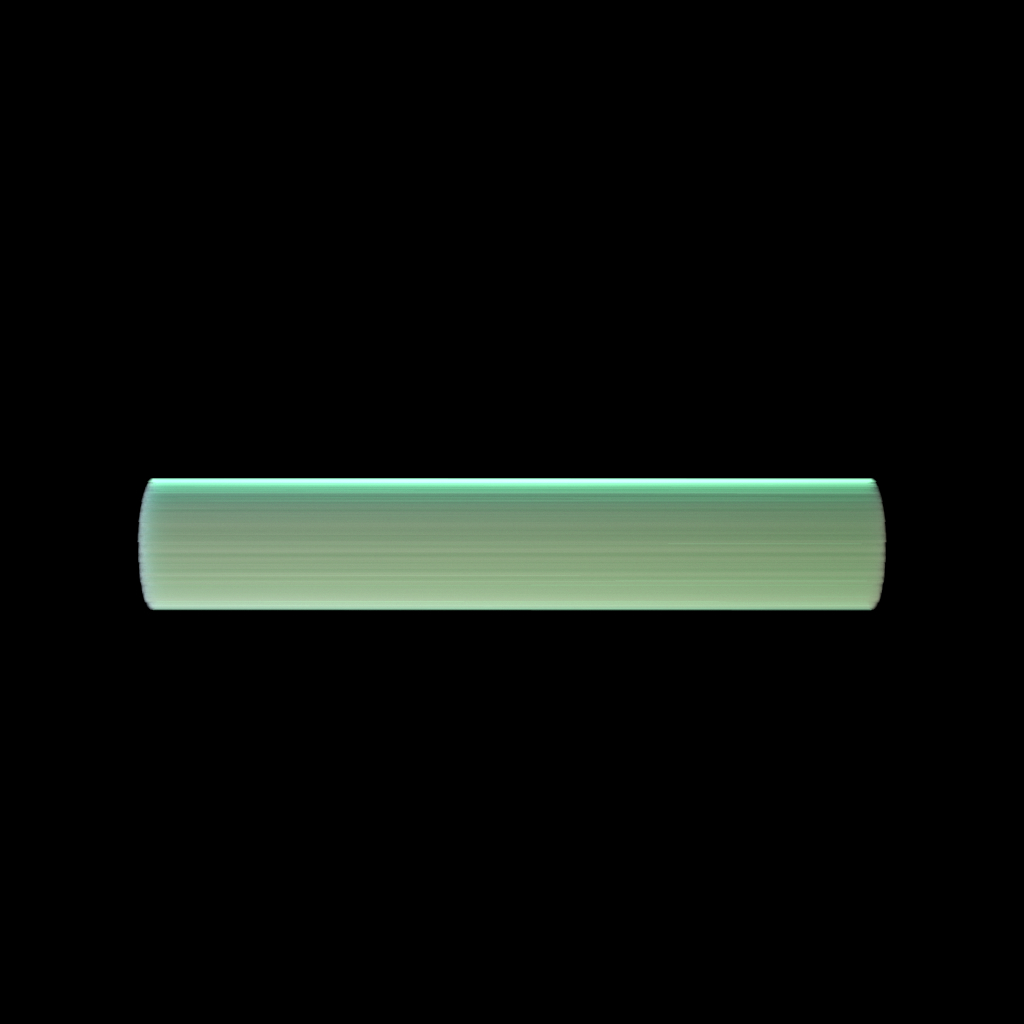}}\hfill%
\rotatebox[origin=c]{90}{\includegraphics[trim={400 340 350 400},clip,height=0.222\linewidth]{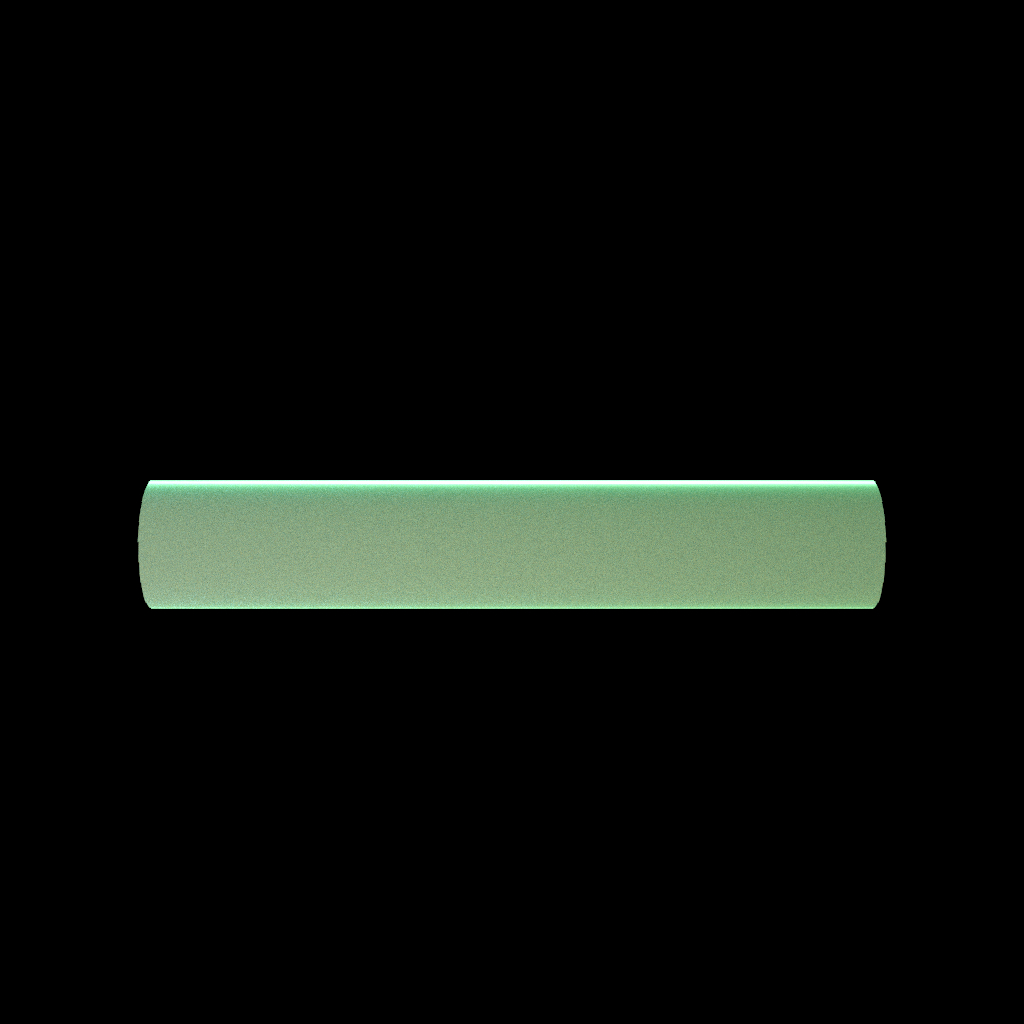}}\hfill%
\rotatebox[origin=c]{90}{\includegraphics[trim={400 340 350 400},clip,height=0.222\linewidth]{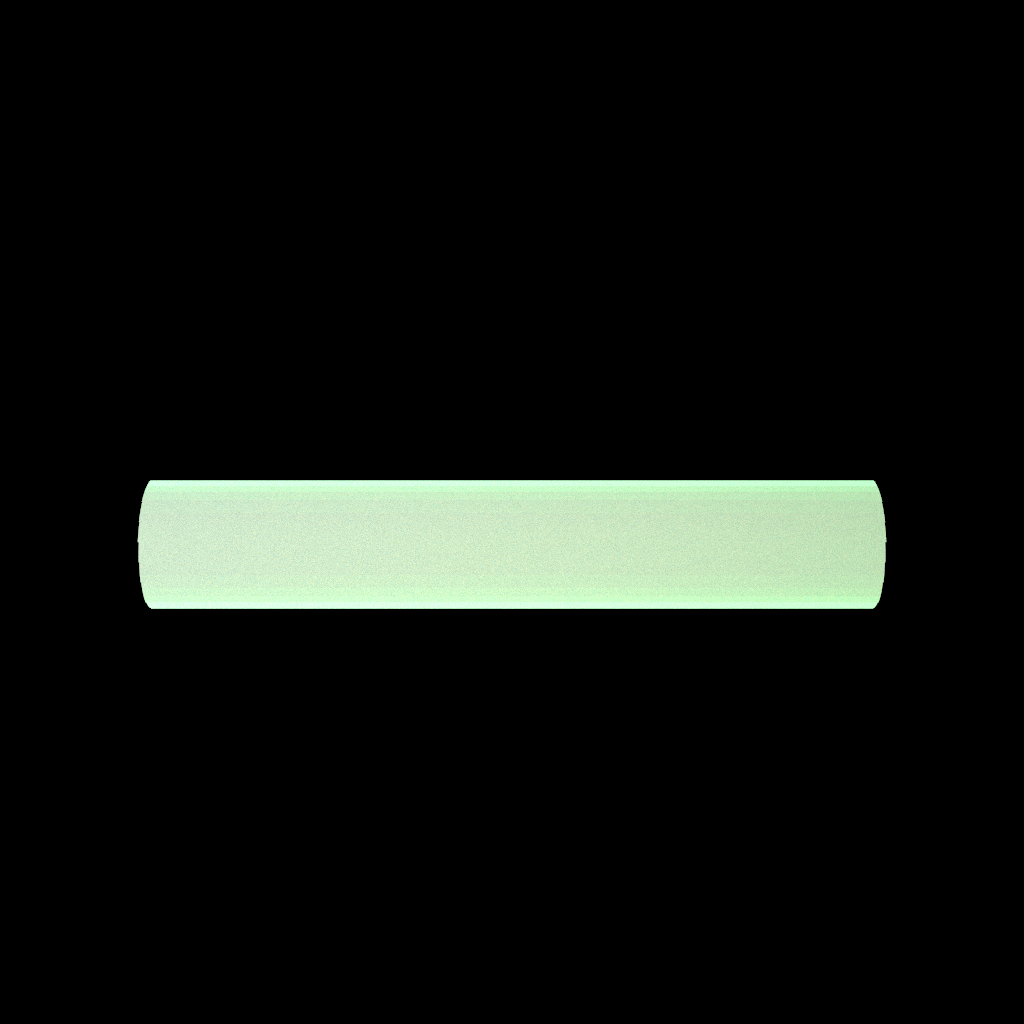}}
{$\quad$~}\\
{$\quad$~}
\rotatebox[origin=c]{90}{\includegraphics[trim={400 340 350 400},clip,height=0.222\linewidth]{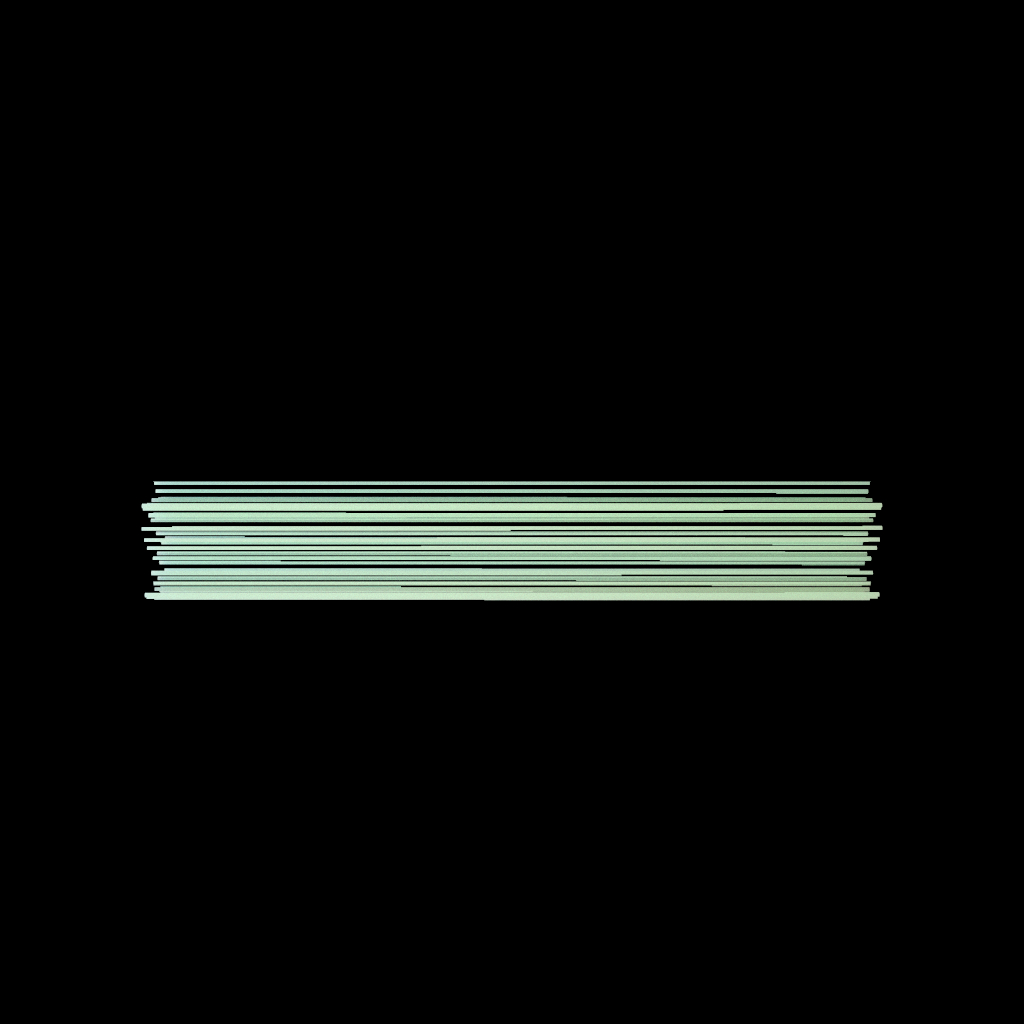}}\hfill%
\rotatebox[origin=c]{90}{\includegraphics[trim={400 340 350 400},clip,height=0.222\linewidth]{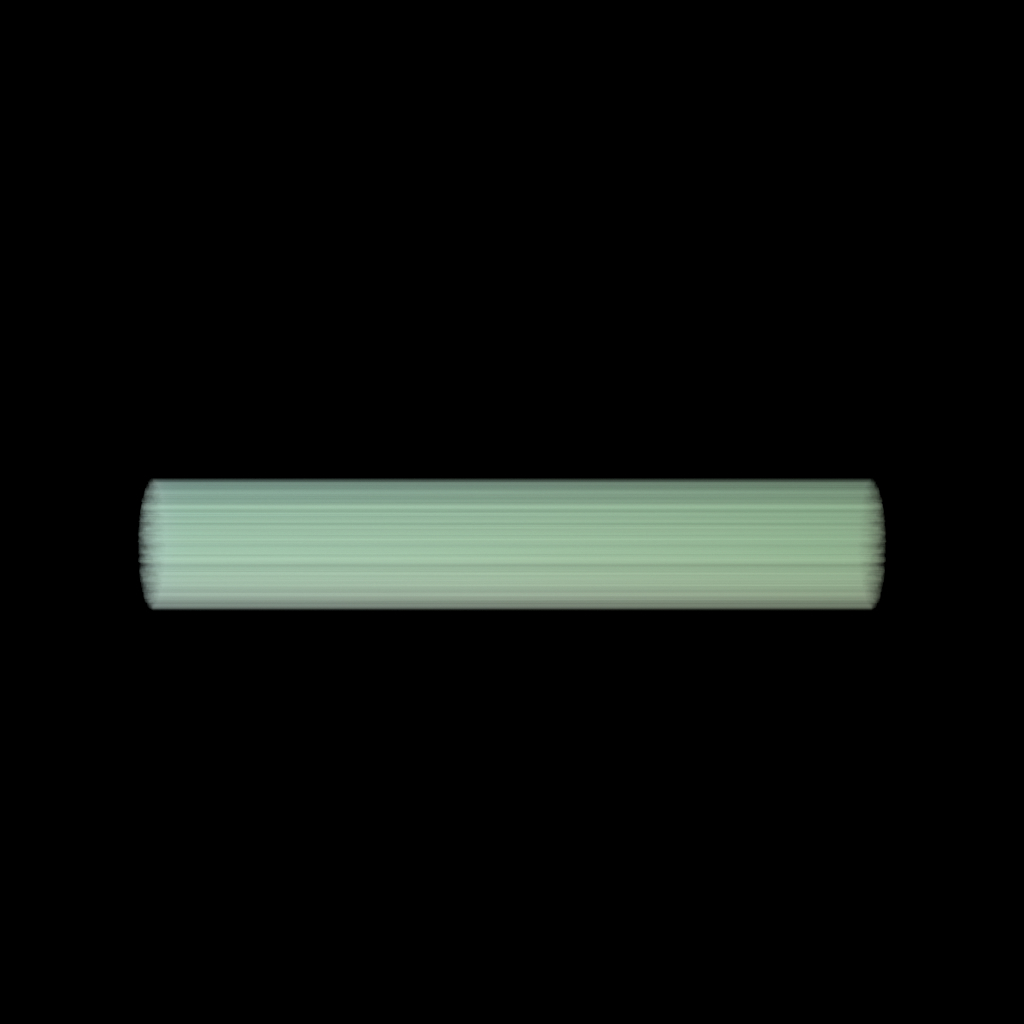}}\hfill%
\rotatebox[origin=c]{90}{\includegraphics[trim={400 340 350 400},clip,height=0.222\linewidth]{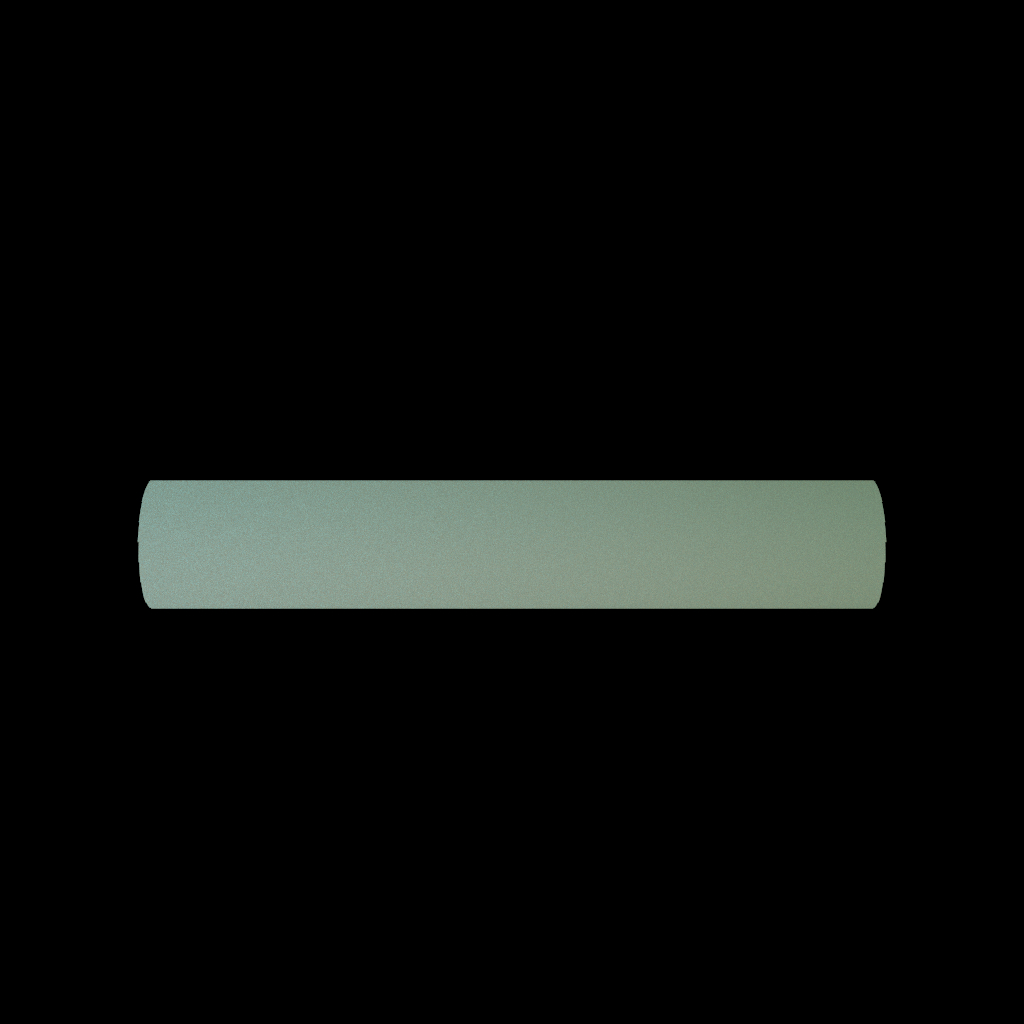}}\hfill%
\rotatebox[origin=c]{90}{\includegraphics[trim={400 340 350 400},clip,height=0.222\linewidth]{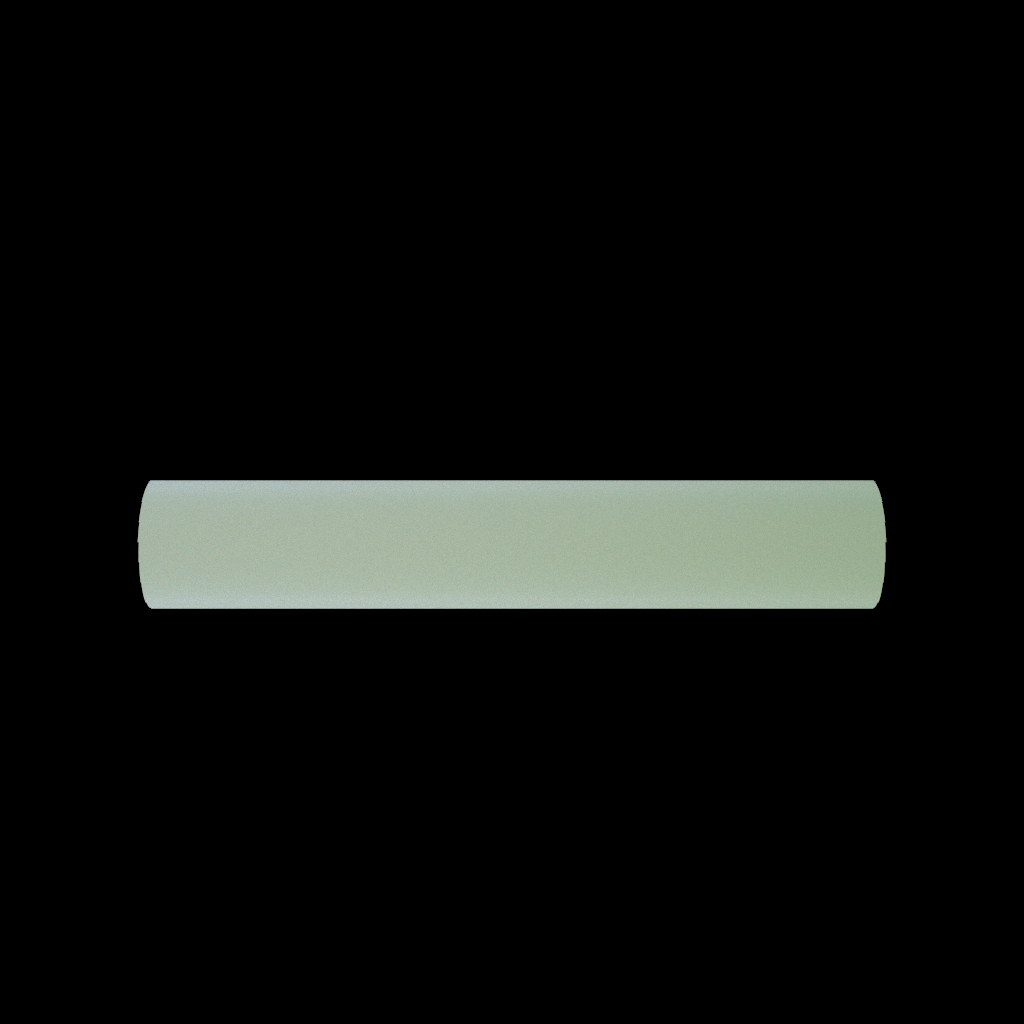}}
{$\quad$~}\\
{$\quad$~}
\rotatebox[origin=c]{90}{\includegraphics[trim={400 340 350 400},clip,height=0.222\linewidth]{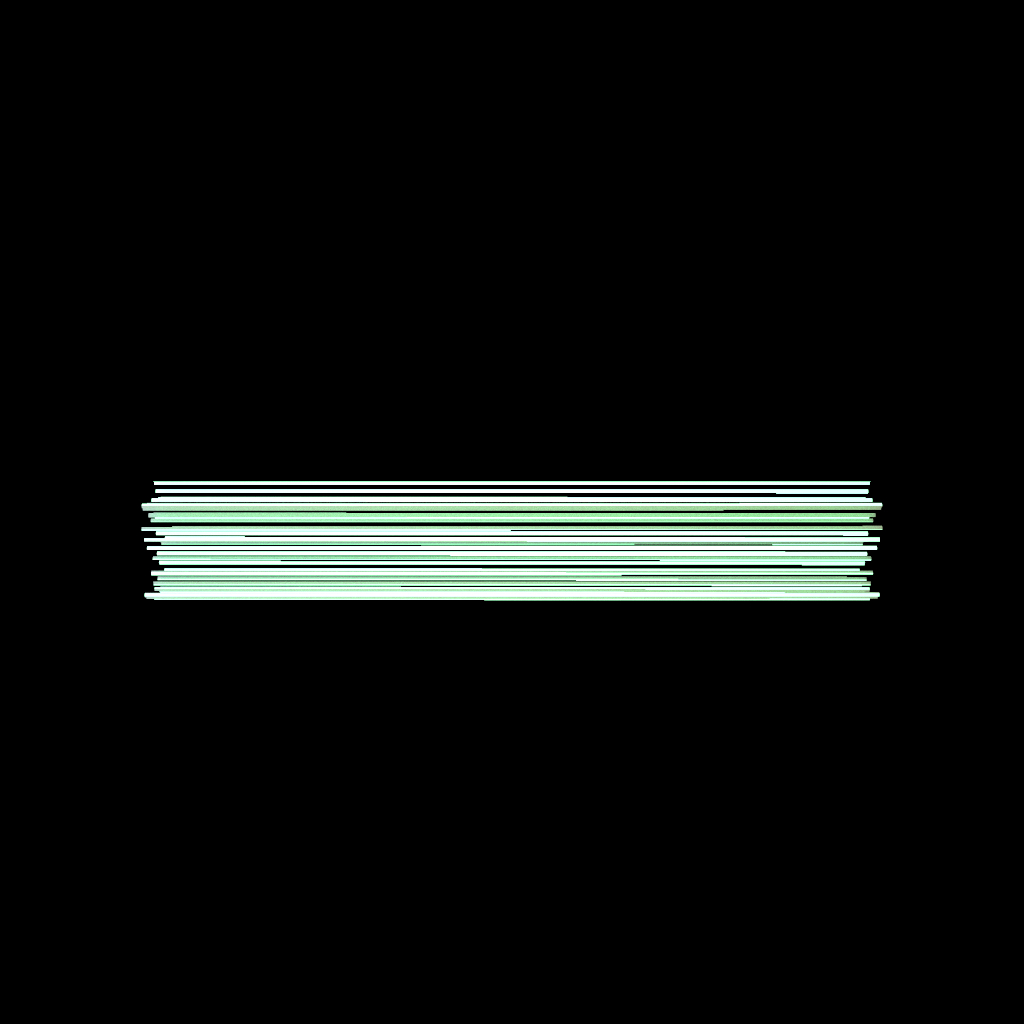}}\hfill%
\rotatebox[origin=c]{90}{\includegraphics[trim={400 340 350 400},clip,height=0.222\linewidth]{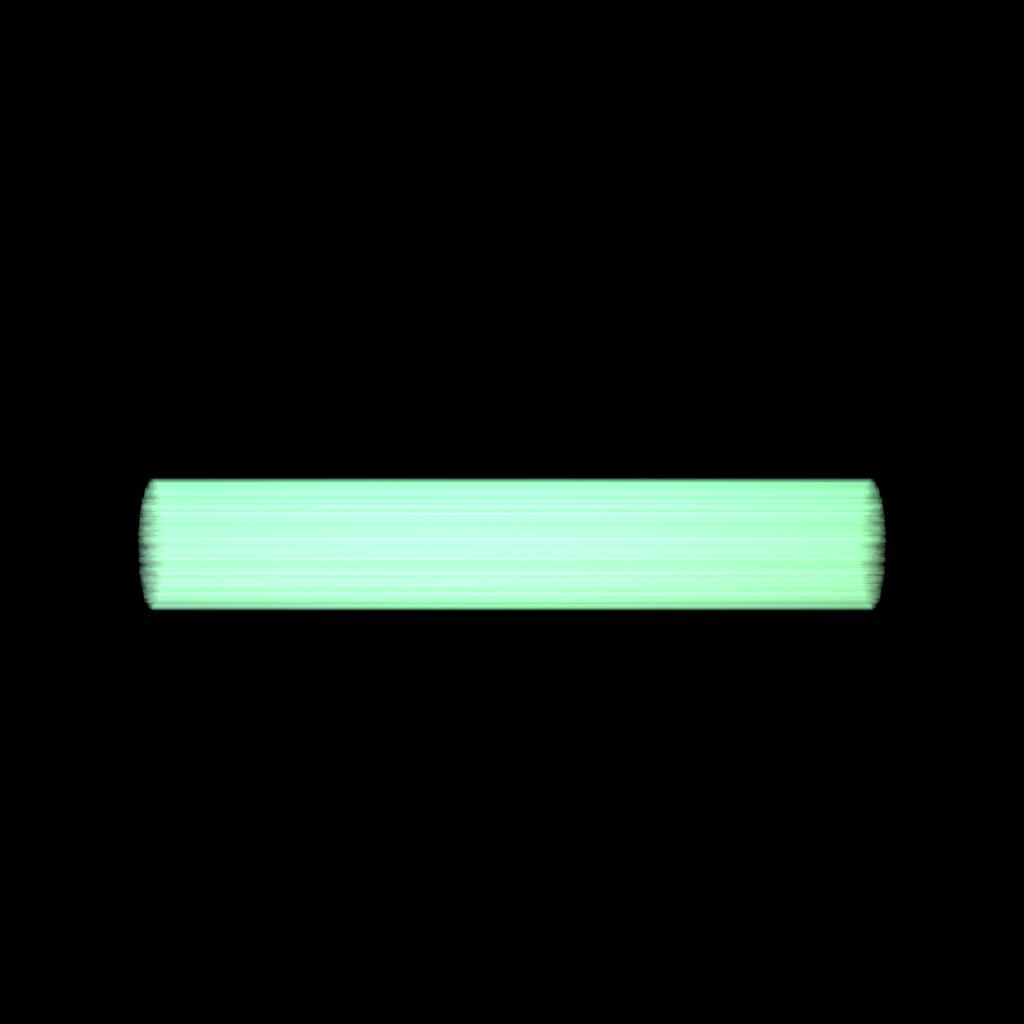}}\hfill%
\rotatebox[origin=c]{90}{\includegraphics[trim={400 340 350 400},clip,height=0.222\linewidth]{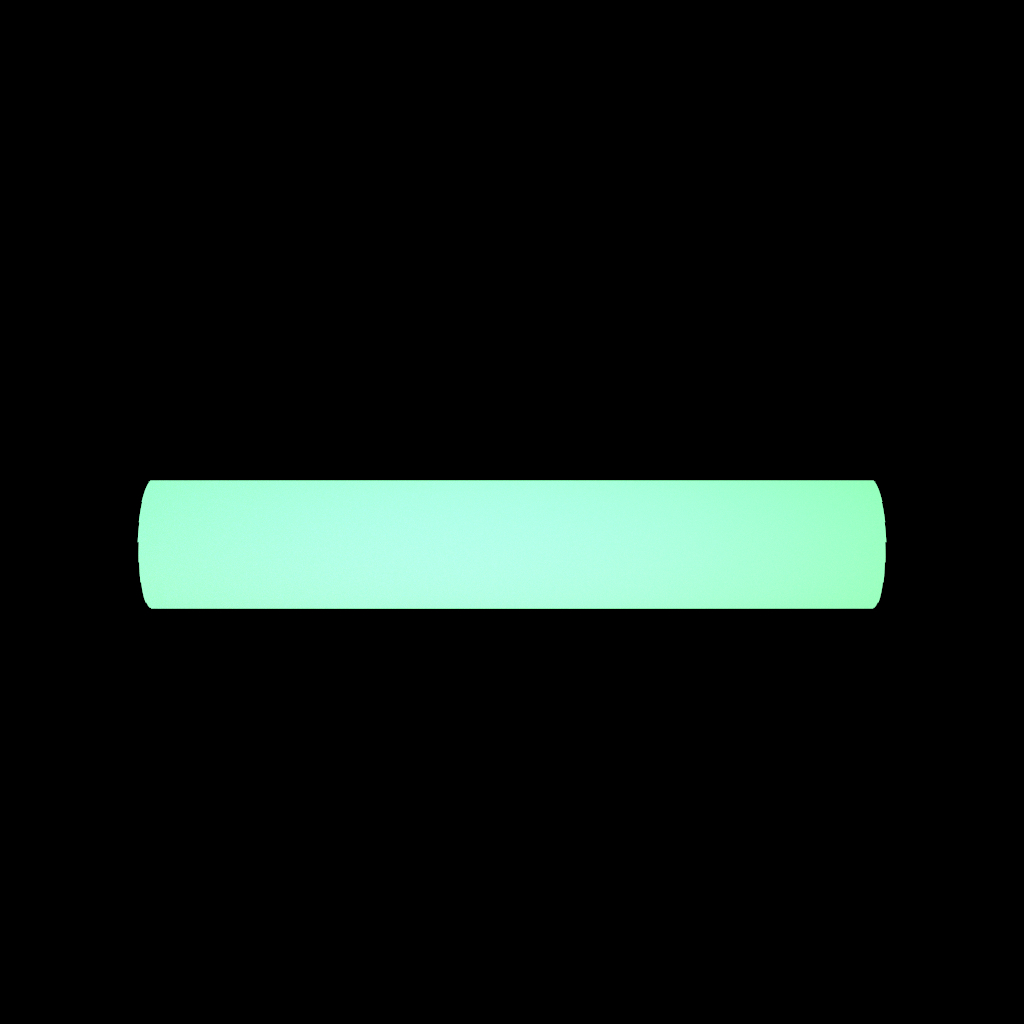}}\hfill%
\rotatebox[origin=c]{90}{\includegraphics[trim={400 340 350 400},clip,height=0.222\linewidth]{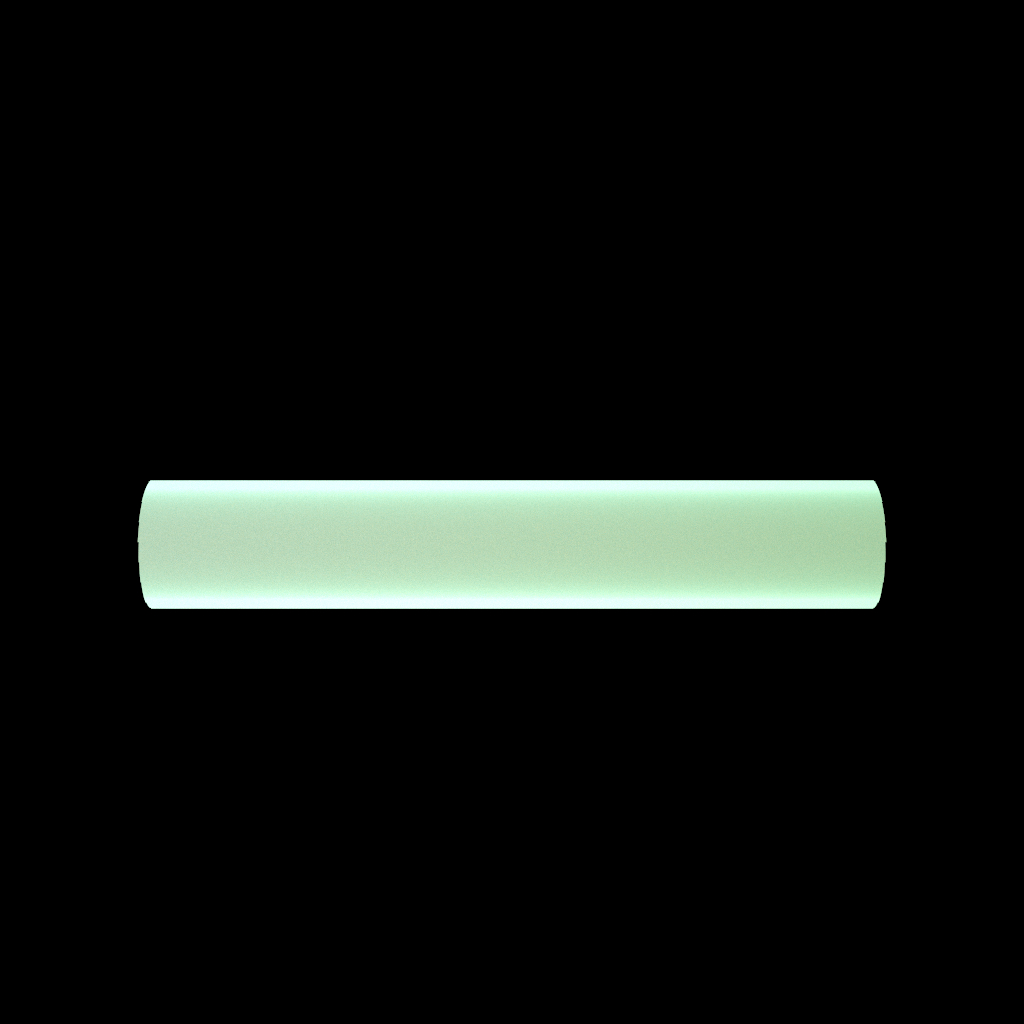}}
{$\quad$~}\\
\vspace{-21em}
\begin{flushleft}
\rotatebox{90}{\small \hspace{4em}Low density \hspace{7em} High density }   
\end{flushleft}
\vspace{-22.3em}
\begin{flushright}
\rotatebox{90}{\small \hspace{3em} Backlit \hspace{3em} Frontlit \hspace{3em} Backlit \hspace{4em} Frontlit }   
\end{flushright}
\vspace{-1.5em}
\begin{flushleft}
{\small \hspace{4em} Ref. \hspace{4em} Avg. ref. \hspace{3.5em} \textbf{Ours} \hspace{2em} \cite{ZhuZJYA23} }   
\end{flushleft}
\vspace{-0.5em}
\caption{Compared to prior aggregated model~\cite{ZhuZJYA23}, ours is closer to avg. ref. that averages 100 different ply instances (ref.) for both frontlit and backlit with high and low hair densities. Both avg. ref. and ref. use single hair BCSDF $f^\text{single}$.}
\label{fig:compareEGSR}
\Description{}
\end{figure}

\subsection{Aggregated Shading Model}

Leveraging the dual scattering, \citet{ZhuZJYA23} proposed an aggregated BCSDF model to approximate light scattering within a ply consisting of $n$ fibers as: 
\begin{align} \label{eq:egsr_bcsdf}
    f^\text{aggregated}(\theta_i, \phi_i, \theta_o, \phi_o) = \sum_{p \in \{R,D,F,B\}}A_p ~ M_p(\theta_h) ~ N_p(\phi),
\end{align}
in which, in addition to reflection $R$ and diffuse $D$ lobes, they introduced forward scattering lobe $F$ and backward scattering lobe $B$ to capture multiple scattering components inside the aggregated ply (see~\autoref{fig:illustration}b). The $F$ lobe represents the distribution of remaining scattered energy that passes forward after multiple scattering inside the ply. The corresponding attenuation $A_F(\theta_d)$ is computed as:
\begin{align}
    A_F(\theta_d) &= \prod_{i=1}^{n} a_F(\theta_d),
\end{align}
where $a_F$ sums the corresponding outgoing radiance toward the front hemisphere, including contributions from $R$, $TT$, and $D$ lobes of a single hair. However, as shown in~\autoref{fig:compareEGSR} second column, their aggregated model does not match the reference well, especially under backlit conditions. Particularly, their $F$ lobe represents the remaining energy after the ray passes through all the fibers in the light path. However, their $R$, $B$, and $D$ lobes still exhibit a diffuse azimuthal distribution, which should be attenuated after multiple scattering. The double-counting leads to higher brightness. Additionally, in sparse fiber distribution, $TT$ lobe exhibits a completely different azimuthal distribution compared to $D$ and $R$ lobes, and it cannot be represented by a single $F$ lobe adequately. 

\section{Our Aggregated Shading Model} \label{sec:bcsdf}

We propose an improved aggregated BCSDF model with three novel enhancements, including 1) modified lobe decomposition strategy, 2) shadowing-masking term $S(\phi, \theta_d)$, and 3) elliptical cross-section handling, to match the reference accurately. Our aggregated BCSDF is defined as:
\begin{align} \label{eq:our_bcsdf}
f^\text{ours}(\theta_i, \phi_i, \theta_o, \phi_o) = S(\phi, \theta_d) \sum_{p \in \{\widehat{R},\widehat{TT},\widehat{D},\widehat{B}\}}A_p M_p(\theta_h) N_p(\phi), 
\end{align}
We consider our aggregated model to characterize lobes of the last intersected hair inside an aggregated geometry. Let $n$ be the total number of hairs intersected by a ray that carries incident radiance. If the ray leaves from the back hemisphere of the aggregated hair, $n$ always equals $1$, meaning that $\widehat{R}$, $\widehat{D}$, and $\widehat{TT}$ are exactly the same as their counterparts in the single hair BCSDF. Conversely, if the ray leaves from the front hemisphere, it must have traversed $n - 1$ hairs already. Following the concept of dual scattering, in our model, the original $F$ lobe in~\autoref{eq:egsr_bcsdf} is distributed to $\widehat{R}$, $\widehat{D}$, and $\widehat{TT}$ lobes, while $\widehat{B}$ lobe only approximates outgoing radiance toward the back hemisphere from the shading point.

\paragraph{Modified $\widehat{R}$, $\widehat{D}$, and $\widehat{TT}$ lobes}
The $\widehat{R}$, $\widehat{D}$, and $\widehat{TT}$ lobes in our model describe the scattering distribution of a hair for incident radiance that arrives at the shading point after being attenuated by $n-1$ preceding hairs. Similar to global scattering, the aggregated attenuation of each lobe is the product of the attenuation per penetrated hair along the scattering path, expressed as:
\begin{align}
A_p &= a_F^{n - 1} a_p, \quad p \in \{\widehat{R},\widehat{D},\widehat{TT}\} .
\end{align}
The longitudinal and azimuthal scattering functions of $\widehat{R}$, $\widehat{D}$, and $\widehat{TT}$ lobes in~\autoref{eq:our_bcsdf} are defined as:
\begin{alignat}{3}
    M_{\widehat{R}}    &= G(\bar{\beta}_{\widehat{R}}^M,~\theta_h), && N_{\widehat{R}} ~= \frac{1}{2\pi}, \nonumber \\
    M_{\widehat{TT}} &= G(\bar{\beta}_{\widehat{TT}}^M, \theta_h), \quad && N_{\widehat{TT}} = G(\bar{\beta}_{\widehat{TT}}^N, \phi - \pi), \\ 
    M_{\widehat{D}}    &= \frac{1}{\pi}, && N_{\widehat{D}}  ~= \frac{1}{2\pi}, \nonumber
\end{alignat}
where scattering distributions are modeled by Gaussian functions with wider variance $\bar{\beta}_p^M$ to account for the longitudinal spread due to front scattering. In particular, $\bar{\beta}_p^{M}$ and $\bar{\beta}_{\widehat{TT}}^{N}$ are calculated by accumulating all variances along the light path as:
\begin{equation}
\begin{split}
    \bar{\beta}_{p}^M &= {(n - 1) \bar{\sigma}_f^{2}(\theta_d) + {\beta_{p}^{M}}}, \quad p \in \{\widehat{R},\widehat{TT}\} \\
    \bar{\beta}_{\widehat{TT}}^{N} &= {(n - 1) \bar{\sigma}_{f}^{2}(\theta_d)  + {\beta_{\widehat{TT}}^{N}}}.
\end{split}
\end{equation}
The spread variance ${\bar{\sigma}_f}^2(\theta_d)$ is calculated by accumulating the forward azimuthal variances of $n$ hairs. $\theta_d = \theta_i - \theta_o$ is the relative longitudinal angle.

\paragraph{Modified $\widehat{B}$ lobe}
In our approach, we employ the $\widehat{B}$ lobe to approximate the contribution of multiple scattering near the shading point after the ray has reached the corresponding hair. It is essential to note that, following the assumption of dual scattering that splits multiple scattering into forward and backward components, the forward scattering is distributed among $\widehat{R}$, $\widehat{D}$, and $\widehat{TT}$ lobes. The $\widehat{B}$ lobe is specifically designed to account for the outgoing path toward the back hemisphere at the shading point, as illustrated in~\autoref{fig:illustration}c.

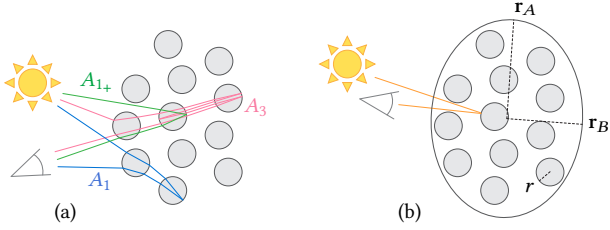
\begin{figure}[ht!]
\centering
\includegraphics[width=0.9\linewidth]{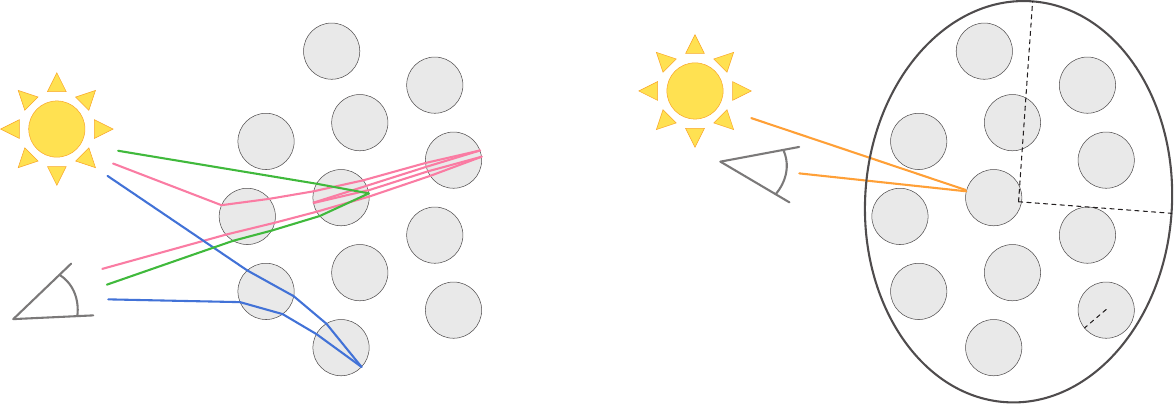}\vspace{-0.3em}
\put(-190,50){\small\textcolor{mygreen}{$A_{1_+}$}}
\put(-187,12){\small\textcolor{myblue}{$A_1$}}
\put(-128,40){\small\textcolor{mypink}{$A_3$}}
\put(-26,78){\small\textcolor{black}{$\rr_A$}}
\put(2,34){\small\textcolor{black}{$\rr_B$}}
\put(-22,10){\small\textcolor{black}{$r$}}
\put(-200,0){\small\textcolor{black}{(a)}}
\put(-70,0){\small\textcolor{black}{(b)}}\vspace{0.5em}
\caption{Illustrations of light pass in aggregated cross-section: (a) the blue $A_1$ and pink $A_3$ paths are considered in the traditional local scattering with one and three backward scattering events by $B$ lobe. The green path illustrates our added $A_{1_+}$ paths, which encounter two deflections before returning to the shading hair. (b) when the incident and outgoing directions are close, the observed hair along the incident direction may not be occluded by others along the outgoing direction. All hairs with radius $r$ are contained by an elliptical cross-section, with major axis $\rr_A$ and minor axis $\rr_B$. }
\Description{}
\label{fig:ds}
\end{figure}

To address the scenario that hair distribution is relatively sparse within the aggregate geometry, allowing light to easily traverse the ``first'' layer of hairs and reach the hair behind (green path in~\autoref{fig:ds}a), we introduce an additional path series, denoted as $A_{1_+}$, for the backscattering light, to supplement existing paths with one $A_1$ and three $A_3$ backward scattering events (denoted by $\texttt{B}$), and the paths with even number of forward scattering events (denoted by $\texttt{F}$). This series represents the scenario where the light hits the hair behind the currently shaded hair first, undergoes an even number of forward scattering events and one backward scattering event, and then hits the back side of the current hair. This path series can be described as $\texttt{F}^{i+1}\texttt{BF}^{i}$, where $i \geq 0$ denotes the number of hairs passed through behind the currently hit hair. The rationale behind considering the $A_{1_+}$ path series is that, when the hair density is relatively low, there is a higher probability that some hairs behind the currently hit hair get illuminated through gaps, significantly contributing to the overall appearance of the outer hair. The corresponding attenuation is computed as follows:
\begin{align}
    A_{1_+} = d_{1_+} a_{B}(\theta_d) \sum_{i=1}^{n} a_F^{2i-1}(\theta_d),
\end{align}
where $d_{1_+}$ accounts for the possibility of the contribution of such paths coming from the side and rear directions. In practice, we approximate it using a constant value as
$d_{1_+} = 0.6$.

The attenuation $A_{\widehat{B}}$, longitudinal $M_{\widehat{B}}$ and azimuthal $N_{\widehat{B}}$ distributions are as follows:
\begin{align}
    A_{\widehat{B}} = A_{1_+} + A_1 + A_3, \quad
    M_{\widehat{B}} = G(\bar{\beta}_{\widehat{B}}^M, \theta_h), \quad
    N_{\widehat{B}} = \frac{I_B(\phi)}{\pi},
\end{align}
where the longitudinal variance $\bar{\beta}_{\widehat{B}}^M$ is the averaged accumulated variance along each possible path, weighted by its attenuation. Since the $R$ and $D$ lobes of a single hair exhibit a diffuse distribution in the azimuthal direction, our aggregated $N_{\widehat{B}}$ is modeled as a uniform distribution over the back hemisphere.

\paragraph{Estimation of hair number $n$.} 
In order to represent a collection of hairs using a thick cylinder with an elliptical cross-section, it is important to estimate the number of hairs a ray intersects inside the aggregated cylinder for computing attenuation and scattering distribution. We define the unitless hair density $\rho$ as the ratio of the area occupied by the hairs to the cross-sectional area of the aggregated cylinder, given by $\rho = {n_\text{total} r^2} / {(|\rr_A| |\rr_B|)}$, where $n_\text{total}$ is the number of hairs inside, $r$ is the single hair radius, and $\rr_A$ and $\rr_B$ are the major and minor axes of the cross-section. Given the ray travel distance $l$, the number of intersecting hairs $n$ can be calculated as $n = \rho l \sqrt{\frac{n_\text{total}}{\pi |\rr_A| |\rr_B|}}$, where the last term estimates number of intersecting hairs per unit distance.

\paragraph{Shadowing-masking $S(\phi, \theta_d)$}
For a cluster of hair strands, even if $\omega_i$ and $\omega_o$ fall within the same hemisphere, it does not guarantee that the hair seen from the view direction can also be seen from the light direction. In other words, when the relative azimuthal angle $\phi \in \left(-\frac{\pi}{2}, \frac{\pi}{2}\right)$, self-shadowing situations can occur within a thick aggregated geometry.

We propose a shadowing-masking term to compensate for self-shadowing for $\phi \in \left(-\frac{\pi}{2}, \frac{\pi}{2}\right)$ by categorizing shadowing at the shading point into three categories: shadowing by a single hair, shadowing by multiple hairs, and no shadow. We approximate the possibility of the shadow ray from a shading point needing to pass through a \underline{s}ingle hair to reach the light source as:
\begin{align}
P_\text{s} &= (1 - \rho) \bigl(1 - \cos(\phi)\bigr), 
\end{align}
where $\cos(\phi)$ is used to approximate the difference between the incident and outgoing directions in the azimuthal plane. As shown in~\autoref{fig:ds}b, hairs located further from the incident direction are more likely to be occluded by other strands along the light direction. The $1-\rho$ term accounts for the fact that the shadow-masking effect becomes less prominent with denser hairs, as a thick fiber better approximates the cluster of hairs.

The shading point can also be occluded by \underline{m}ultiple hairs along the shadow ray. We approximate the probability as
\begin{align}
P_\text{m} &= (1 - \rho)~c_\text{m},
\end{align}
where $c_\text{m}$ is a constant value empirically determined ($c_\text{m} = 0.2$ for all our experiments).
With $P_s$ and $P_m$, we can determine the probability of the shading point reached directly by light with \underline{n}o-shadow as:
\begin{align}
P_\text{n} &= 1 - P_\text{m} - P_\text{s}.
\end{align}
Thus, the shadowing-masking $S(\phi, \theta_d)$ term is formulated as:
\begin{align}
S(\phi, \theta_d) = P_\text{s}(\phi) a_F(\theta_d) + P_\text{n} \mathds{1}, \label{eq:shadowmasking}
\end{align}
where $\mathds{1}$ represents a three-dimensional vector of ones. Based on the dual scattering assumption, the attenuation needs to be multiplied by $a_F$, as the light ray must undergo forward scattering once to reach the actual shading hair. Note that no lighting calculation is performed for $P_\text{m}$, given the assumption that its contribution to lighting is negligible.

\section{Real-time Rendering Pipeline} \label{sec:rtpipeline}

This section outlines the construction process of the hair hierarchy and describes how we utilize it in rendering. 

\subsection{Initialization} \label{sec:lod_init}

Given a strand-based hair model with tens of thousands of \emph{single hairs}, we build the hair hierarchy in four stages: 1) selecting a number of guide hairs for simulation, 2) clustering hair strands into $n_L$ levels, 3) fitting the ellipses for thick hair cross-sections, and 4) computing interpolation weights. We denote the levels in the hierarchy by $L_0$ (coarsest) through $L_{n_L-1}$ (finest). 

\paragraph{1. Guide hair selection}
Initially, a Catmull–Rom spline is fitted with $n_\text{c}$ control points for each hair strand. Following~\cite{Wang09}, we use k-means clustering to create a user-defined number of hair clusters. The distance metric between two single hairs $i$ and $j$ is defined as the sum of distances between corresponding control points along the two single hairs, expressed as $D(i, j) = \sum_k^{n_\text{c}} | \pp^{i,k} - \pp^{j,k} |^2$, where $\pp^{i,k}$ denotes the position of the $k$-th control point along the hair $i$. Within each cluster, the hair with the smallest average distance to all other hairs is selected as the guide hair, which serves as the basis for the physical-based simulation and is used to animate each single hair via linear skinning~\cite{Somasundaram2015} in runtime. Note that the cluster used for guide hair selection in this stage would not be discarded afterward.

\paragraph{2. Hair strand clustering}
After identifying guide hairs, each single hair locates three guide hairs that are closest to it. Single hairs sharing the same set of guide hairs are clustered together to form the coarsest level $L_0$. The subsequent levels, from $L_1$ to $L_{n_L-1}$, are directly clustered from $L_0$. While we do not maintain strict bounding relations between consecutive levels, we ensure that the cross-sectional area of the thick hairs at each level is approximately four times larger than that of the previous level.


\paragraph{3. Cross-section fitting}
To construct a thick elliptical hair from a cluster of single hairs, we first create $n_\text{c}$ sets of control points where each set $\mathcal{P}^k$ contains all single hairs' $k$-th control points. For each set $\mathcal{P}^k$, we compute the average position of all control points $\overline{\pp}_\text{cs}^k$ and normalized average tangent $\overline{\tt}^k$ to define the center of the ellipse. Principal Component Analysis (PCA) is then employed to fit an ellipsoid for $\mathcal{P}^k$, after which we discard the axis closest to $\overline{\tt}^k$. Finally, we identify four \emph{corner control points} ${\pp}_{0,1,2,3}^k$ from $\mathcal{P}^k$ that are closest to the four endpoints of the remaining two axes. 

\paragraph{4. Interpolation weight computing}
In hair skinning, three guide hairs form a triangular prism for linear skinning. 
We compute and record interpolation weight for every single hair control point, as well as for every thick hair cross-section center $\overline{\pp}_\text{cs}^k$ and corner control points ${\pp}_{0,1,2,3}^k$.

\paragraph{5. LoD level initialization}
We compute the initial LoD levels for the starting view in rendering. Using a user-specified screen-space hair width threshold $\epsilon_w$, we determine the LoD level for each cluster within $L_0$ as follows: we examine the LoD level from coarsest to the finest until we find the first level where the maximum thick hair width in the cluster is less than $\epsilon_w$. At runtime, we adjust the LoD levels starting from the level used in the previous frame, making this process much more efficient. Note that the LoD selection is performed on a per-segment basis.




\subsection{Runtime}

\label{sec:runtime}

With guide hairs dynamically simulated at runtime, our rendering framework uses the simulated control points of the guide hairs and the precomputed interpolation weights to determine the positions and shapes of the hairs at desired LoD levels (we update quantities like $\overline{\pp}_\text{cs}^k$ and ${\pp}_{0,1,2,3}^k$ by interpolation using these positions and weights). We dynamically compute the thick hair widths to determine the proper LoD levels and calculate density and lobes in our aggregated BCSDF.




\paragraph{Hair width calculation}
To ensure that thick hairs selected by the width-based LoD metric adequately cover the corresponding single hairs, we dynamically compute the hair widths as follows. 
First, the four corner control points ${\pp}_{0,1,2,3}^k$ are projected to the plane defined by $\overline{\pp}_\text{cs}^k$'s position and its tangent. Then, we measure the largest screen-space distances (on the $xy$ plane) and the largest depth differences (along the $z$ axis) between these projected points as $l_\text{w}$ and $l_\text{t}$ as hair width and thickness, respectively. The corresponding density and estimated hair number are computed as $\rho = {n_\text{total} r^2} / {(l_\text{w} l_\text{t})}$ for shadow-masking in \autoref{eq:shadowmasking} and $n = \rho l \sqrt{\frac{n_\text{total}}{\pi l_\text{w} l_\text{t}}}$ for all lobes in \autoref{eq:our_bcsdf}.

\paragraph{Dynamic LoD selection}
During runtime, we select the LoD level for each cluster in $L^0$. The goal of our LoD selection is to ensure that all thick hairs rendered have widths that are just below the screen-space hair width threshold $\epsilon_w$. Checking through all LoD levels can be expensive at runtime. Therefore, we propose an approximately correct, deferred LoD selection that reuses information between frames.
For each cluster in every frame, we render its thick hairs at the current LoD level (the LoD level for the first frame is determined during initialization) and record the maximum width, denoted as $w_{\text{cur}}$. We also calculate the maximum width at one coarser level, denoted as $w_{\text{cur}_{-}}$. Using these two widths, we determine the LoD level $L$ for the next frame as:
\begin{align}\label{eq:lod_select}
L =
\begin{cases}
L + 1, & \text{if }  w_\text{cur} \hspace{0.5em}> \epsilon_w,  \\
L,     & \text{if } w_\text{cur} \hspace{0.5em}\leq \epsilon_w < w_{\text{cur}_{-}} \\
L - 1, & \text{if } w_{\text{cur}_{-}} \leq \epsilon_w.
\end{cases},
\end{align}

\paragraph{Runtime pipeline}
With the geometry generated for the given LoD level, we proceed with the conventional rendering pipeline, which includes three passes:

\paragraph{1. Shadow pass}
Head and hair are rendered to generate a shadow map and deep opacity map~\cite{ZinkeYWK08}, respectively.

\paragraph{2. Shading pass}
The width for each thick hair is calculated dynamically in the vertex shader. The maximum hair width for each cluster is recorded in a buffer using atomicMax. The tessellation shader subdivides hair strands based on their distance from the camera, and the geometry shader extends them into camera-facing strips. The BCSDF described in~\autoref{sec:bcsdf} is employed in the fragment shader for shading. 

\paragraph{3. Compute pass}
The maximum hair width in each cluster at one coarser level is computed to determine the LoD for the next frame using the selection strategy (\autoref{eq:lod_select}). Note that this compute pass only invokes the vertex shader. 

Note that our runtime pipeline does not explicitly handle thick hairs overlapping, as hairs are rendered as camera-oriented strips for both the shading and deep opacity map passes.

\section{RESULTS} \label{sec:result}

\begin{figure}[ht]
\centering
\newcommand{\figcap}[1]{\begin{minipage}{0.33\linewidth}\centering#1\end{minipage}}
\includegraphics[trim={75 0 75 150},clip,width=0.333\linewidth]{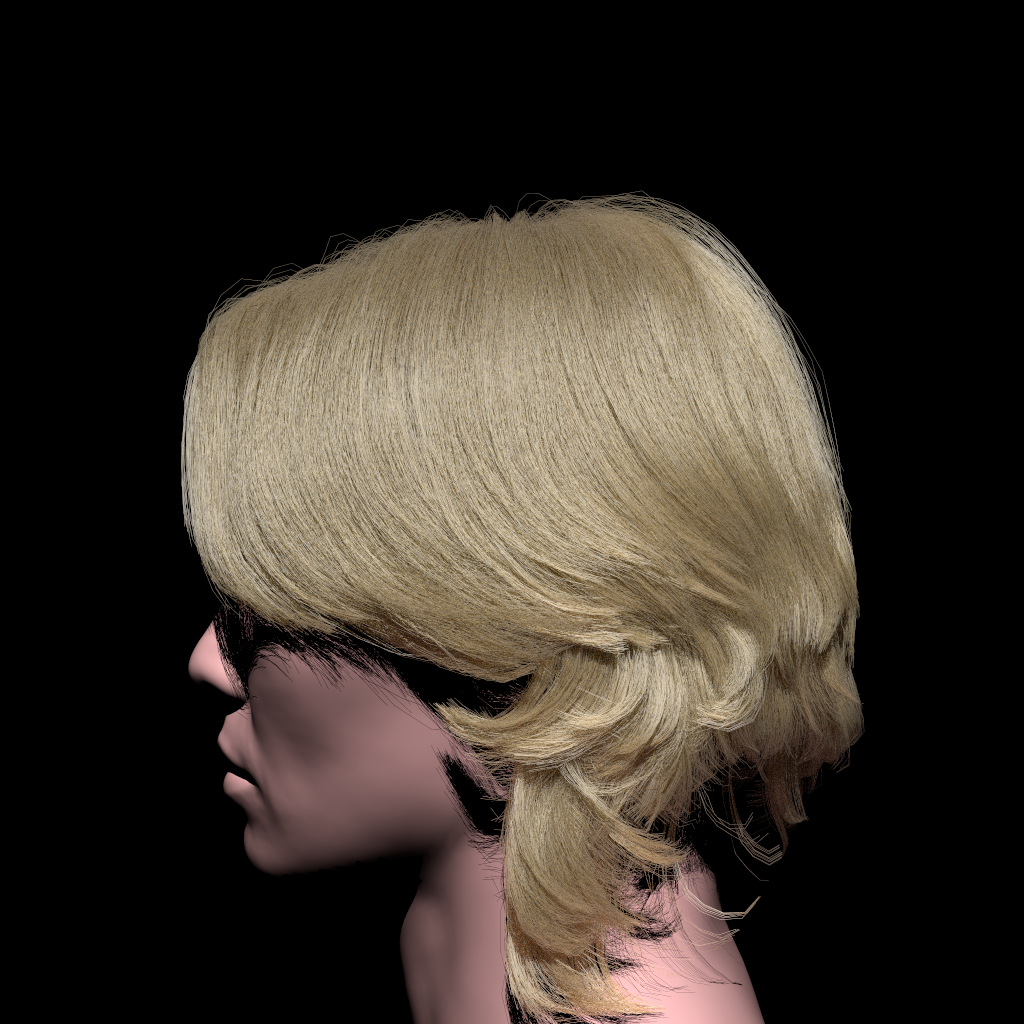}\hfill%
\includegraphics[trim={75 0 75 150},clip,width=0.333\linewidth]{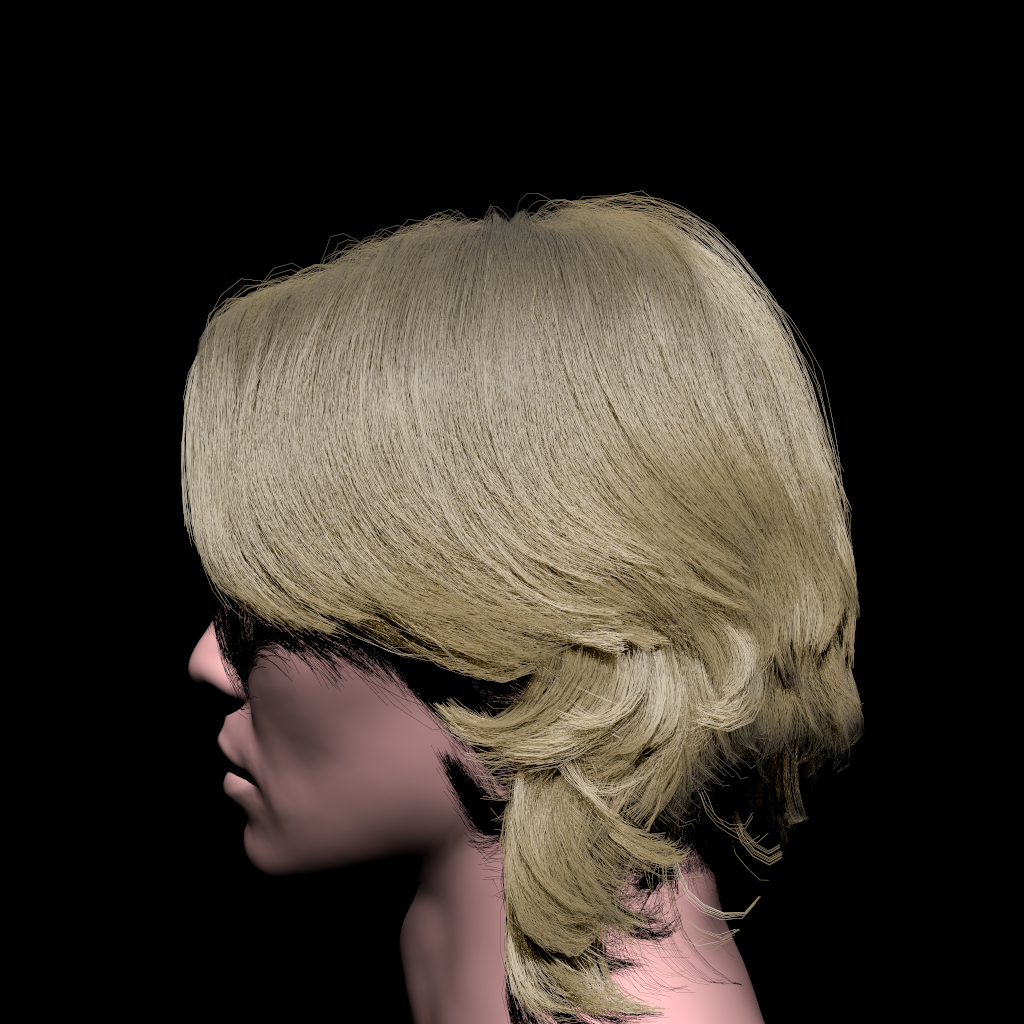}\hfill%
\includegraphics[trim={75 0 75 150},clip,width=0.333\linewidth]{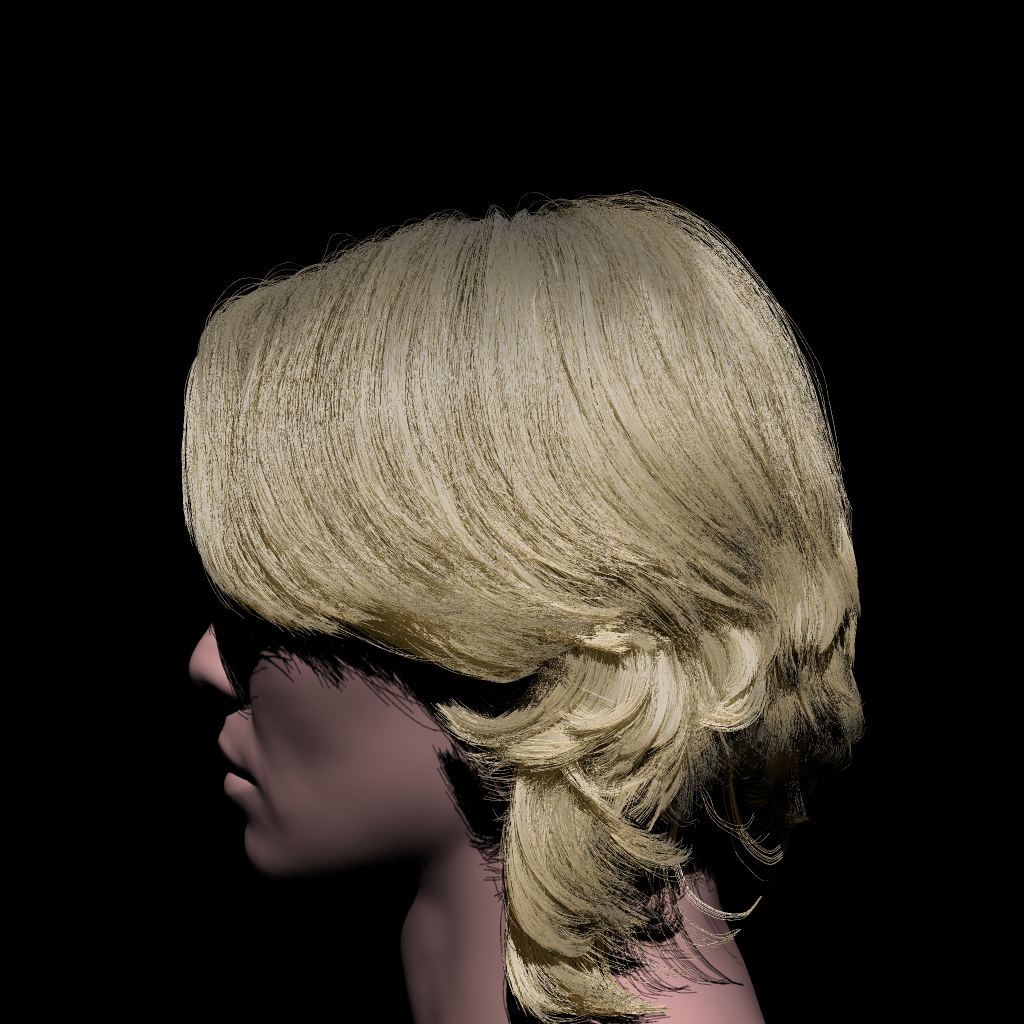}\\ \vspace{-1.5em}
\figcap{\footnotesize \hspace{8em} \color{white}{6 mins}}\hfill%
\figcap{\footnotesize \hspace{8.7em} \color{white}{\textbf{1.6 s}}}\hfill%
\figcap{\footnotesize \hspace{8em} \color{white}{\textbf{10 ms}}}\\
\figcap{\footnotesize Offline path tracing (CPU)}\hfill%
\figcap{\footnotesize \textbf{Ours, ray shooting (CPU)}}\hfill%
\figcap{\footnotesize \textbf{Ours, raster (GPU)}}\\
\caption{Our approach (offline ray shooting and real-time OpenGL rasterizer) achieves close results to the path tracing reference, while our real-time implementation is 36000$\times$ faster for close view.}
\label{fig:cpuvsgpu}
\Description{}
\end{figure}

\begin{figure*}[ht]
\centering
\newcommand{\figcap}[1]{\begin{minipage}{0.11\linewidth}\centering#1\end{minipage}}
\newcommand{\figcapleft}[1]{\begin{minipage}{0.11\linewidth}#1\end{minipage}}
\figcap{~ }\hfill%
\figcap{ Near view}\hfill%
\figcapleft{(60\%)}\hfill%
\figcap{~ }\hfill%
\figcap{ Middle view}\hfill%
\figcapleft{(20\%) }\hfill%
\figcap{~ }\hfill%
\figcap{ Far view}\hfill%
\figcapleft{ (2\%) }\vspace{0.em}\\
\figcap{\small Full w/ PT }\hfill%
\figcap{\small \textbf{LoD w/ ours}}\hfill%
\figcap{\small Full w/ DS }\hfill%
\figcap{\small Full w/ PT }\hfill%
\figcap{\small \textbf{LoD w/ ours}}\hfill%
\figcap{\small Full w/ DS }\hfill%
\figcap{\small Full w/ PT }\hfill%
\figcap{\small \textbf{LoD w/ ours}}\hfill%
\figcap{\small Full w/ DS }\vspace{0.1em}\\
\includegraphics[trim={0 0 0 0},clip,width=0.11\linewidth]{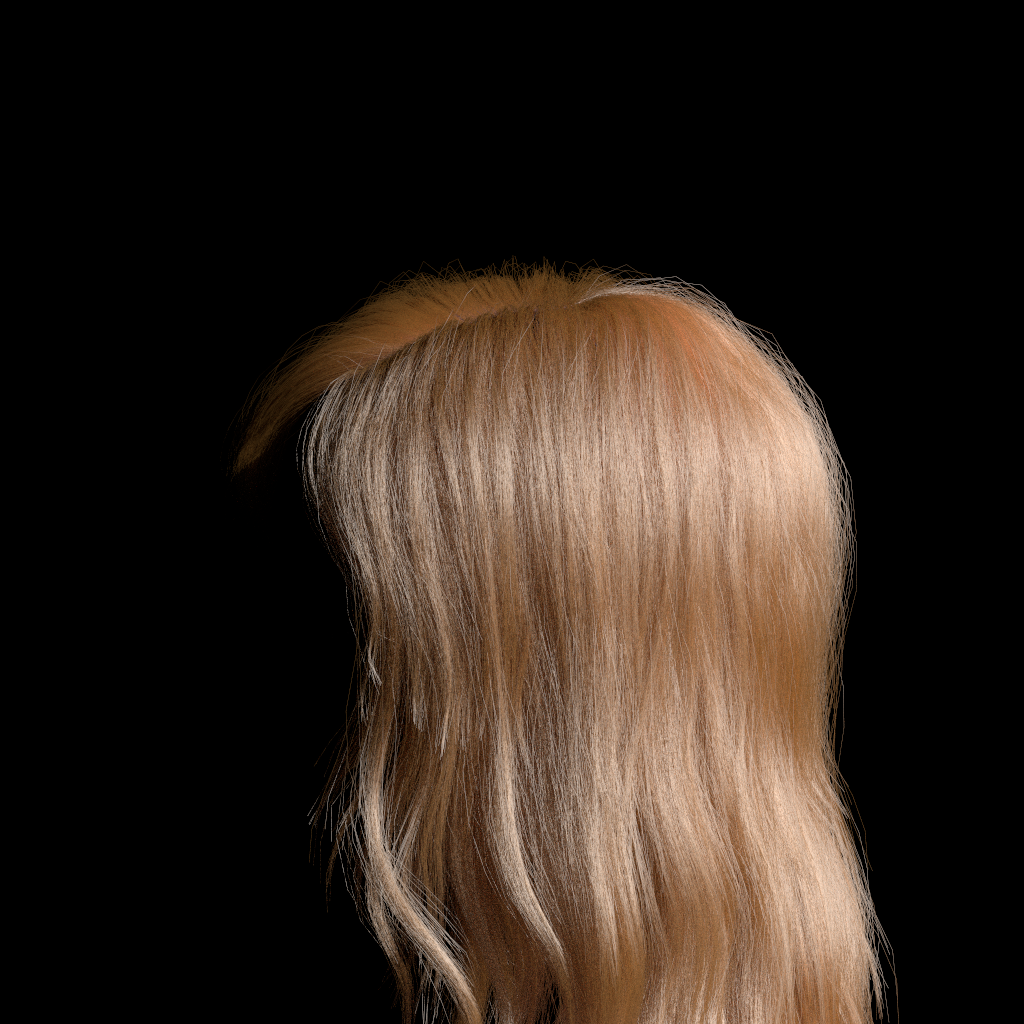}\hfill%
\includegraphics[trim={0 0 0 0},clip,width=0.11\linewidth]{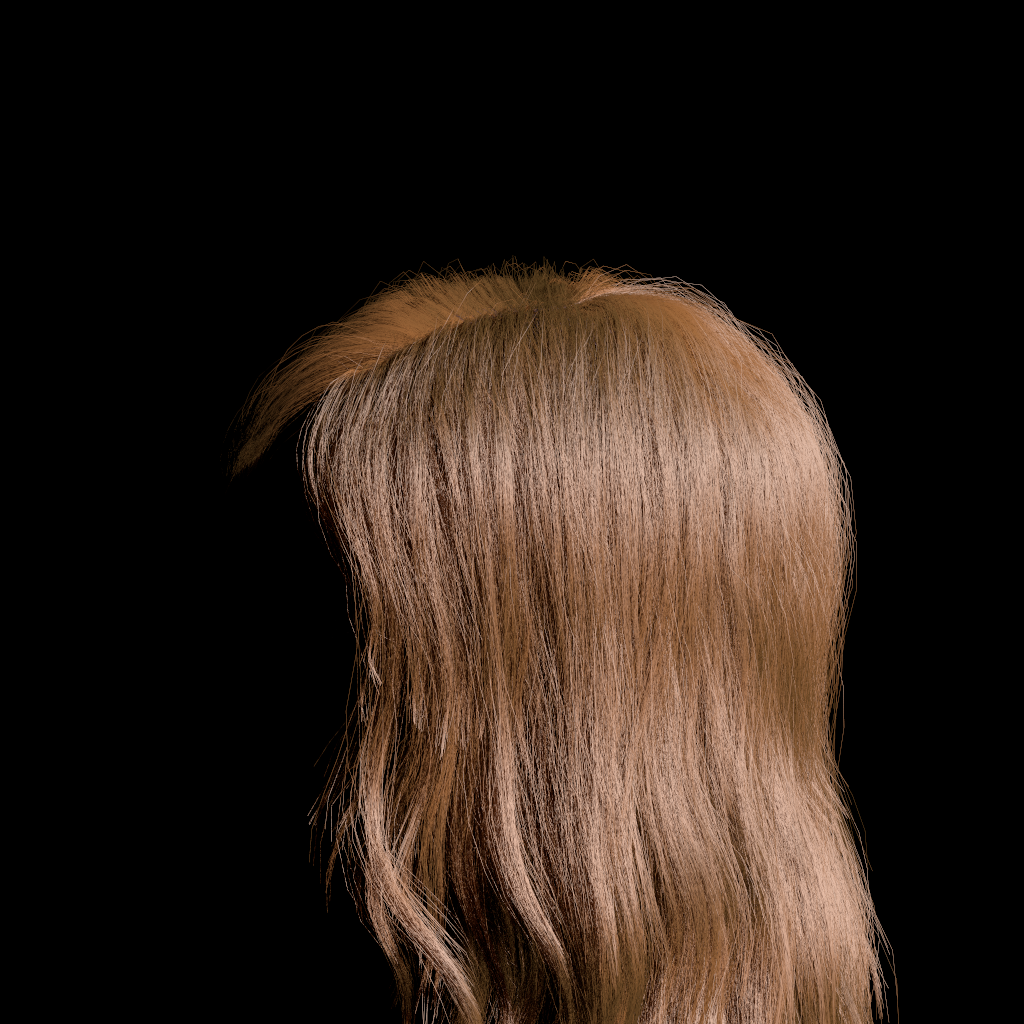}\hfill%
\includegraphics[trim={0 0 0 0},clip,width=0.11\linewidth]{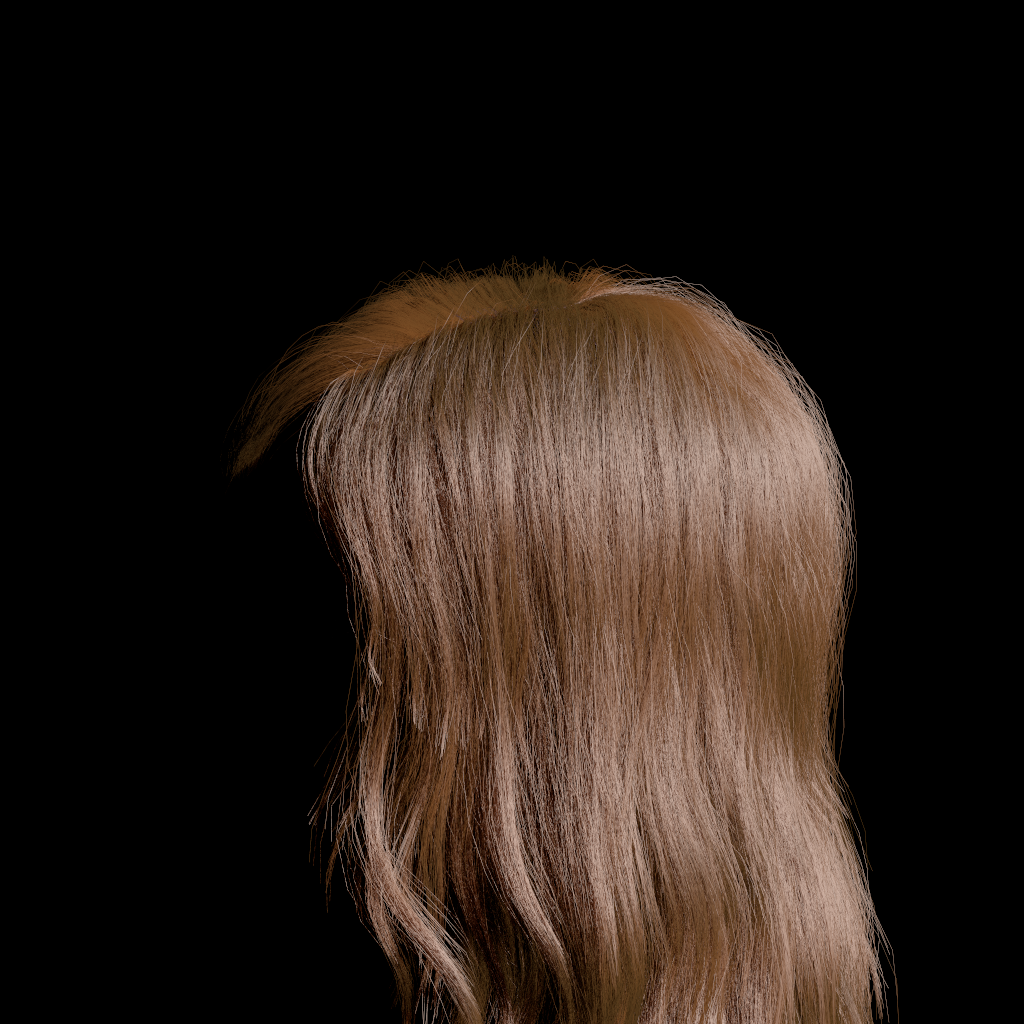}\hfill%
\includegraphics[trim={325 325 325 325},clip,width=0.11\linewidth]{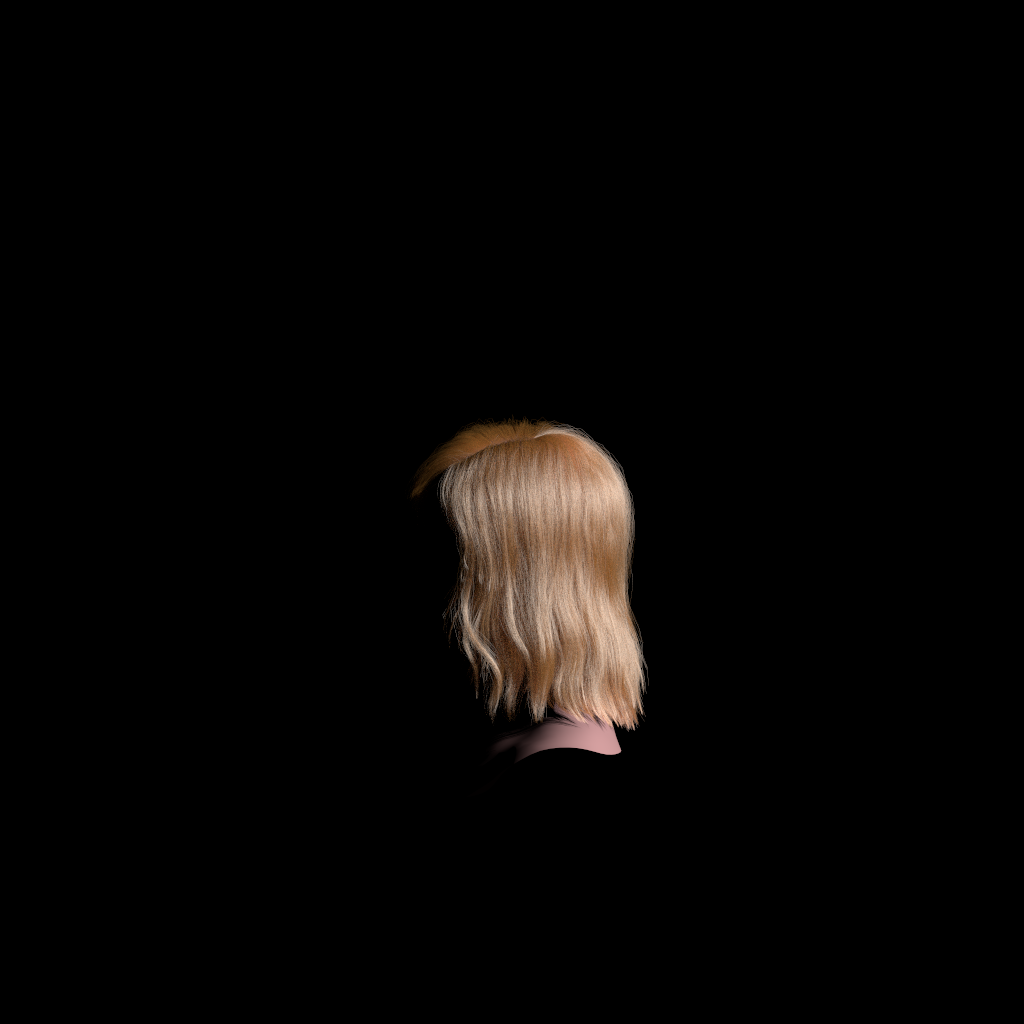}\hfill%
\includegraphics[trim={325 325 325 325},clip,width=0.11\linewidth]{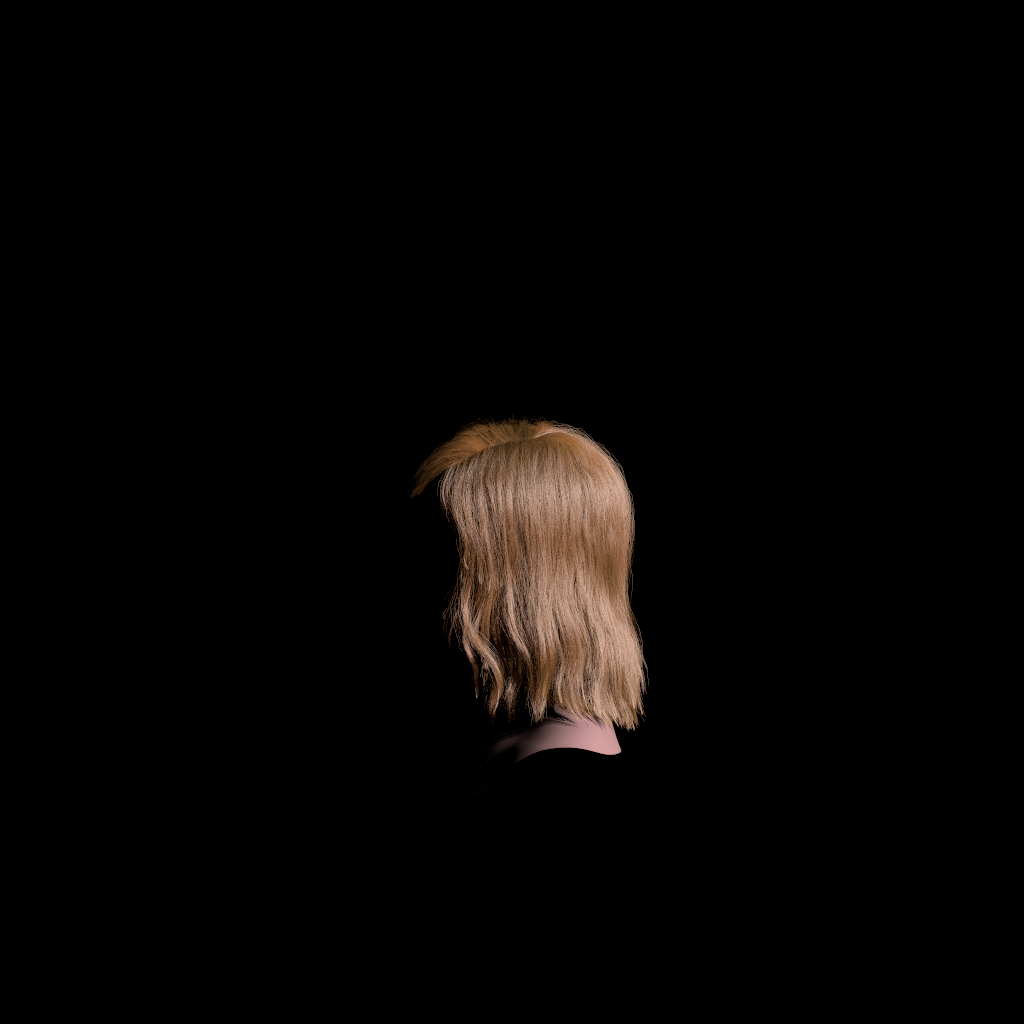}\hfill%
\includegraphics[trim={325 325 325 325},clip,width=0.11\linewidth]{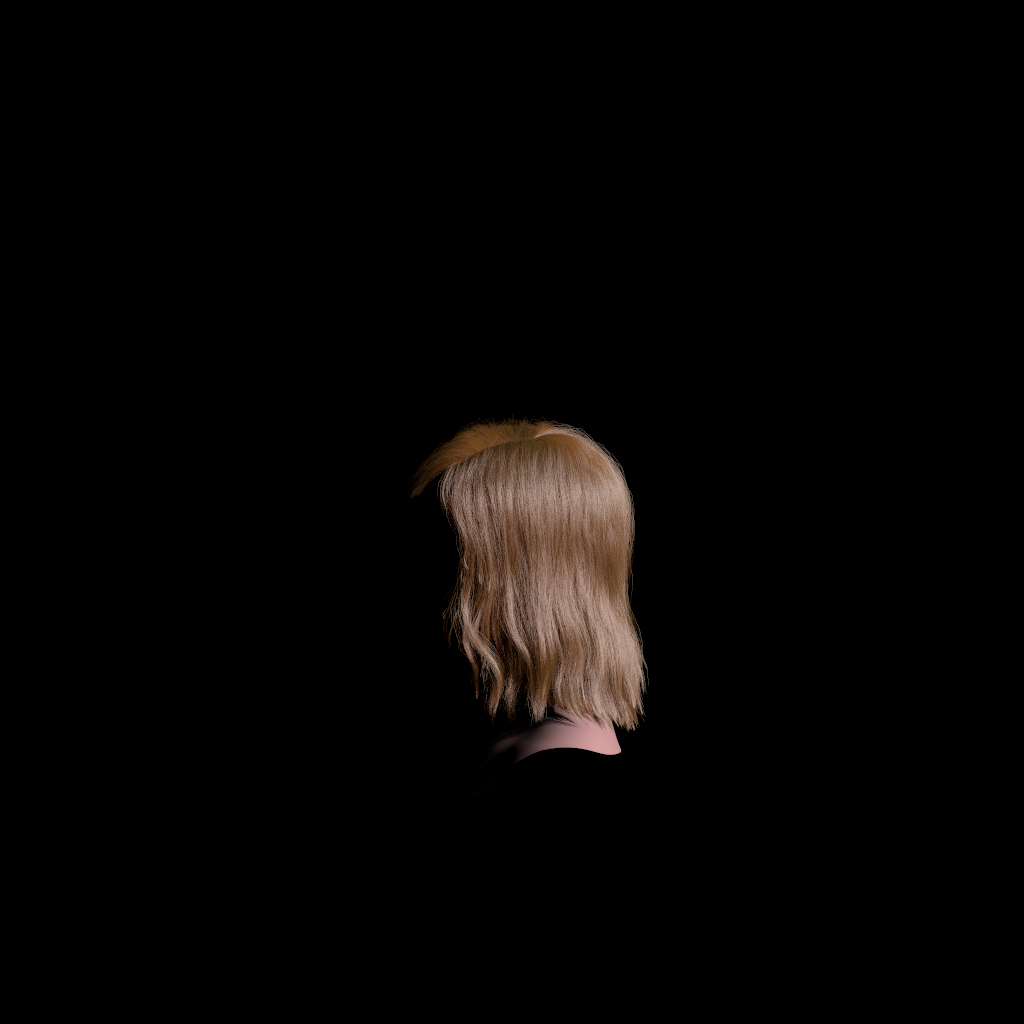}\hfill%
\includegraphics[trim={450 450 450 450},clip,width=0.11\linewidth]{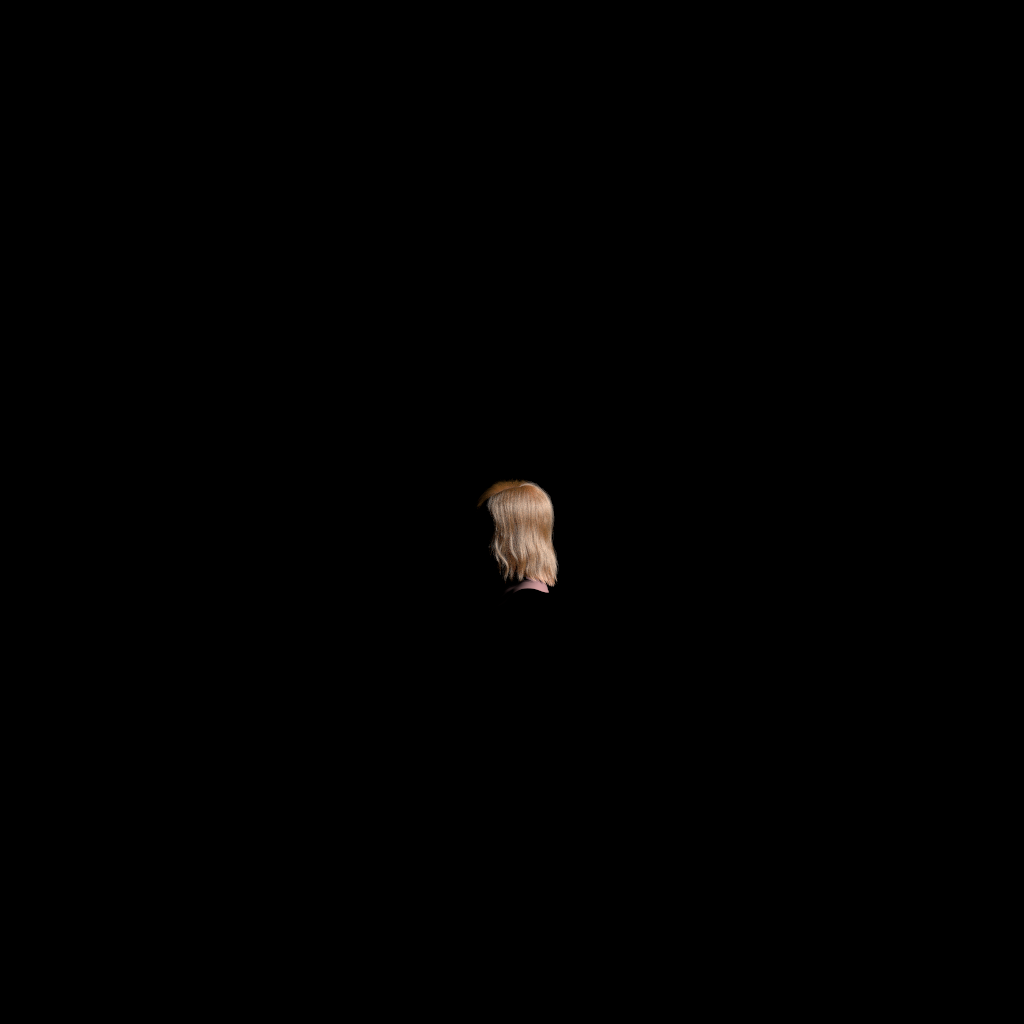}\hfill%
\includegraphics[trim={450 450 450 450},clip,width=0.11\linewidth]{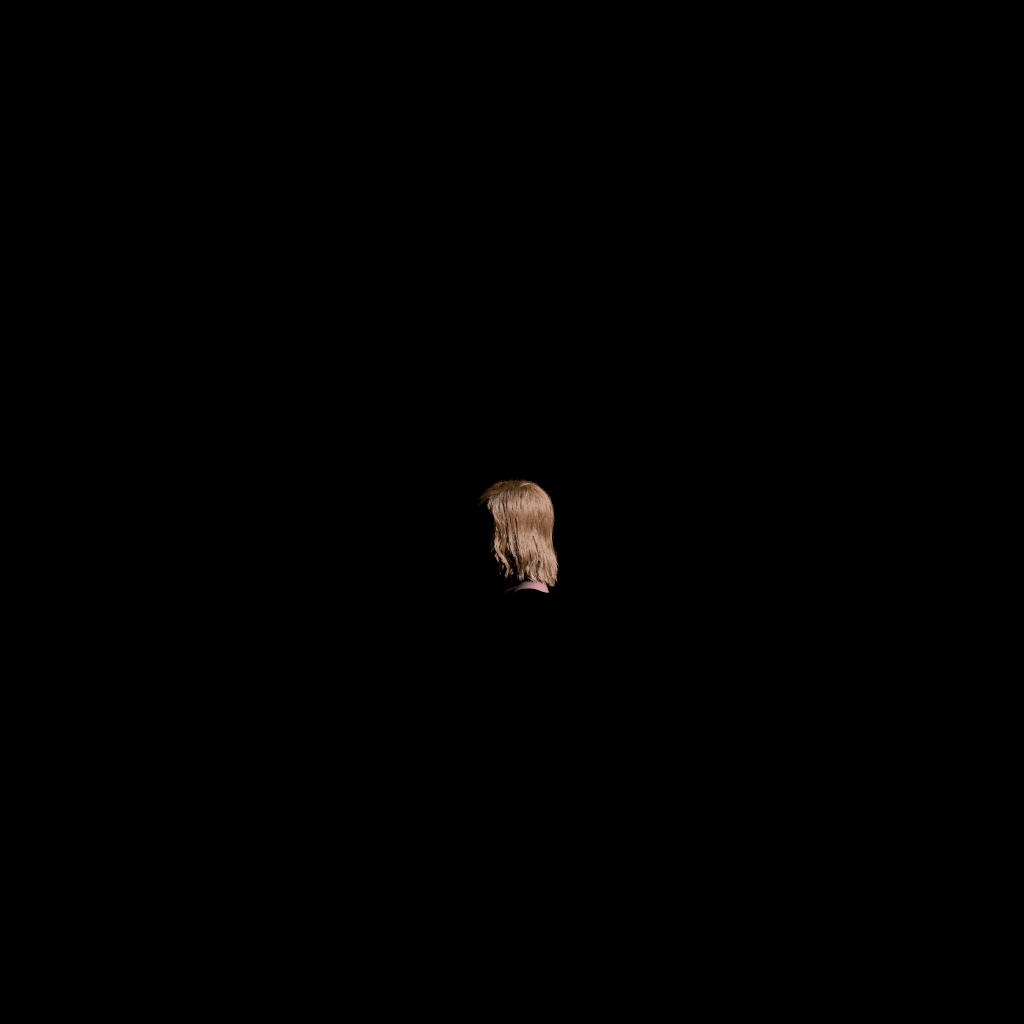}\hfill%
\includegraphics[trim={450 450 450 450},clip,width=0.11\linewidth]{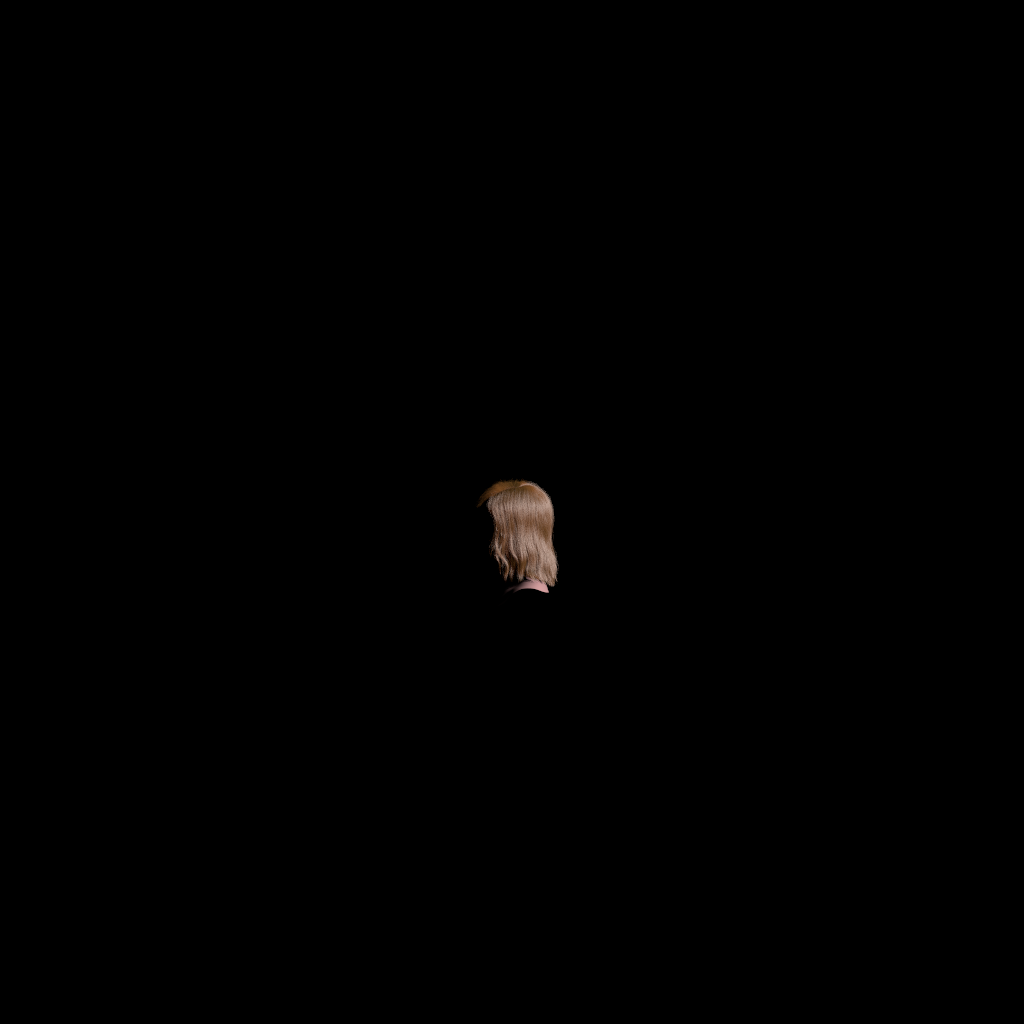}\\
\includegraphics[trim={0 0 0 0},clip,width=0.11\linewidth]{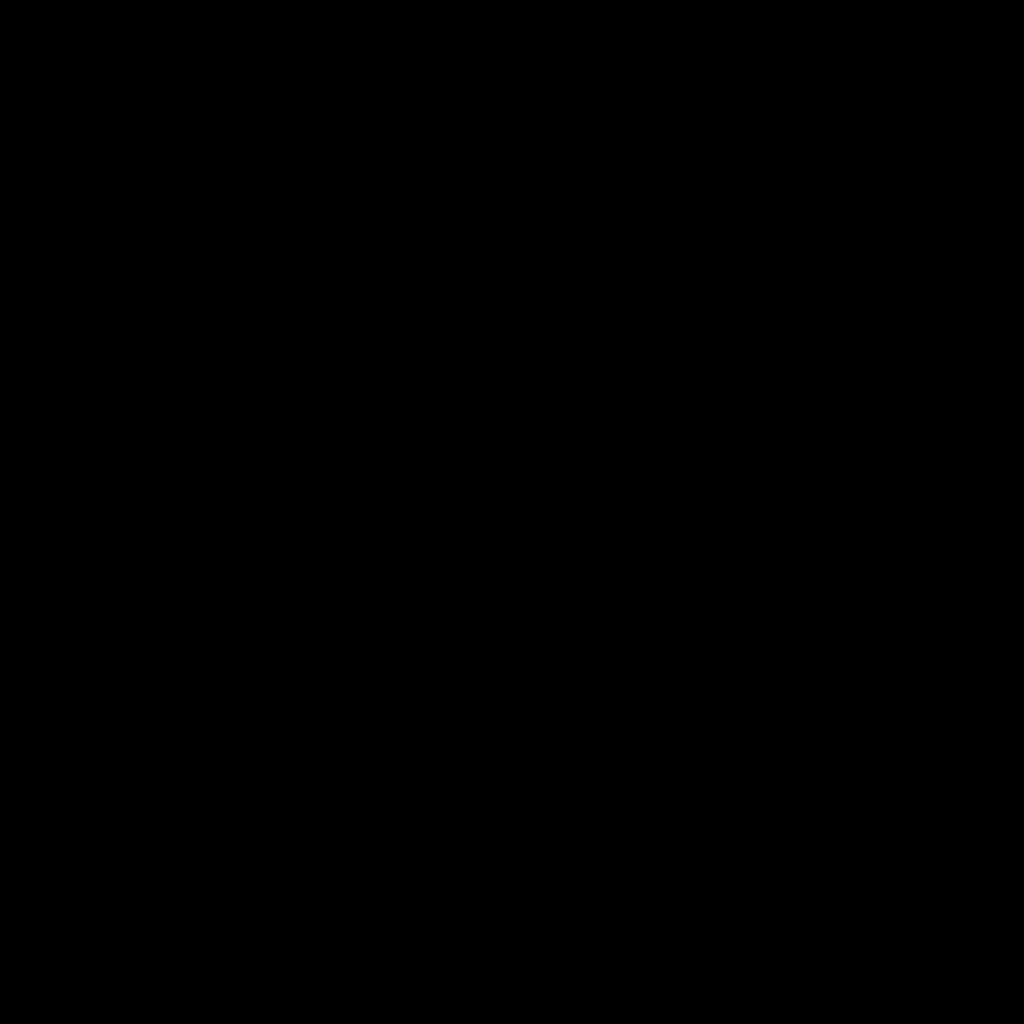}\hfill%
\includegraphics[trim={0 0 0 0},clip,width=0.11\linewidth]{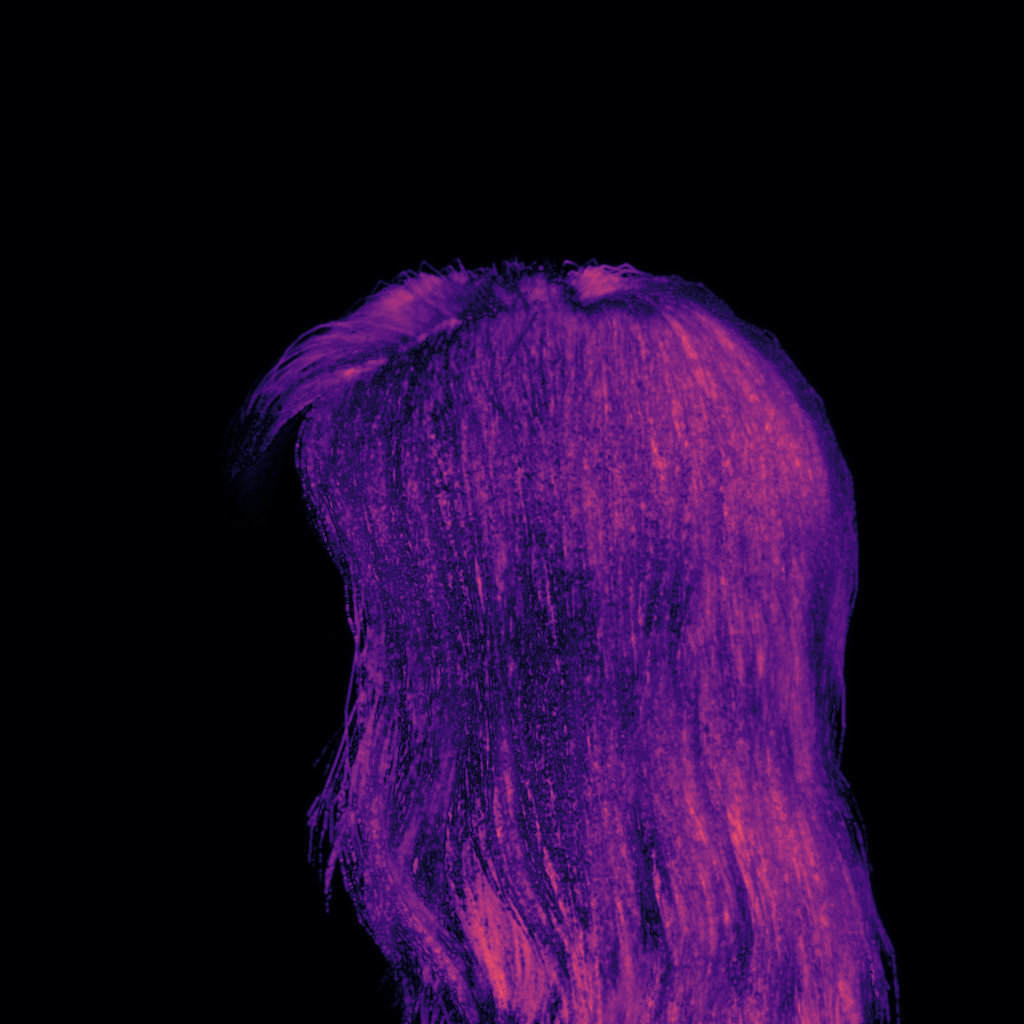}\hfill%
\includegraphics[trim={0 0 0 0},clip,width=0.11\linewidth]{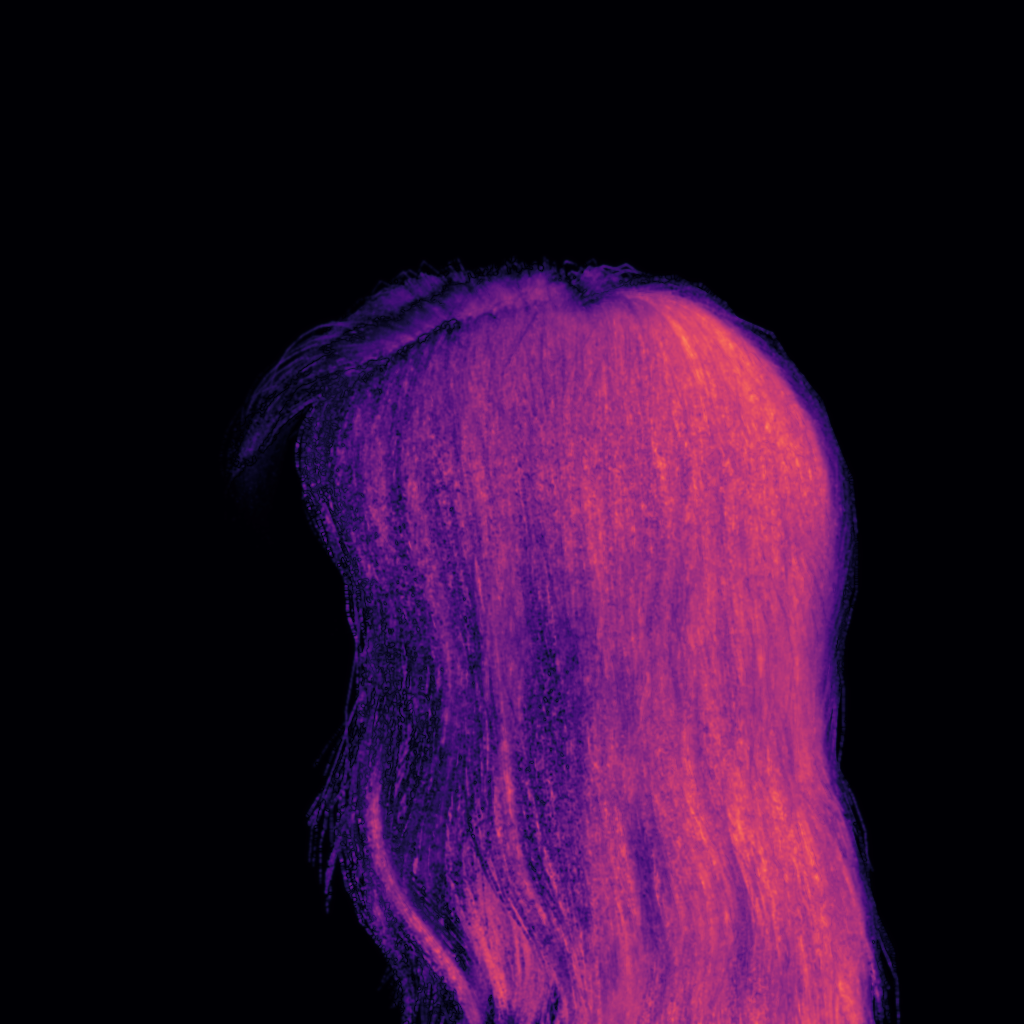}\hfill%
\includegraphics[trim={0 0 0 0},clip,width=0.11\linewidth]{FIG/FIG_SIGA/main_result/hadley/gt/0.png}\hfill%
\includegraphics[trim={325 325 325 325},clip,width=0.11\linewidth]{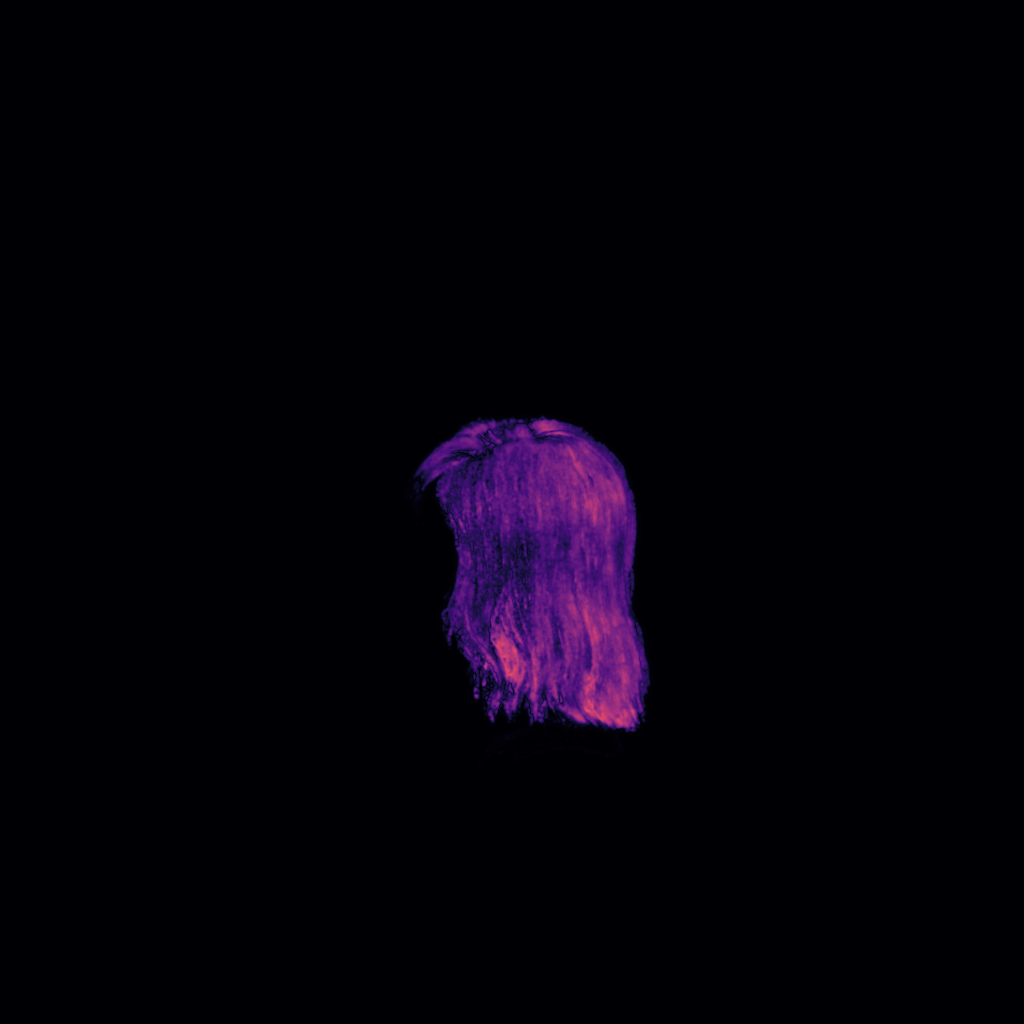}\hfill%
\includegraphics[trim={325 325 325 325},clip,width=0.11\linewidth]{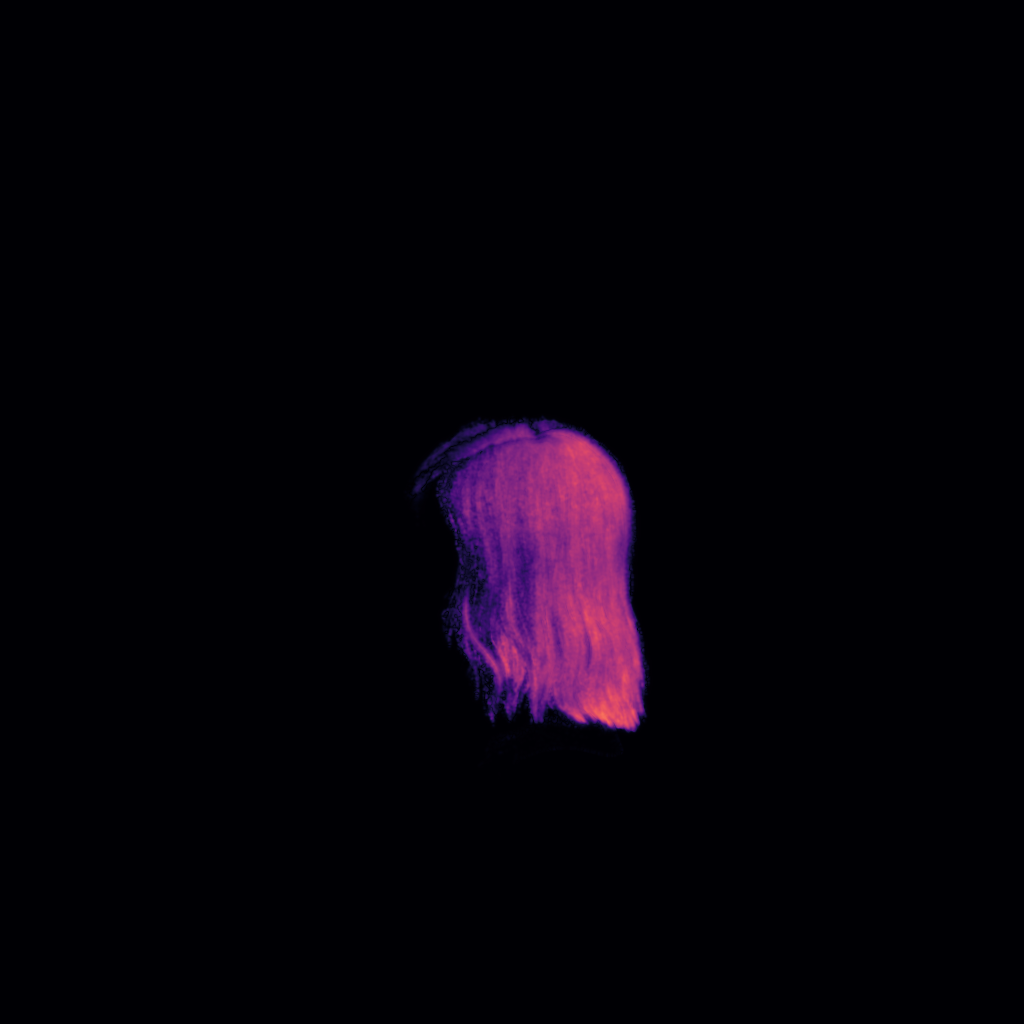}\hfill%
\includegraphics[trim={0 0 0 0},clip,width=0.11\linewidth]{FIG/FIG_SIGA/main_result/hadley/gt/0.png}\hfill%
\includegraphics[trim={450 450 450 450},clip,width=0.11\linewidth]{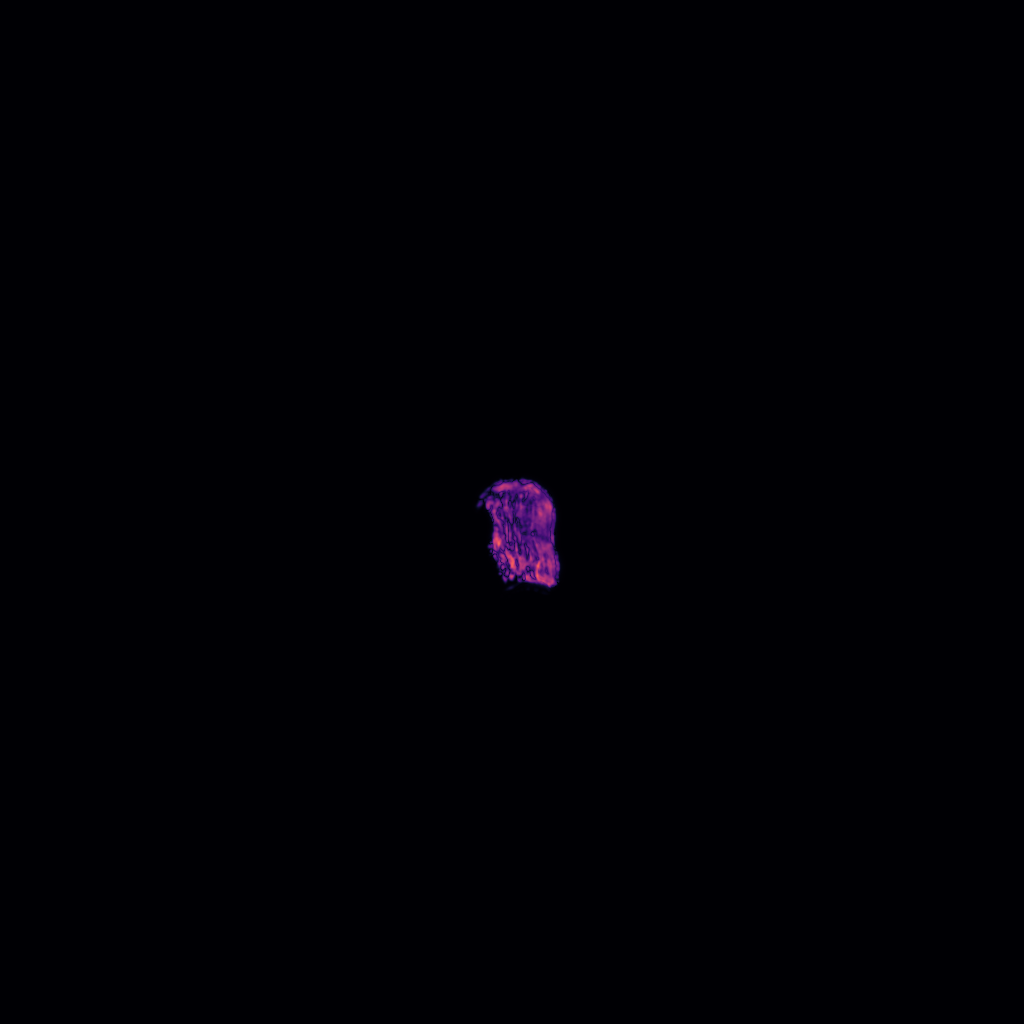}\hfill%
\includegraphics[trim={450 450 450 450},clip,width=0.11\linewidth]{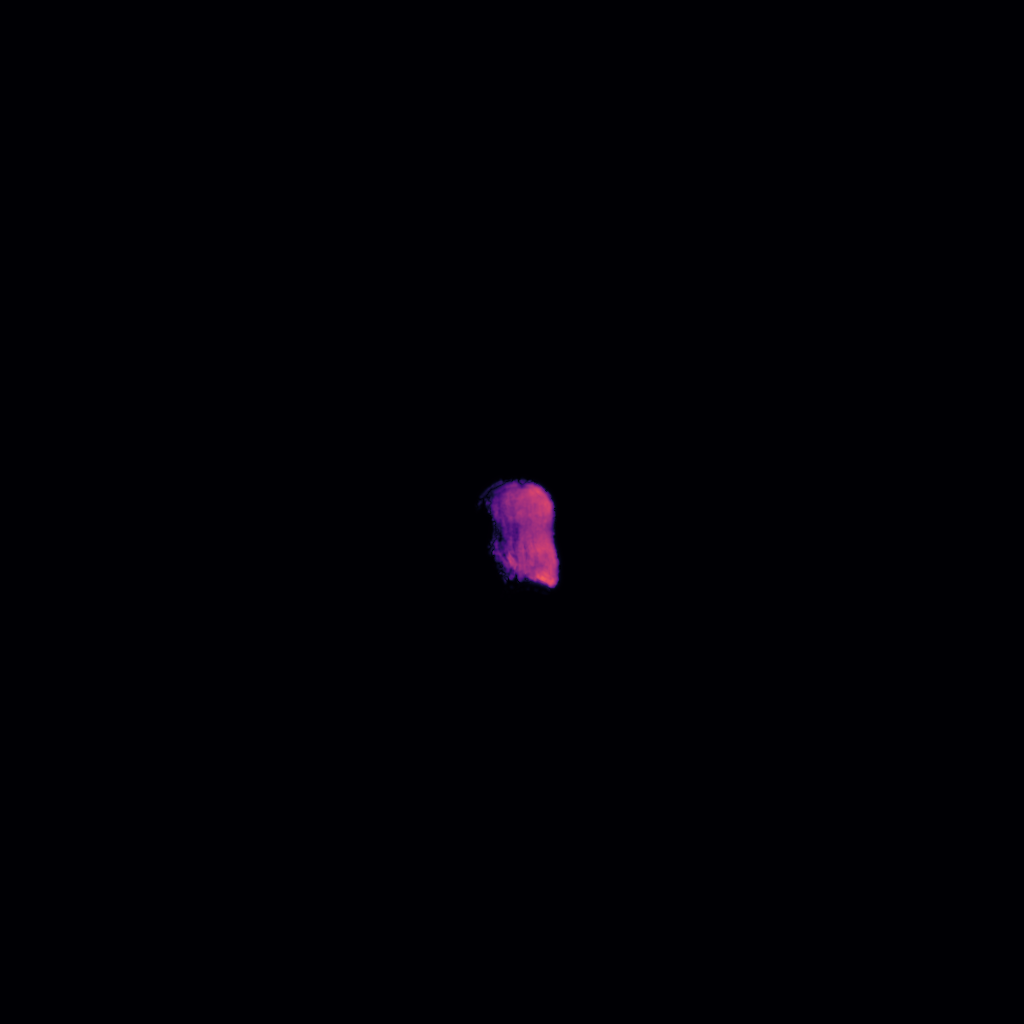}\\
\includegraphics[trim={0 0 0 0},clip,width=0.11\linewidth]{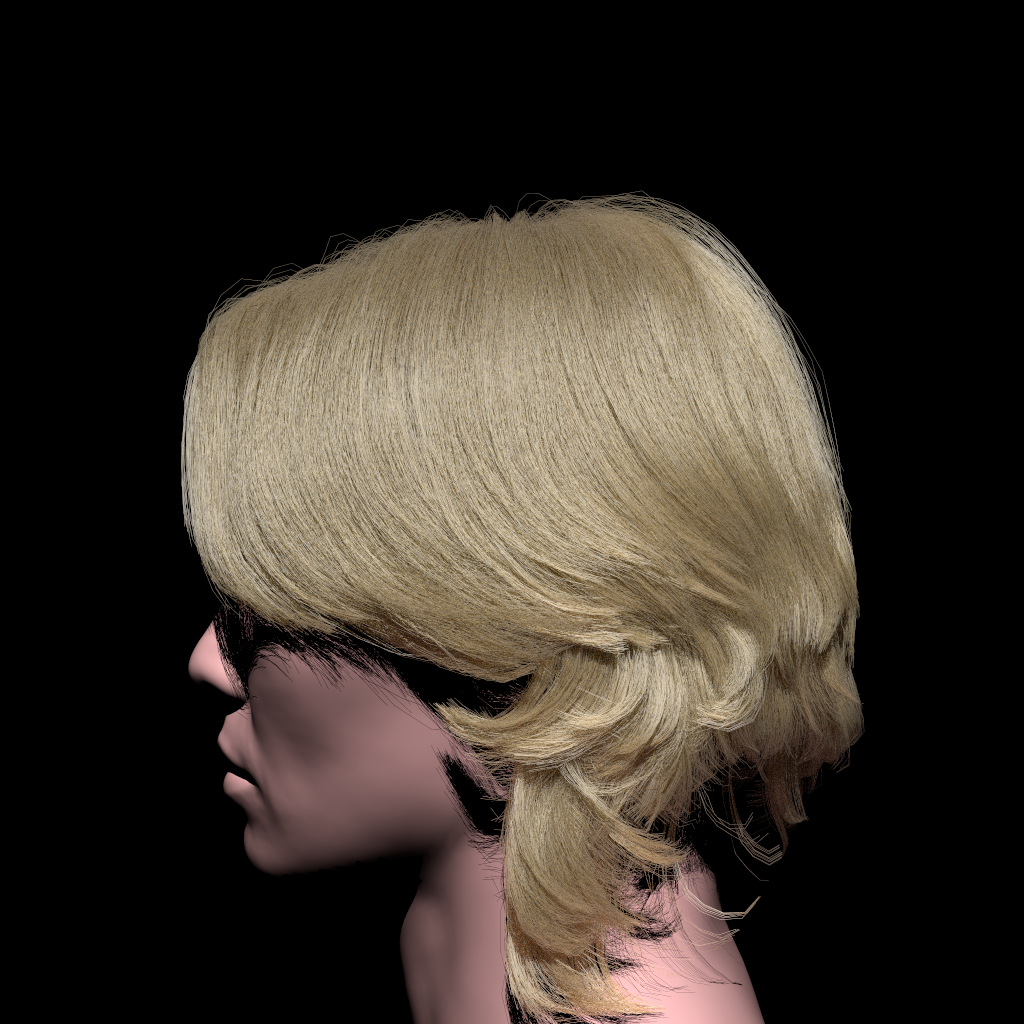}\hfill%
\includegraphics[trim={0 0 0 0},clip,width=0.11\linewidth]{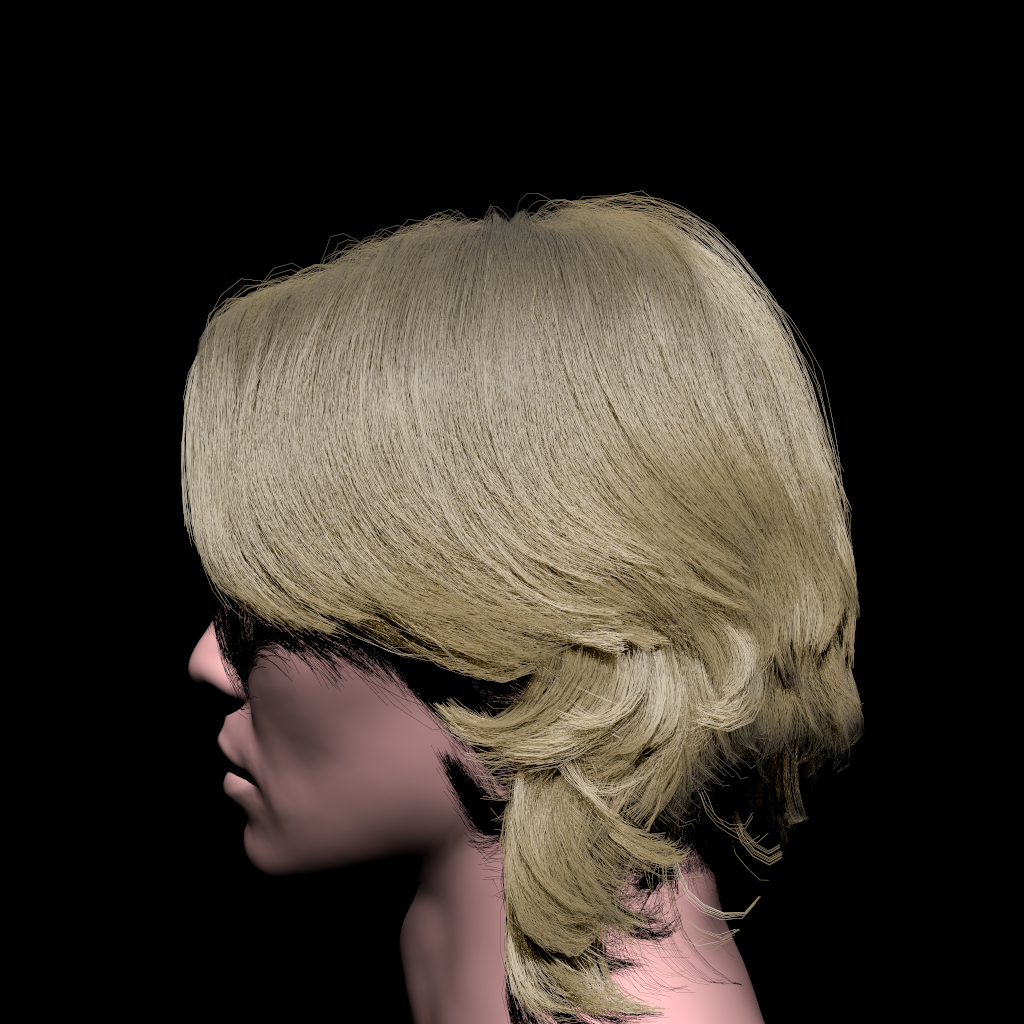}\hfill%
\includegraphics[trim={0 0 0 0},clip,width=0.11\linewidth]{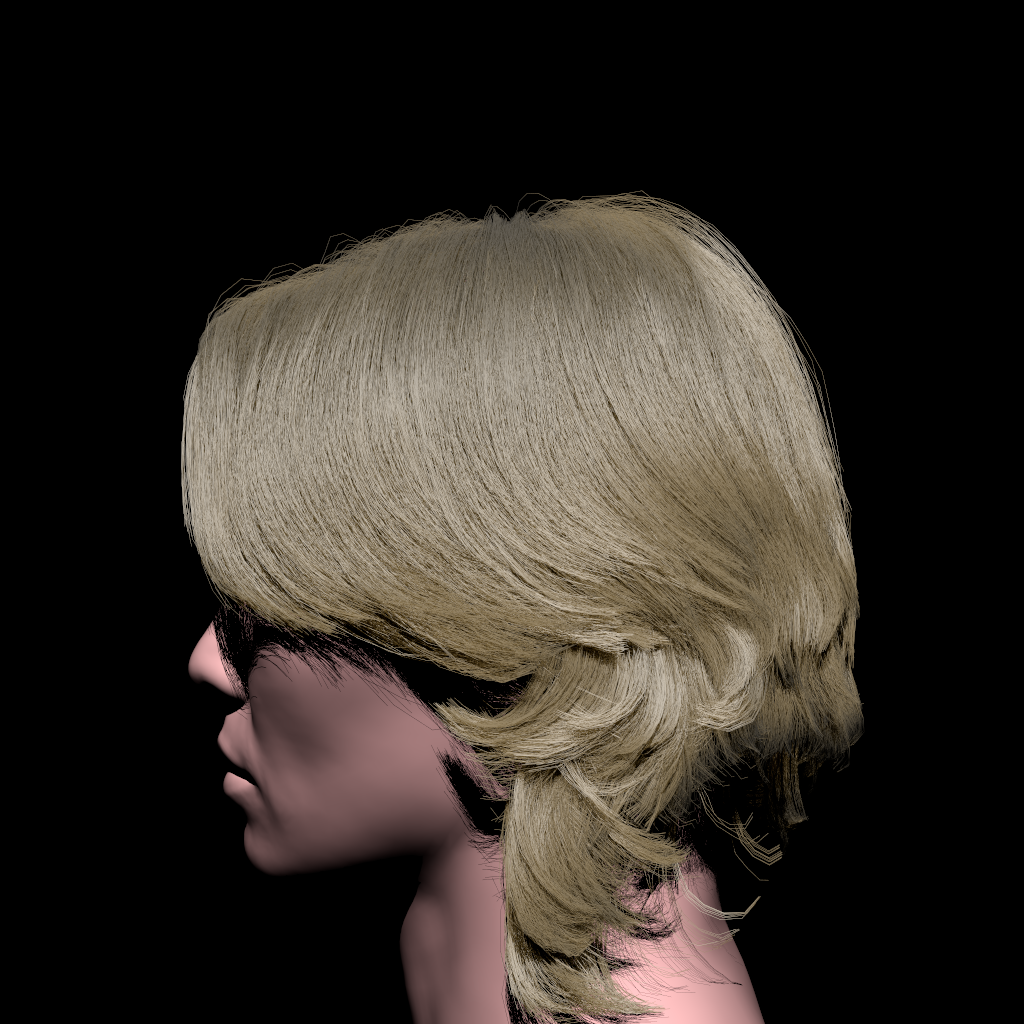}\hfill%
\includegraphics[trim={325 325 325 325},clip,width=0.11\linewidth]{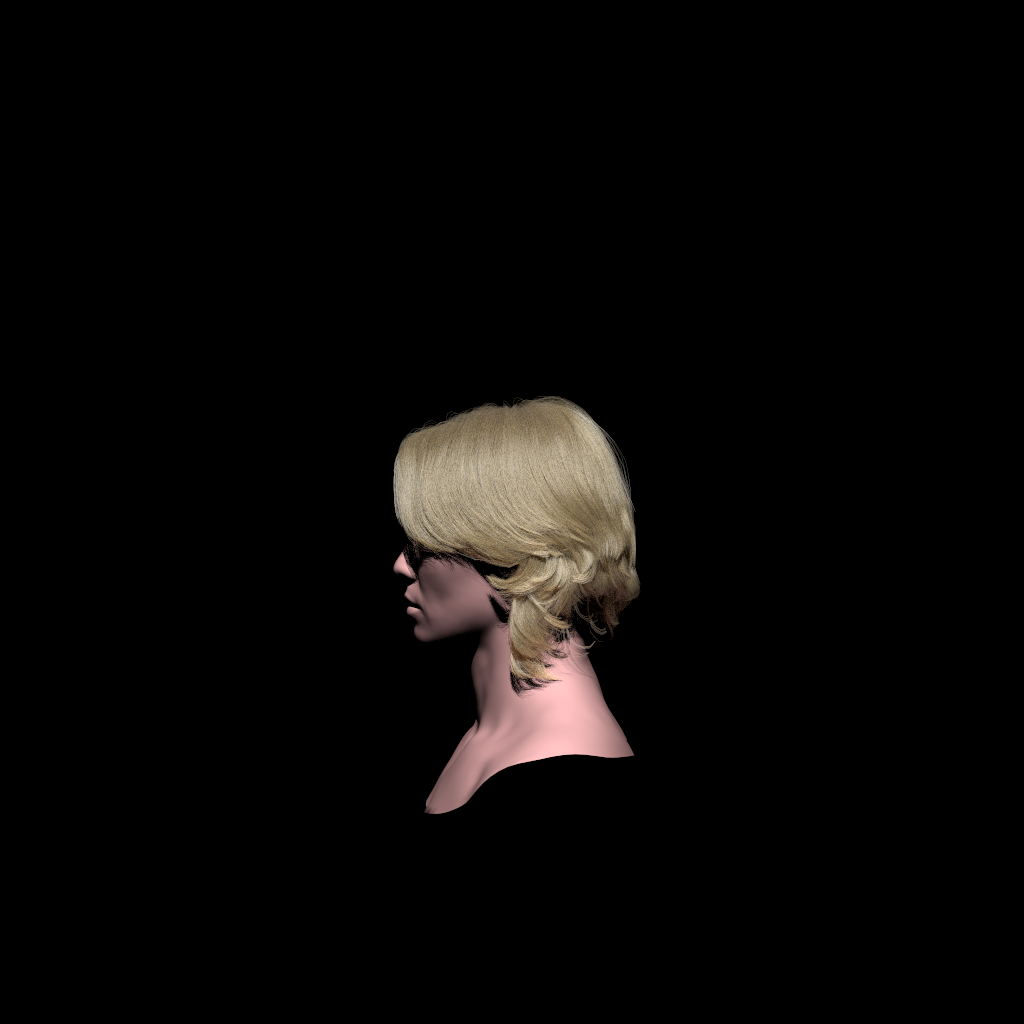}\hfill%
\includegraphics[trim={325 325 325 325},clip,width=0.11\linewidth]{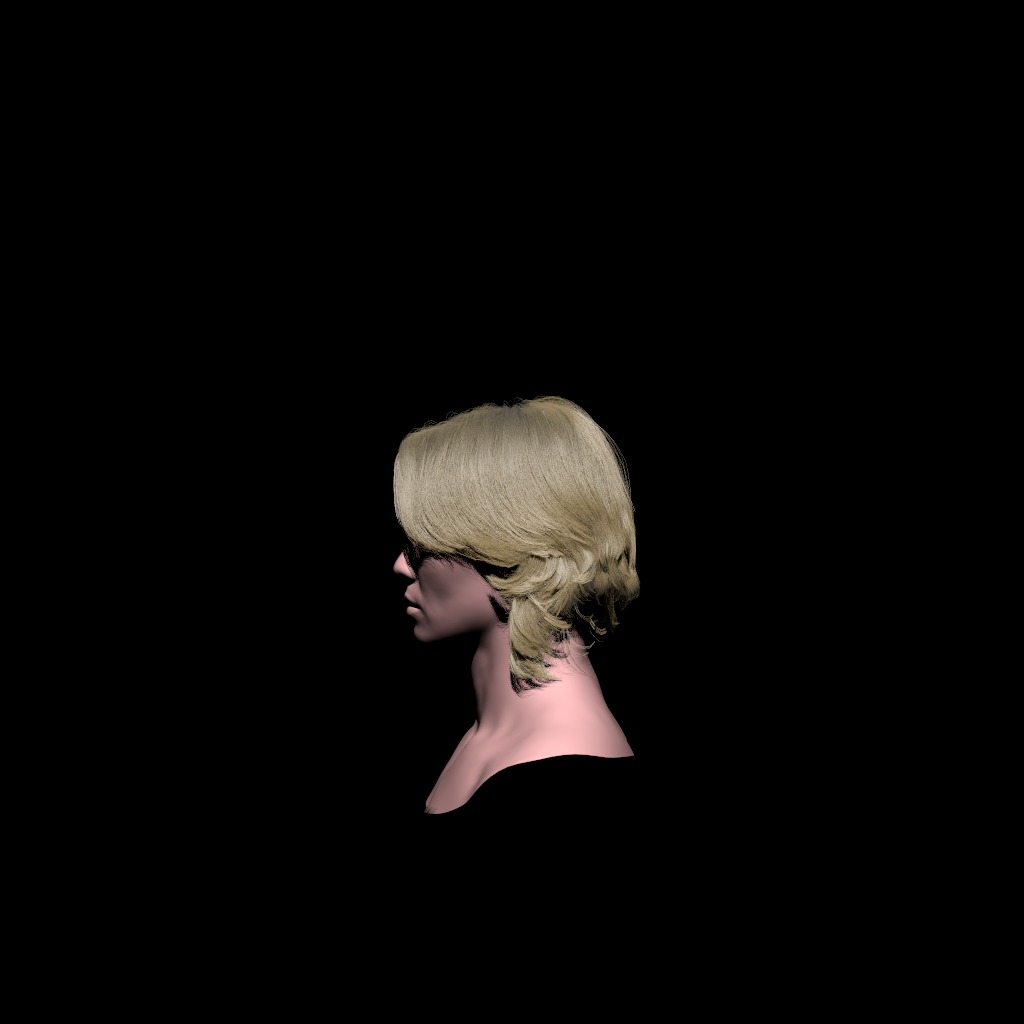}\hfill%
\includegraphics[trim={325 325 325 325},clip,width=0.11\linewidth]{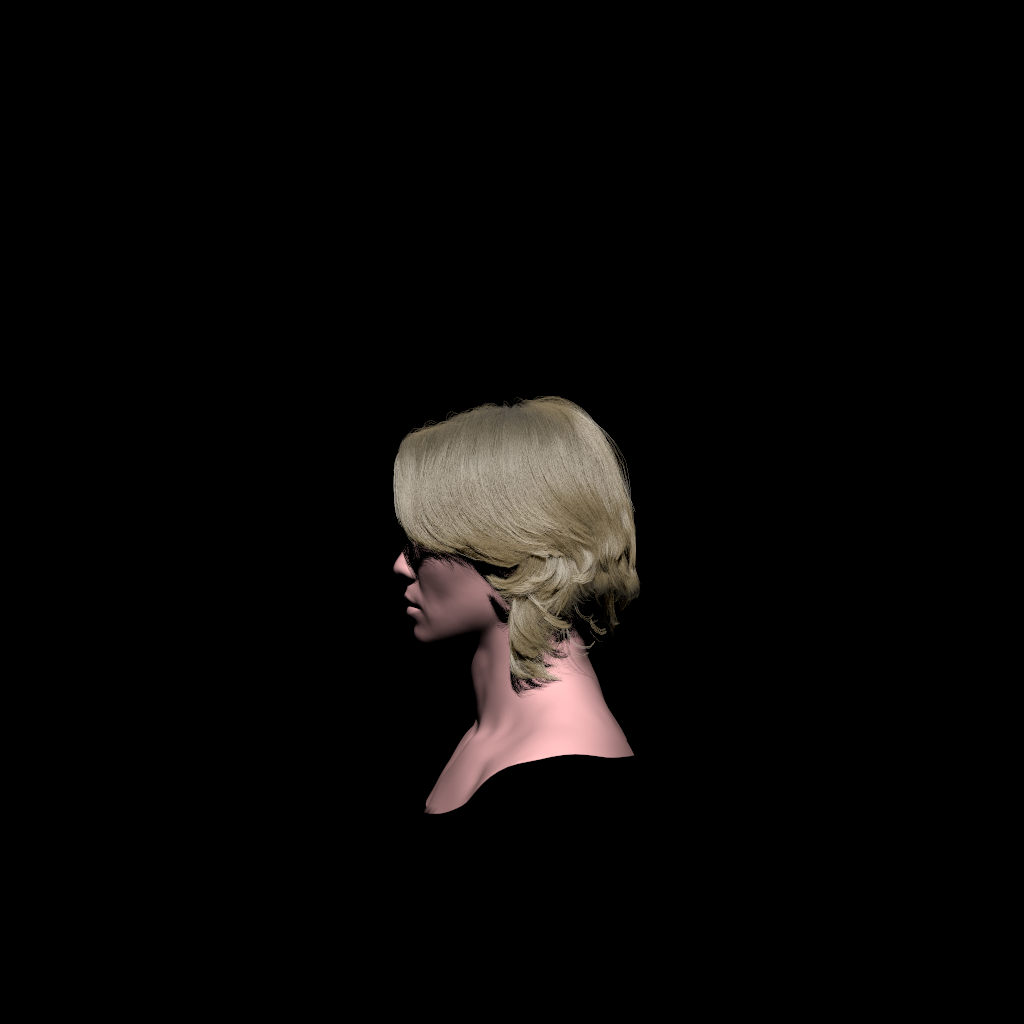}\hfill%
\includegraphics[trim={450 450 450 450},clip,width=0.11\linewidth]{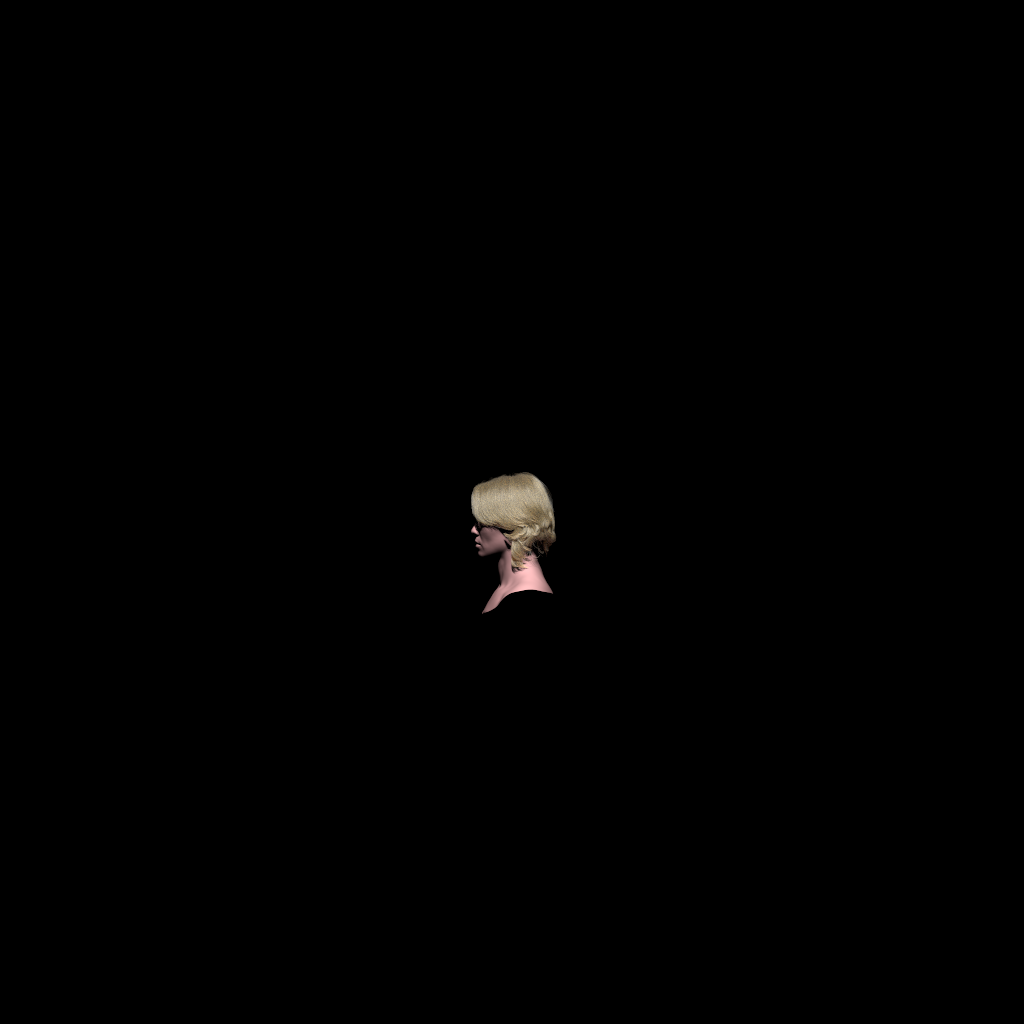}\hfill%
\includegraphics[trim={450 450 450 450},clip,width=0.11\linewidth]{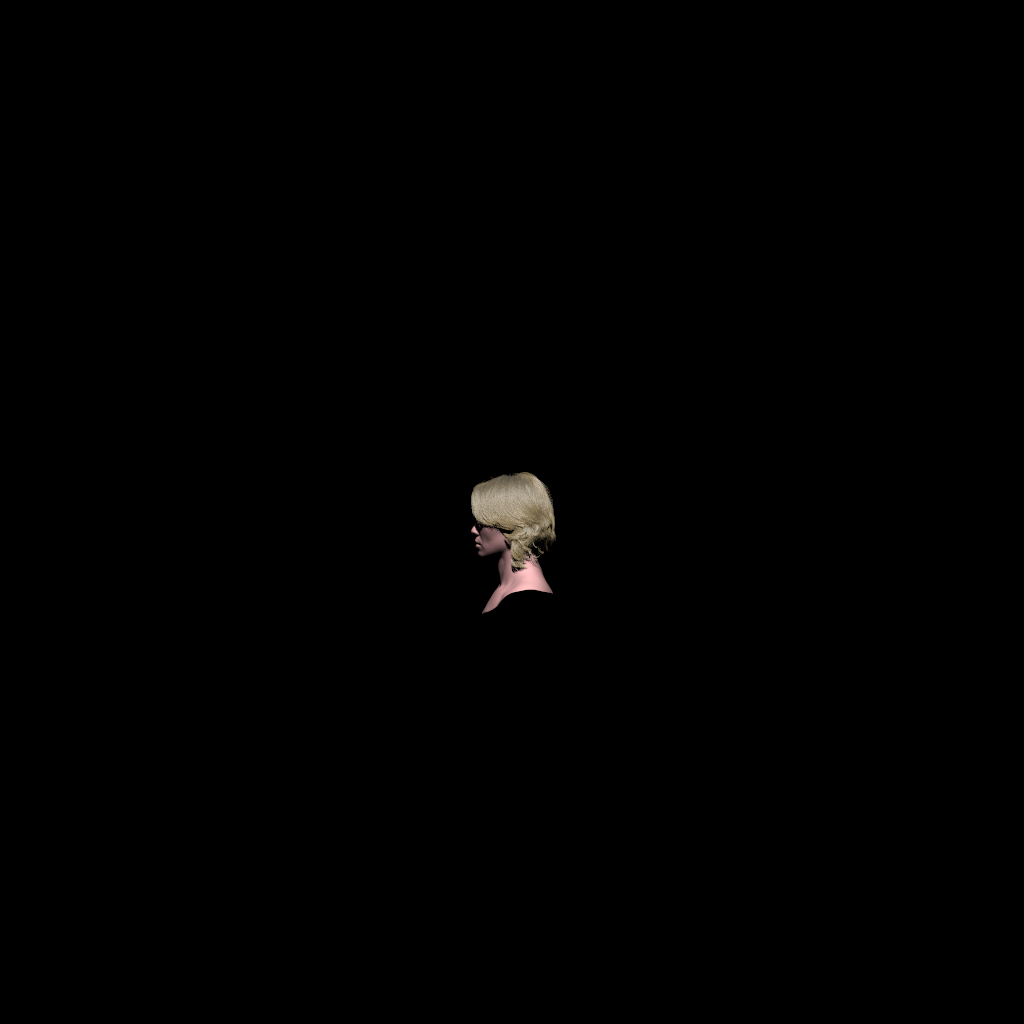}\hfill%
\includegraphics[trim={450 450 450 450},clip,width=0.11\linewidth]{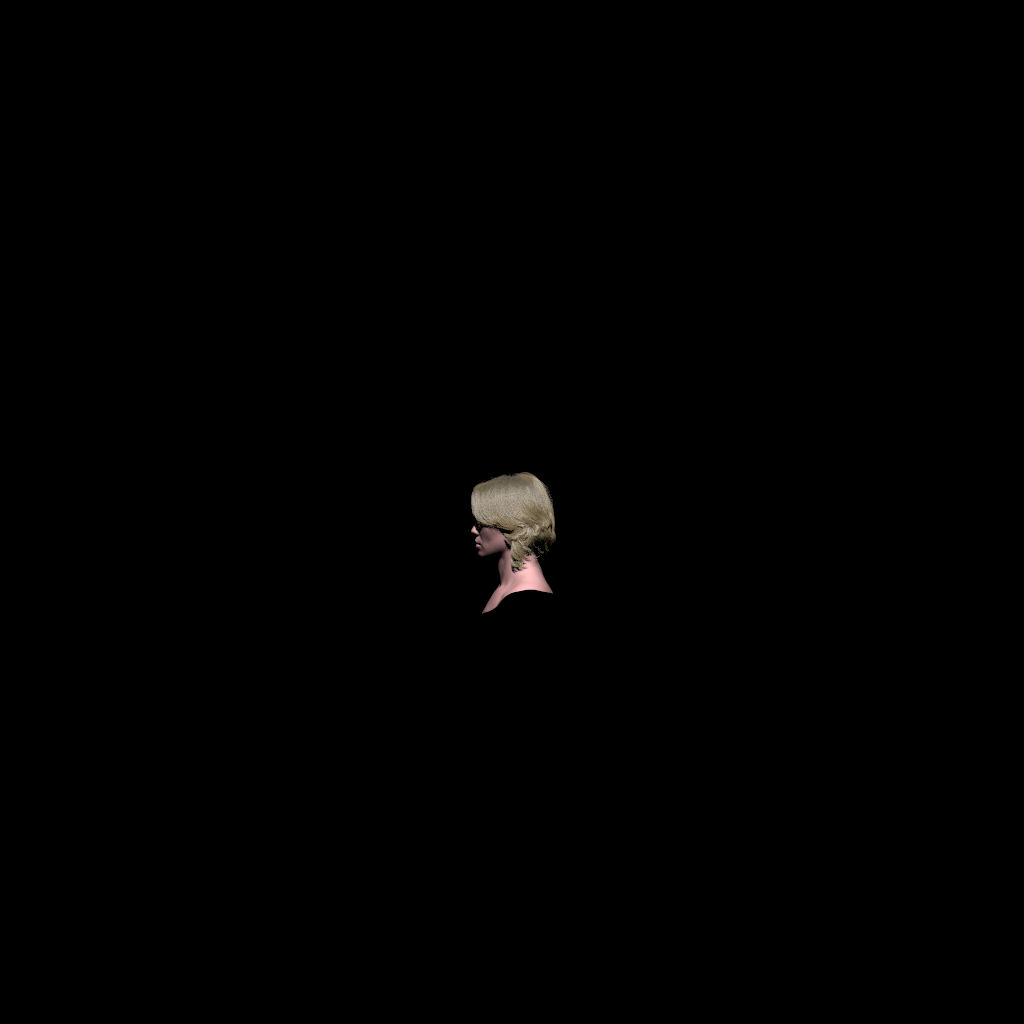}\\
\includegraphics[trim={0 0 0 0},clip,width=0.11\linewidth]{FIG/FIG_SIGA/main_result/hadley/gt/0.png}\hfill%
\includegraphics[trim={0 0 0 0},clip,width=0.11\linewidth]{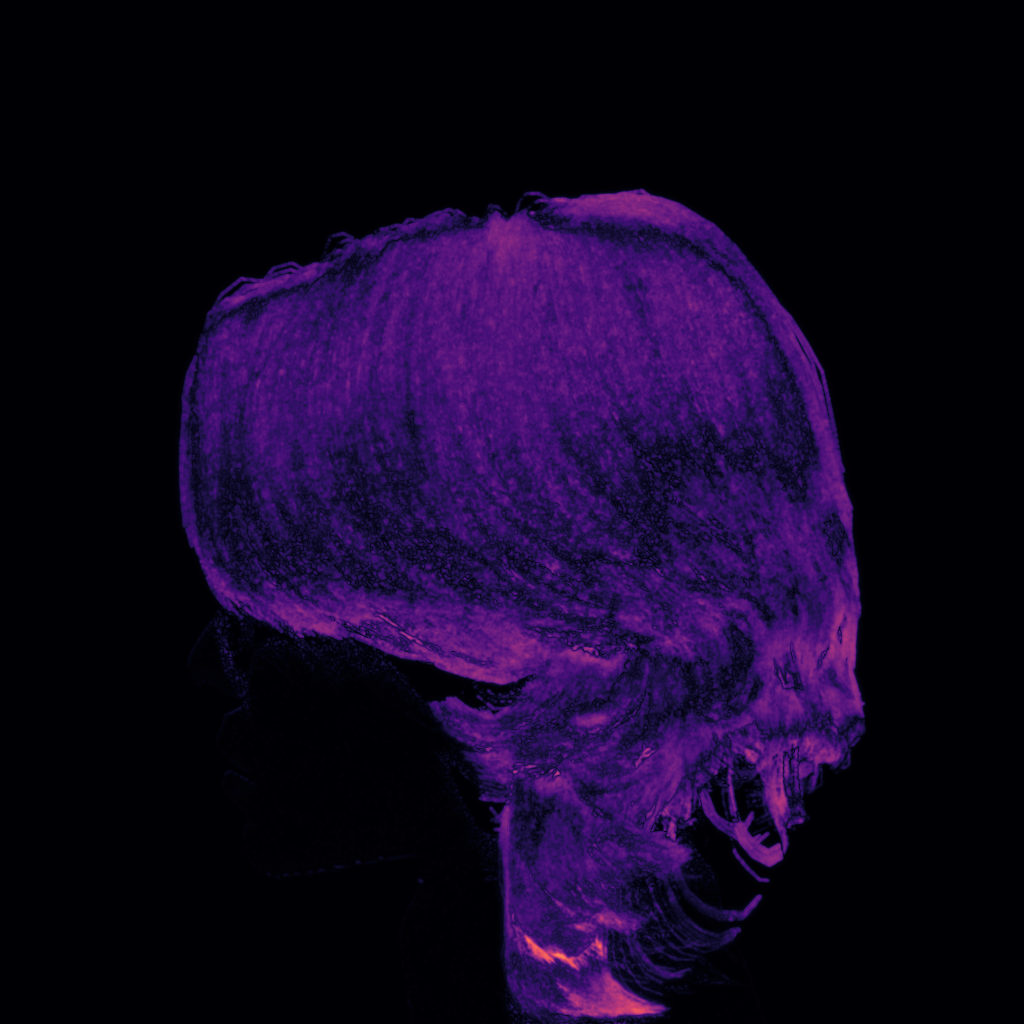}\hfill%
\includegraphics[trim={0 0 0 0},clip,width=0.11\linewidth]{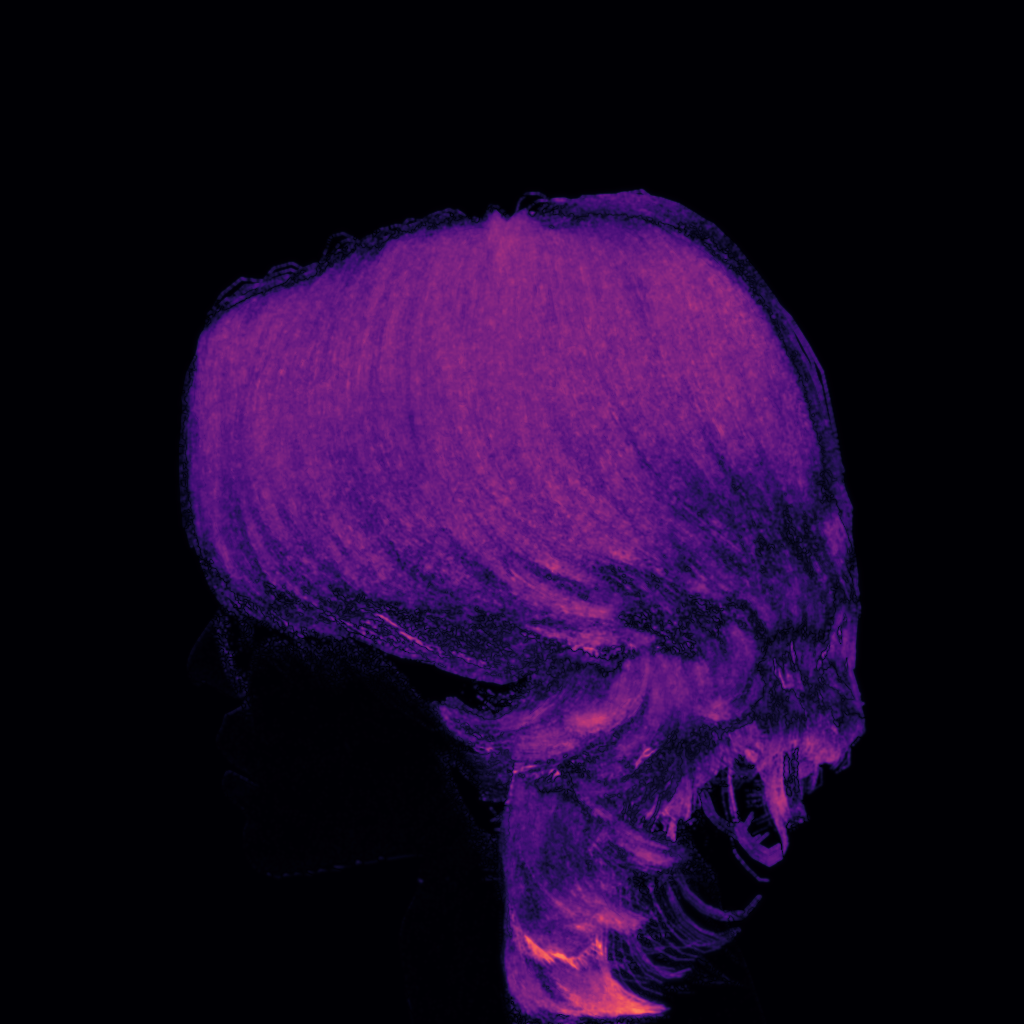}\hfill%
\includegraphics[trim={0 0 0 0},clip,width=0.11\linewidth]{FIG/FIG_SIGA/main_result/hadley/gt/0.png}\hfill%
\includegraphics[trim={325 325 325 325},clip,width=0.11\linewidth]{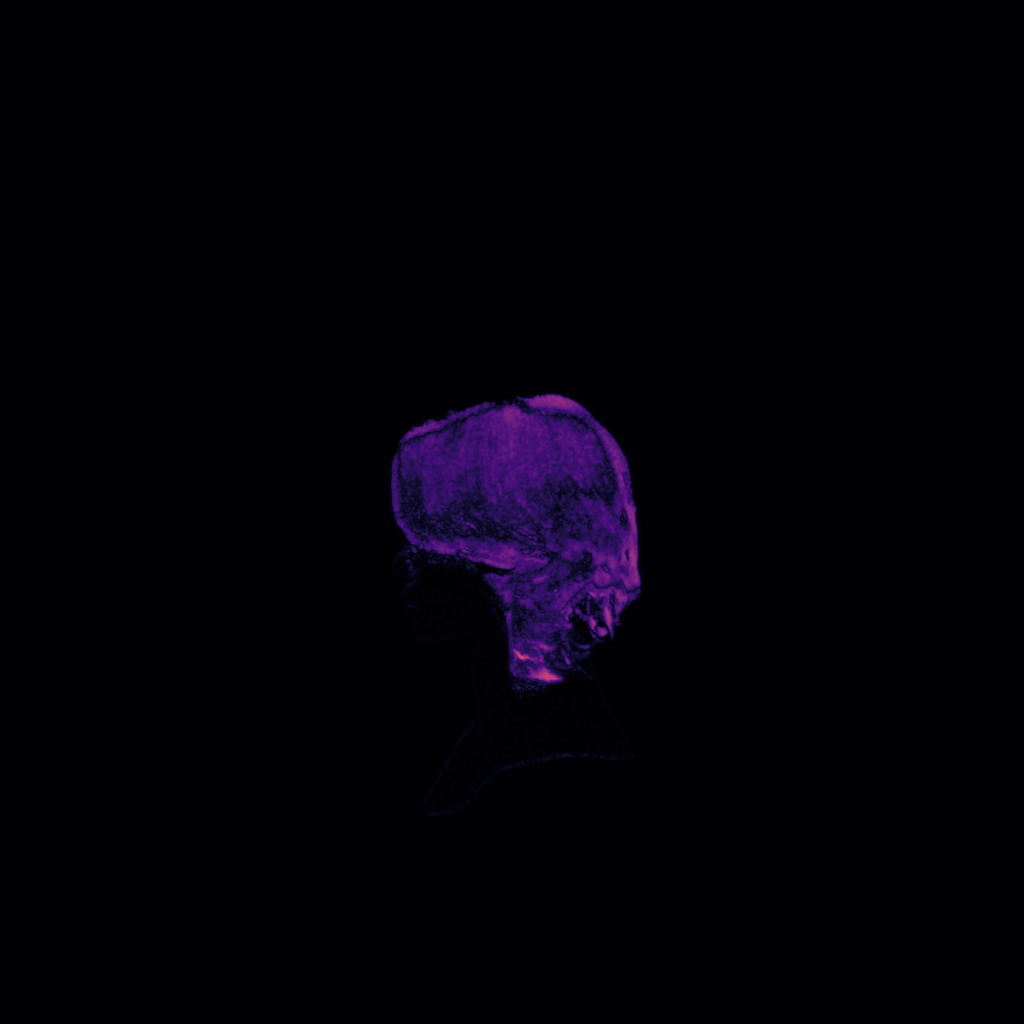}\hfill%
\includegraphics[trim={325 325 325 325},clip,width=0.11\linewidth]{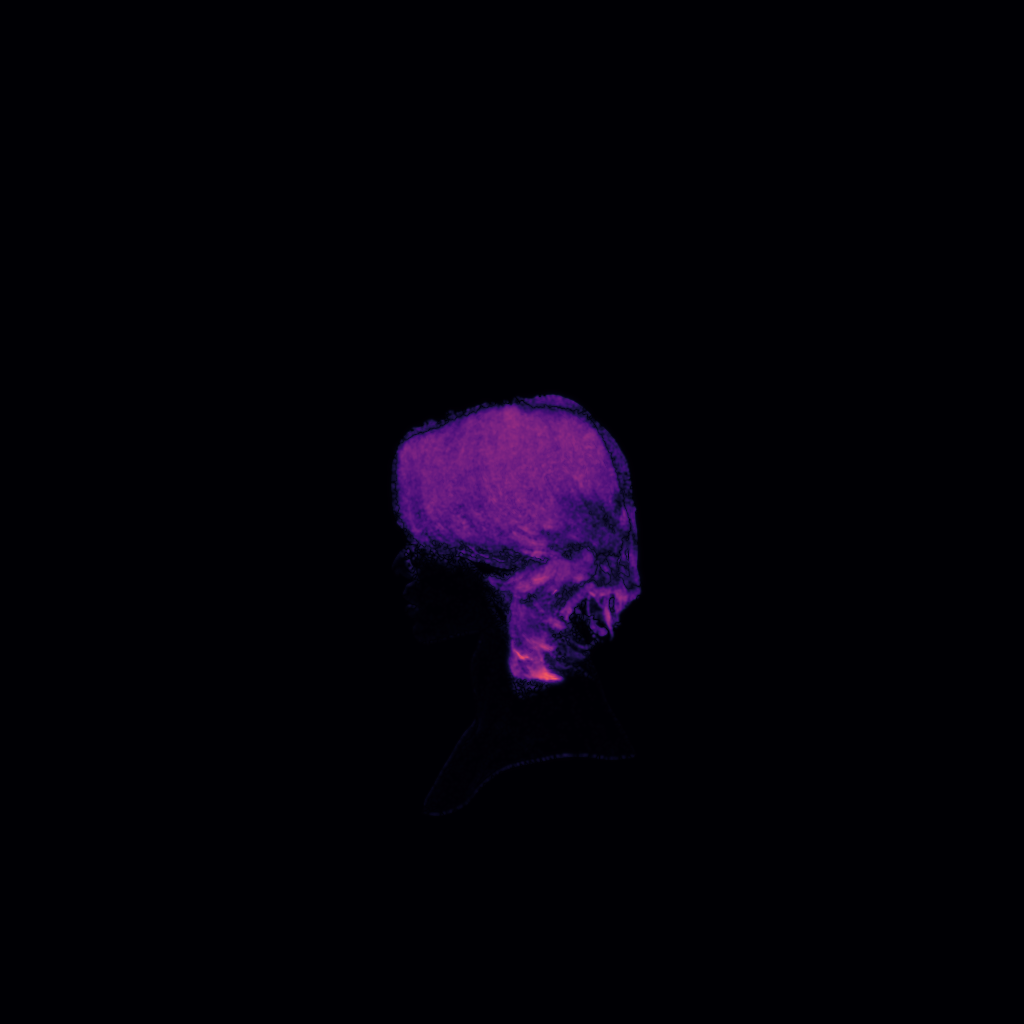}\hfill%
\includegraphics[trim={0 0 0 0},clip,width=0.11\linewidth]{FIG/FIG_SIGA/main_result/hadley/gt/0.png}\hfill%
\includegraphics[trim={450 450 450 450},clip,width=0.11\linewidth]{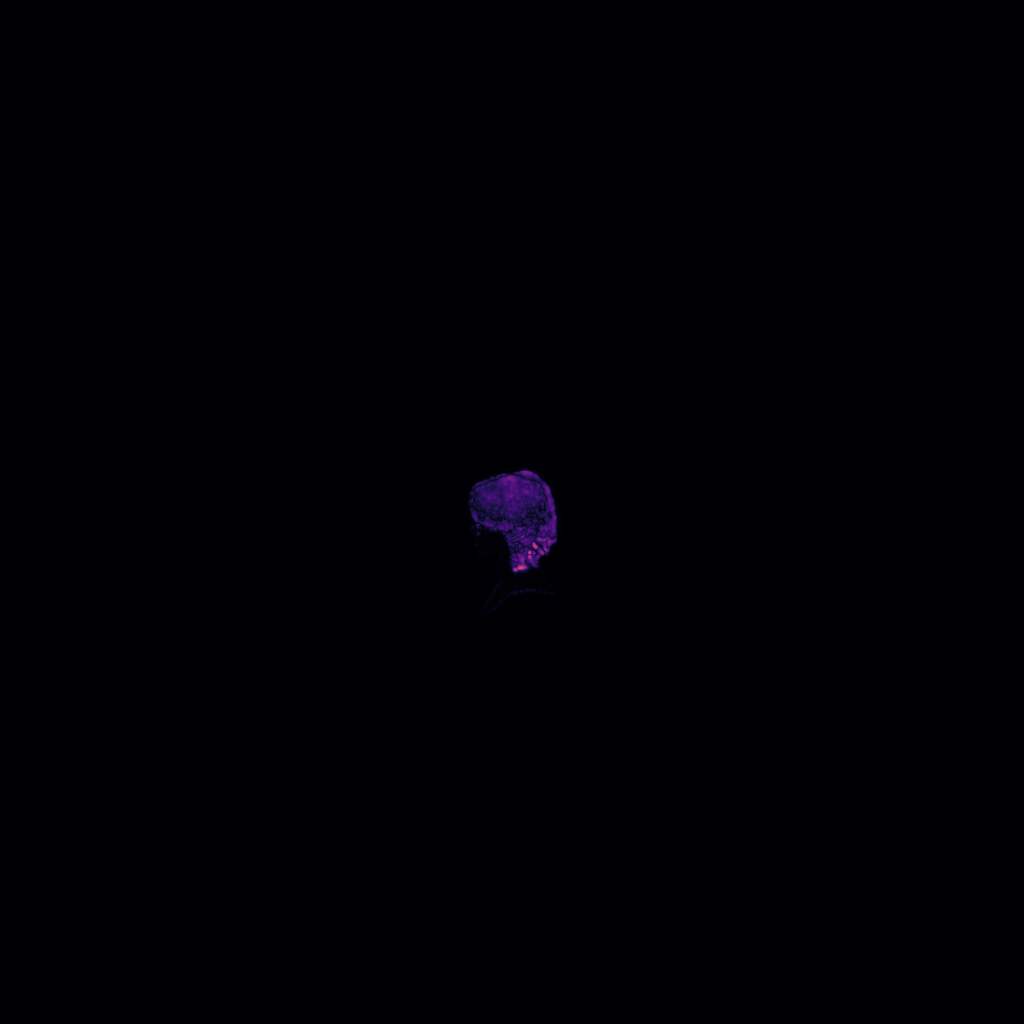}\hfill%
\includegraphics[trim={450 450 450 450},clip,width=0.11\linewidth]{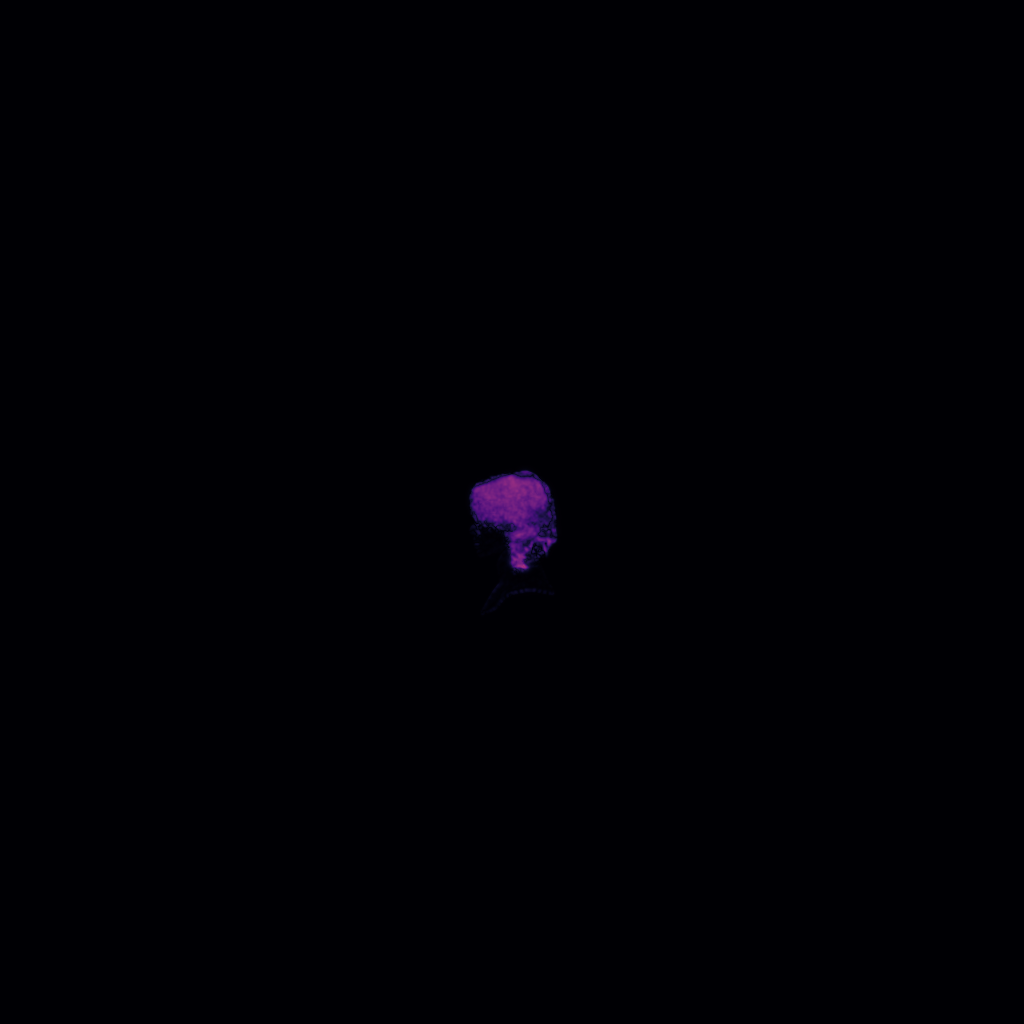}\\
\includegraphics[trim={180 260 180 100},clip,width=0.11\linewidth]{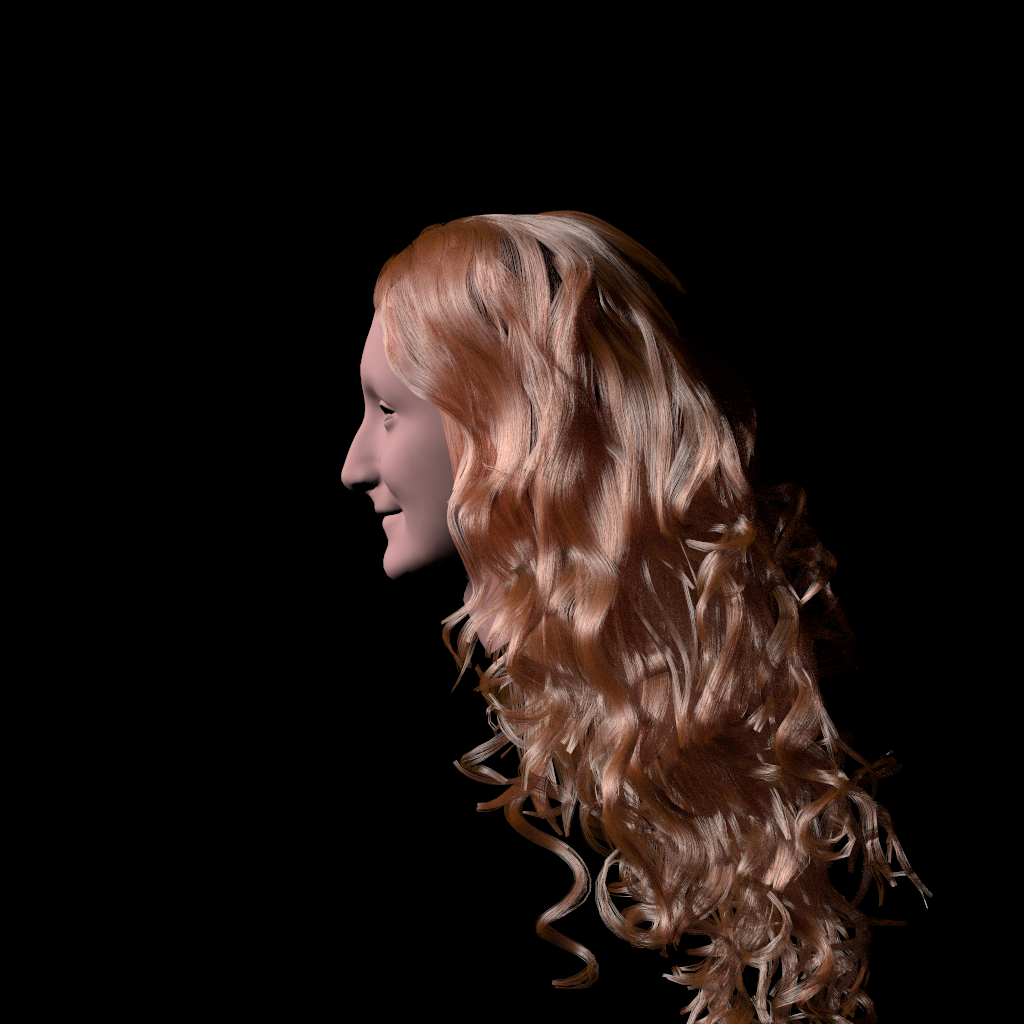}\hfill%
\includegraphics[trim={180 260 180 100},clip,width=0.11\linewidth]{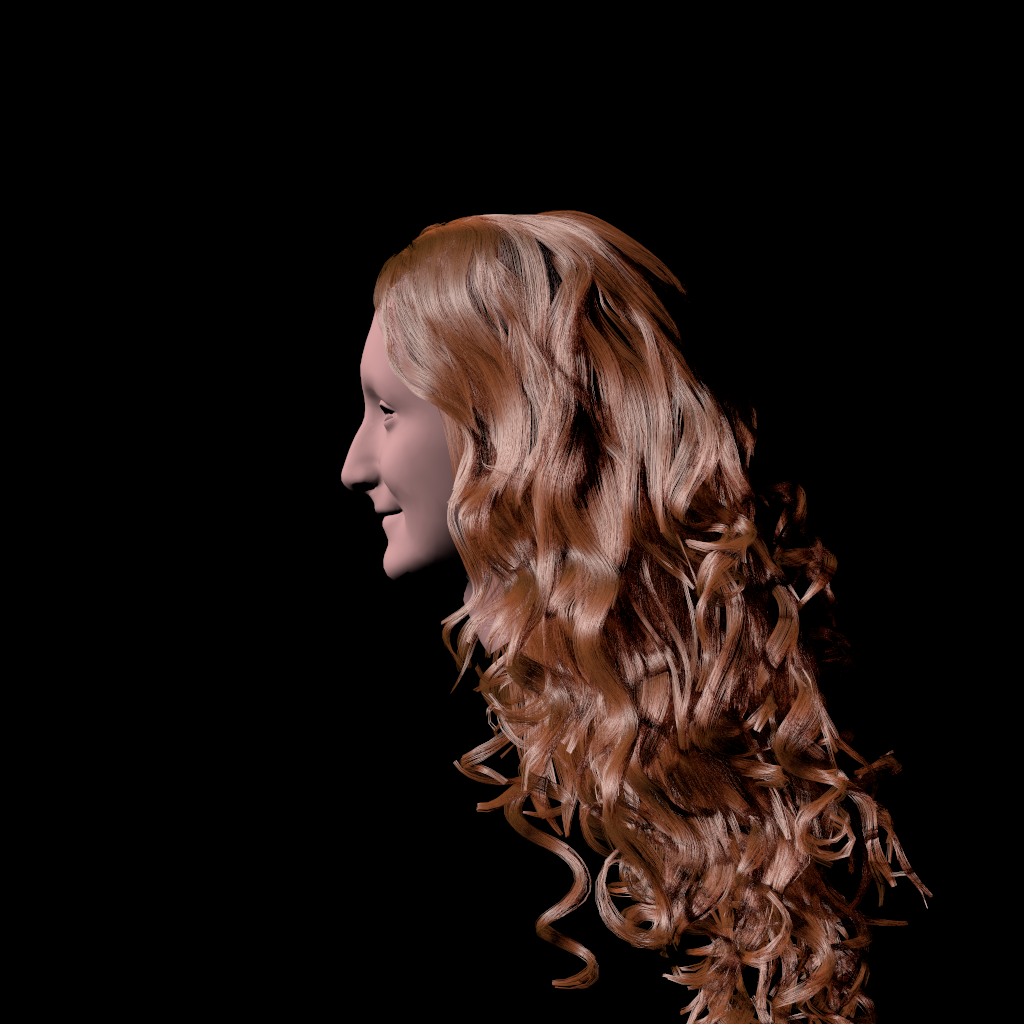}\hfill%
\includegraphics[trim={180 260 180 100},clip,width=0.11\linewidth]{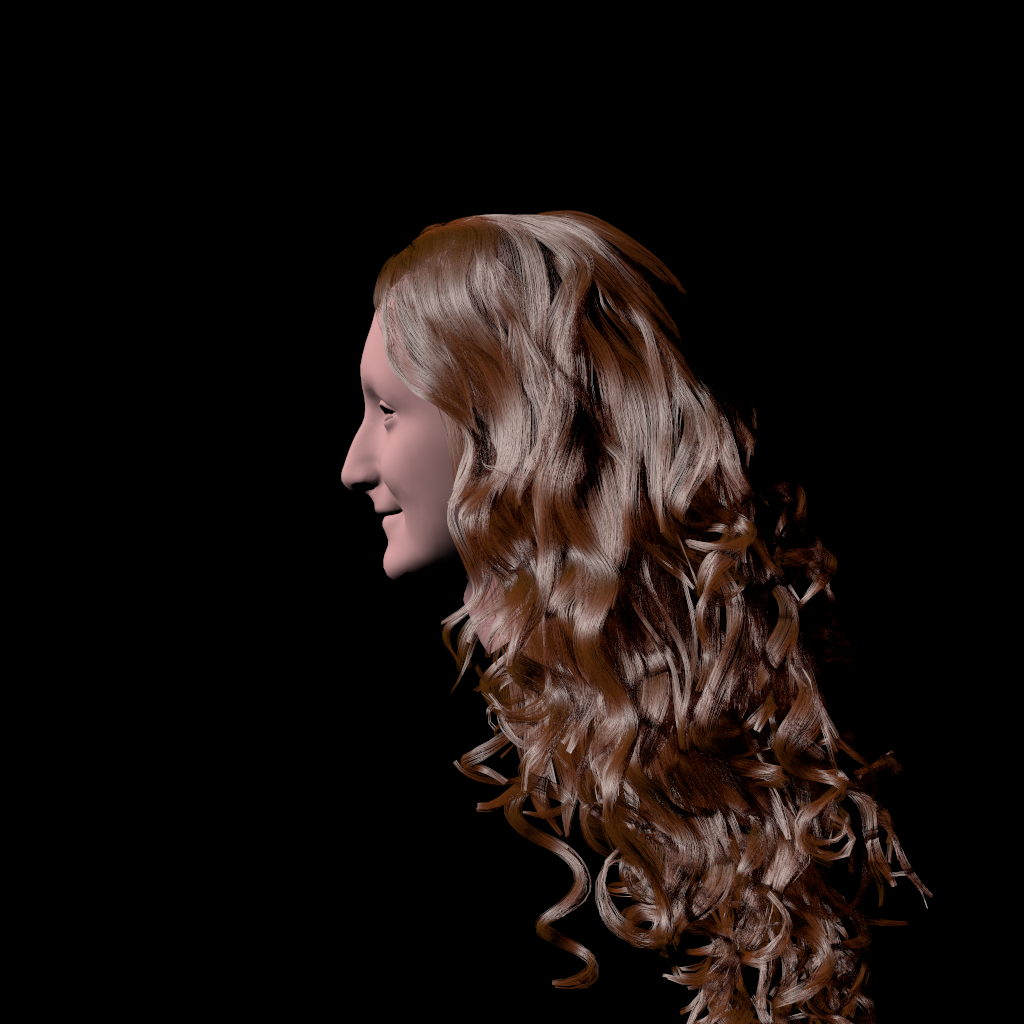}\hfill%
\includegraphics[trim={350 390 350 310},clip,width=0.11\linewidth]{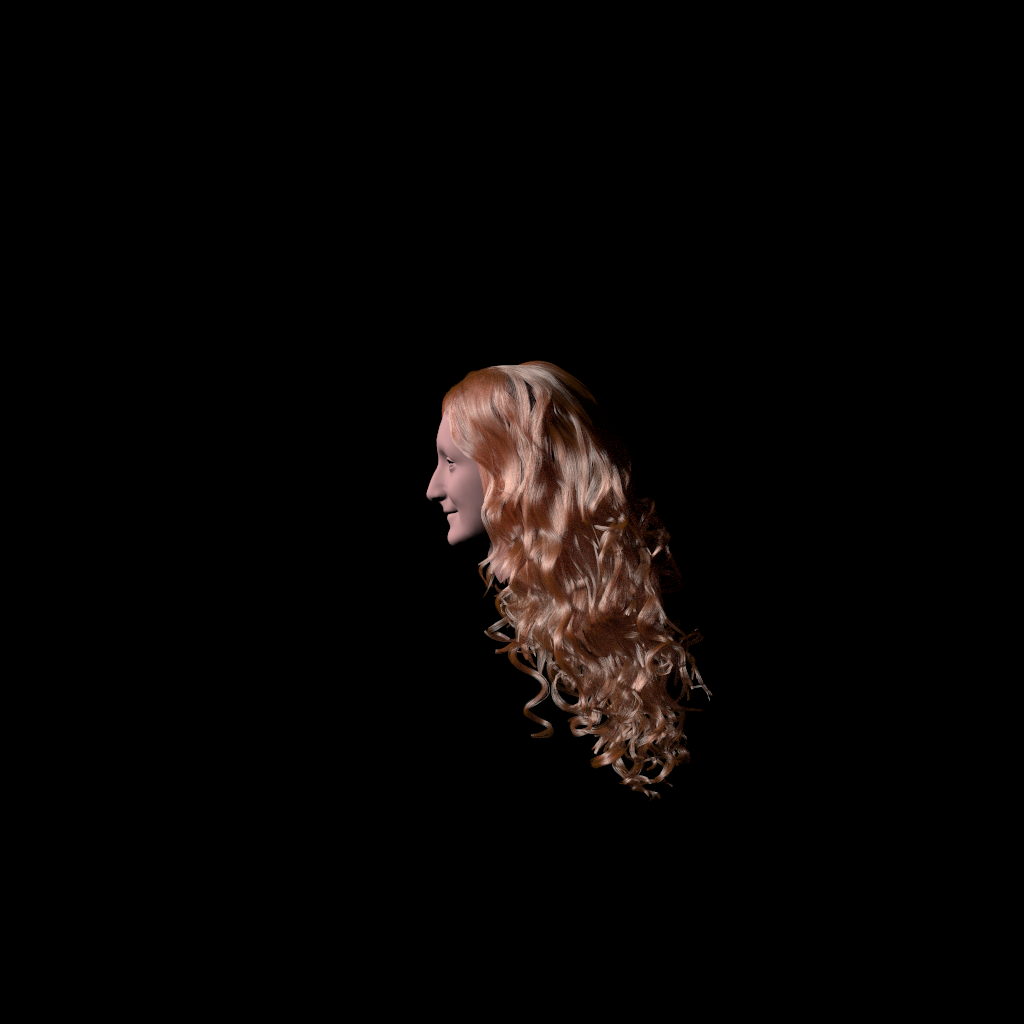}\hfill%
\includegraphics[trim={350 390 350 310},clip,width=0.11\linewidth]{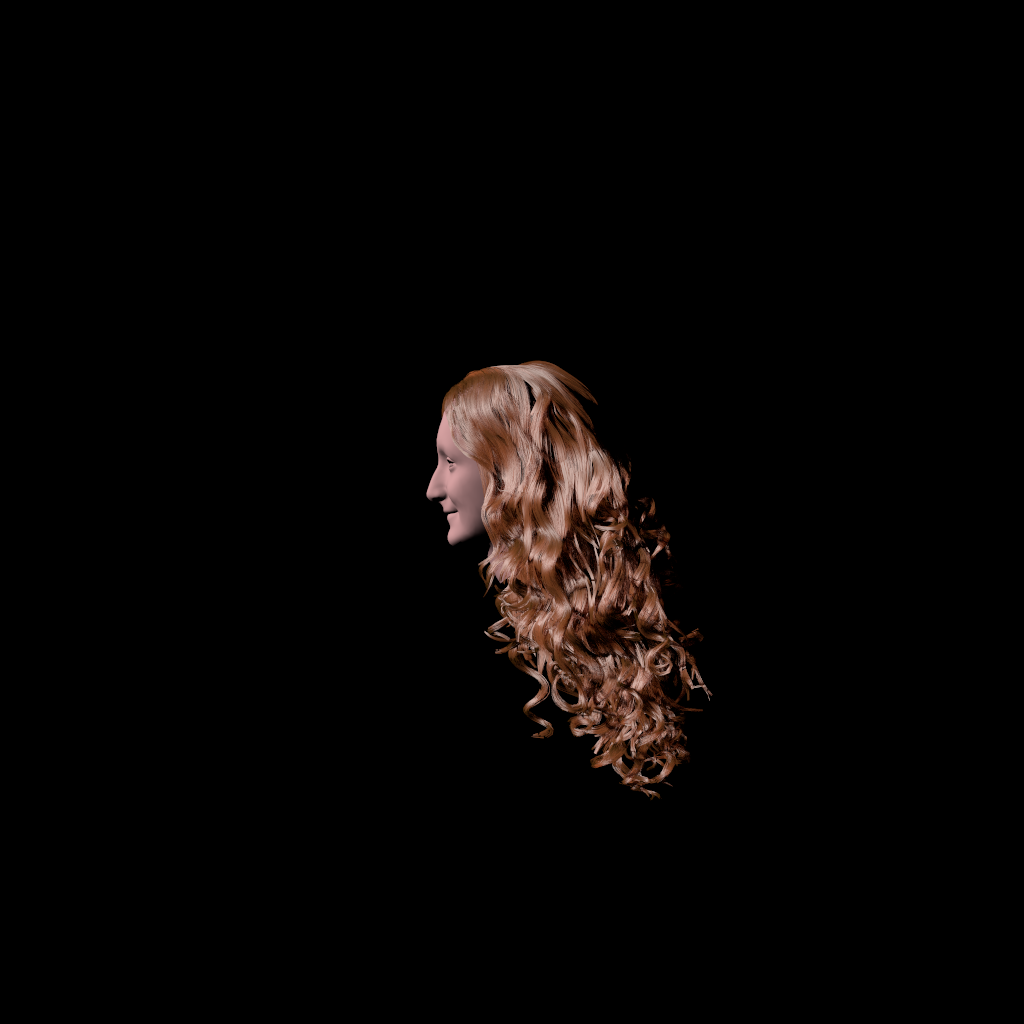}\hfill%
\includegraphics[trim={350 390 350 310},clip,width=0.11\linewidth]{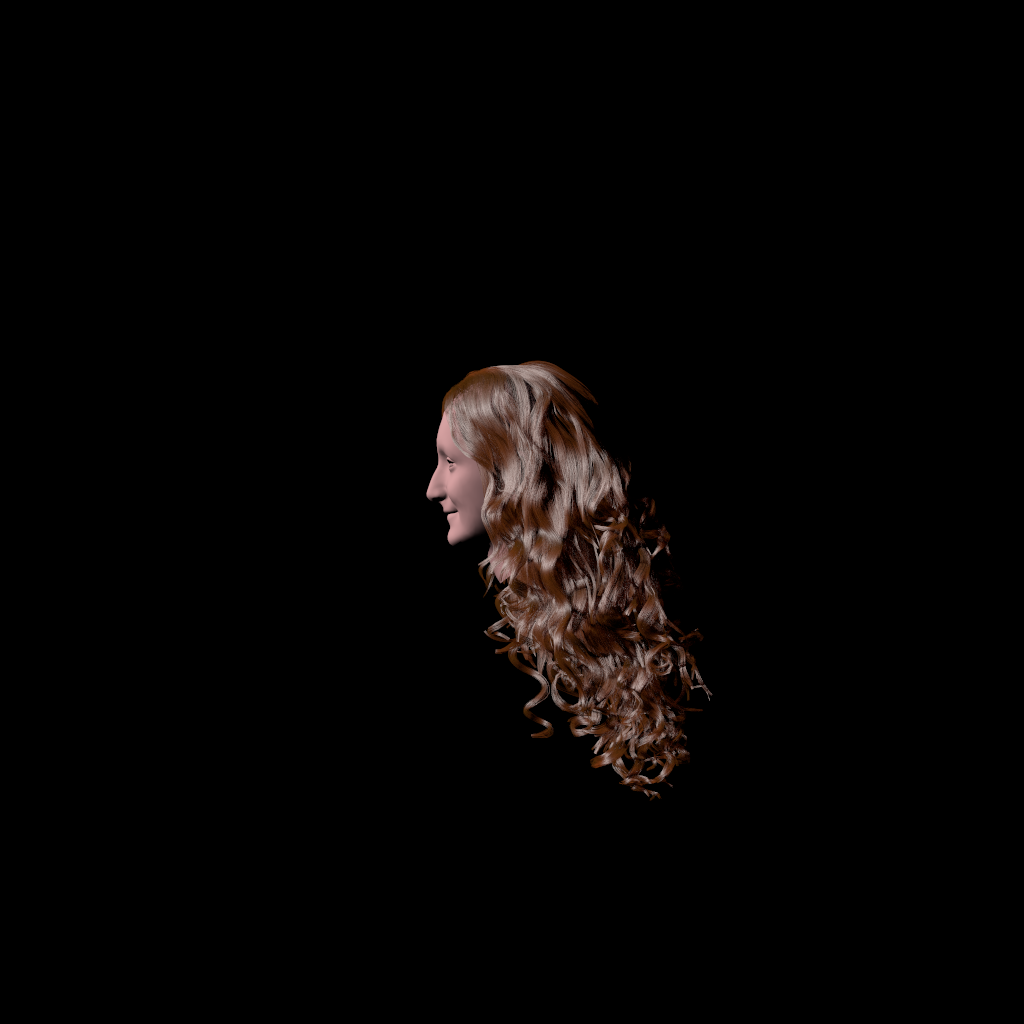}\hfill%
\includegraphics[trim={450 470 450 430},clip,width=0.11\linewidth]{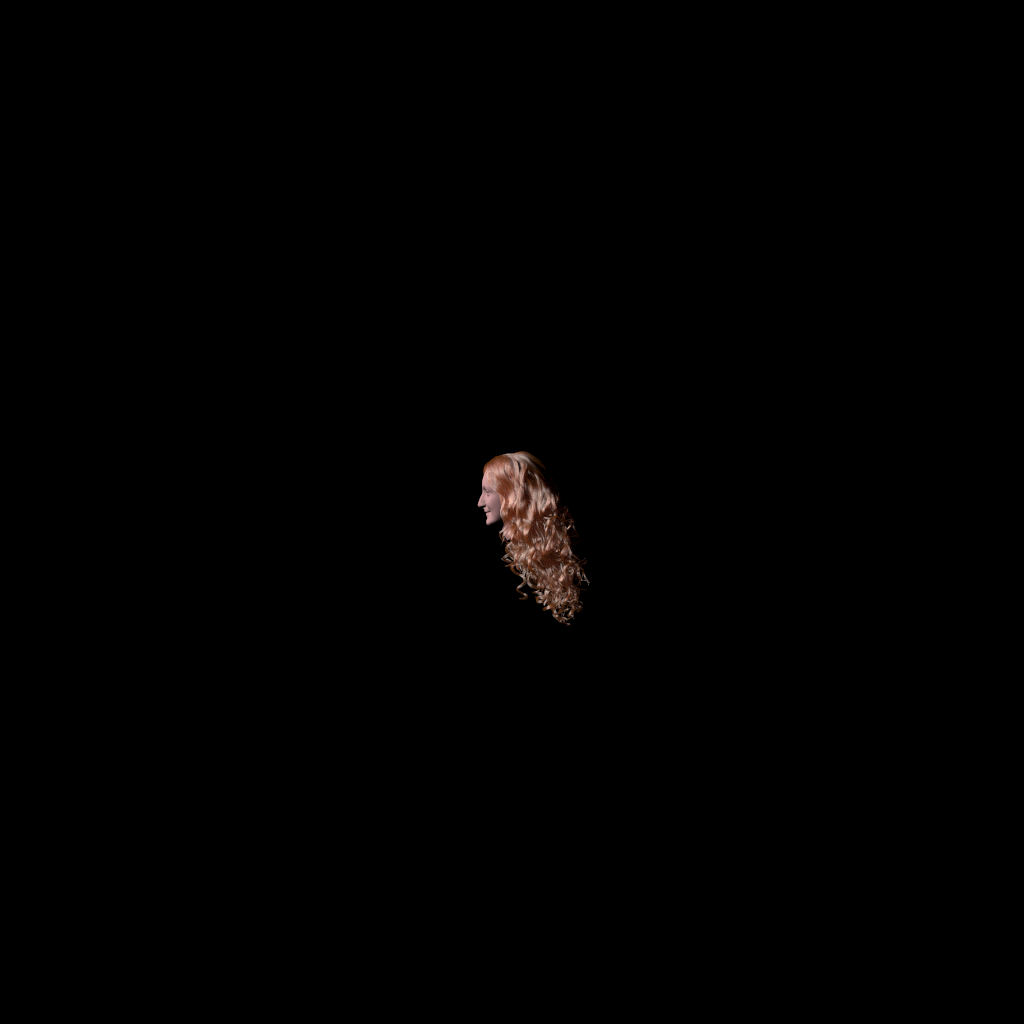}\hfill%
\includegraphics[trim={450 470 450 430},clip,width=0.11\linewidth]{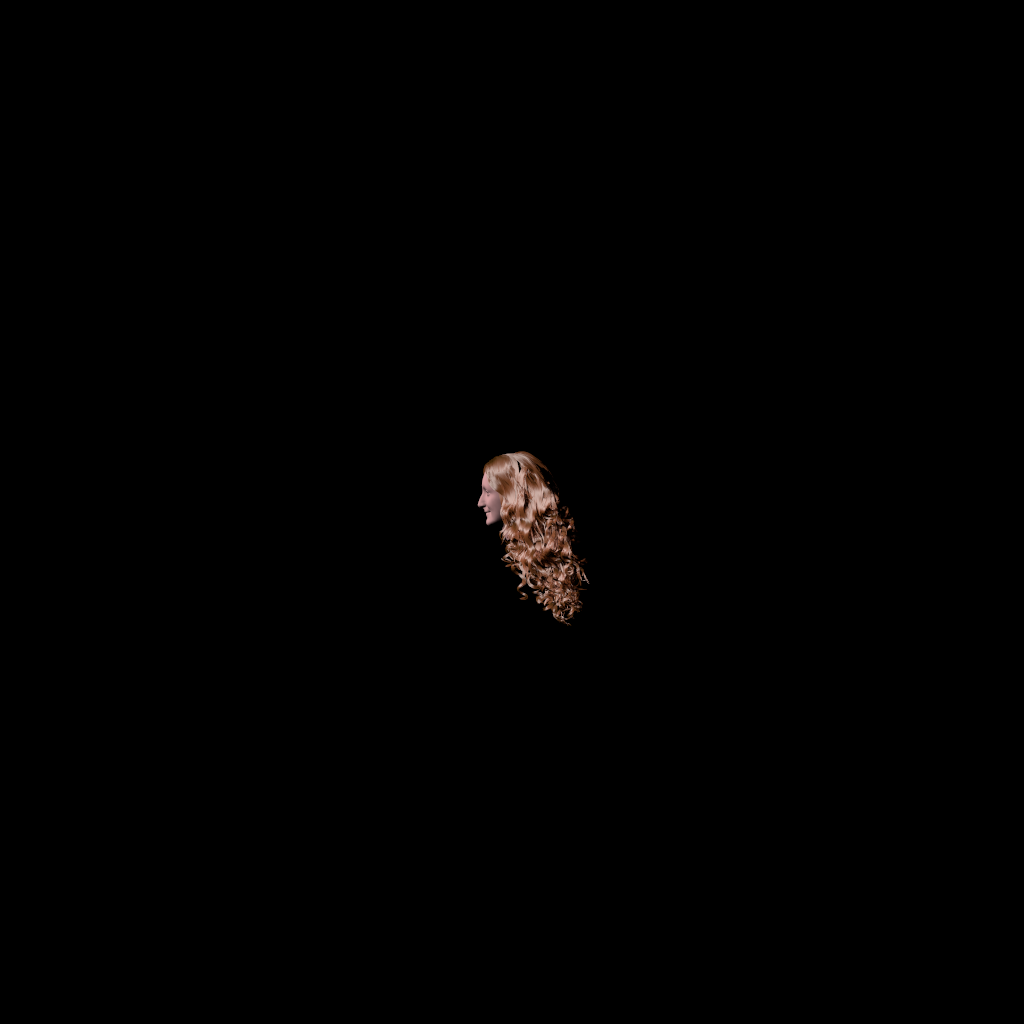}\hfill%
\includegraphics[trim={450 470 450 430},clip,width=0.11\linewidth]{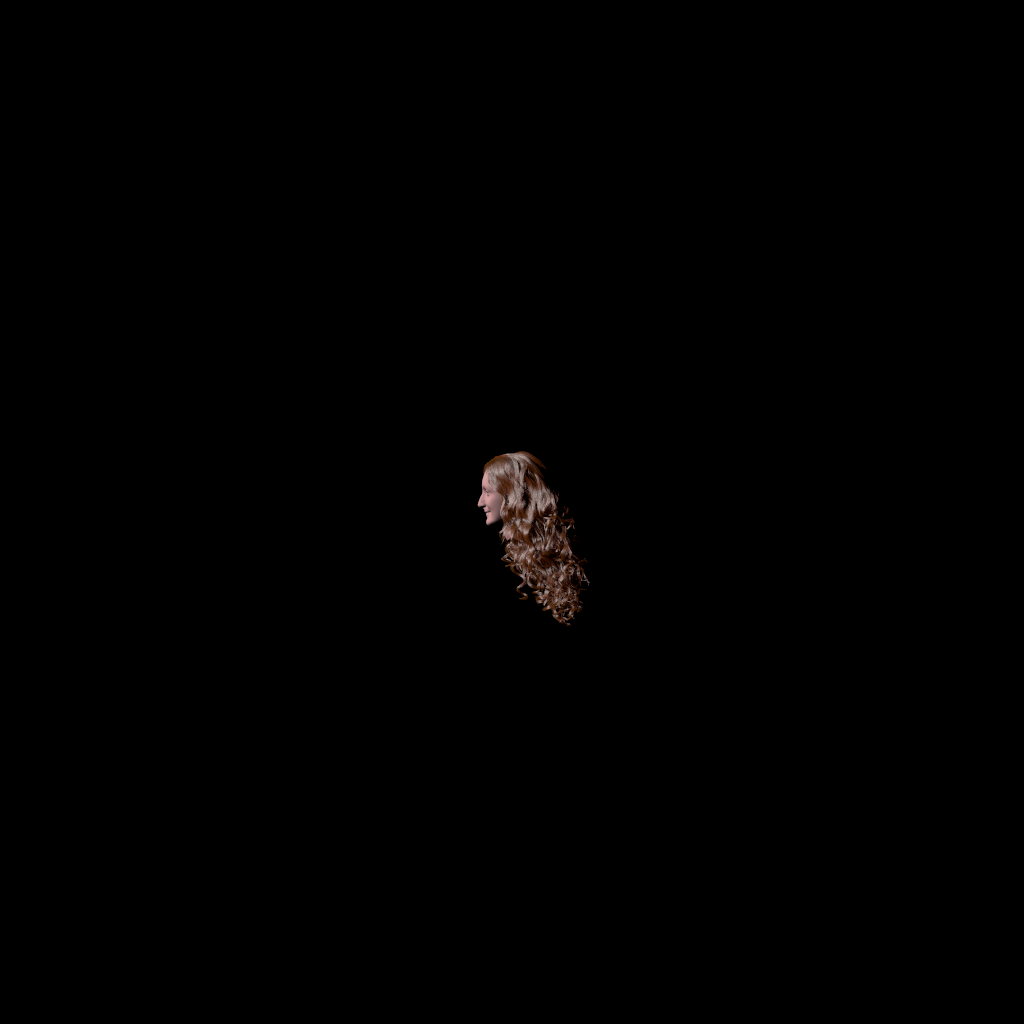}\\
\includegraphics[trim={0 0 0 0},clip,width=0.11\linewidth]{FIG/FIG_SIGA/main_result/hadley/gt/0.png}\hfill%
\includegraphics[trim={180 260 180 100},clip,width=0.11\linewidth]{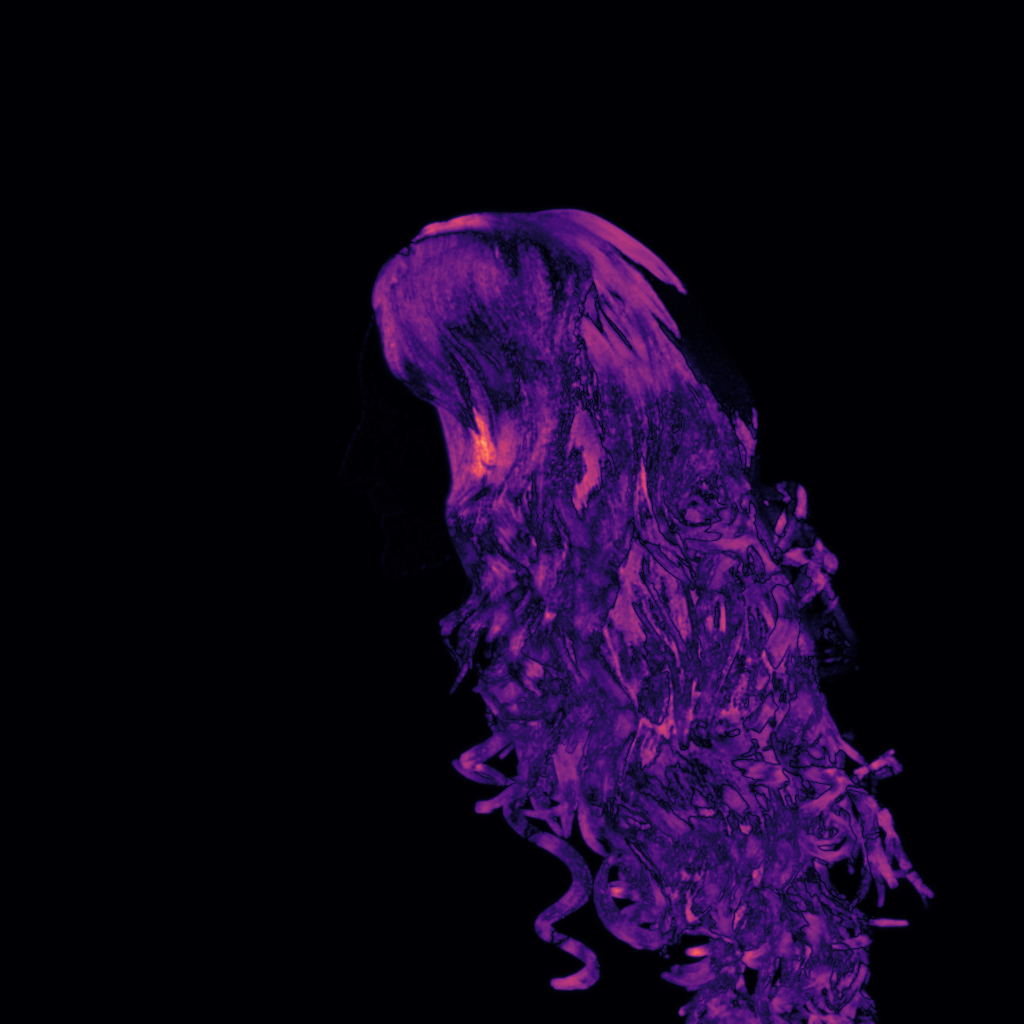}\hfill%
\includegraphics[trim={180 260 180 100},clip,width=0.11\linewidth]{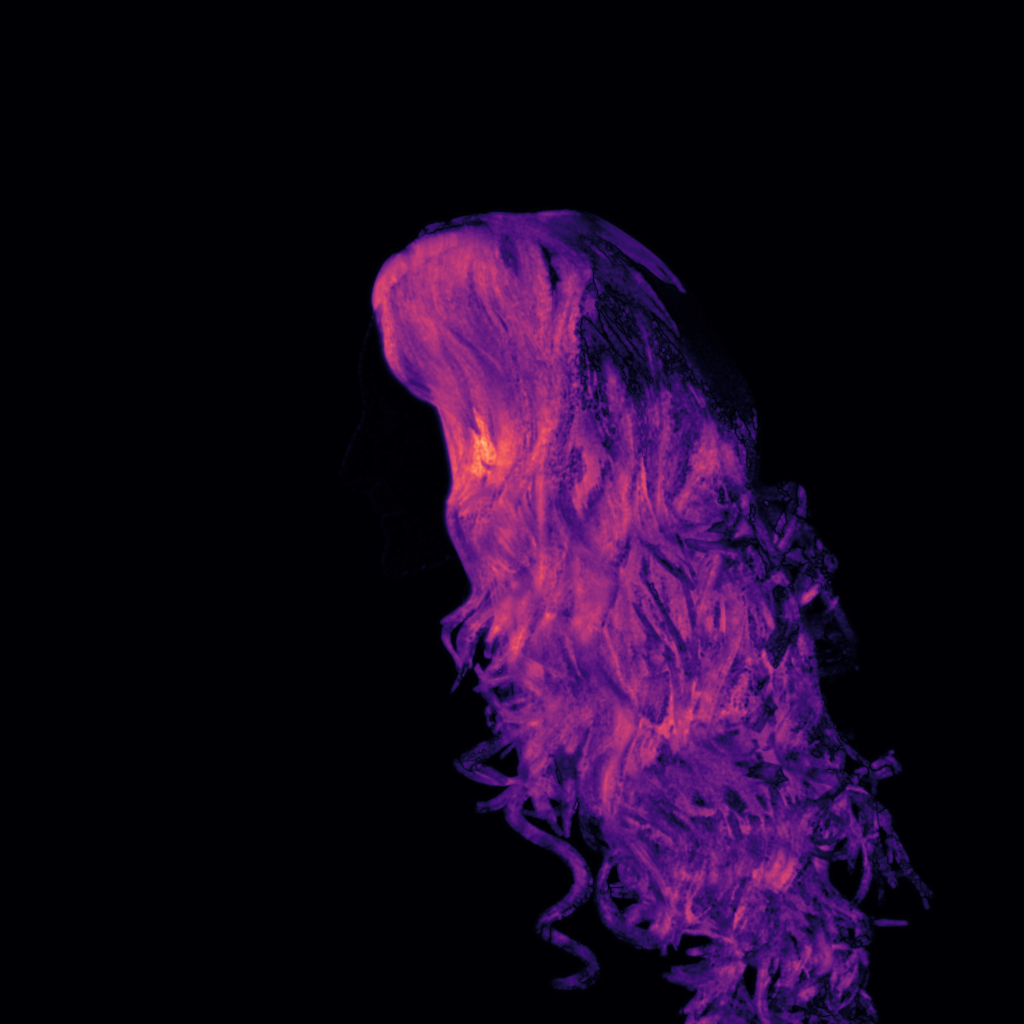}\hfill%
\includegraphics[trim={0 0 0 0},clip,width=0.11\linewidth]{FIG/FIG_SIGA/main_result/hadley/gt/0.png}\hfill%
\includegraphics[trim={350 390 350 310},clip,width=0.11\linewidth]{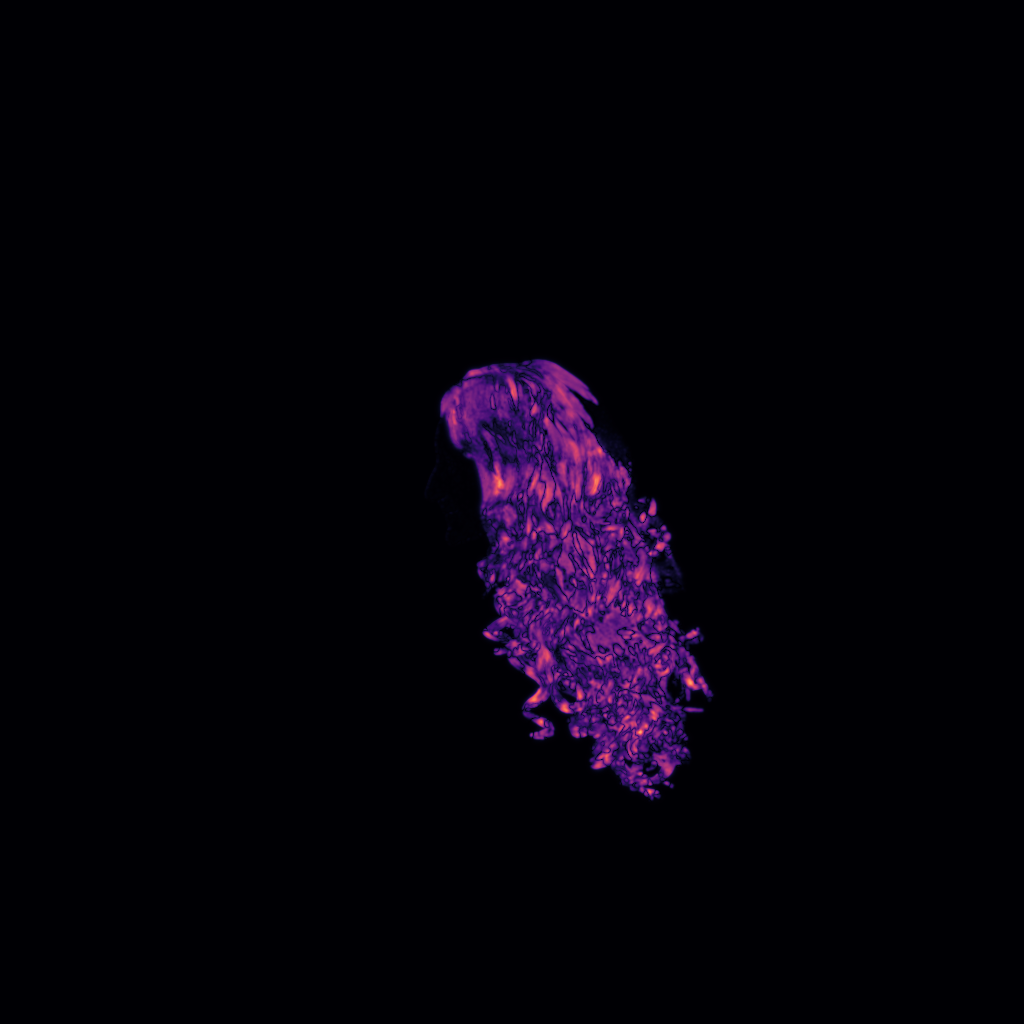}\hfill%
\includegraphics[trim={350 390 350 310},clip,width=0.11\linewidth]{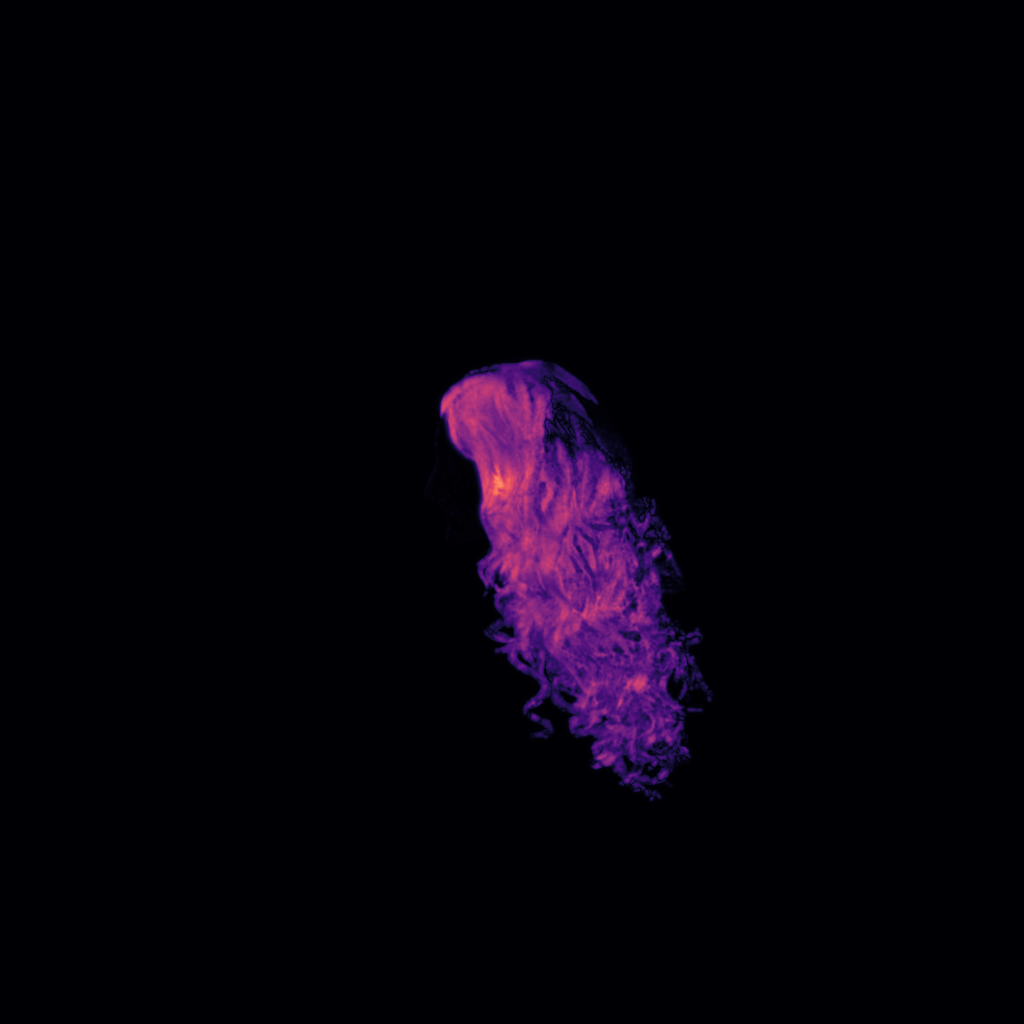}\hfill%
\includegraphics[trim={0 0 0 0},clip,width=0.11\linewidth]{FIG/FIG_SIGA/main_result/hadley/gt/0.png}\hfill%
\includegraphics[trim={450 470 450 430},clip,width=0.11\linewidth]{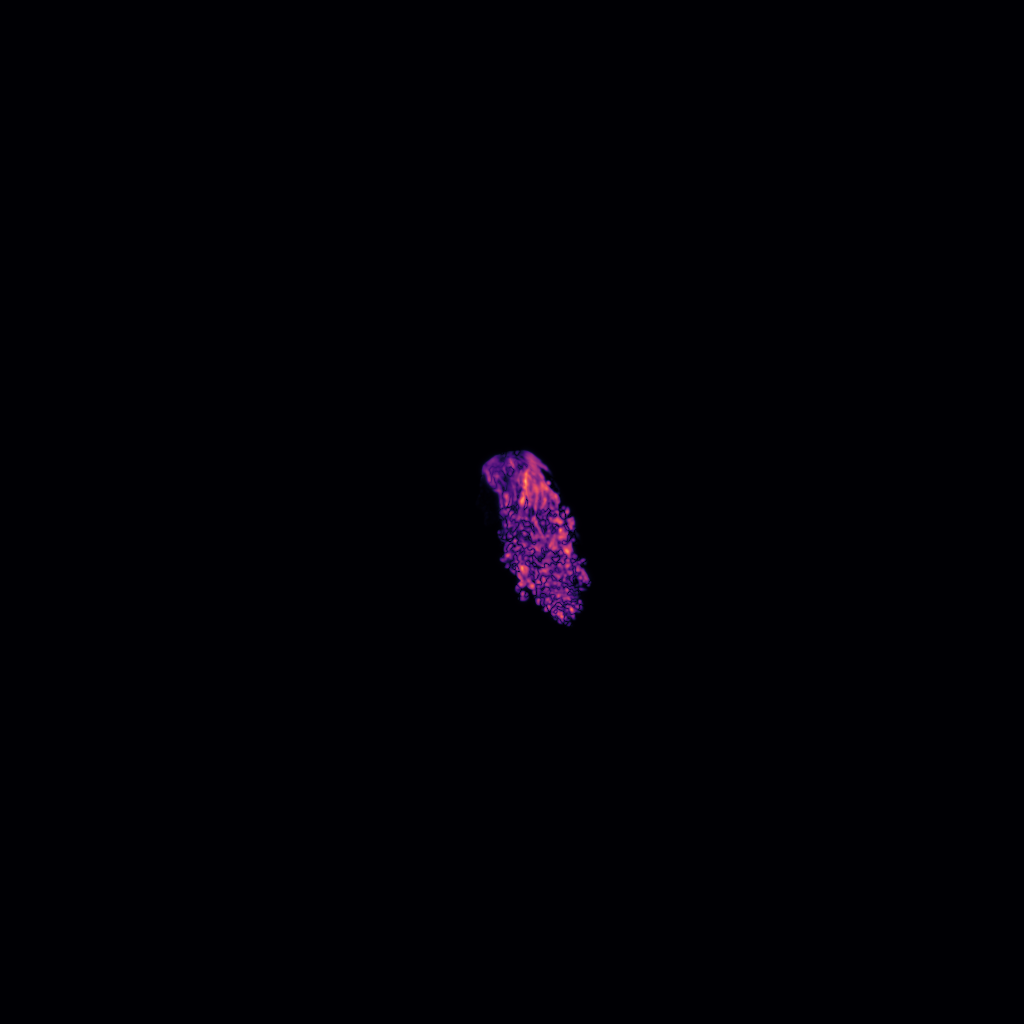}\hfill%
\includegraphics[trim={450 470 450 430},clip,width=0.11\linewidth]{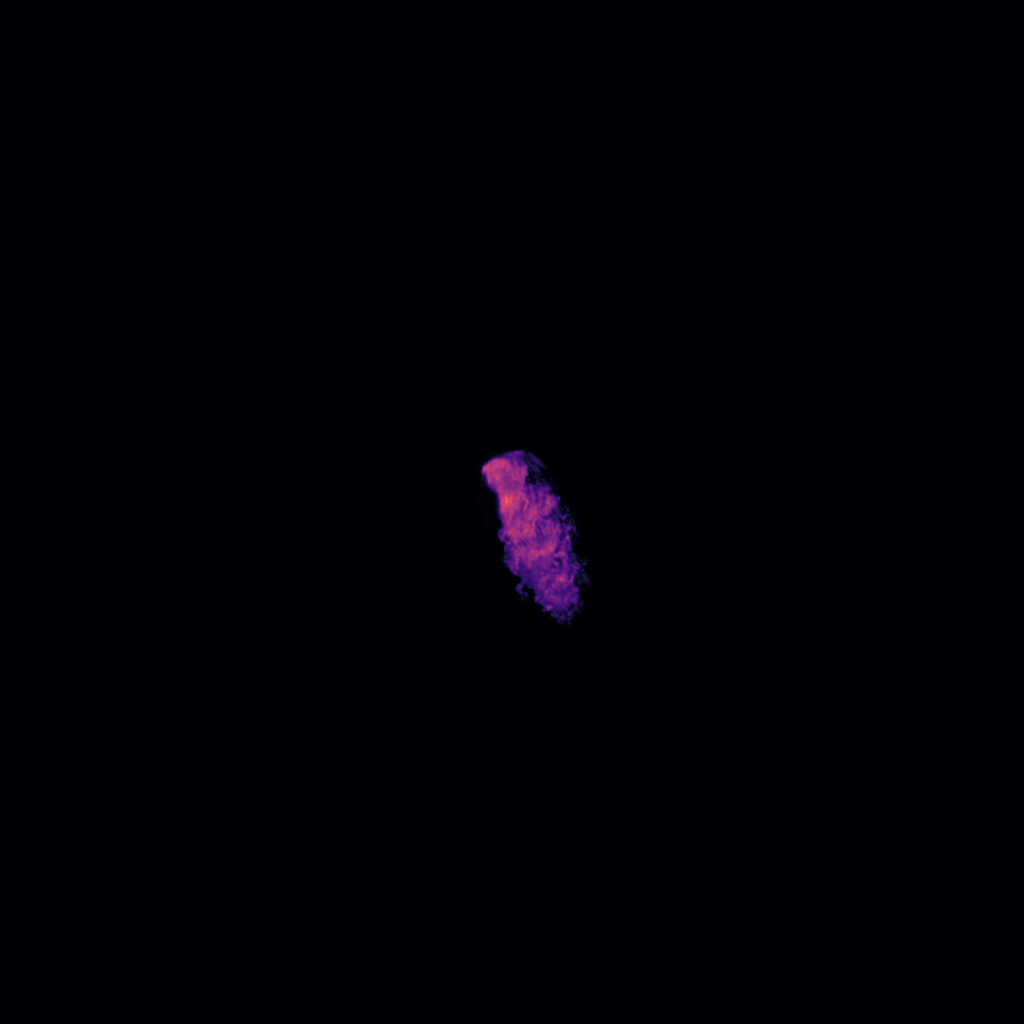}\\

\includegraphics[trim={0 0 0 0},clip,width=0.11\linewidth]{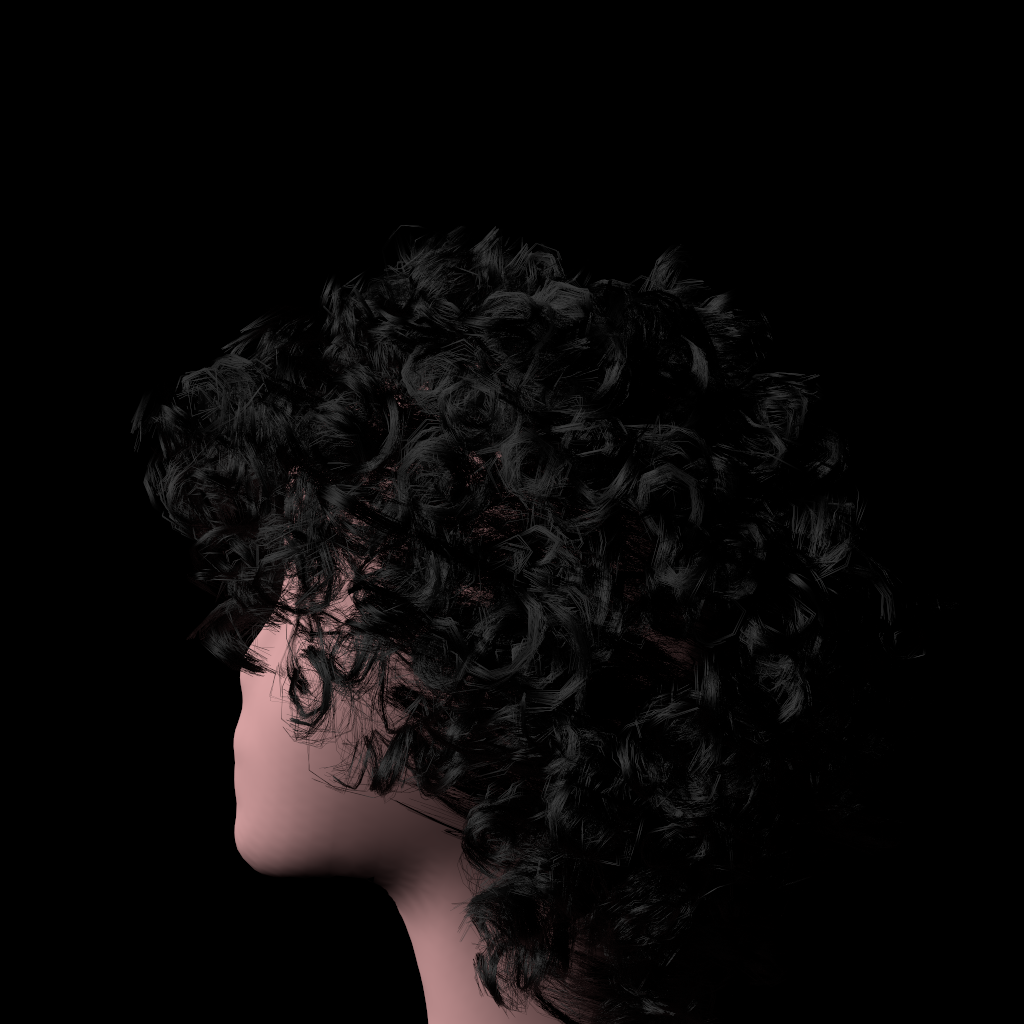}\hfill%
\includegraphics[trim={0 0 0 0},clip,width=0.11\linewidth]{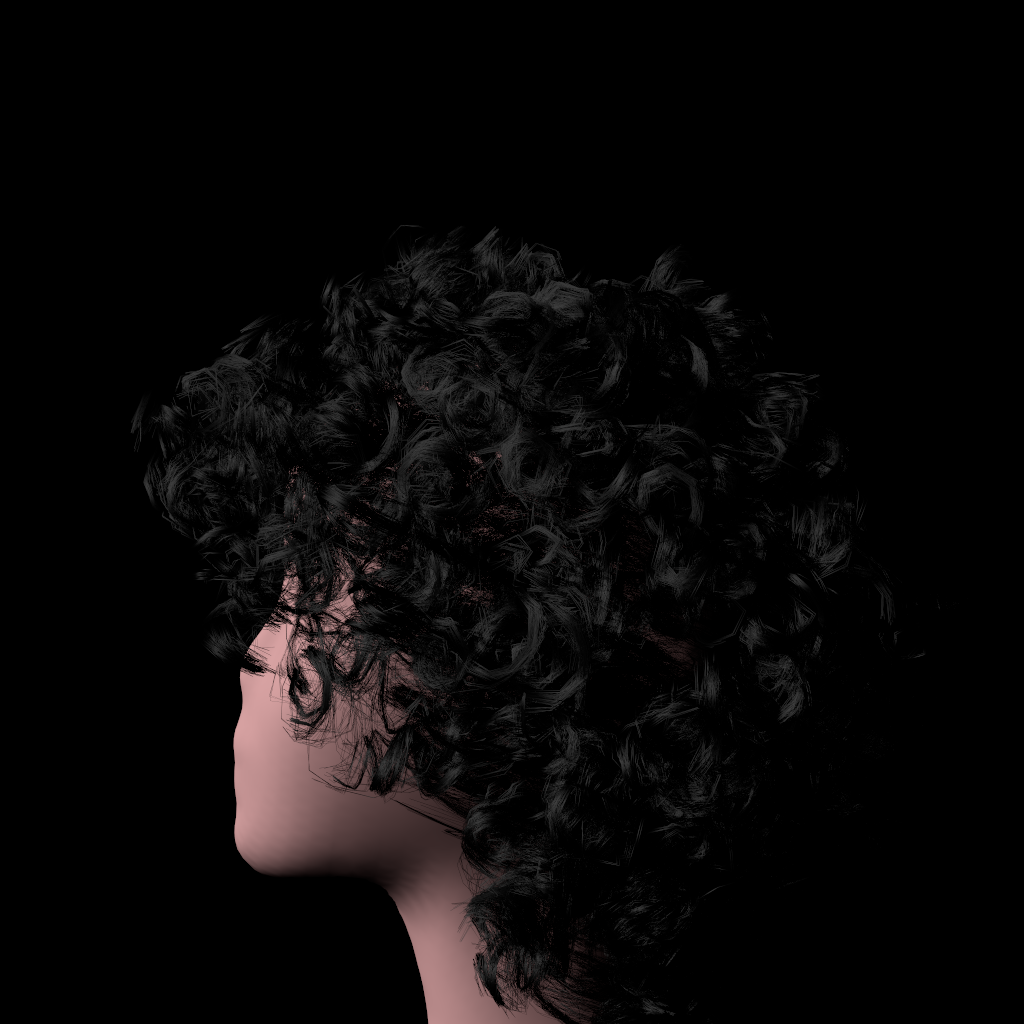}\hfill%
\includegraphics[trim={0 0 0 0},clip,width=0.11\linewidth]{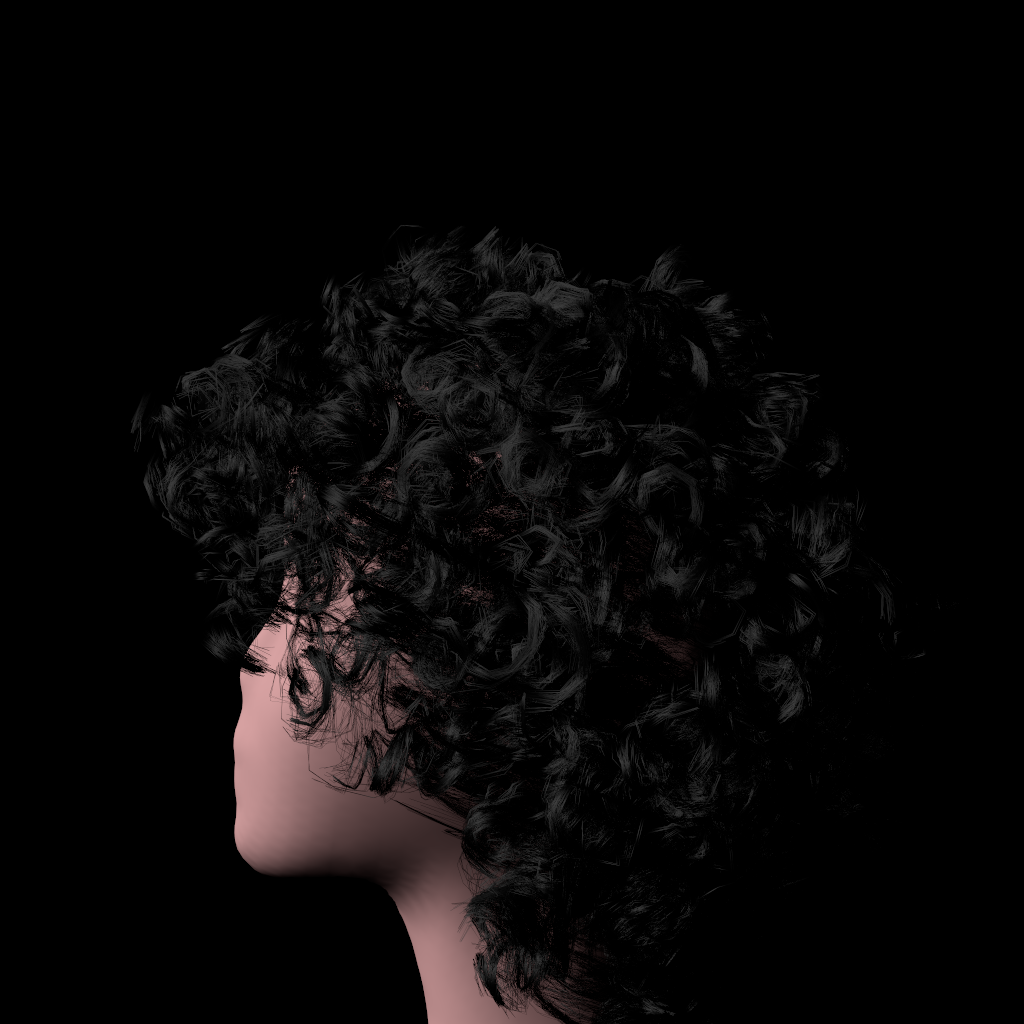}\hfill%
\includegraphics[trim={350 350 350 350},clip,width=0.11\linewidth]{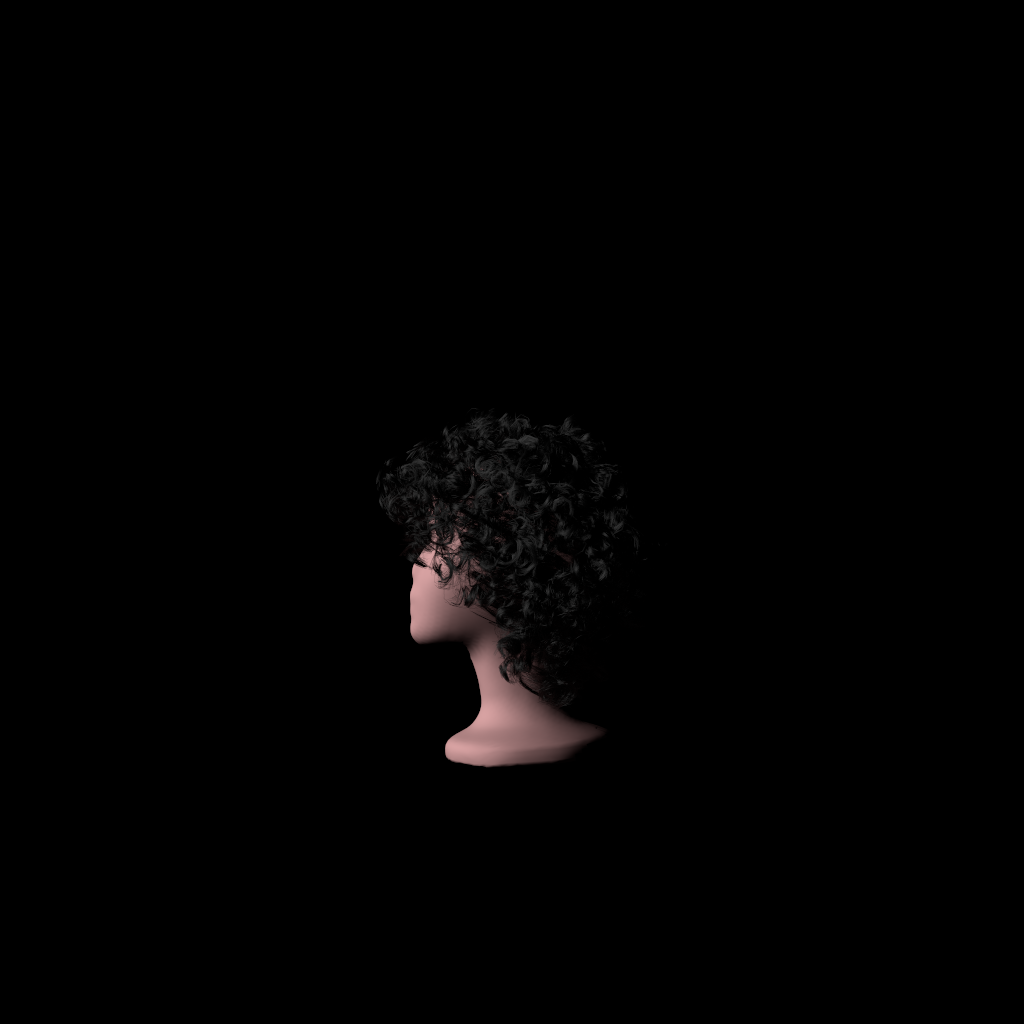}\hfill%
\includegraphics[trim={350 350 350 350},clip,width=0.11\linewidth]{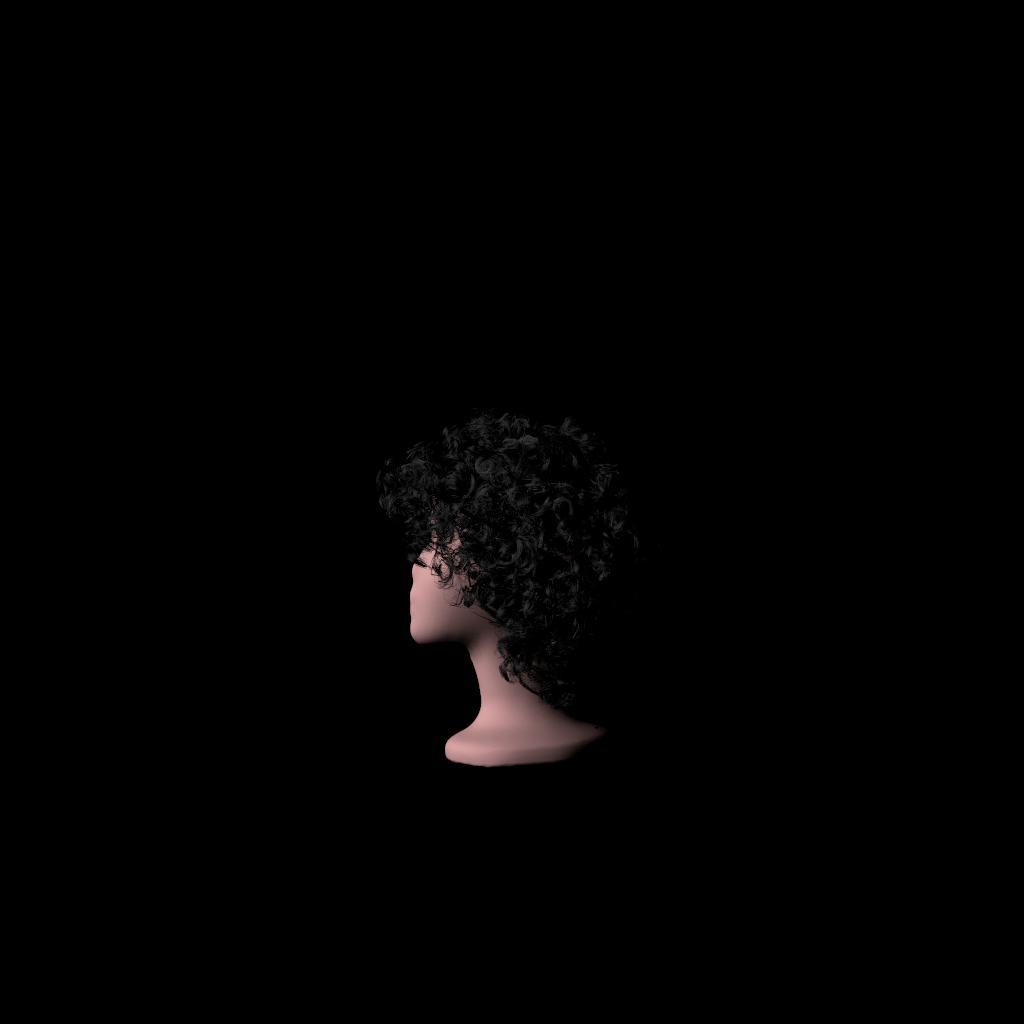}\hfill%
\includegraphics[trim={350 350 350 350},clip,width=0.11\linewidth]{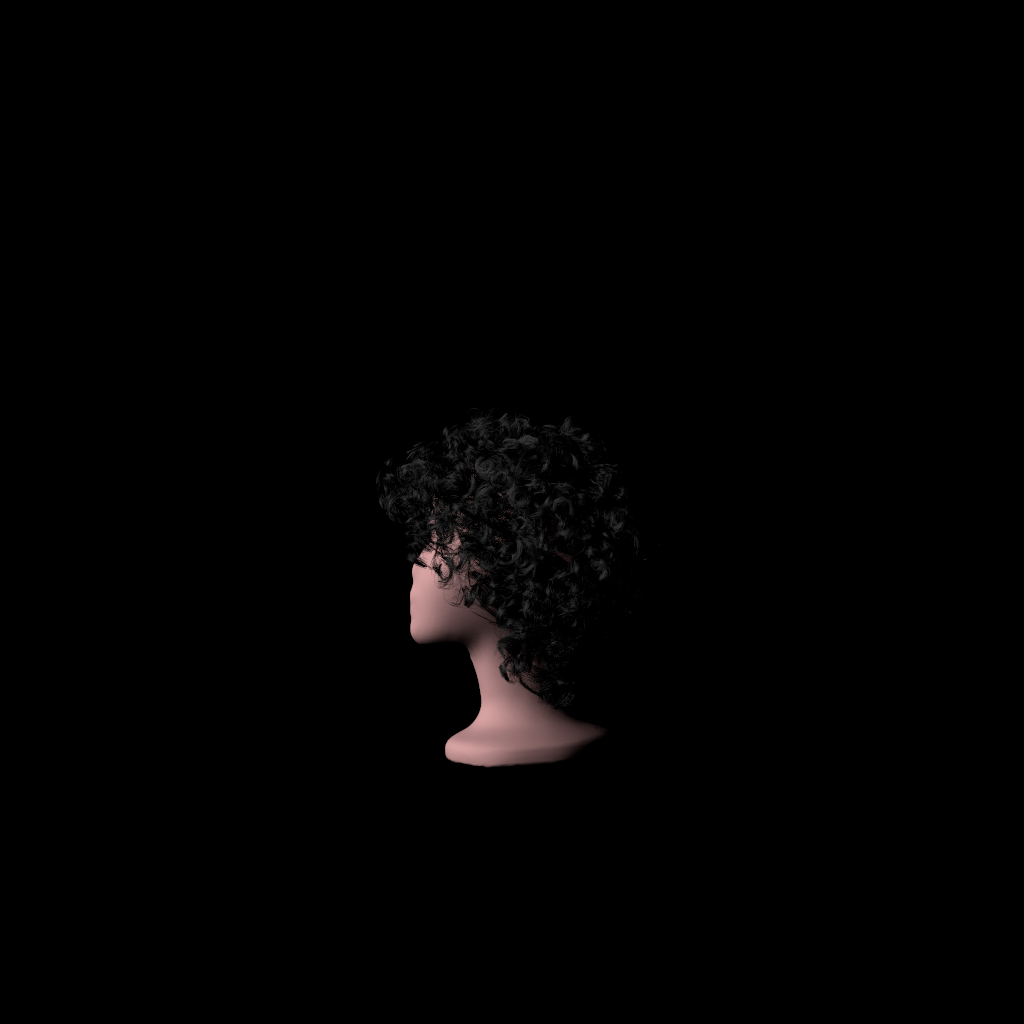}\hfill%
\includegraphics[trim={450 450 450 450},clip,width=0.11\linewidth]{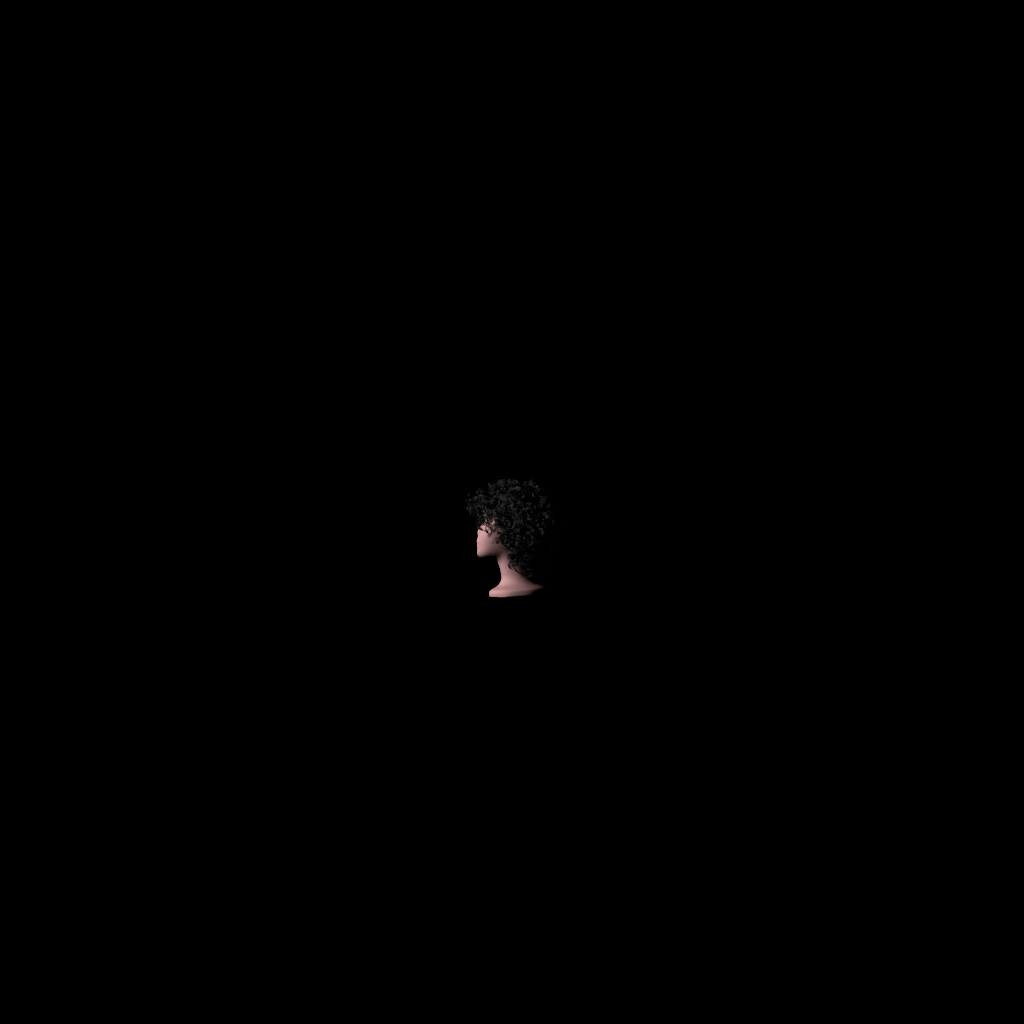}\hfill%
\includegraphics[trim={450 450 450 450},clip,width=0.11\linewidth]{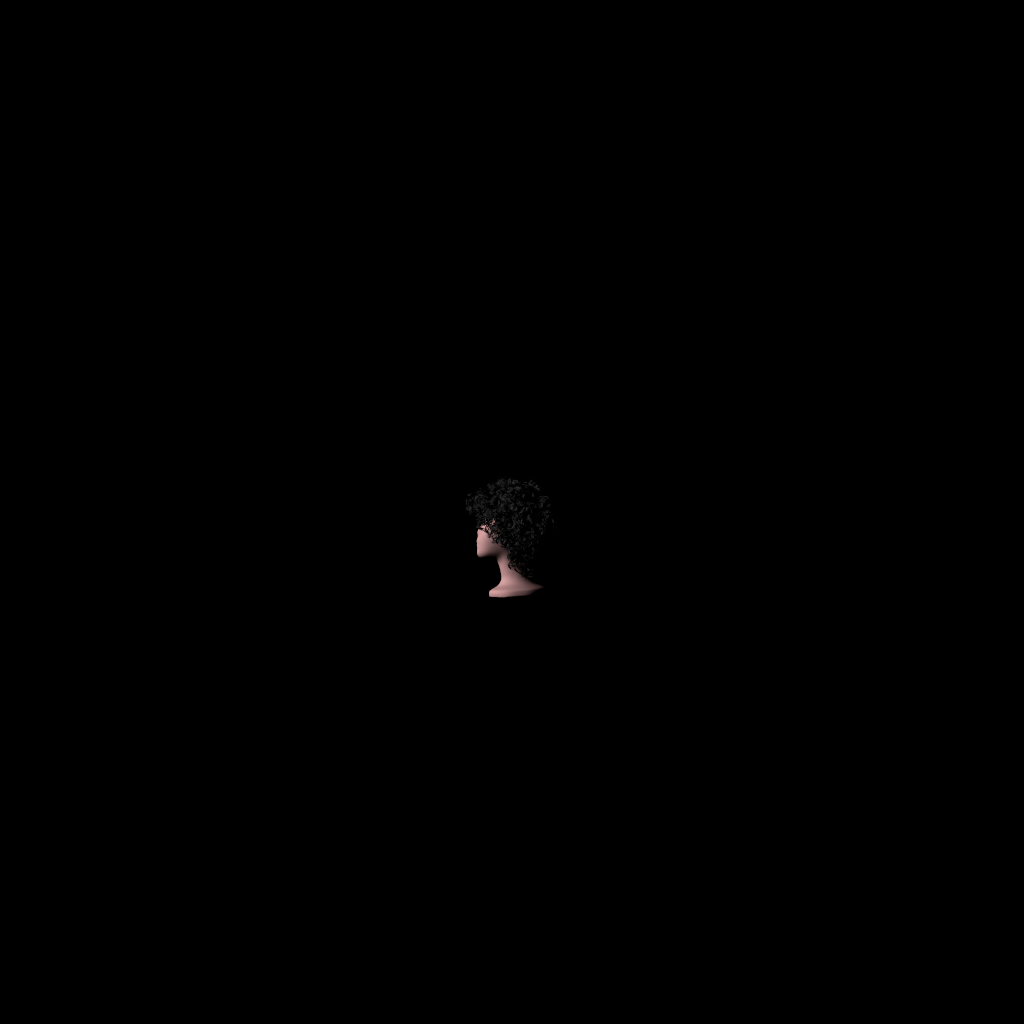}\hfill%
\includegraphics[trim={450 450 450 450},clip,width=0.11\linewidth]{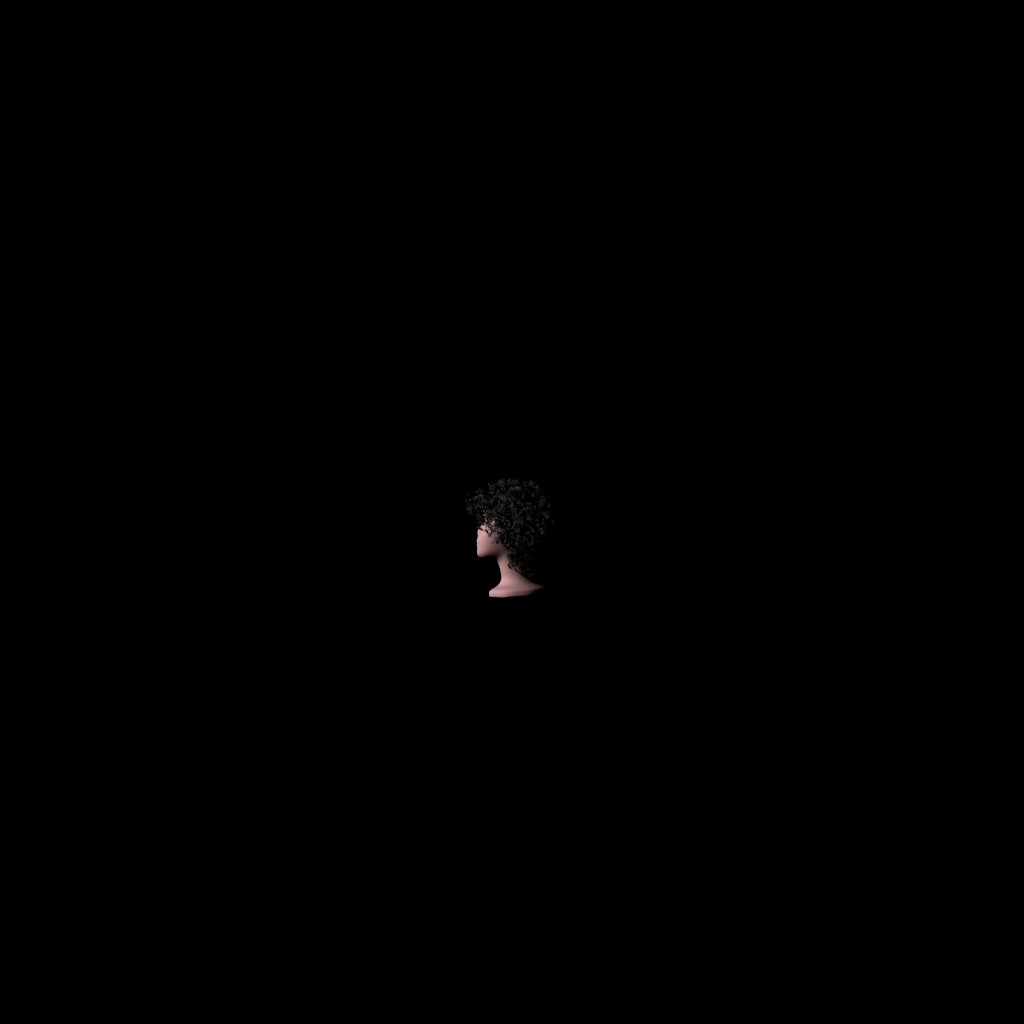}\\
\includegraphics[trim={0 0 0 0},clip,width=0.11\linewidth]{FIG/FIG_SIGA/main_result/hadley/gt/0.png}\hfill%
\includegraphics[trim={0 0 0 0},clip,width=0.11\linewidth]{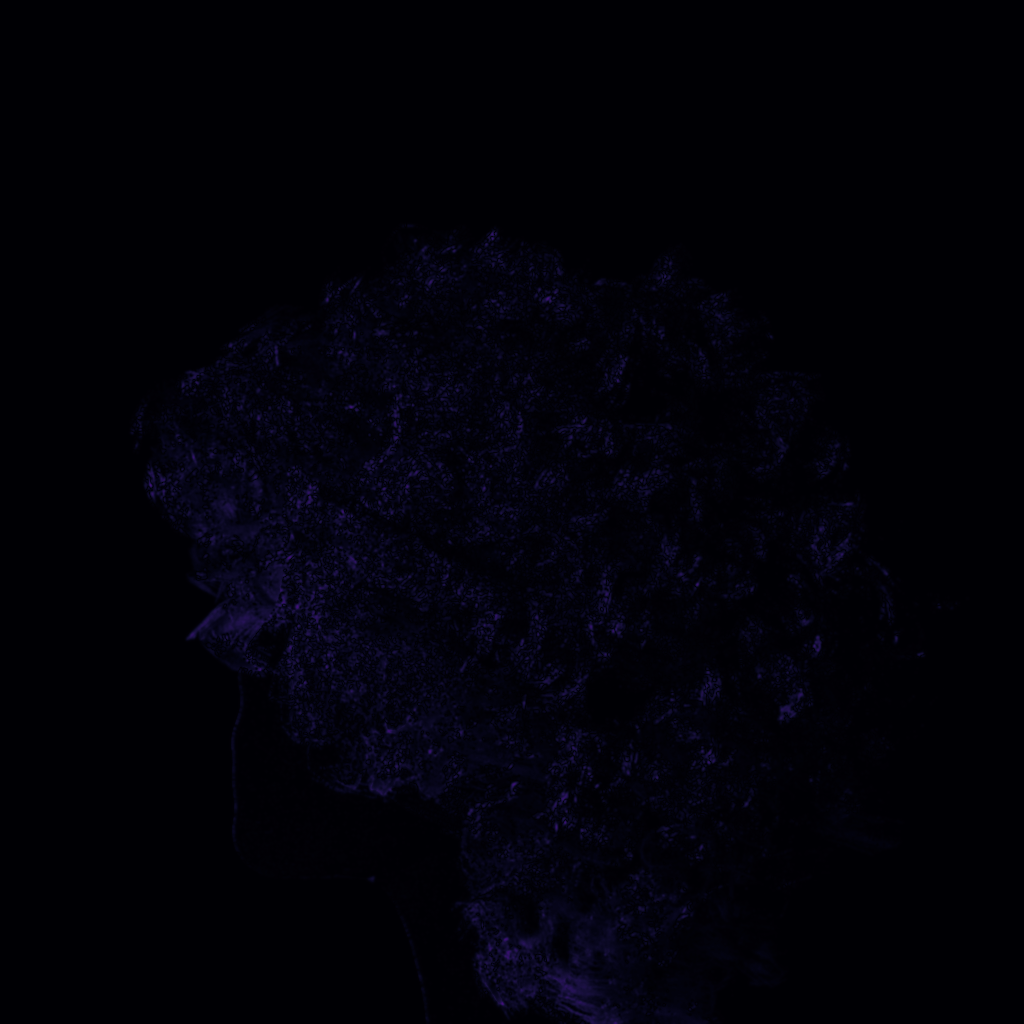}\hfill%
\includegraphics[trim={0 0 0 0},clip,width=0.11\linewidth]{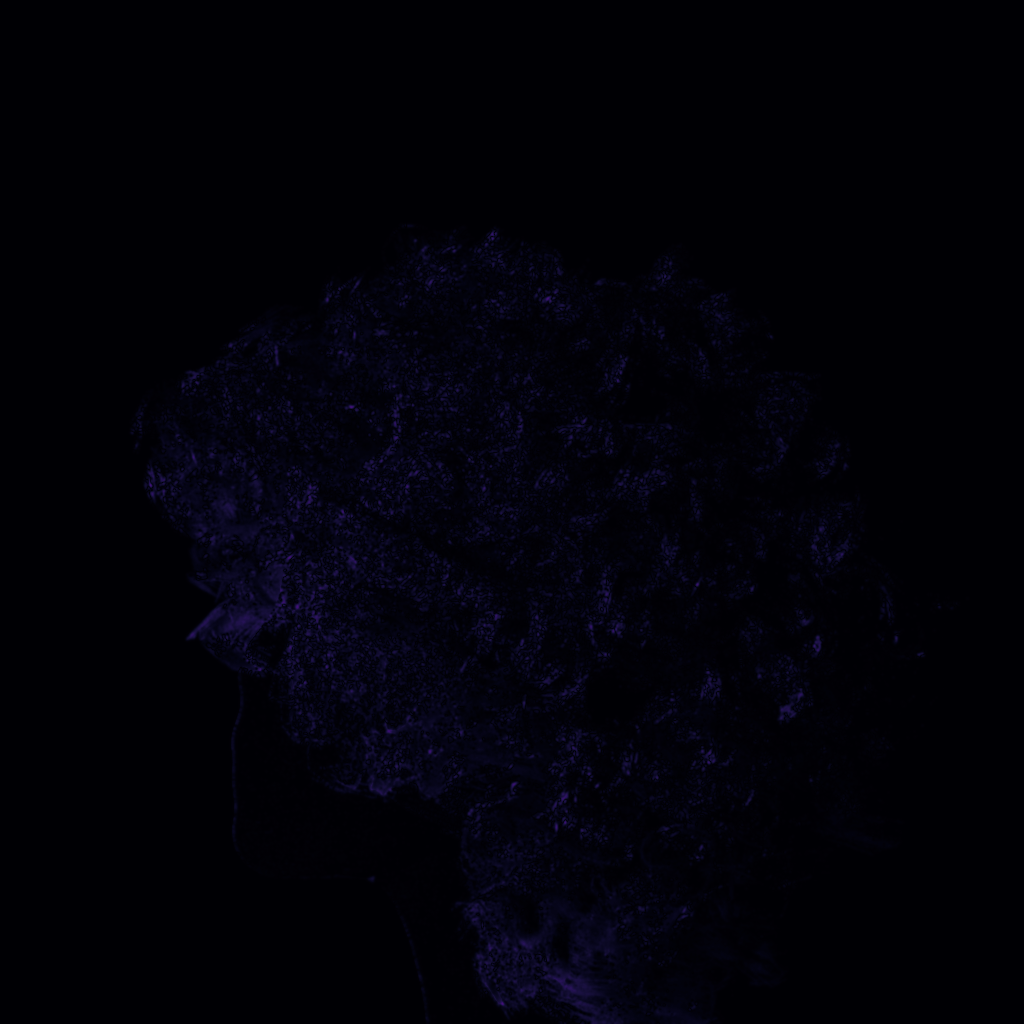}\hfill%
\includegraphics[trim={0 0 0 0},clip,width=0.11\linewidth]{FIG/FIG_SIGA/main_result/hadley/gt/0.png}\hfill%
\includegraphics[trim={350 350 350 350},clip,width=0.11\linewidth]{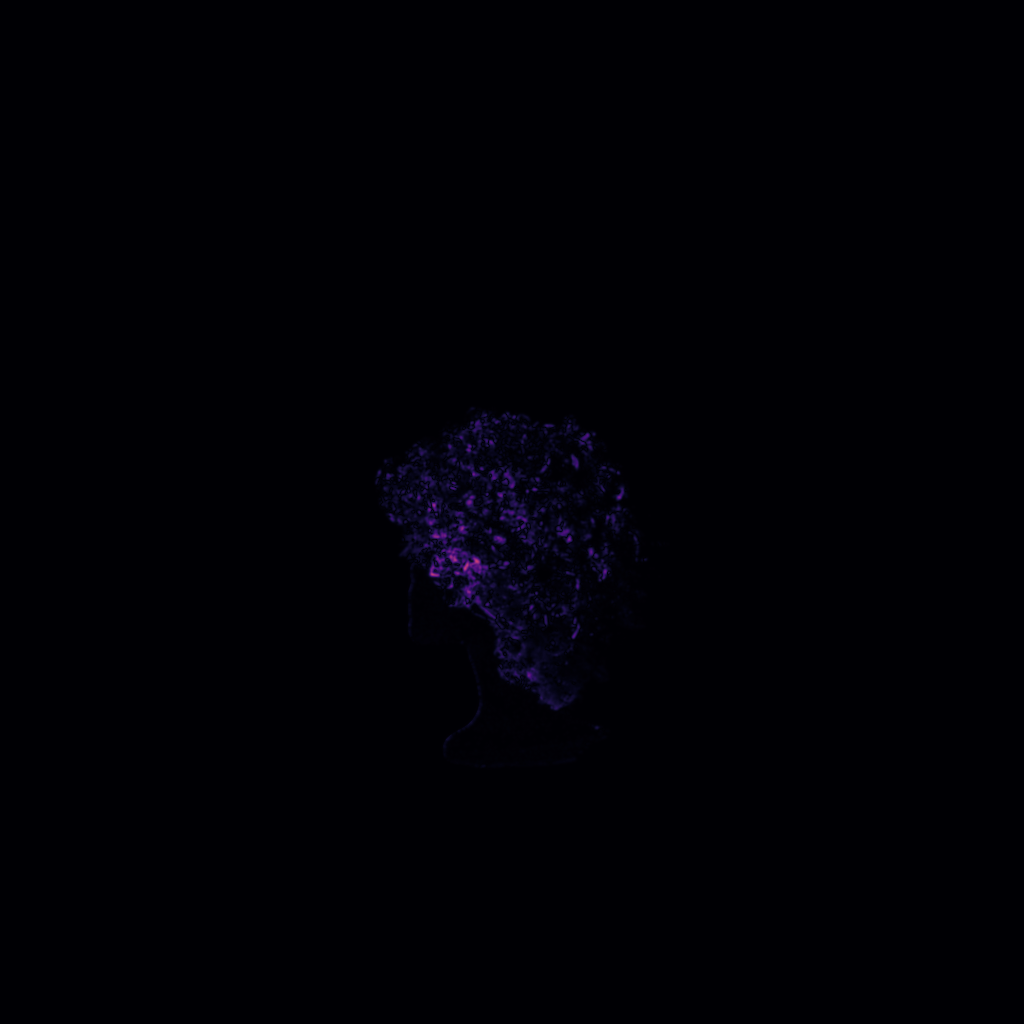}\hfill%
\includegraphics[trim={350 350 350 350},clip,width=0.11\linewidth]{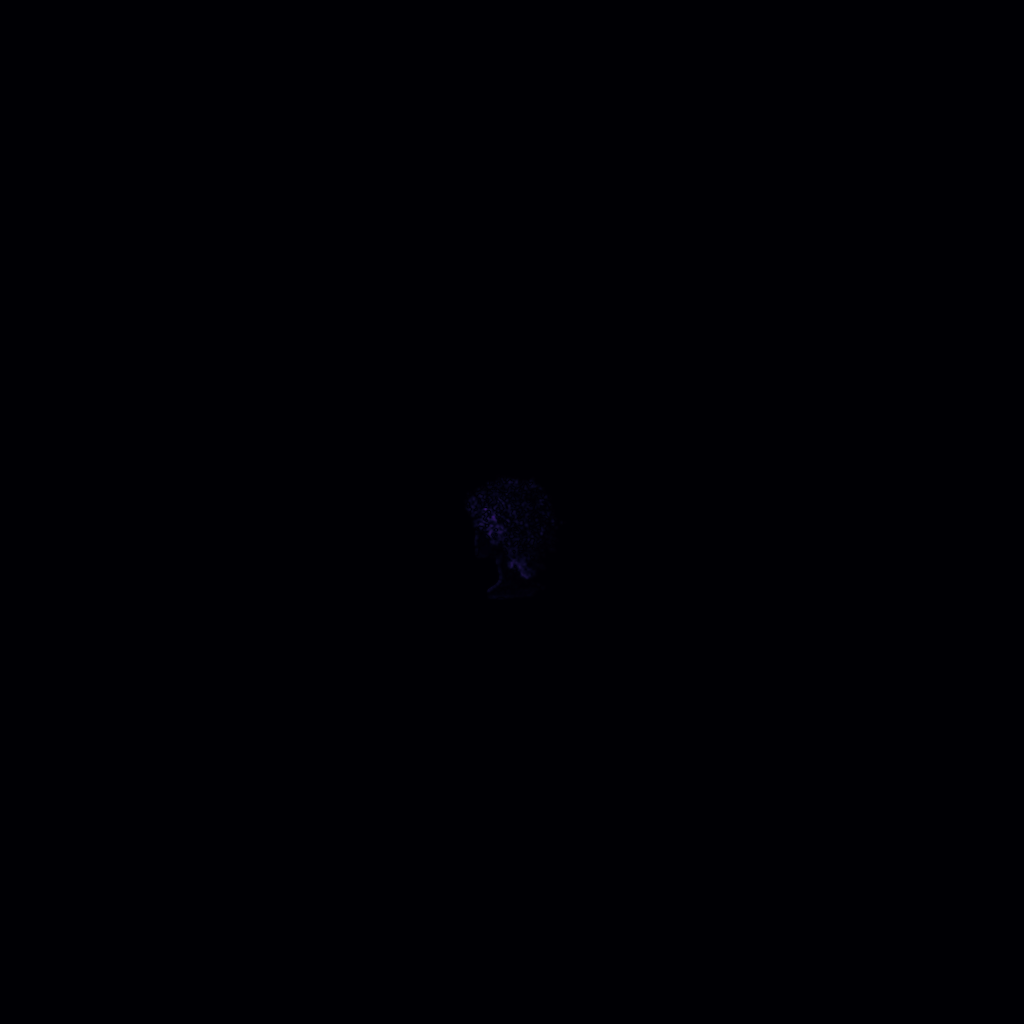}\hfill%
\includegraphics[trim={0 0 0 0},clip,width=0.11\linewidth]{FIG/FIG_SIGA/main_result/hadley/gt/0.png}\hfill%
\includegraphics[trim={450 450 450 450},clip,width=0.11\linewidth]{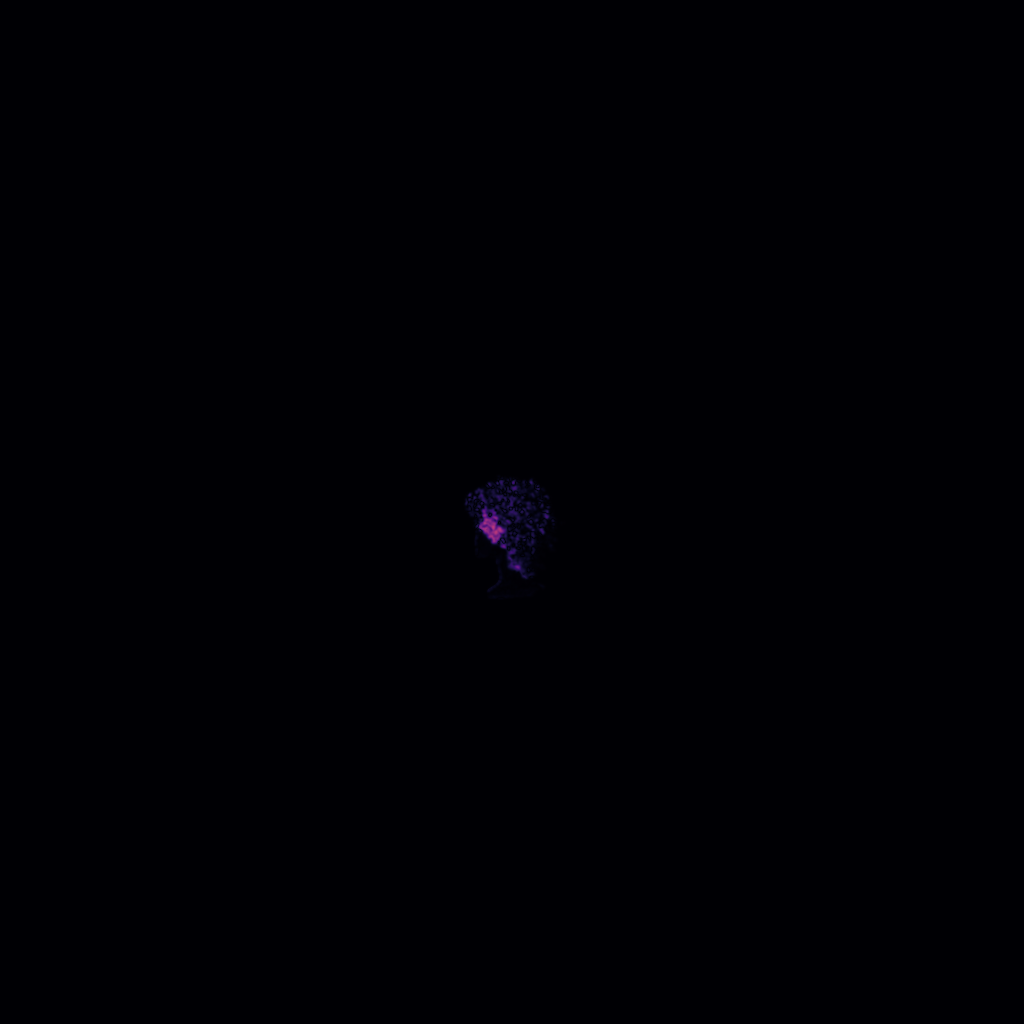}\hfill%
\includegraphics[trim={450 450 450 450},clip,width=0.11\linewidth]{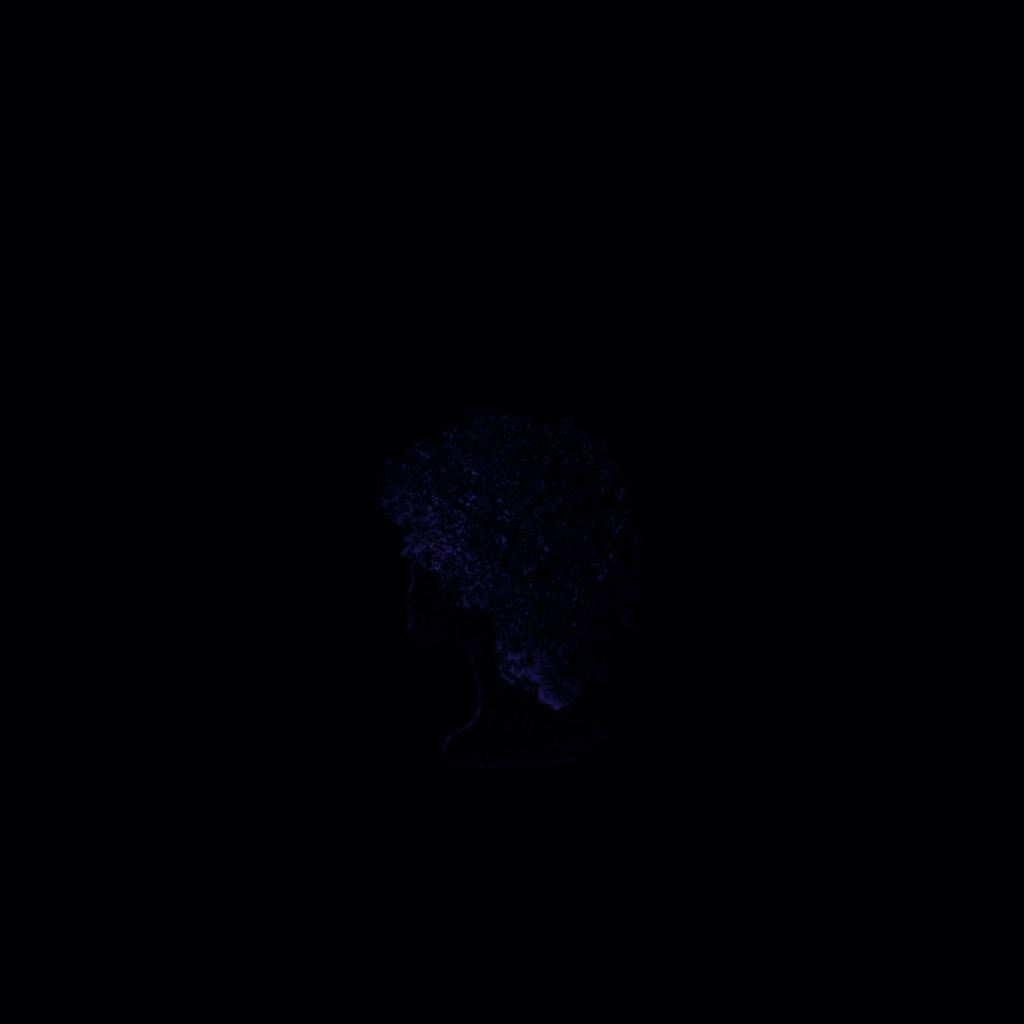}\\

\vspace{-51.1em}
\figcap{\small \color{myRed}{970K} }\hfill%
\figcap{\small \color{myRed}{930K} }\hfill%
\figcap{\small \color{myRed}{970K} }\hfill%
\figcap{\small \color{myRed}{970K} }\hfill%
\figcap{\small \color{myRed}{790K} }\hfill%
\figcap{\small \color{myRed}{970K} }\hfill%
\figcap{\small \color{myRed}{970K} }\hfill%
\figcap{\small \color{myRed}{250K} }\hfill%
\figcap{\small \color{myRed}{970K} }\\
\vspace{5em}
\figcap{\small \color{myCyan}{~ } }\hfill%
\figcap{\small \color{myCyan}{0.090} }\hfill%
\figcap{\small \color{myCyan}{0.128} }\hfill%
\figcap{\small \color{myCyan}{~ } }\hfill%
\figcap{\small \color{myCyan}{0.110} }\hfill%
\figcap{\small \color{myCyan}{0.140} }\hfill%
\figcap{\small \color{myCyan}{~ } }\hfill%
\figcap{\small \color{myCyan}{0.132} }\hfill%
\figcap{\small \color{myCyan}{0.160} }\\
\vspace{5.1em}
\figcap{\small \color{myRed}{1050K} }\hfill%
\figcap{\small \color{myRed}{1010K} }\hfill%
\figcap{\small \color{myRed}{1050K} }\hfill%
\figcap{\small \color{myRed}{1050K} }\hfill%
\figcap{\small \color{myRed}{857K} }\hfill%
\figcap{\small \color{myRed}{1050K} }\hfill%
\figcap{\small \color{myRed}{1050K} }\hfill%
\figcap{\small \color{myRed}{418K} }\hfill%
\figcap{\small \color{myRed}{1050K} }\\
\vspace{5.1em}
\figcap{\small \color{myCyan}{~ } }\hfill%
\figcap{\small \color{myCyan}{0.066} }\hfill%
\figcap{\small \color{myCyan}{0.086} }\hfill%
\figcap{\small \color{myCyan}{~ } }\hfill%
\figcap{\small \color{myCyan}{0.062} }\hfill%
\figcap{\small \color{myCyan}{0.082} }\hfill%
\figcap{\small \color{myCyan}{~ } }\hfill%
\figcap{\small \color{myCyan}{0.065} }\hfill%
\figcap{\small \color{myCyan}{0.086} }\\
\vspace{5.15em}
\figcap{\small \color{myRed}{1900K} }\hfill%
\figcap{\small \color{myRed}{1856K} }\hfill%
\figcap{\small \color{myRed}{1900K} }\hfill%
\figcap{\small \color{myRed}{1900K} }\hfill%
\figcap{\small \color{myRed}{1535K} }\hfill%
\figcap{\small \color{myRed}{1900K} }\hfill%
\figcap{\small \color{myRed}{1900K} }\hfill%
\figcap{\small \color{myRed}{574K} }\hfill%
\figcap{\small \color{myRed}{1900K} }\\
\vspace{5.1em}
\figcap{\small \color{myCyan}{~ } }\hfill%
\figcap{\small \color{myCyan}{0.053} }\hfill%
\figcap{\small \color{myCyan}{0.070} }\hfill%
\figcap{\small \color{myCyan}{~ } }\hfill%
\figcap{\small \color{myCyan}{0.125} }\hfill%
\figcap{\small \color{myCyan}{0.141} }\hfill%
\figcap{\small \color{myCyan}{~ } }\hfill%
\figcap{\small \color{myCyan}{0.141} }\hfill%
\figcap{\small \color{myCyan}{0.137} }\\
\vspace{5.15em}
\figcap{\small \color{myRed}{1600K} }\hfill%
\figcap{\small \color{myRed}{1510K} }\hfill%
\figcap{\small \color{myRed}{1600K} }\hfill%
\figcap{\small \color{myRed}{1600K} }\hfill%
\figcap{\small \color{myRed}{983K} }\hfill%
\figcap{\small \color{myRed}{1600K} }\hfill%
\figcap{\small \color{myRed}{1600K} }\hfill%
\figcap{\small \color{myRed}{330K} }\hfill%
\figcap{\small \color{myRed}{1600K} }\\
\vspace{5.1em}
\figcap{\small \color{myCyan}{~ } }\hfill%
\figcap{\small \color{myCyan}{0.009} }\hfill%
\figcap{\small \color{myCyan}{0.009} }\hfill%
\figcap{\small \color{myCyan}{~ } }\hfill%
\figcap{\small \color{myCyan}{0.023} }\hfill%
\figcap{\small \color{myCyan}{0.010} }\hfill%
\figcap{\small \color{myCyan}{~ } }\hfill%
\figcap{\small \color{myCyan}{0.026} }\hfill%
\figcap{\small \color{myCyan}{0.013} }\\

\vspace{5.5em}
\caption{Four hairstyles, ranging from straight to curly hair, under near, middle, and far views. Compared to path tracing with full hair geometry (full w/ PT), our approach (LoD w/ ours) utilizing reduced hair geometry through level-of-detail techniques can outperform existing dual scattering employing full geometry (full w/ DS) in both memory size and appearance. The \textnormal{\reflectbox{F}LIP} error of ours increases for the far view due to the geometry simplification. {\color{myRed}{Pink}} and {\color{myCyan}{cyan}} denote the number of segments and \textnormal{\reflectbox{F}LIP} error of screen-space bounding box of hairs. 
}
\label{fig:main}
\Description{}
\end{figure*}

We conduct all experiments on a system with an AMD Ryzen ThreadRipper 3970X 32-core CPU, 256 GB of memory, and an NVIDIA RTX 3090 GPU. We implement our method (aggregated BCSDF + LoD strategy) on both a CPU renderer and a GPU renderer (\autoref{fig:cpuvsgpu}). In contrast to employing full hair geometry in path tracing (PT) and dual scattering (DS) \cite{ZinkeYWK08}, we utilize reduced hair geometry achieved through our hair LoD solution. Specifically, we group individual hair strands into 256 clusters and fit each cluster with thick hairs across 5-6 LoD levels, depending on the total number of hair strands. Each thick hair is represented by a B-Spline with 16 control points. For sufficient granularity, each thick hair segment is assigned an individual LoD level. 


To validate the accuracy of our proposed method, we use a CPU render implemented in Embree~\cite{Embree} with 20 samples per pixel (spp). Similar to \cite{ZinkeYWK08}, we use ray shooting to find all intersecting hairs along the shadow path to compute the transmittance for dual scattering. The CPU reference is generated by path tracing the full hair geometry with up to 70 ray bounces and 1024 spp. The hair BCSDF parameters are chosen generally consistent with the values provided by \citet{MarschnerJCWH03}. In particular, we use $M_R = $ 0.08 - 0.15, $M_{TT} = $0.12 - 0.3, and $N_{TT} = $0.2 - 0.4. We further adjust attenuation to match the real-life hair color, as $A_R = $ (0.03,0.03,0.03) - (0.25,0.25,0.25), $A_{TT} = $ (0.25,0.17,0.07) - (0.38,0.25,0.18), $A_D = $ (0.2,0.11,0.1) - (0.40,0.21,0.08). 
%
%

To further evaluate our real-time performance and rendering quality, we implement our framework and counterpart full hair geometry and hair cards with DS using OpenGL 4, in which the shadow computation in all methods uses deep opacity maps. 

We use a LoD threshold of $\epsilon_w = 2$ pixels for all our experiments. The output image resolution for all results is 1024$\times$1024. The initialization of our method takes from 5 to 10 minutes, depending on the number of hair strands, with the majority of the time spent on hair clustering. 
To assess the impact of our LoD solution on rendering accuracy, we present results from near, middle, and far views with 60\%, 20\%, and 2\% screen occupancy ratios, respectively. Note that our ray tracer produces a slightly different estimation of hair screen width than our rasterizer, which causes the hair count in the two systems to be different at the same view. 


\paragraph{Hairstyles} 
In \autoref{fig:main}, we evaluate our approach on Four hairstyles, ranging from straight to curly. Two of these hairstyles are sourced from Unreal Metahumans~\cite{unrealengine}, and two are captured from the real-world examples~\cite{Derouet-Jourdan:2013,Shen2023}. For the four hairstyles, we use different hair colors: brown, blonde, red, and black. 
Compared to using full hair geometry with DS, our method consistently delivers higher accuracy with lower \FLIP errors for all close views and middle and far views across the first three hairstyles. The additional light path series $A_{1_+}$ introduced by our method significantly enhances shading accuracy. However, as the viewing distance increases, the error of our method increases slightly due to geometric simplification. 



\begin{figure}[ht]
\centering
\newcommand{\figcap}[1]{\begin{minipage}{0.31\linewidth}\centering#1\end{minipage}}
\hspace{0.04\linewidth}\hfill%
\figcap{\small Near view (60\%)}\hfill%
\figcap{\small Middle view (20\%)}\hfill%
\figcap{\small Far view (2\%)}\vspace{.4em}
\\\hspace{0.04\linewidth}\hfill%
\includegraphics[trim={0   0   0   0  },clip,width=0.315\linewidth]{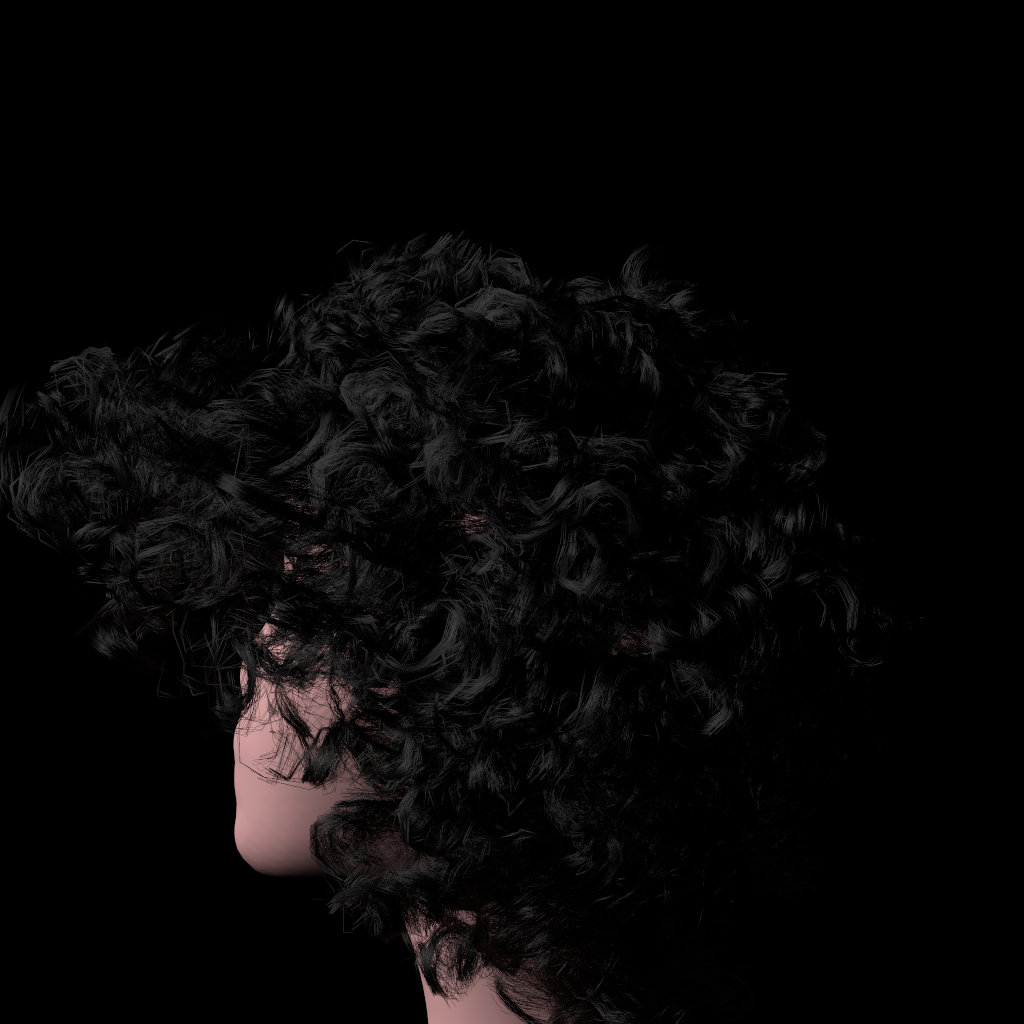}\hfill%
\includegraphics[trim={325 325 325 325},clip,width=0.315\linewidth]{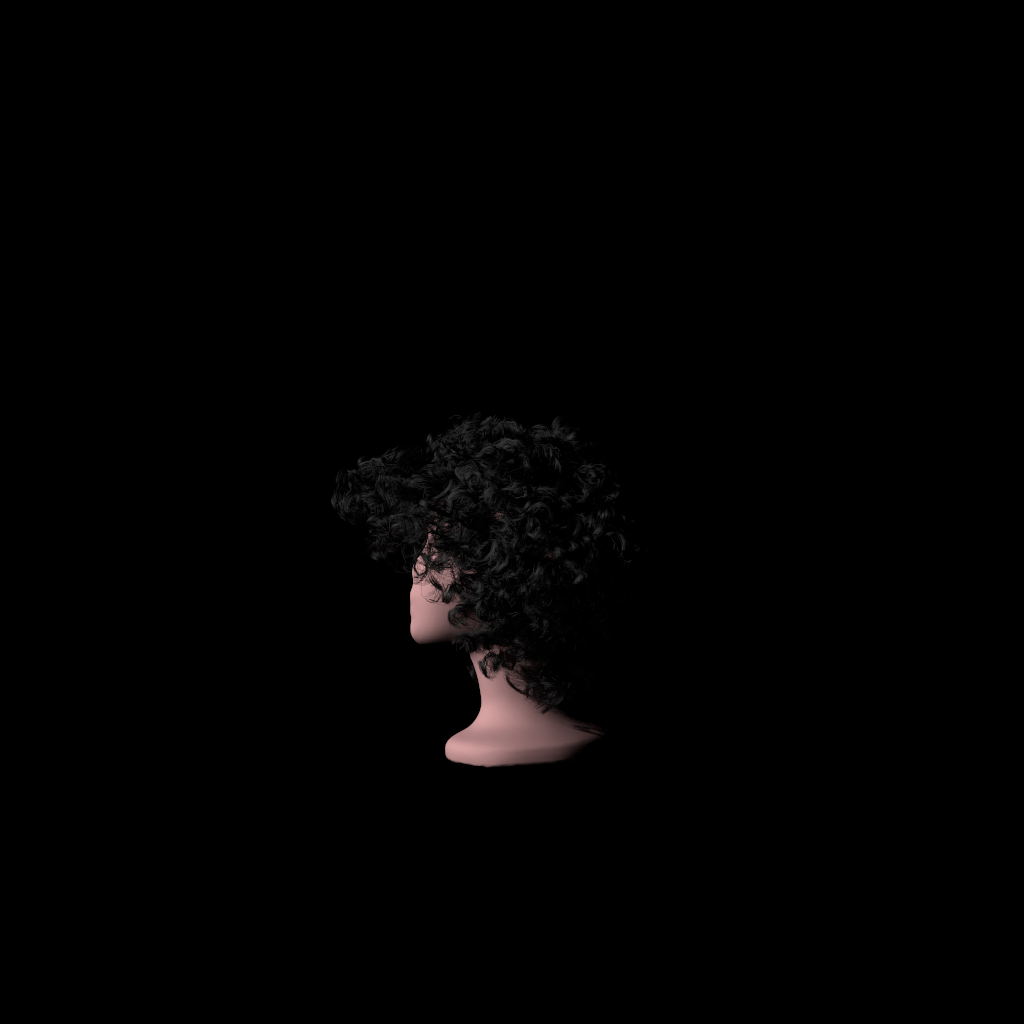}\hfill%
\includegraphics[trim={450 450 450 450},clip,width=0.315\linewidth]{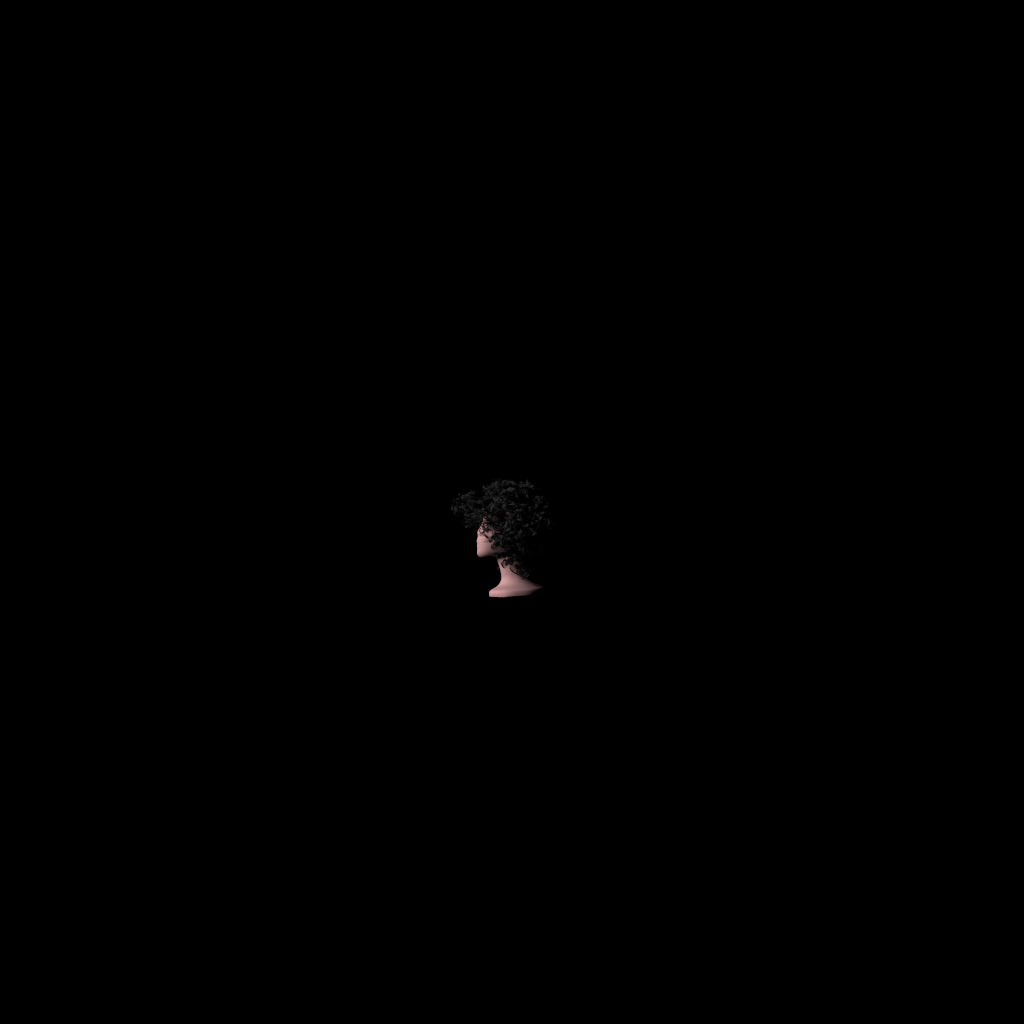}\\
\hspace{0.04\linewidth}\hfill%
\includegraphics[trim={0   0   0   0  },clip,width=0.315\linewidth]{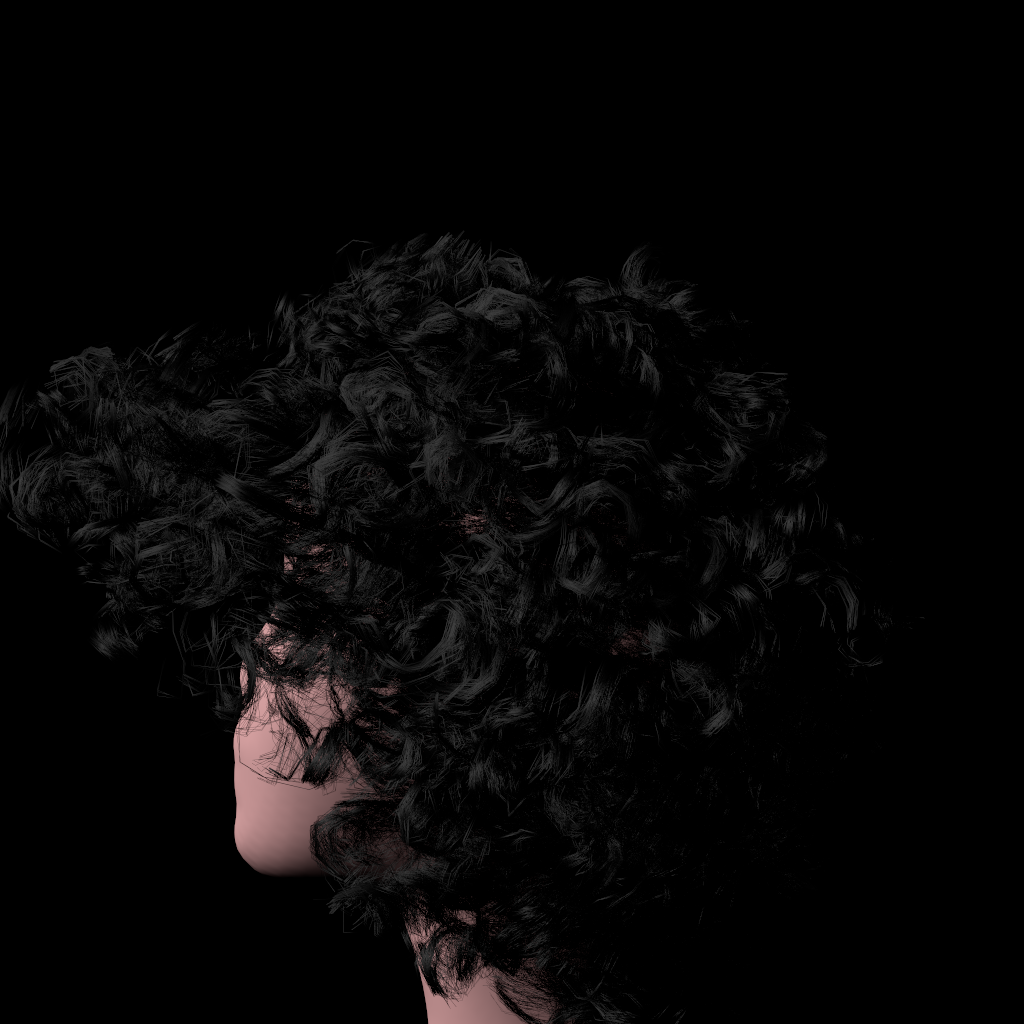}\hfill
\includegraphics[trim={325 325 325 325},clip,width=0.315\linewidth]{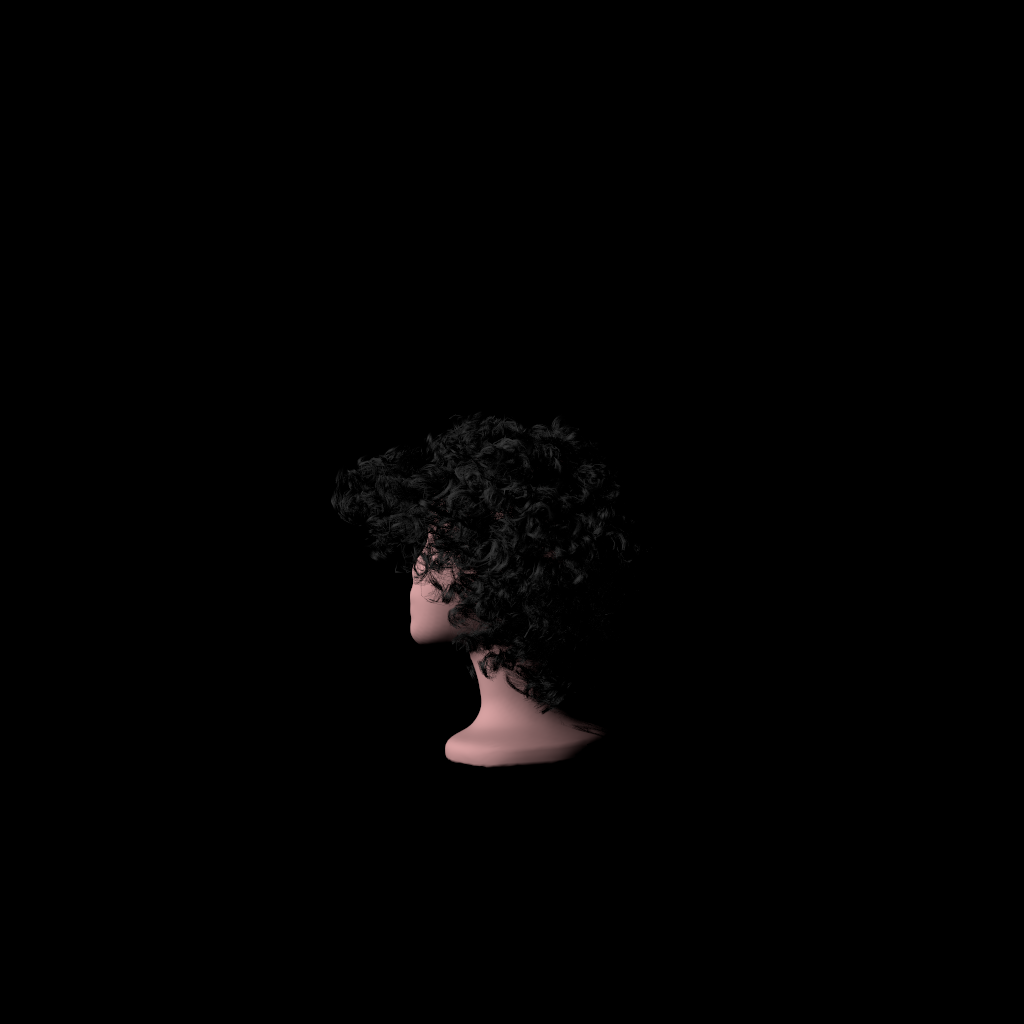}\hfill
\includegraphics[trim={450 450 450 450},clip,width=0.315\linewidth]{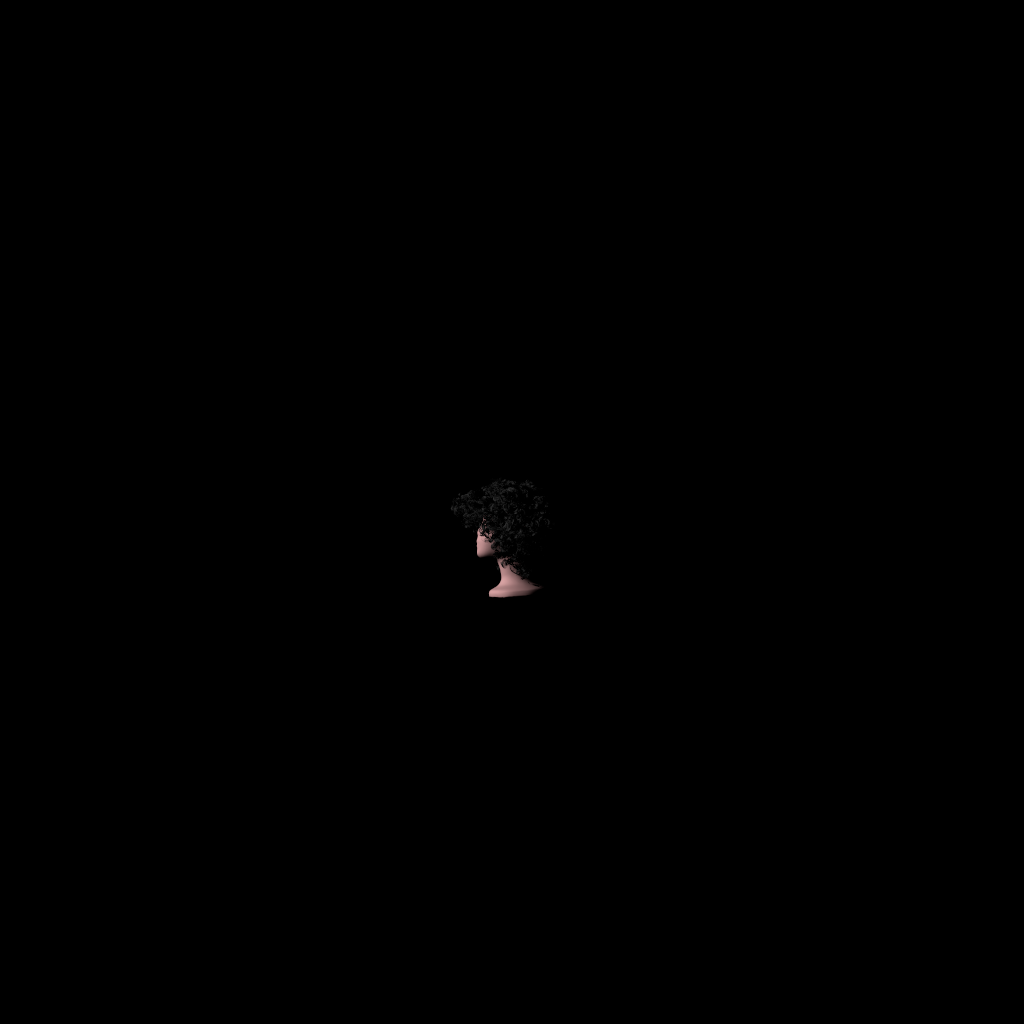}
\\
\hspace{0.04\linewidth}\hfill%
\includegraphics[trim={0 0 0 0},        clip,width=0.315\linewidth]{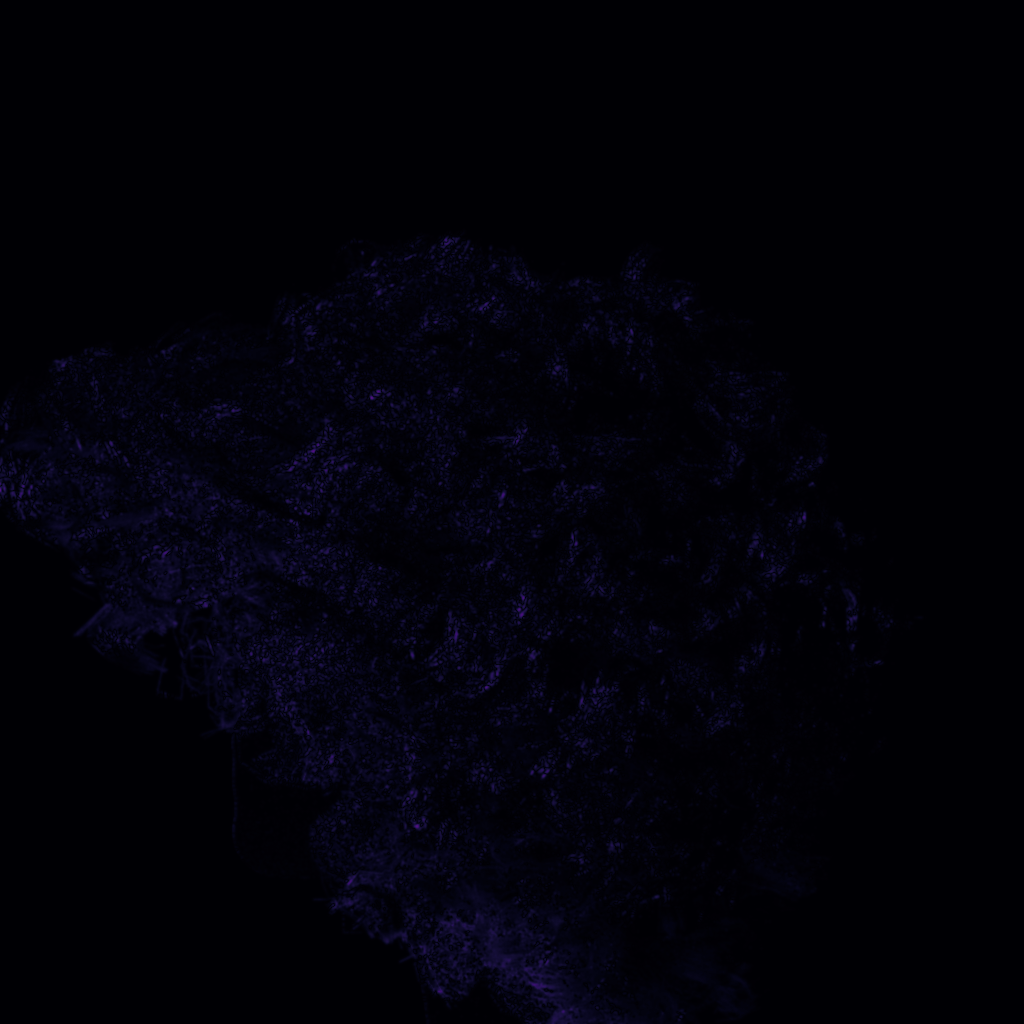}\hfill%
\includegraphics[trim={325 325 325 325},clip,width=0.315\linewidth]{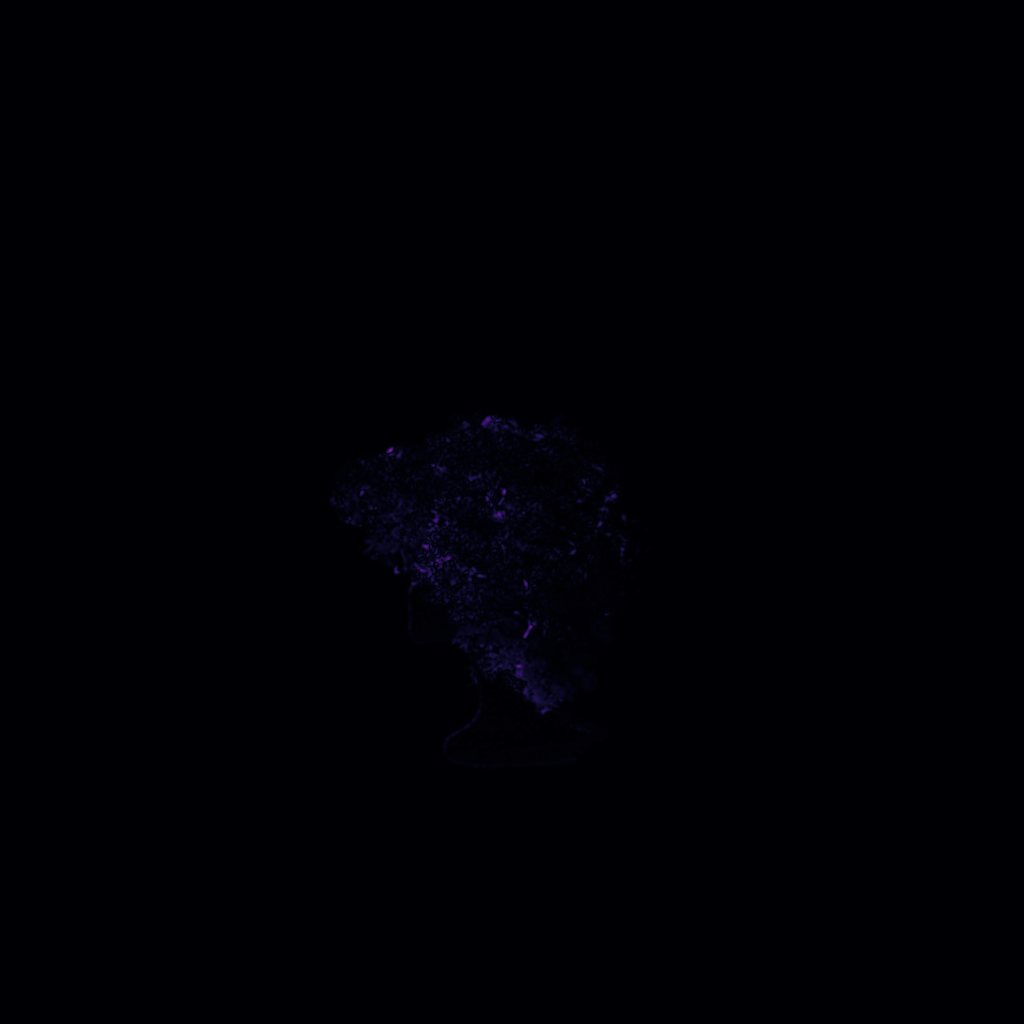}\hfill%
\includegraphics[trim={450 450 450 450},clip,width=0.315\linewidth]{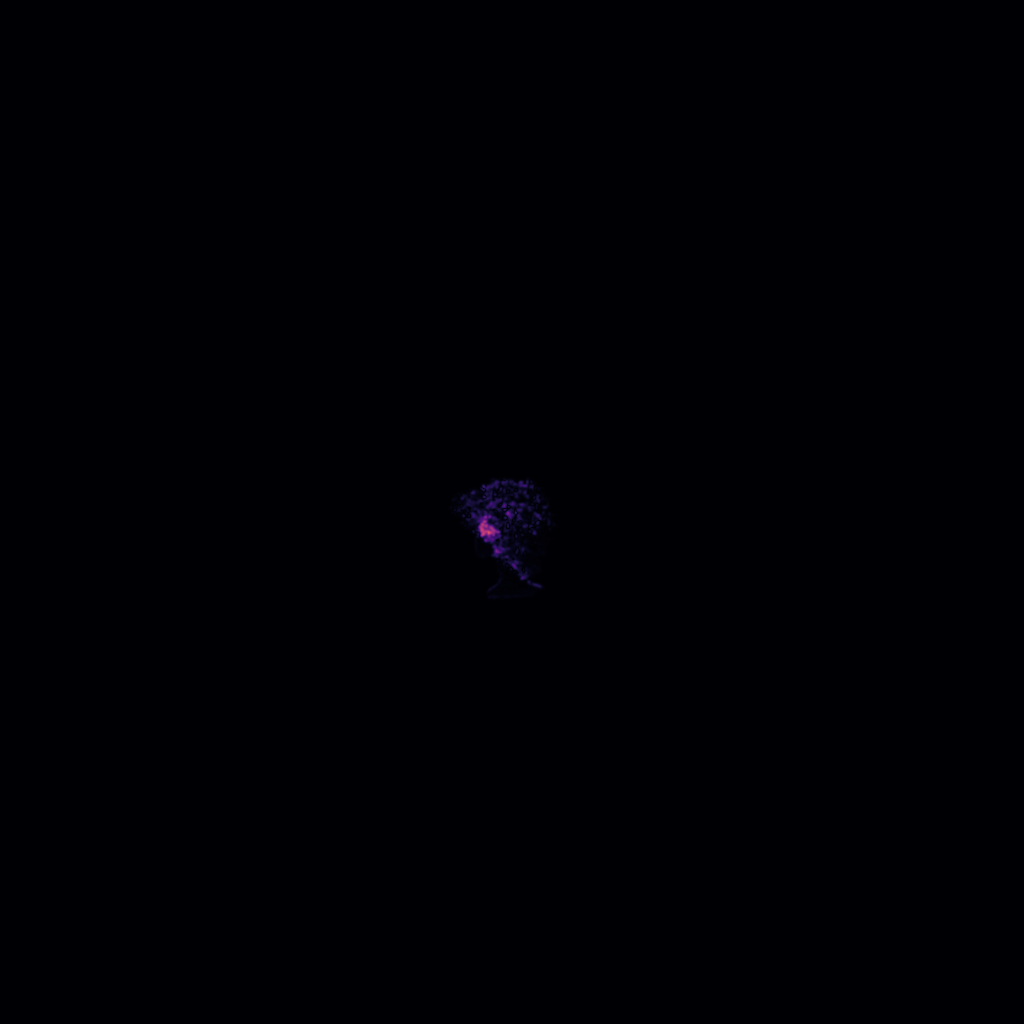}
\\\hspace{0.04\linewidth}\hfill%
\includegraphics[trim={0   0   0   0  },clip,width=0.315\linewidth]{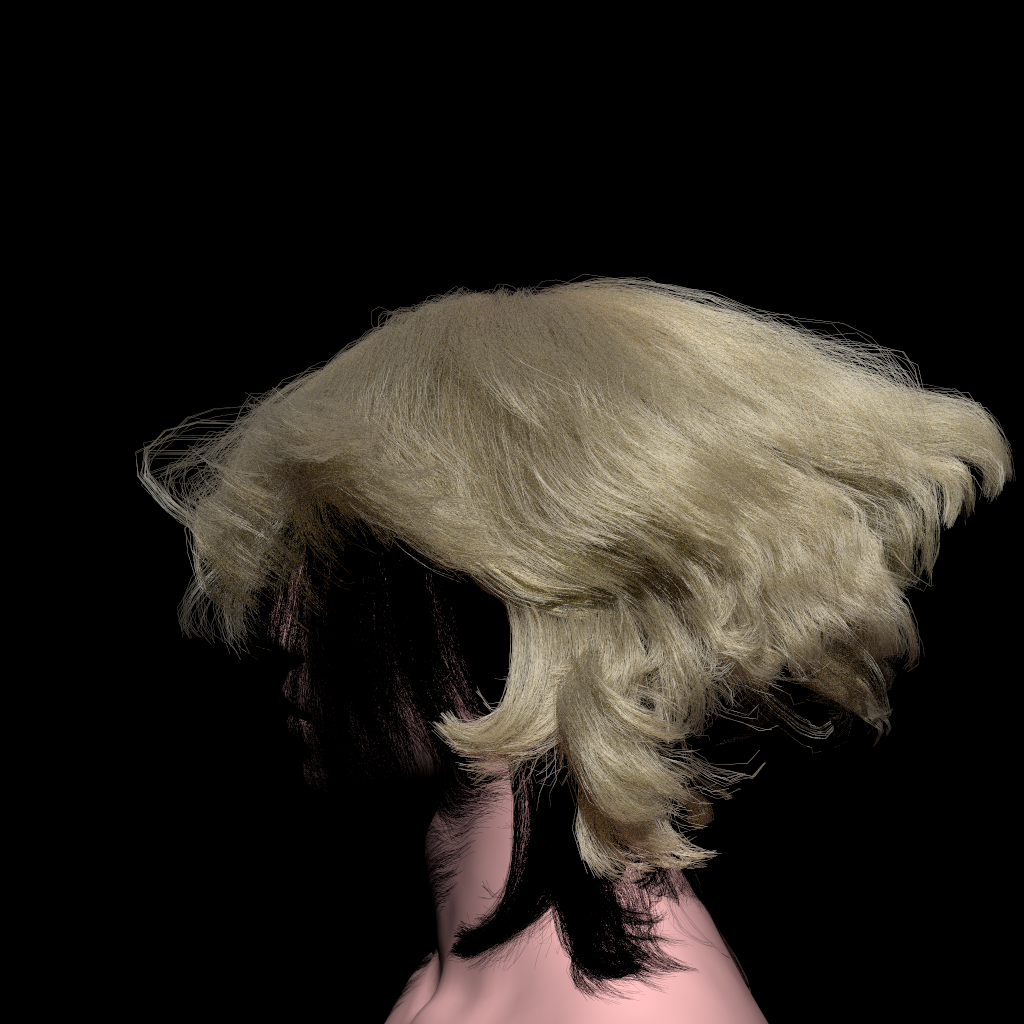}\hfill%
\includegraphics[trim={309 309 309 309},clip,width=0.315\linewidth]{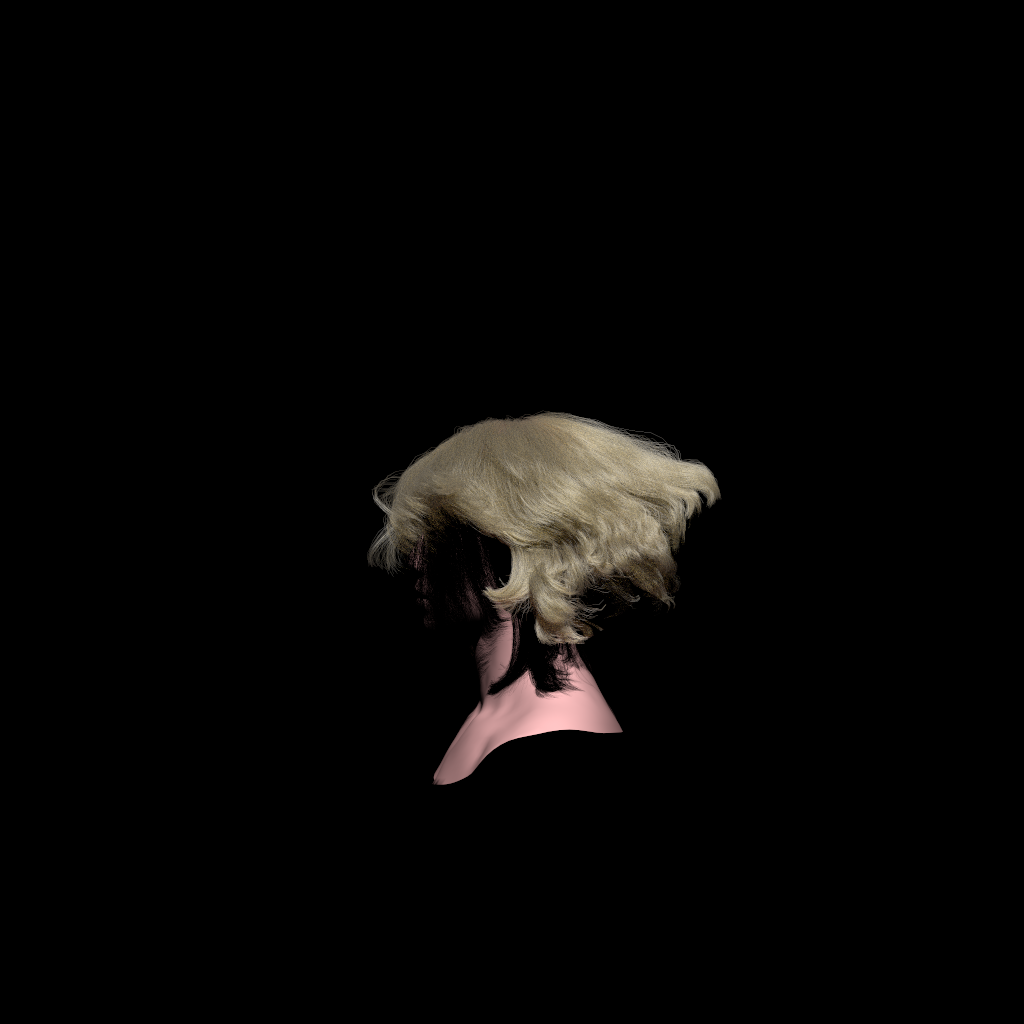}\hfill%
\includegraphics[trim={435 435 435 435},clip,width=0.315\linewidth]{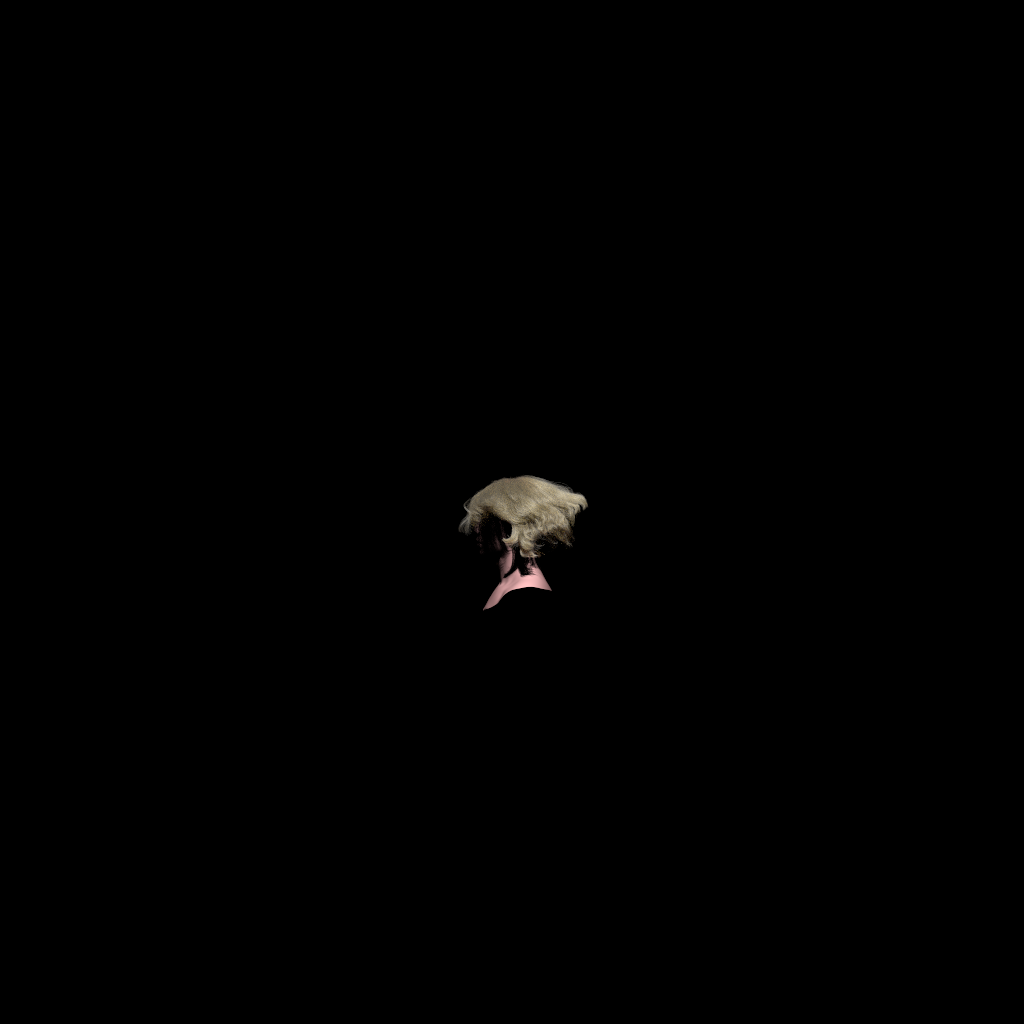}\\
\hspace{0.04\linewidth}\hfill%
\includegraphics[trim={0   0   0   0  },clip,width=0.315\linewidth]{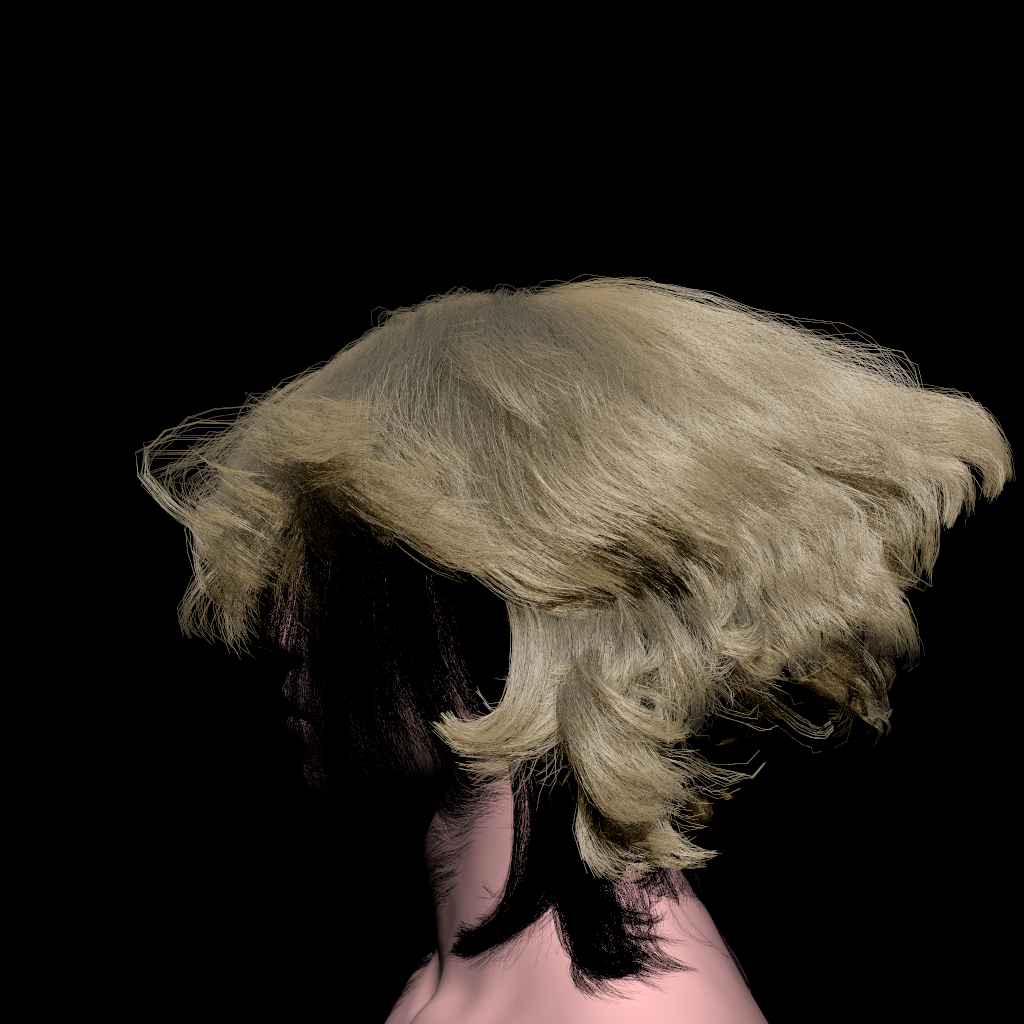}\hfill
\includegraphics[trim={309 309 309 309},clip,width=0.315\linewidth]{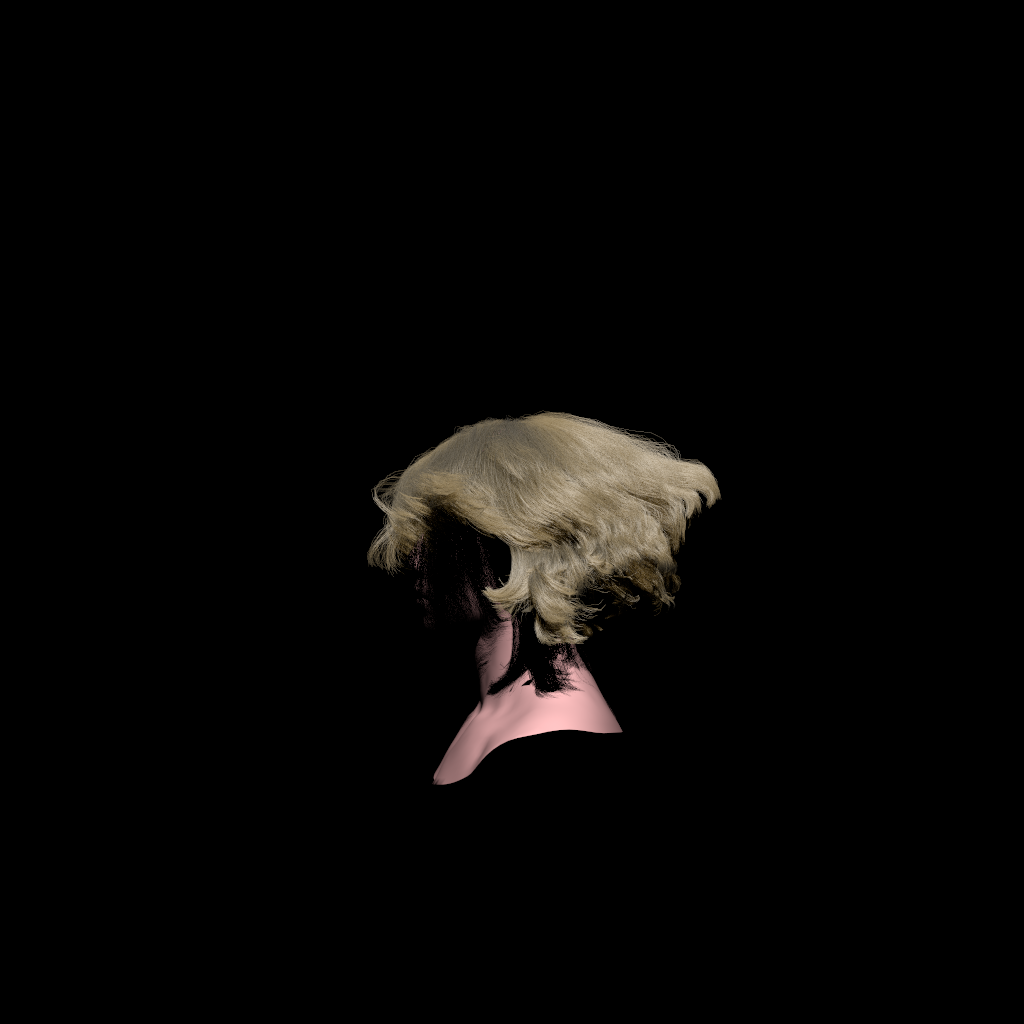}\hfill
\includegraphics[trim={435 435 435 435},clip,width=0.315\linewidth]{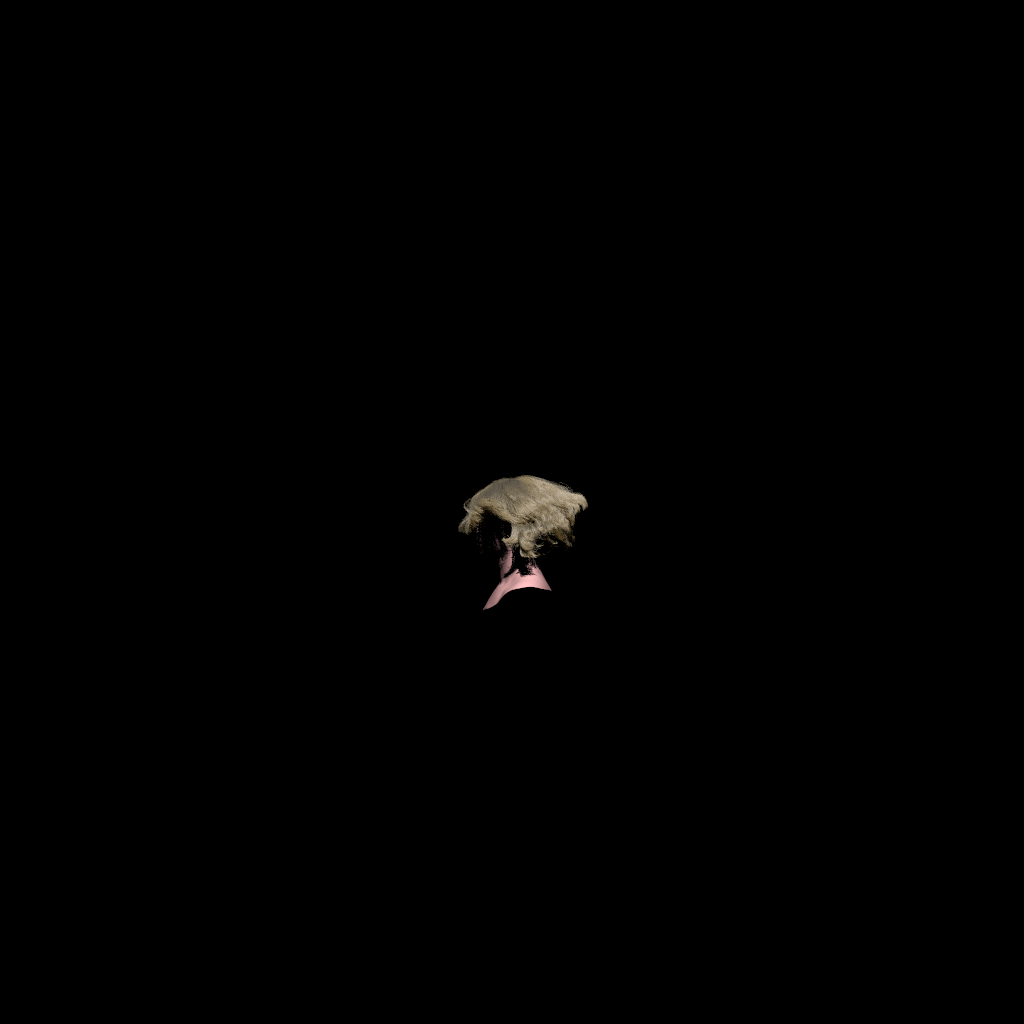}
\\
\hspace{0.04\linewidth}\hfill%
\includegraphics[trim={0 0 0 0},        clip,width=0.315\linewidth]{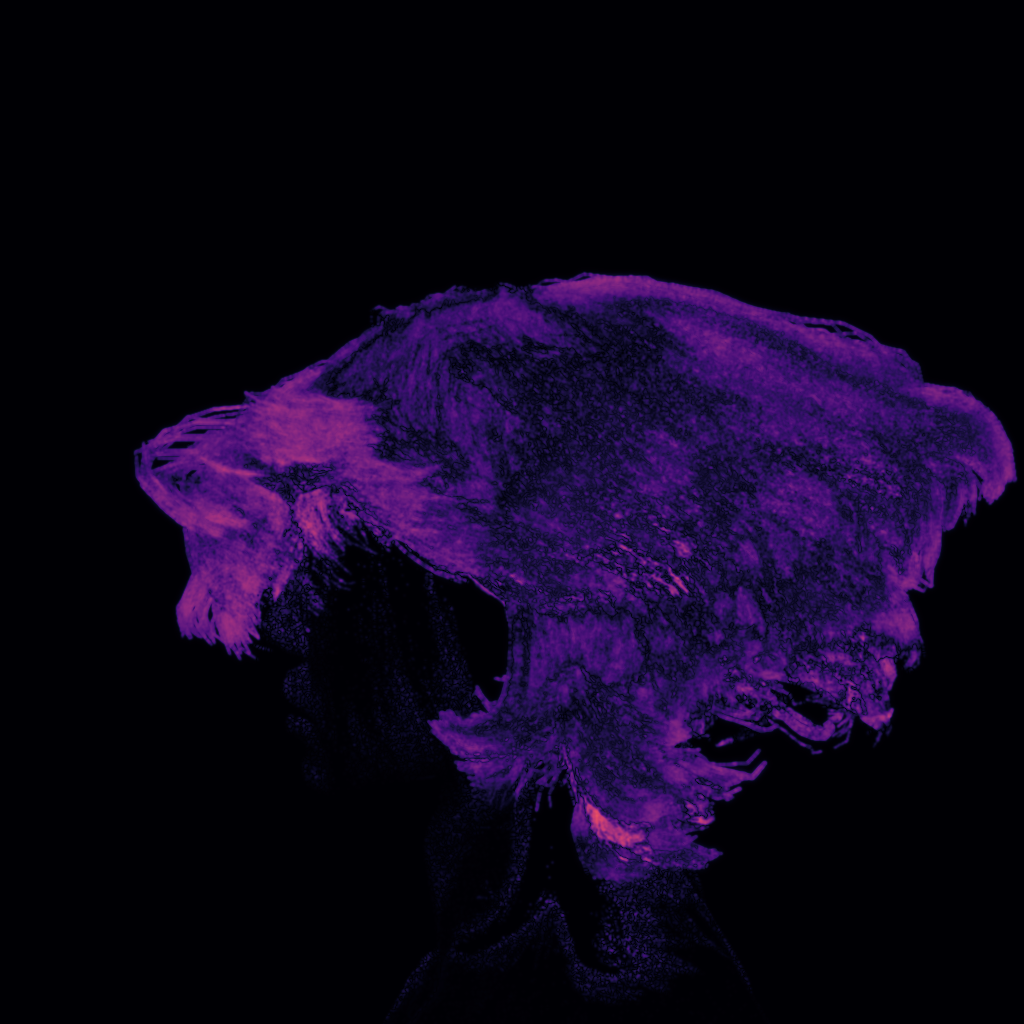}\hfill%
\includegraphics[trim={309 309 309 309},clip,width=0.315\linewidth]{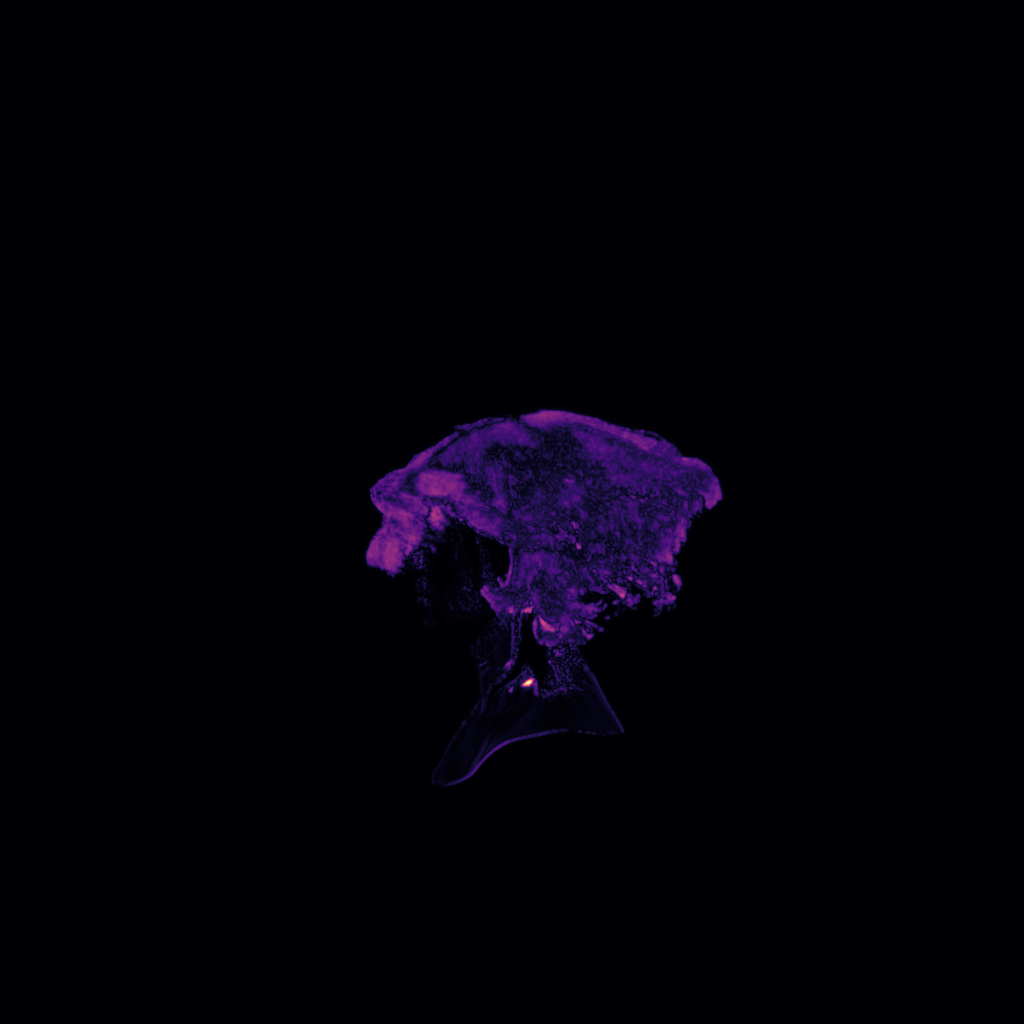}\hfill%
\includegraphics[trim={435 435 435 435},clip,width=0.315\linewidth]{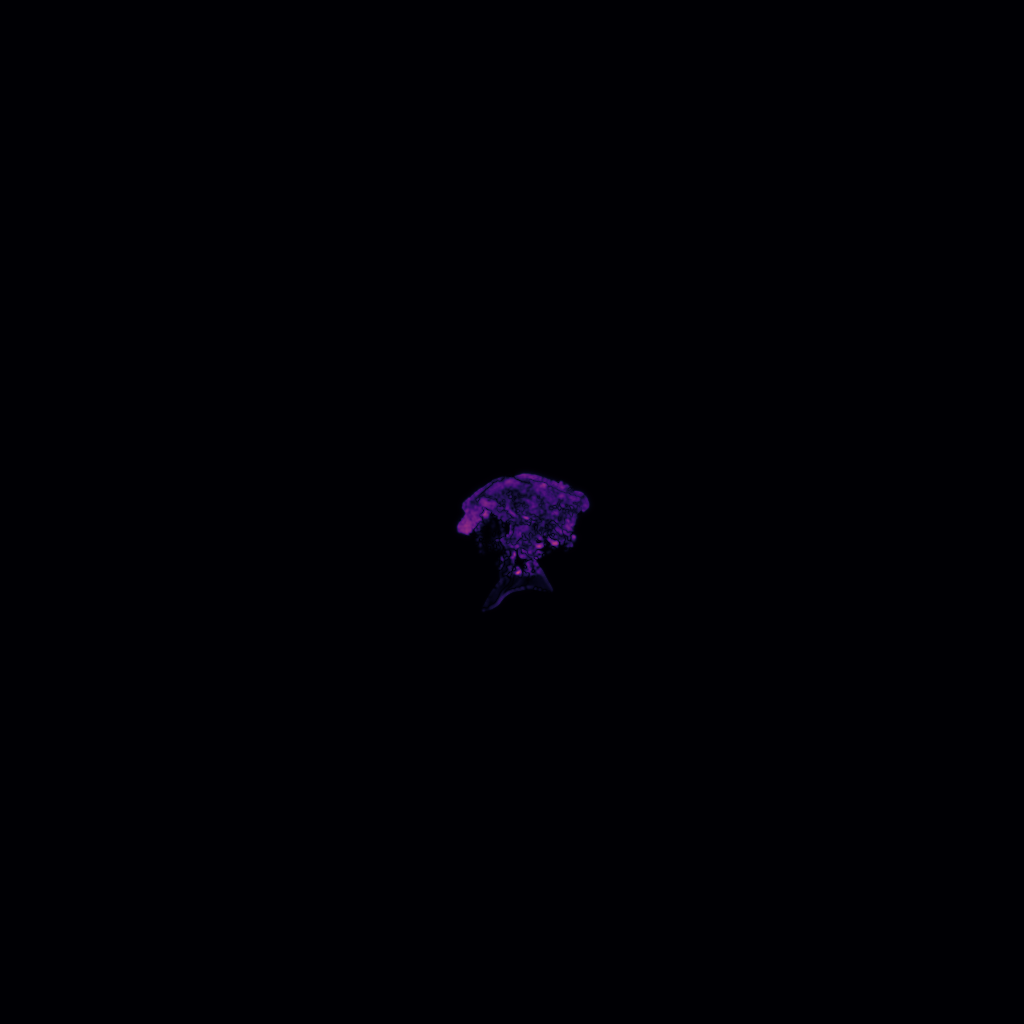}\\
\vspace{-52em}
\begin{flushleft}
{\small \hspace{8em} \color{myRed}{$1600$K}  \hspace{6.7em} \color{myRed}{$1600$K} \hspace{6.7em} \color{myRed}{$1600$K}}
\end{flushleft}
\vspace{7em}
\begin{flushleft}
{\small \hspace{8em} \color{myRed}{$1480$K}  \hspace{7.1em} \color{myRed}{$960$K} \hspace{7.1em} \color{myRed}{$300$K}}
\end{flushleft}
\vspace{7em}
\begin{flushleft}
{\small \hspace{8.4em} \color{myCyan}{0.010}  \hspace{7.1em} \color{myCyan}{0.012} \hspace{7.1em} \color{myCyan}{0.023}}
\end{flushleft}
\vspace{7em}
\begin{flushleft}
{\small \hspace{8em} \color{myRed}{$1050$K}  \hspace{6.7em} \color{myRed}{$1050$K} \hspace{6.7em} \color{myRed}{$1050$K}}
\end{flushleft}
\vspace{7em}
\begin{flushleft}
{\small \hspace{8.4em} \color{myRed}{$930$K}  \hspace{7.1em} \color{myRed}{$840$K} \hspace{7.1em} \color{myRed}{$650$K}}
\end{flushleft}
\vspace{7em}
\begin{flushleft}
{\small \hspace{8.4em} \color{myCyan}{0.053}  \hspace{7.1em} \color{myCyan}{0.054} \hspace{7.1em} \color{myCyan}{0.052}}
\end{flushleft}
\vspace{-44.5em}
\begin{flushleft}{
\rotatebox{90}{\small \hspace{2em} \reflectbox{F}LIP error \hspace{2.7em} \textbf{Ours (ray shooting)} \hspace{1em} Offline path tracing \hspace{3.em} \reflectbox{F}LIP error \hspace{2.7em} \textbf{Ours (ray shooting)} \hspace{1em} Offline path tracing  }\hfill%
}\end{flushleft}
\caption{Animated hairs from two simulation sequences exhibit varying shapes. {\color{myRed}{Pink}} and {\color{myCyan}{cyan}} denote the number of segments and \textnormal{\reflectbox{F}LIP} error of screen-space bounding box of hairs.}
\label{fig:dynamics}
\Description{}
\end{figure}

\paragraph{Animated hairs}
As shown in~\autoref{fig:dynamics}, our method maintains its rendering accuracy with hair animation. We deployed a strand-based hair simulator~\cite{Hsu2023} to simulate the 256 guide hairs of two hairstyles and derived all hairs via linear skinning. Please see the supplementary video for the complete simulation sequence, where our method allows a smooth transition between different LoD levels in real-time.
Note that the hair LoD level is determined based on its screen space width, which means that in extreme cases of intense hair movement, the finest LoD level may be required. However, in our experiments, even during frames with the most intense motion, our method achieves a significant level of simplification while maintaining nearly identical visual quality to that of the full geometry.

\begin{figure}[th]
\centering
\includegraphics[trim={0 0 0 0},clip,width=\linewidth]{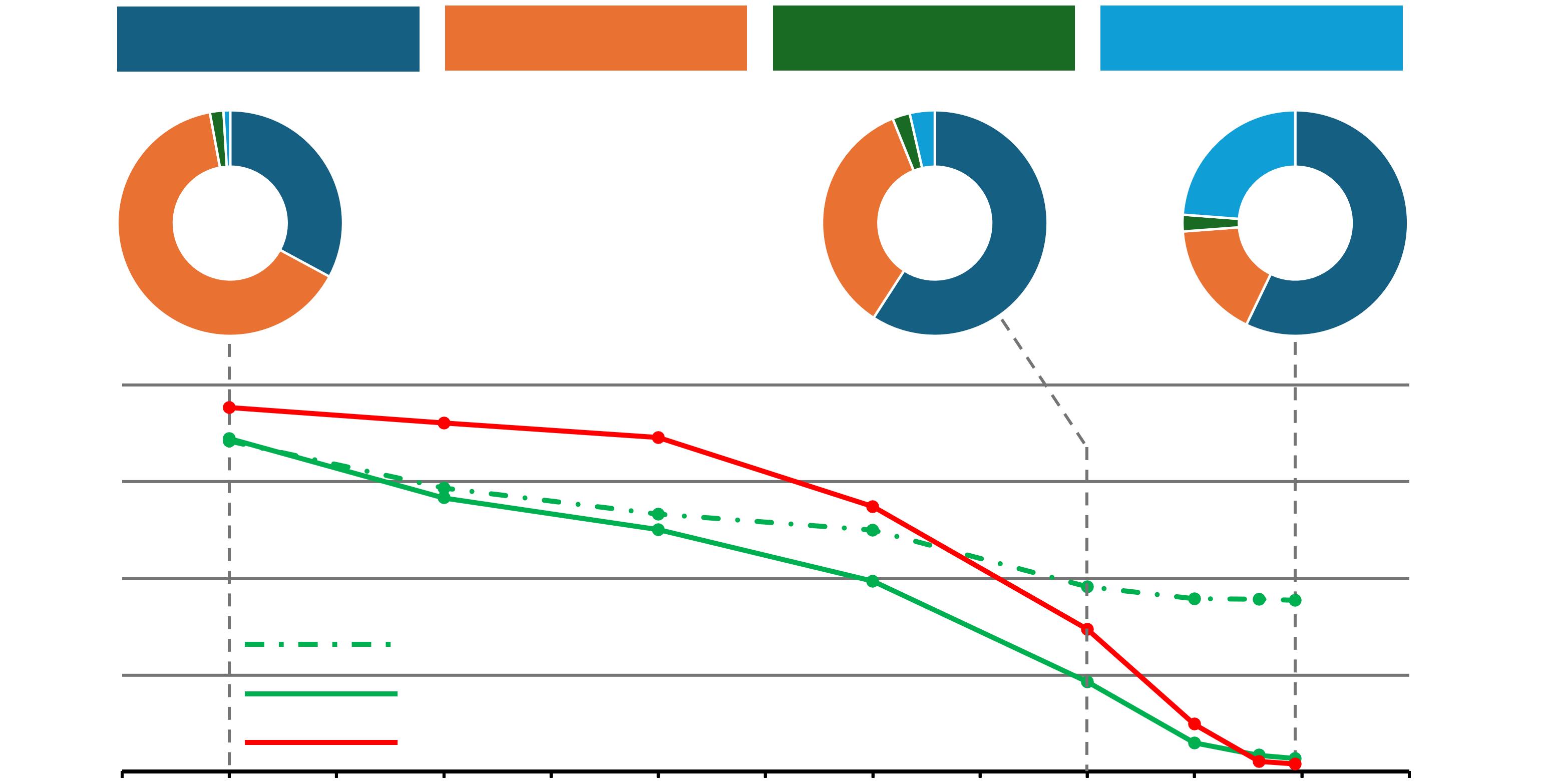}  
\put(-221,114){\small \color{white}{Shadow pass}}
\put(-170,114){\small \color{white}{Shading pass}}
\put(-121,114){\small \color{white}{Compute pass}}
\put(-58,114) {\small \color{white}{Others}}
\put(-214,86){\small {10.3}}
\put(-101,86){\small {2.8}}
\put(-45, 86){\small {0.4}}
\put(-245,61) {\small \color{mygreen}{12 ms}}
\put(-244,45) {\small \color{mygreen}{ 9 ms}}
\put(-244,30) {\small \color{mygreen}{ 6 ms}}
\put(-244,15) {\small \color{mygreen}{ 3 ms}}
\put(-20, 61) {\small \color{red}{2.0}M}
\put(-20, 45) {\small \color{red}{1.5}M}
\put(-20, 30) {\small \color{red}{1.0}M}
\put(-20, 15) {\small \color{red}{0.5}M}
\put(-179, 20) {\small Full geometry time}
\put(-179, 11) {\small Our time}
\put(-179, 4)  {\small Our \#triangles}
\vspace{-0.5em}
\begin{flushleft}{\small Screen \% \hspace{0.em} 50\% \hspace{2.1em} 40\% \hspace{2.1em} 30\% \hspace{2.1em} 20\% \hspace{2.1em} 10\% \hspace{2.2em} 0\% }\end{flushleft}
\caption{ A performance visualization of the scene in \autoref{fig:haircard}. As the screen occupation percentage decreases from 50\% to 0.3\%, our GPU rasterizer progressively outperforms rasterizing full hair geometry, achieving a slight overhead of 0.99$\times$ (10.3ms vs. 10.2ms) to a significant improvement of 13$\times$ (0.4ms vs. 5.3ms). The pie charts provide a breakdown of timings at three screen percentages of hairs.  }

\label{fig:performance}
\Description{}
\end{figure}

\begin{figure}[th]
\centering
\newcommand{\figcap}[1]{\begin{minipage}{0.31\linewidth}\centering#1\end{minipage}}
\hspace{0.04\linewidth}\hfill%
\figcap{\small Near view (50\%)}\hfill%
\figcap{\small Middle view (10\%)}\hfill%
\figcap{\small Far view (0.3\%)}\vspace{.4em}
\\
\hspace{0.04\linewidth}\hfill%
\includegraphics[trim={0   0   0   0  },clip,width=0.315\linewidth]{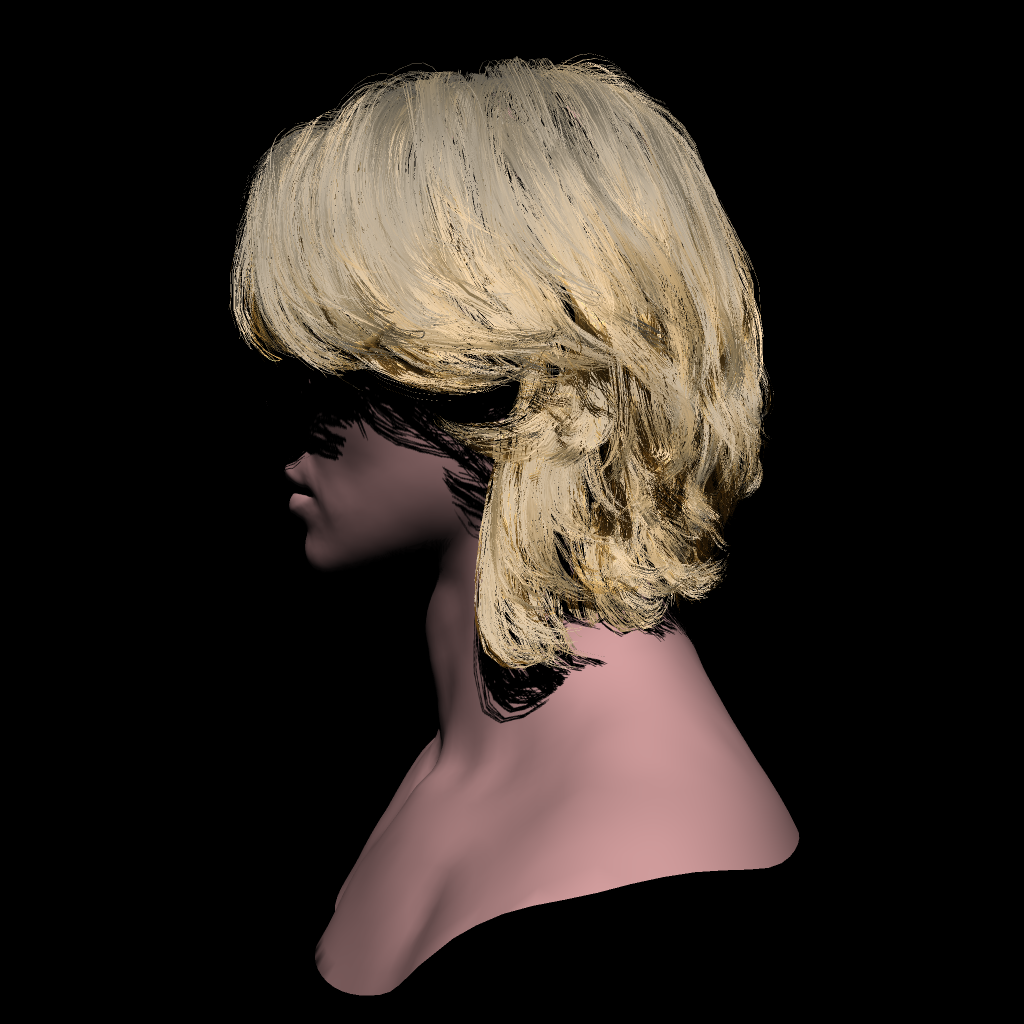}\hfill%
\includegraphics[trim={350 350 350 350},clip,width=0.315\linewidth]{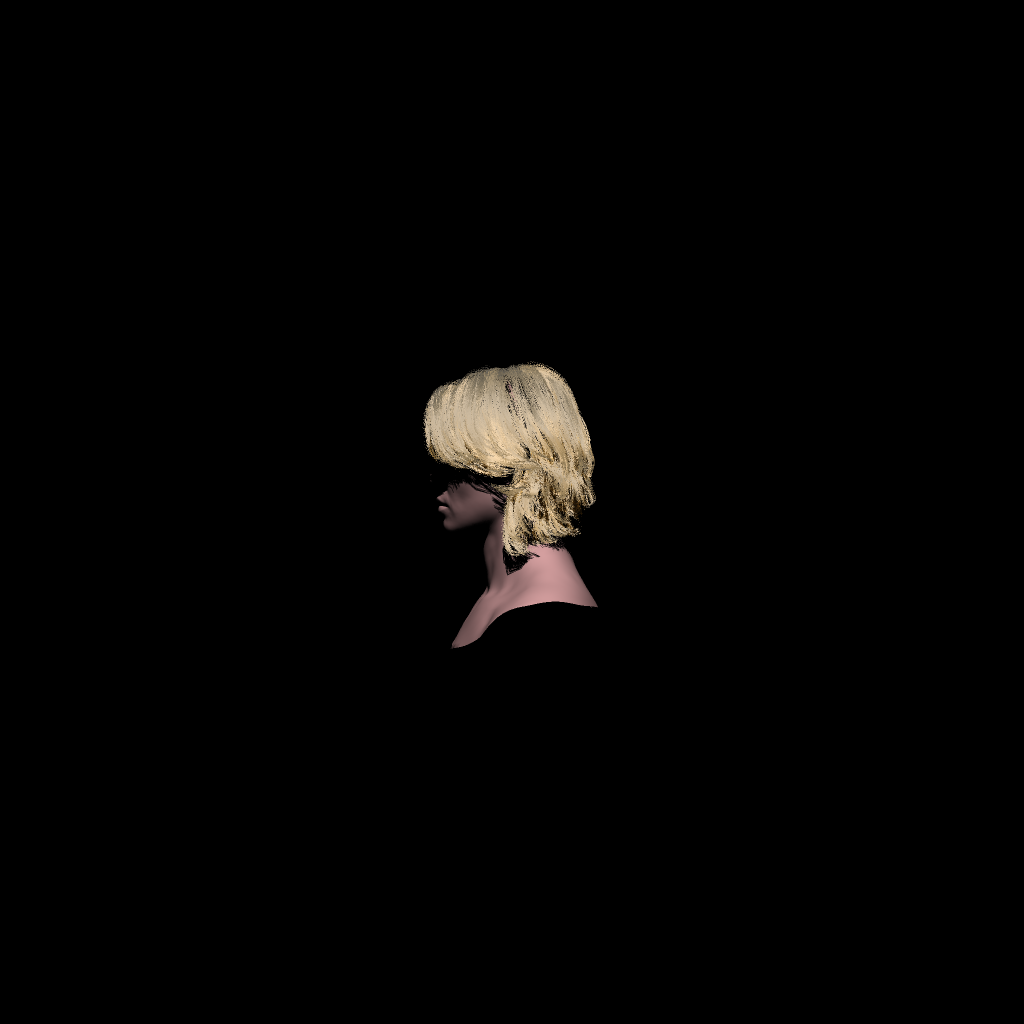}\hfill%
\includegraphics[trim={473 473 473 473},clip,width=0.315\linewidth]{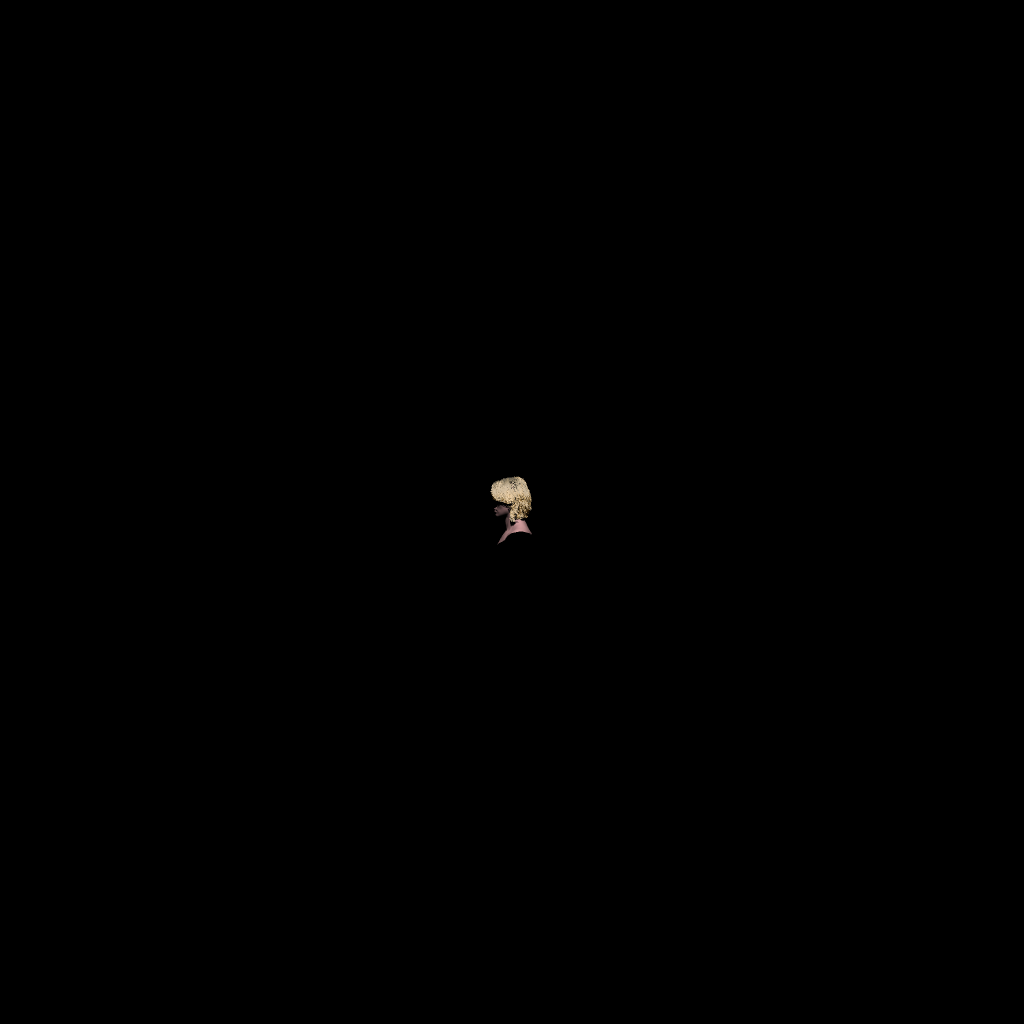}\\
\hspace{0.04\linewidth}\hfill%
\includegraphics[trim={0   0   0   0  },clip,width=0.315\linewidth]{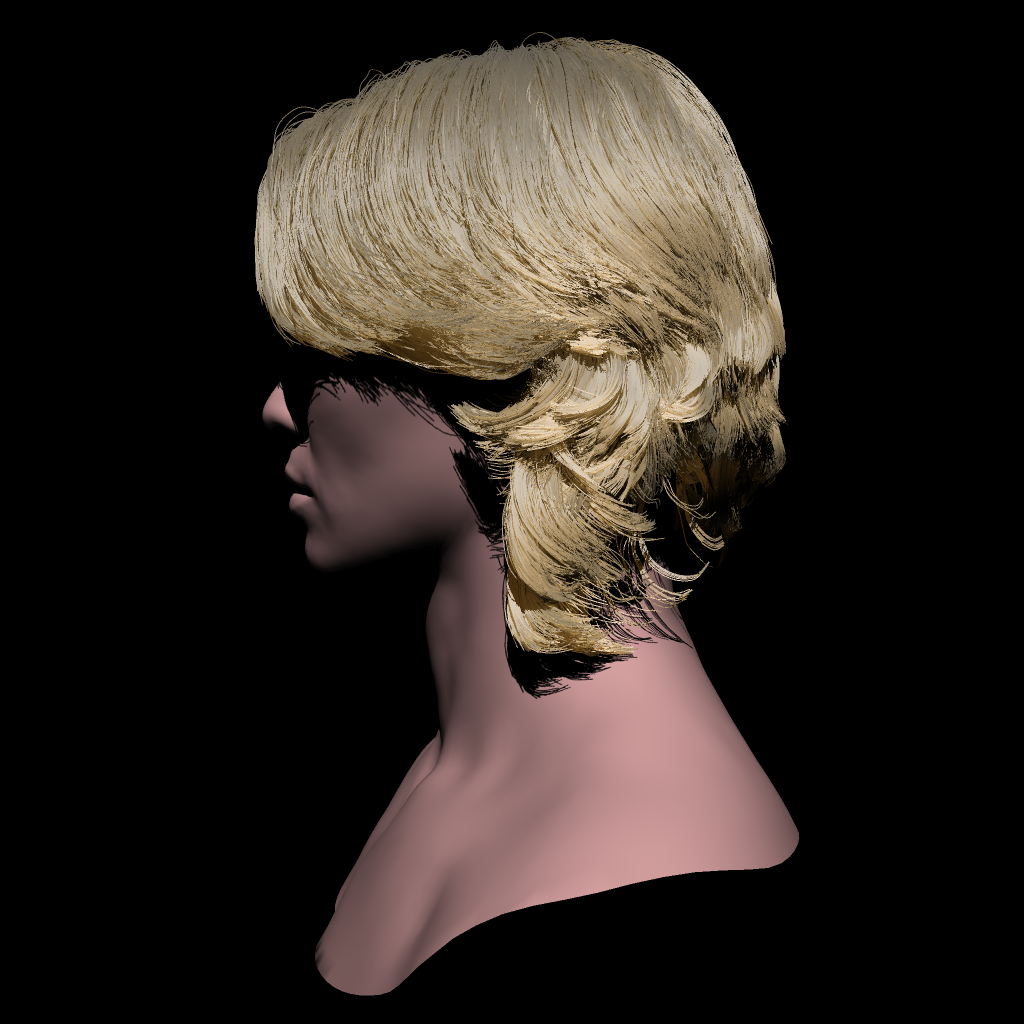}\hfill%
\includegraphics[trim={350 350 350 350},clip,width=0.315\linewidth]{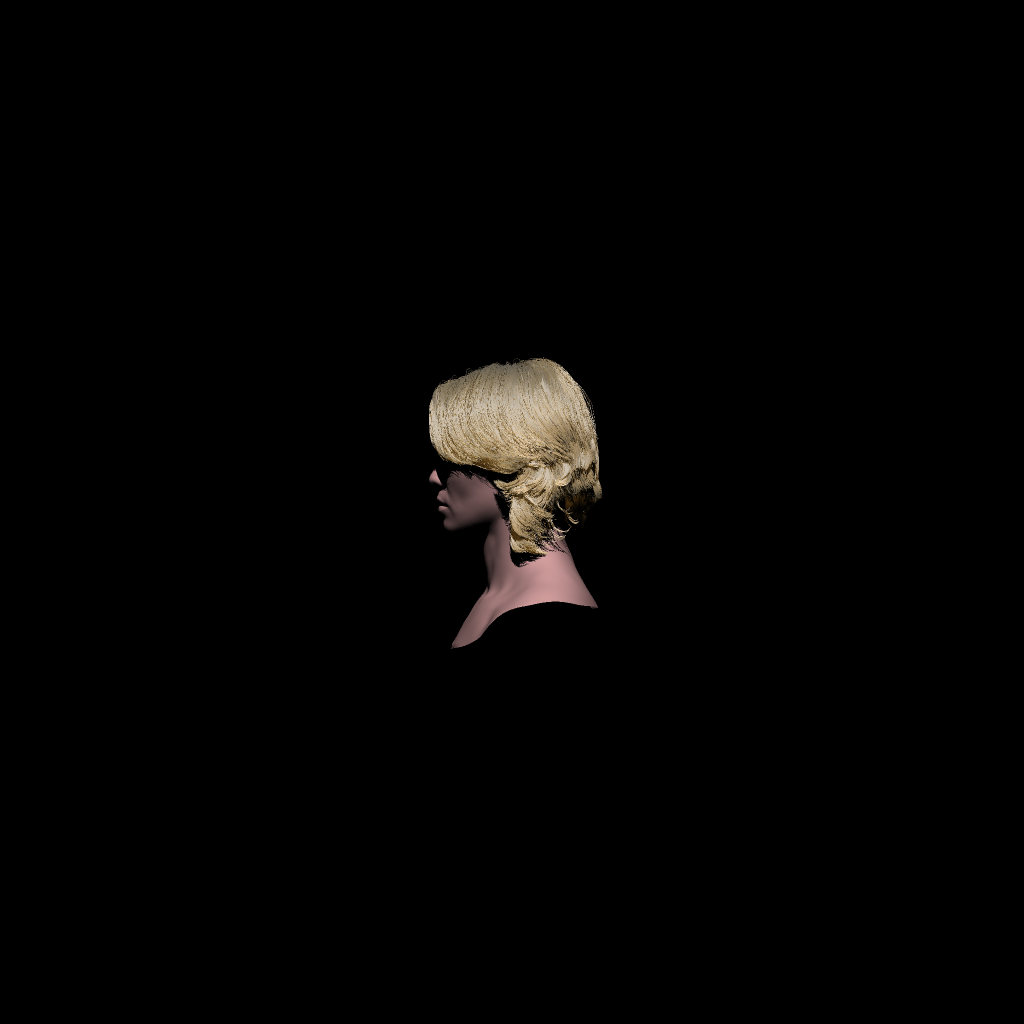}\hfill%
\includegraphics[trim={473 473 473 473},clip,width=0.315\linewidth]{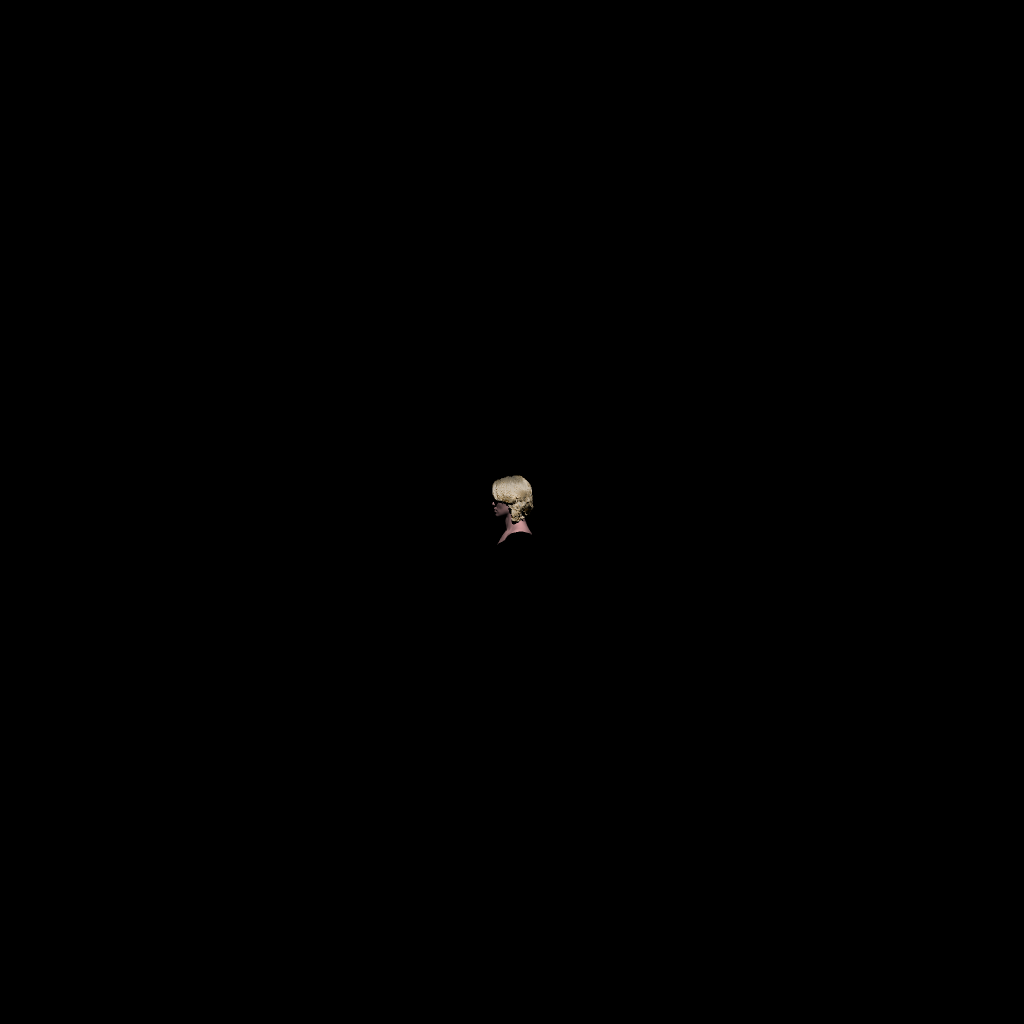}\\
\hspace{0.04\linewidth}\hfill%
\includegraphics[trim={0   0   0   0  },clip,width=0.315\linewidth]{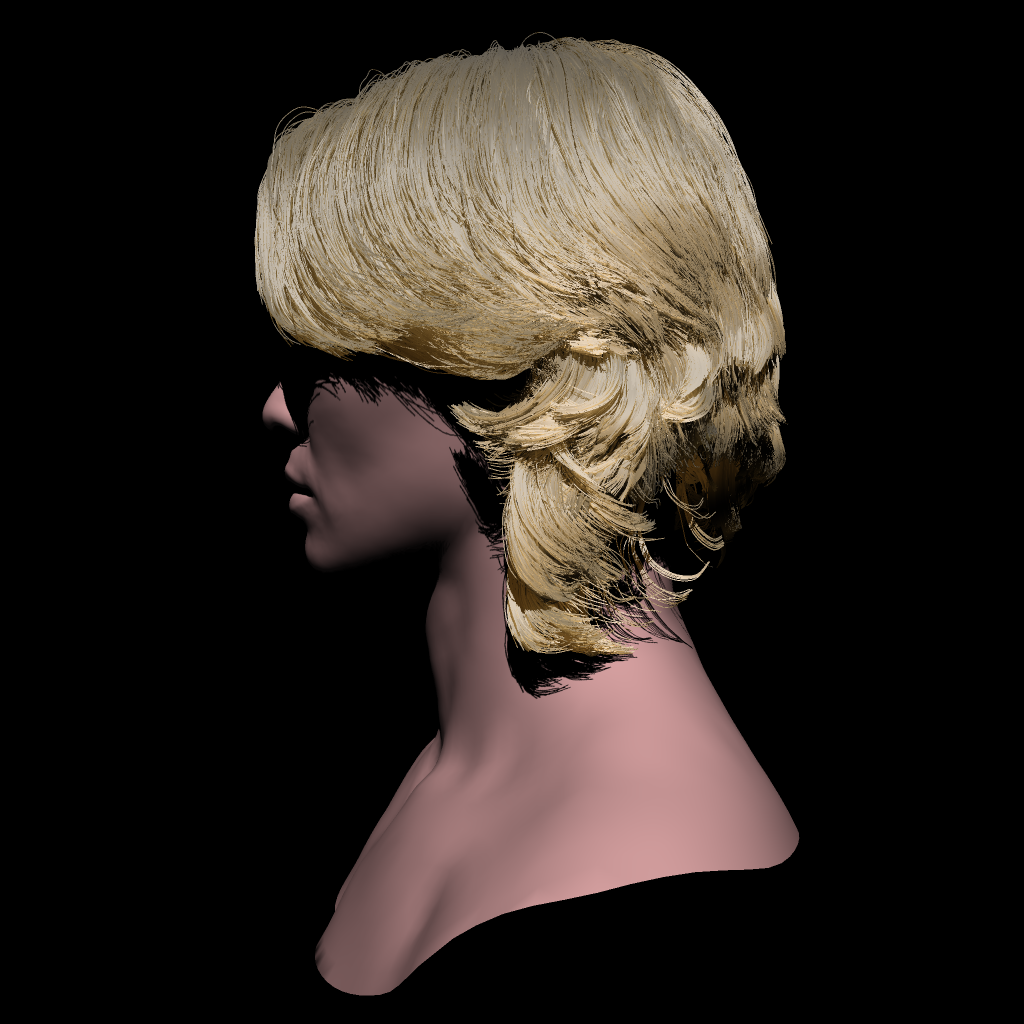}\hfill%
\includegraphics[trim={350 350 350 350},clip,width=0.315\linewidth]{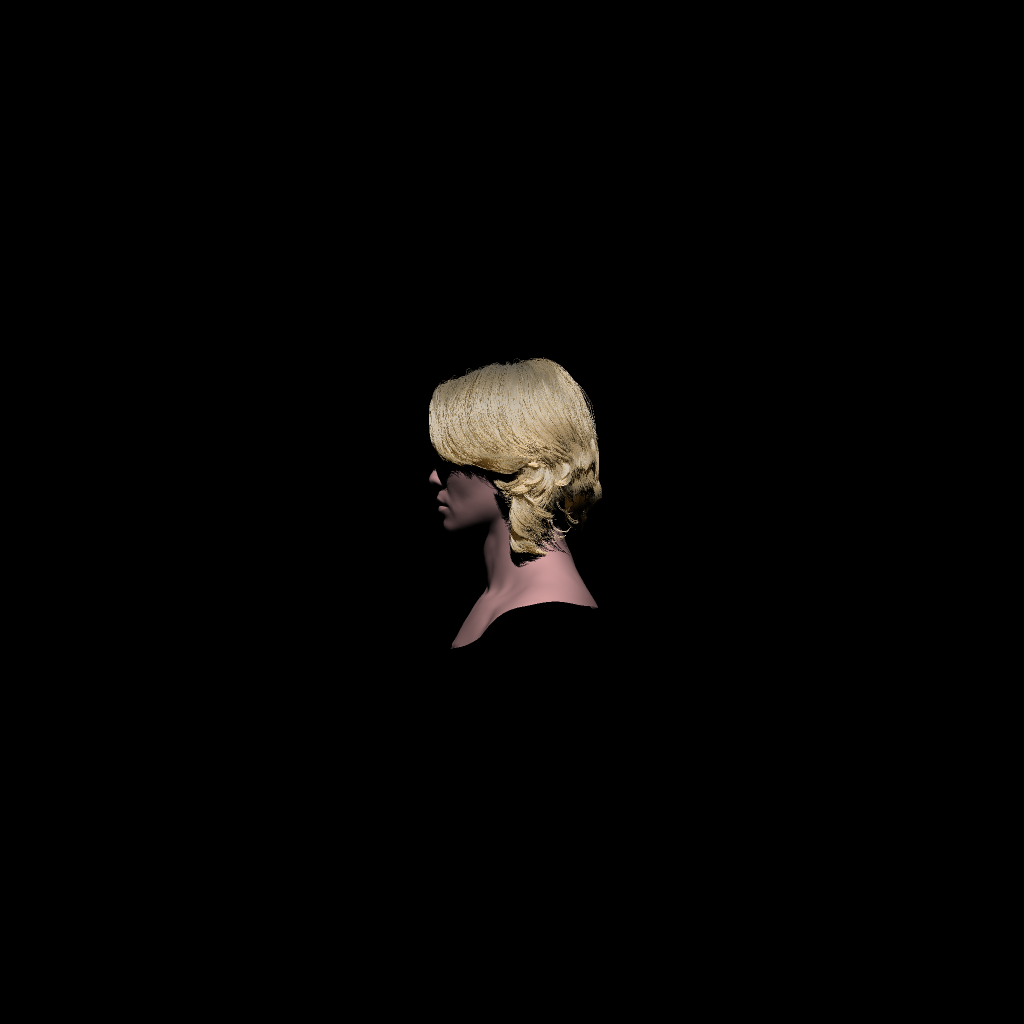}\hfill%
\includegraphics[trim={473 473 473 473},clip,width=0.315\linewidth]{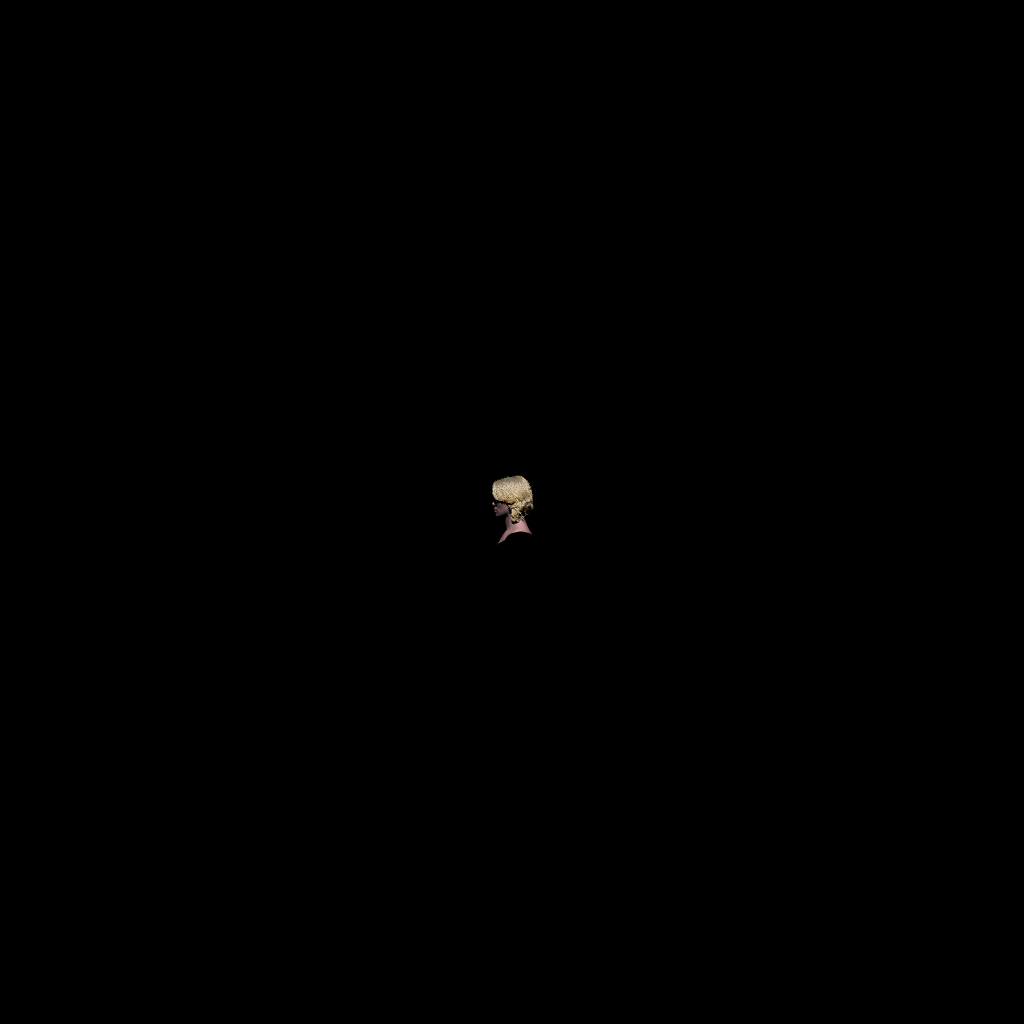}\\
\vspace{-18.6em}
\hspace{0.04\linewidth}\hfill%
\figcap{\small \color{white}{$60$K triangles}}\hfill%
\figcap{\small \color{white}{$30$K triangles}}\hfill%
\figcap{\small \color{white}{$7$K triangles}}\\
\vspace{7.4em}
\hspace{0.04\linewidth}\hfill%
\figcap{\small \color{white}{$1880$K triangles}}\hfill%
\figcap{\small \color{white}{$640$K triangles}}\hfill%
\figcap{\small \color{white}{$70$K triangles}}\\
\vspace{7.4em}
\hspace{0.04\linewidth}\hfill%
\figcap{\small \color{white}{$2100$K triangles}}\hfill%
\figcap{\small \color{white}{$2100$K triangles}}\hfill%
\figcap{\small \color{white}{$2100$K triangles}}\\
\vspace{-26.2em}
\begin{flushleft}
{\small \hspace{7.7em} \color{white}{4.1 ms}  \hspace{6.8em} \color{white}{1.6 ms} \hspace{6.8em} \color{white}{1.3 ms}}
\end{flushleft}
\vspace{7em}
\begin{flushleft}
{\small \hspace{7.3em} \color{white}{10.3 ms}  \hspace{6.8em} \color{white}{2.8 ms} \hspace{6.7em} \color{white}{0.4 ms}}
\end{flushleft}
\vspace{7em}
\begin{flushleft}
{\small \hspace{7.3em} \color{white}{10.3 ms}  \hspace{6.8em} \color{white}{5.8 ms} \hspace{6.7em} \color{white}{5.3 ms}}
\end{flushleft}
\vspace{-18.5em}
\begin{flushleft}{
\rotatebox{90}{\small \hspace{0.25em} {Full geo. (rasterizer)} \hspace{1.25em} \textbf{Ours (rasterizer)} \hspace{1.25em} Hair cards (rasterizer)}\hfill%
}\end{flushleft}
\caption{Full hair strand geometry, our strand-based LoD model, and hair card model at near, middle, and far views. We mark the performance and the number of triangles for hair strips in the image.}
\label{fig:haircard}
\Description{}
\end{figure}



\paragraph{Real-time performance} 
\autoref{fig:performance} illustrates the real-time performance of our GPU rasterizer pipeline. The most demanding scenario occurs with a close view (50\% screen occupation ratio), where only a limited number of single hairs can be represented by thick hairs. In this case, our approach needs to render a similar number of hair segments as the full hair geometry, along with additional compute passes for thick hair width computation. Nonetheless, our approach exhibits only a slight overhead (<1\%). As the camera distance increases, hair strands gradually transition to thicker hairs, reducing the number of segments required for rendering. This leads to a noticeable speedup of $2\times$ for the middle view (10\% of screen) and up to $13\times$ for the far view (0.3\% of screen). The breakdown of single frame time indicates that the percentage of time taken by the shading pass decreases with the camera distance due to the reduction in hair geometry and fewer pixels required to shade. Additionally, our pipeline requires a constant time of about 0.1 ms for hair data organization based on the new LoD level for the next frame rendering.

\paragraph{Comparison to hair cards}
\autoref{fig:haircard} compares our strand-based model with a hair cards model. The hair cards model has 3 LoD levels, featuring 600K, 300K, and 70K triangles, respectively. We manually configure two distances for the hair cards to do the LoD switching and render them using our real-time GPU rasterizer. Despite requiring more triangles in our LoD structure, our approach facilitates a seamless transition between LoD levels and supports strand-based hair simulation. In contrast, the hair card textures at different LoD levels often exhibit inconsistencies, leading to noticeable popping artifacts during the LoD transition. 

Note that we render haircards using an order-independent rasterization, meaning shading only the closest fragment without using alpha blending, for performance purposes. A threshold is set for the coverage texture, and fragments are discarded when their coverage falls below this threshold. Notably, in far-view scenarios, although hair cards have fewer triangles, they necessitate reading the coverage texture to compute accurate shadows. This requirement leads to higher computational costs compared to our method. 
%

\begin{figure}[ht]
\centering
\includegraphics[trim=0 10 0 15, clip, width =\linewidth]{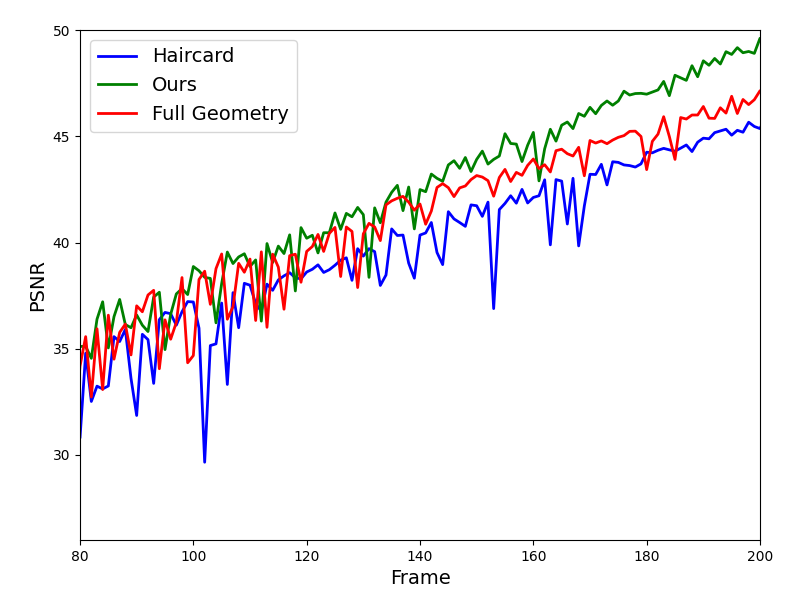}
\caption{Evaluation of transition smoothness. We utilize PSNR to measure the similarity of adjacent frames. Our method (Green) constantly has the highest score compared to full geometry (Red) and hair card (Blue) while avoiding the drastic changes in LoD switching.}
\label{fig:transition}
\Description{}
\end{figure}

Additionally, we measured the smoothness of our transitions by comparing the similarity of adjacent frames using PSNR. As shown in~\autoref{fig:transition}, our method does not experience the drastic changes that occur with hair cards during LoD transitions. Furthermore, due to the appropriate LOD level, the flickering in our real-time rendering is less pronounced than with full geometry. Our PSNR remains consistently at the highest level among the three methods. For a more detailed comparison of LoD transitions, please refer to the supplemental video.

\begin{figure}[ht]
\centering
\newcommand{\figcap}[1]{\begin{minipage}{0.495\linewidth}\centering#1\end{minipage}}
\hspace{0.04\linewidth}\hfill%
\includegraphics[trim={40 22 40 40},clip,height=0.35\linewidth]{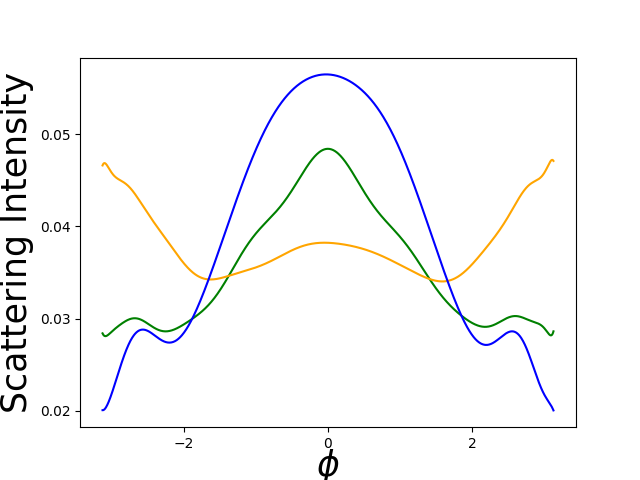}\hfill%
\includegraphics[trim={40 22 40 40},clip,height=0.35\linewidth]{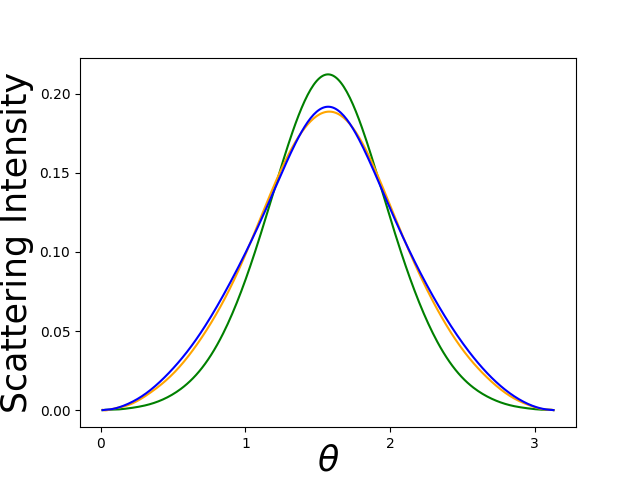}\\
\vspace{-9em}
\begin{flushleft}{
\rotatebox{90}{\small \hspace{1em} {Scattering intensity} }\hfill%
}\end{flushleft}
\figcap{\small \hspace{2em} Azimuthal profile $\phi$ }\hfill%
\figcap{\small \hspace{0.5em} Longitudinal profile $\theta$ }\\
\caption{The average scattering intensity (green) across 100 ply instances with distinct fiber geometries. Our method (blue) offers a more accurate approximation than that of ~\cite{ZhuZJYA23} (orange), particularly in the azimuthal profile. }
\label{fig:scatterplot}
\Description{}
\end{figure}

\paragraph{Comparison to~\cite{ZhuZJYA23}}
We utilize CPU path tracing to render 100 ply instances with diverse fiber geometries and average the rendering outcomes.  Additionally, we illustrate the scattering intensity at the azimuthal and longitudinal planes in~\autoref{fig:scatterplot}. The aggregated model proposed by~\citet{ZhuZJYA23} does not align well with the reference, particularly under backlit conditions, whereas our model provides a more accurate approximation.

\begin{figure}[ht!]
\newcommand{\figcap}[1]{\begin{minipage}{0.33\linewidth}\centering#1\end{minipage}}
{\includegraphics[trim={400 450 350 450},clip,width=0.32\linewidth]{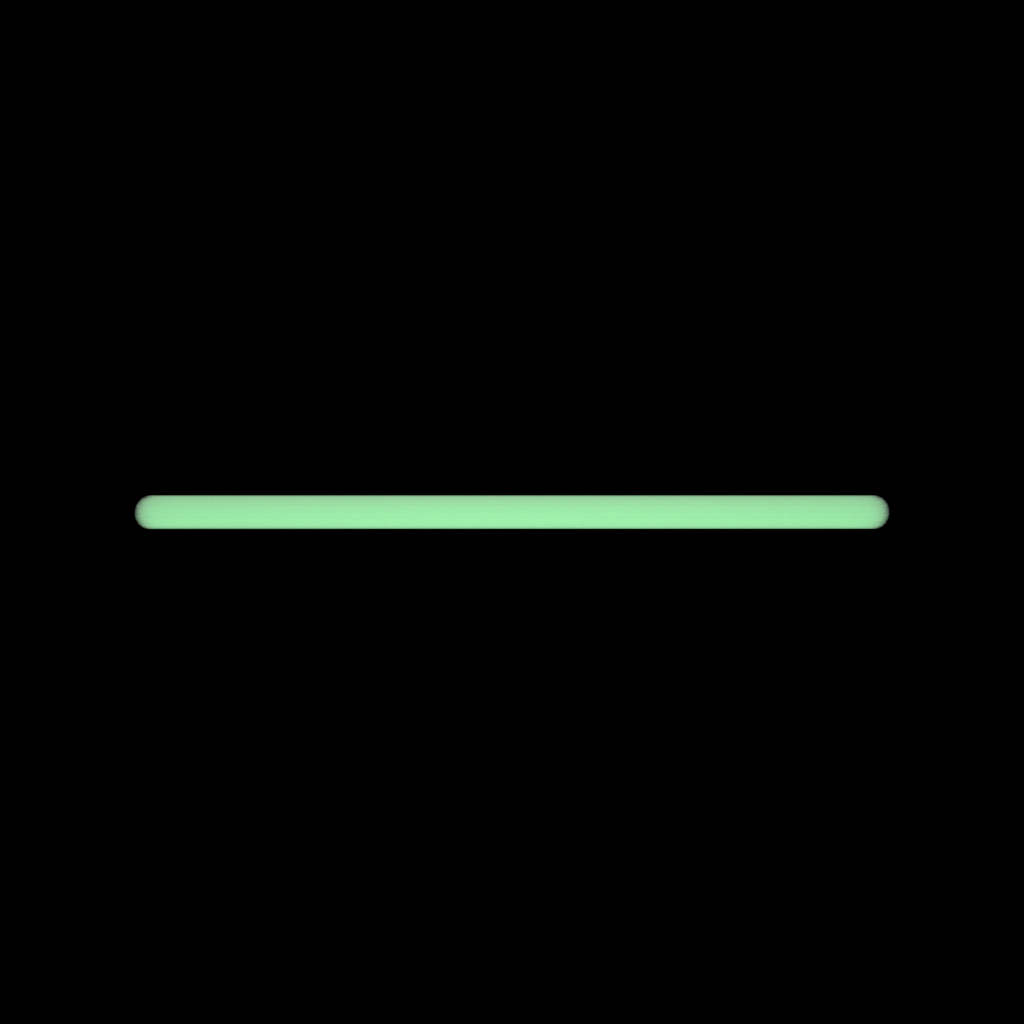}}\hfill%
{\includegraphics[trim={400 450 350 450},clip,width=0.32\linewidth]{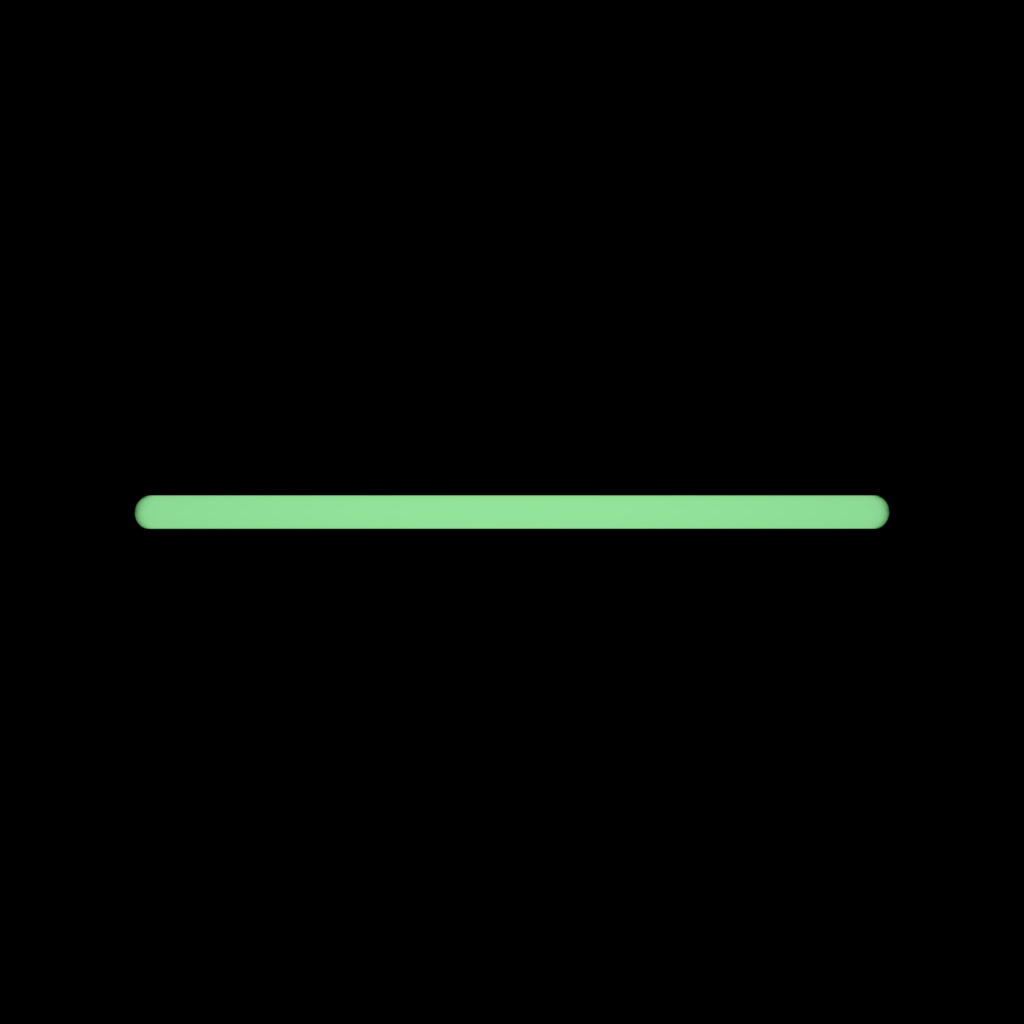}}\hfill%
{\includegraphics[trim={400 450 350 450},clip,width=0.32\linewidth]{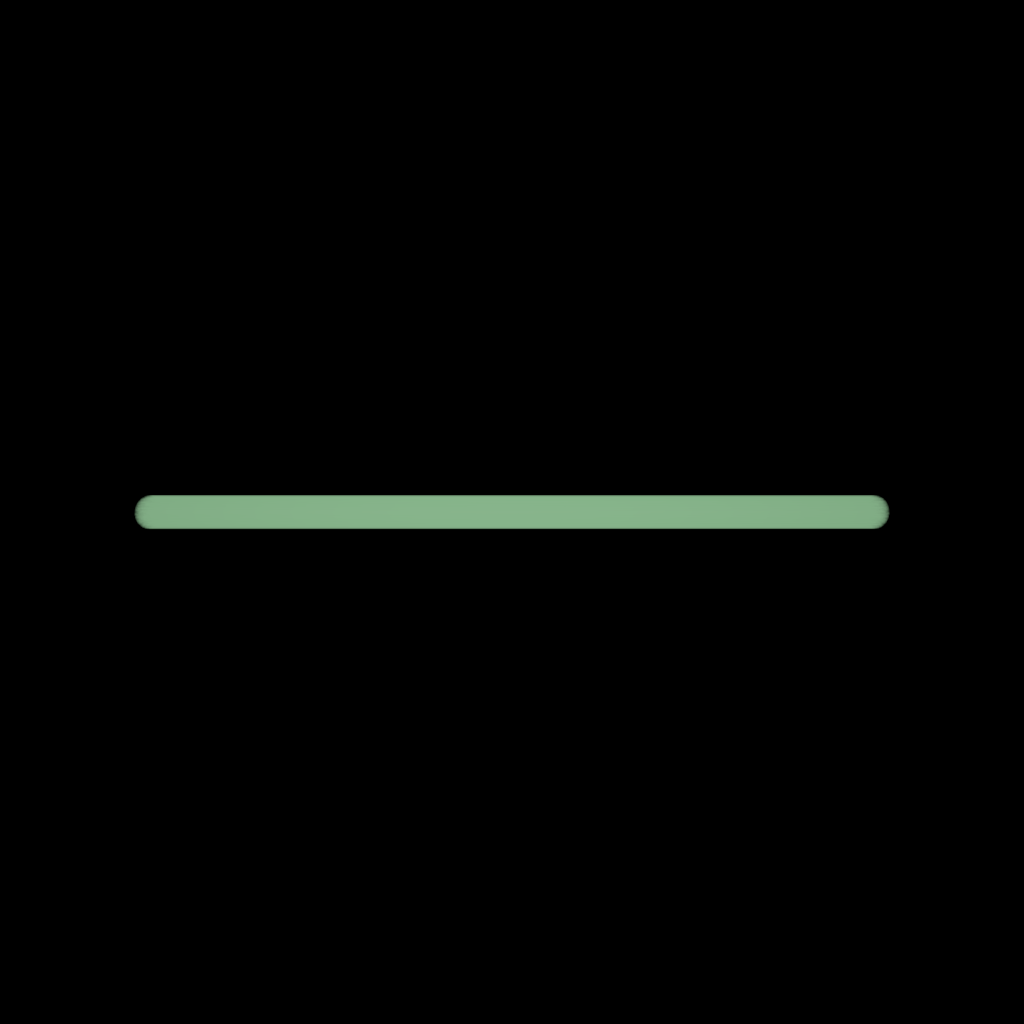}}{$\quad$~}\\
{\includegraphics[trim={400 450 350 450},clip,width=0.32\linewidth]{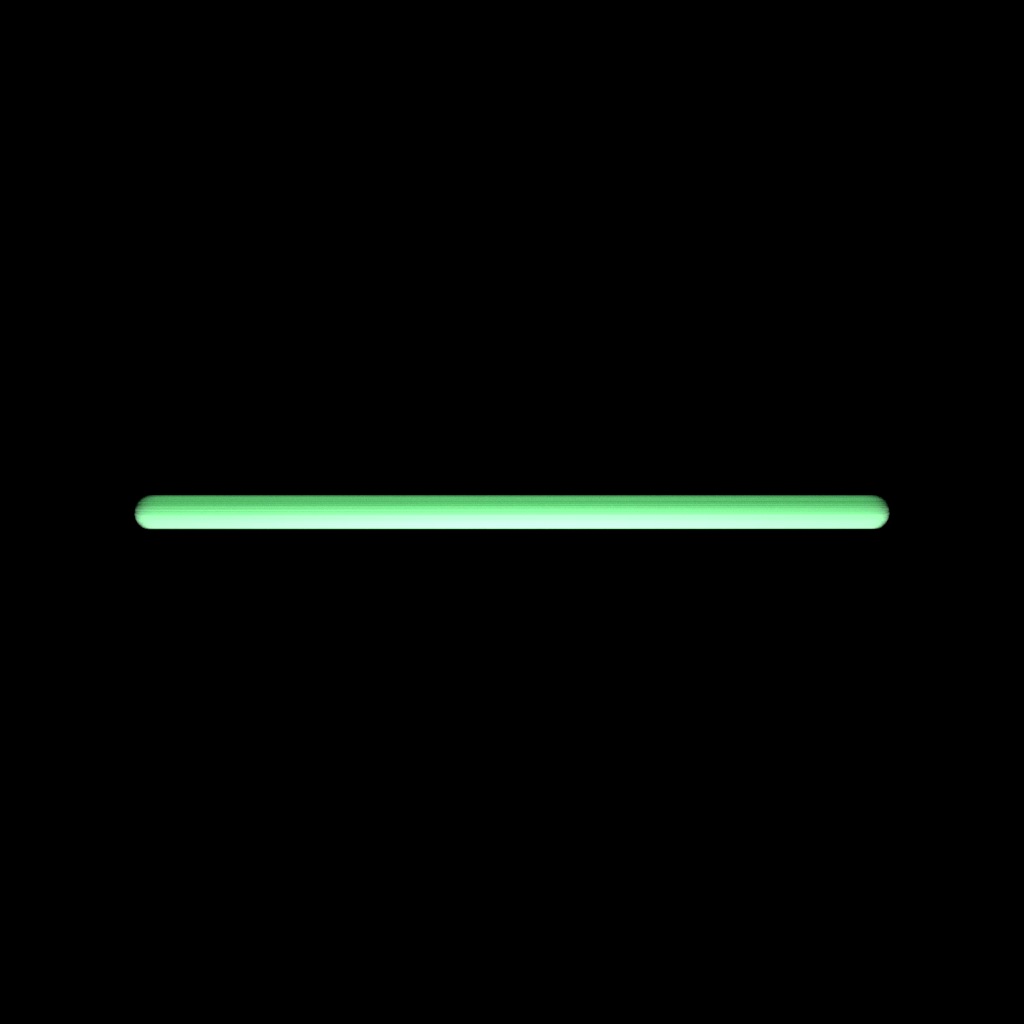}}\hfill%
{\includegraphics[trim={400 450 350 450},clip,width=0.32\linewidth]{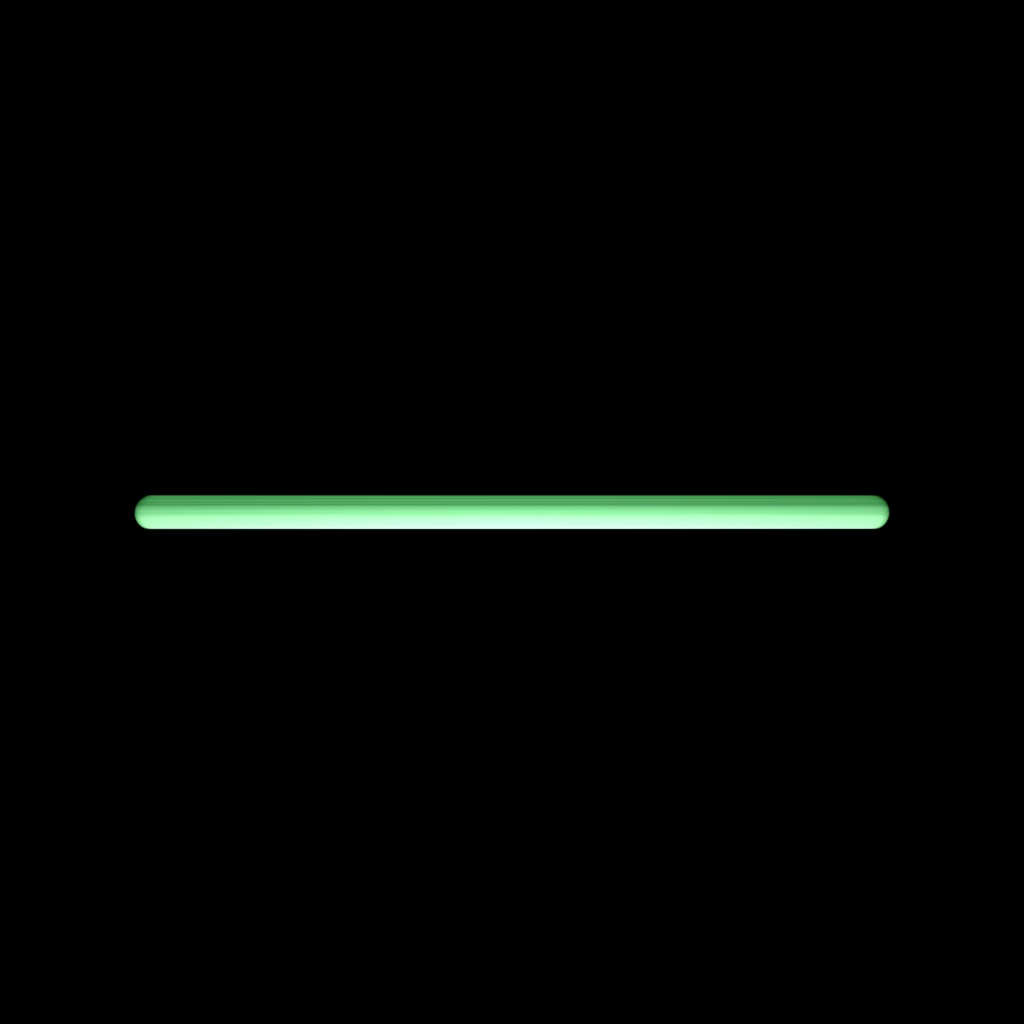}}\hfill%
{\includegraphics[trim={400 450 350 450},clip,width=0.32\linewidth]{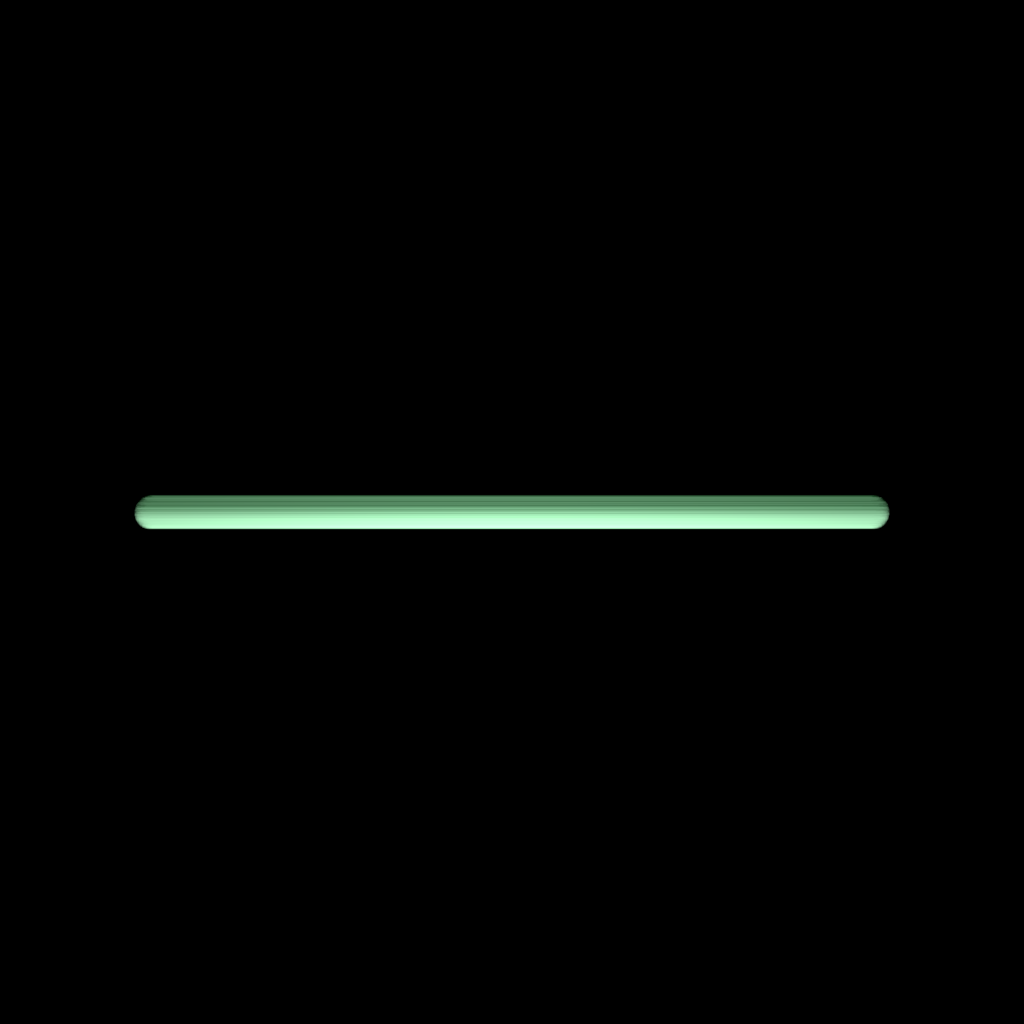}}{$\quad$~}\\
{\includegraphics[trim={400 450 350 450},clip,width=0.32\linewidth]{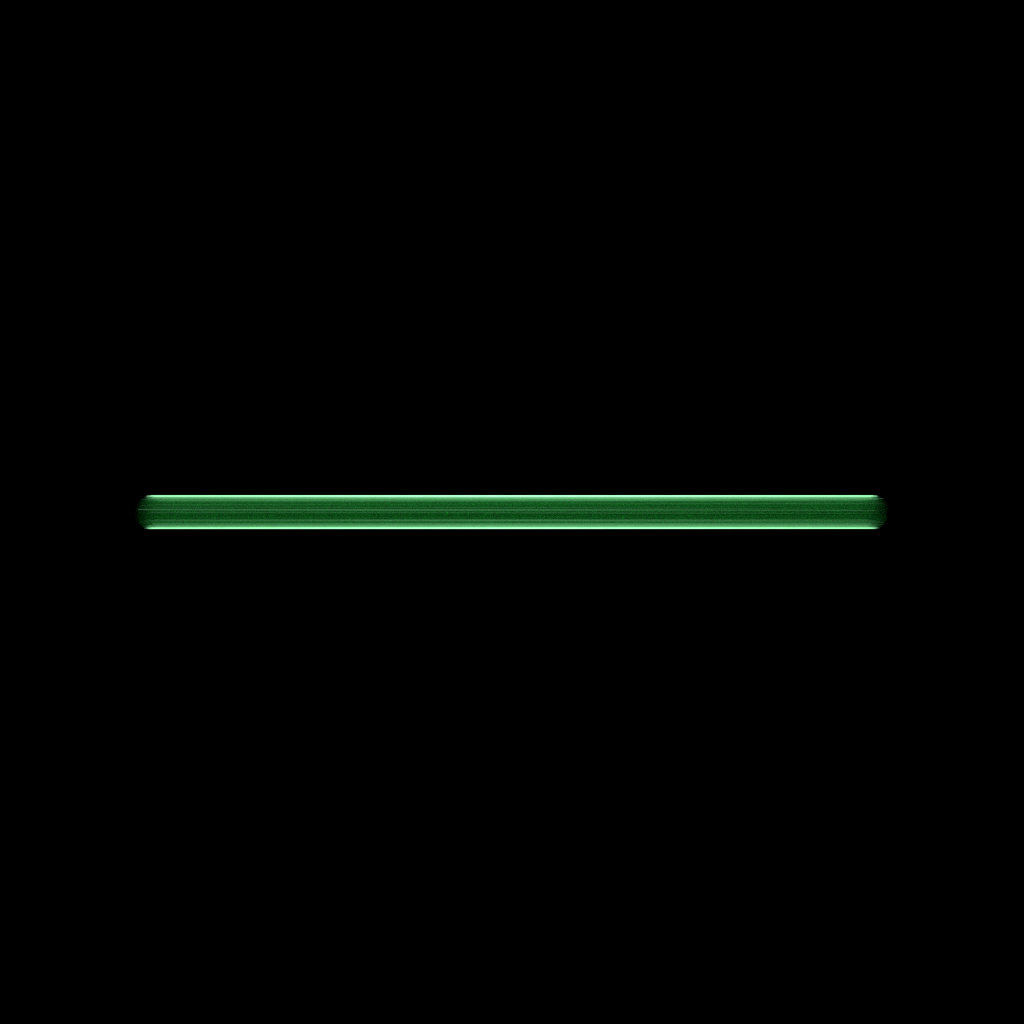}}\hfill%
{\includegraphics[trim={400 450 350 450},clip,width=0.32\linewidth]{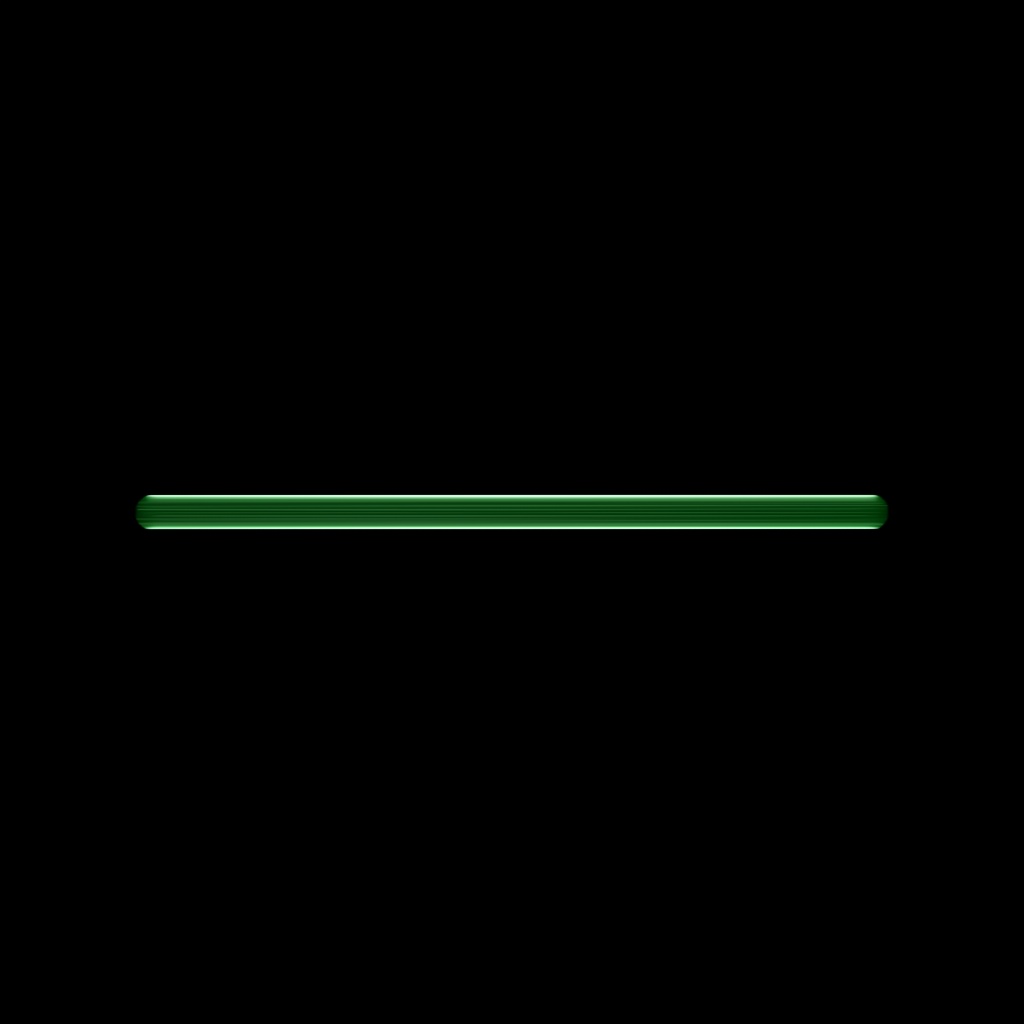}}\hfill%
{\includegraphics[trim={400 450 350 450},clip,width=0.32\linewidth]{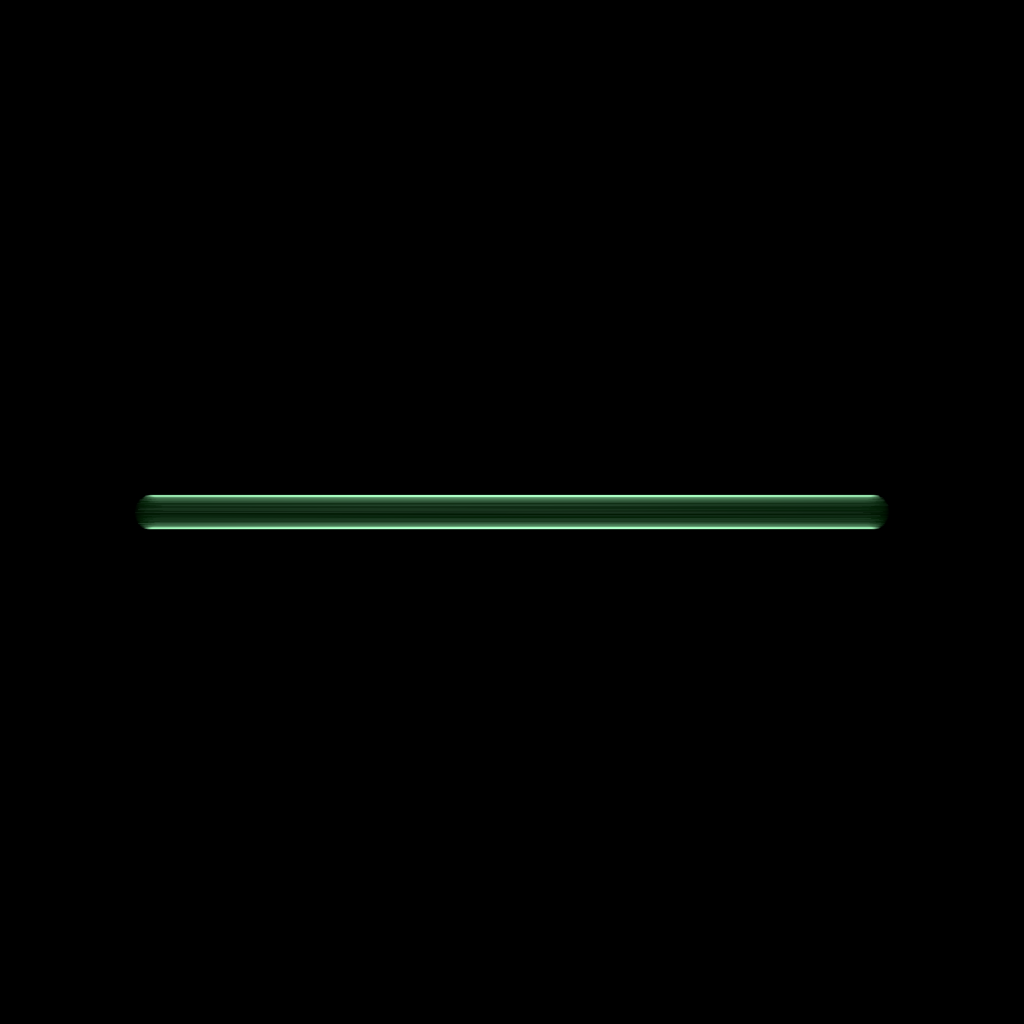}}{$\quad$~}\\
\vspace{-11.5em}
\begin{flushright}
\rotatebox{90}{\small \hspace{0.6em} Backlit \hspace{1.2em} Toplit \hspace{1.2em} Frontlit}   
\end{flushright}
\vspace{-0.5em}
\begin{flushleft}
{\small \hspace{2.75em} Avg. ref. \hspace{5em} \textbf{Avg. DS w/ {$A_{1_{+}}$}} \hspace{4em} Avg. DS }   
\end{flushleft}
\vspace{-0.5em}
\caption{Given a cluster of fibers forming an elliptical cross-section with a major axis to minor axis ratio of $4:1$, $A_{1_+}$ term can compensate for dual scattering to closely match the average reference (avg. ref.). The major axis is pointing to the camera. Avg. indicates averaging rendering results from 100 different fiber instances.}
\label{fig:a1plus}
\Description{}
\end{figure}

\paragraph{Ablation study on $A_{1_{+}}$} 
To assess the influence of $A_{1_{+}}$ path series, we conduct an ablation study involving the averaging of rendering results from 100 different fiber instances. Each instance is rendered with single BCSDF with full single fiber geometries using CPU path tracing. The fiber cluster forms an elliptical cross-section with a major axis to minor axis ratio of $4:1$. \autoref{fig:a1plus} demonstrate that the $A_{1_+}$ term effectively compensates for dual scattering, resulting in close alignment with the average reference (avg. ref.).

\begin{figure}[ht!]
\newcommand{\figcap}[1]{\begin{minipage}{0.495\linewidth}\centering#1\end{minipage}}
\includegraphics[trim=75 0 75 150,clip, width=0.499\linewidth]{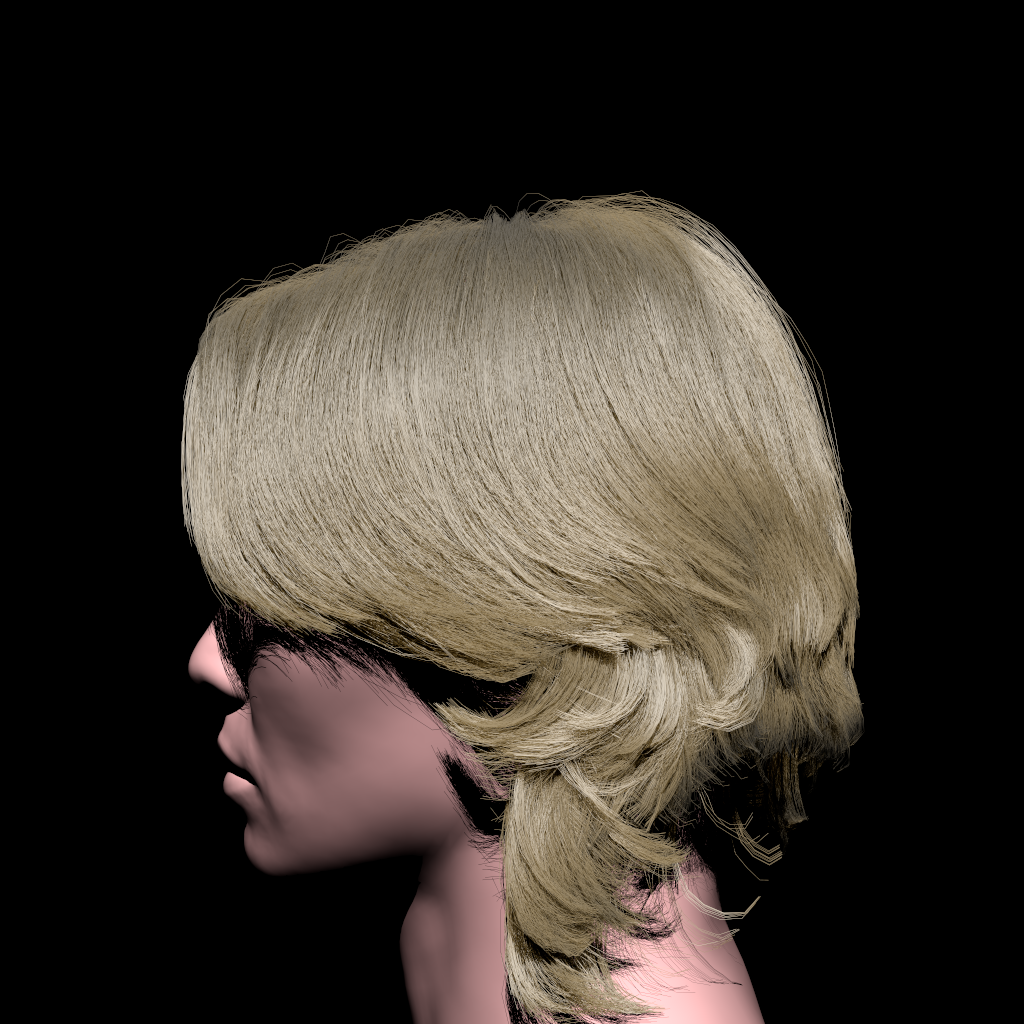}\hfill
\includegraphics[trim=75 0 75 150,clip, width=0.499\linewidth]{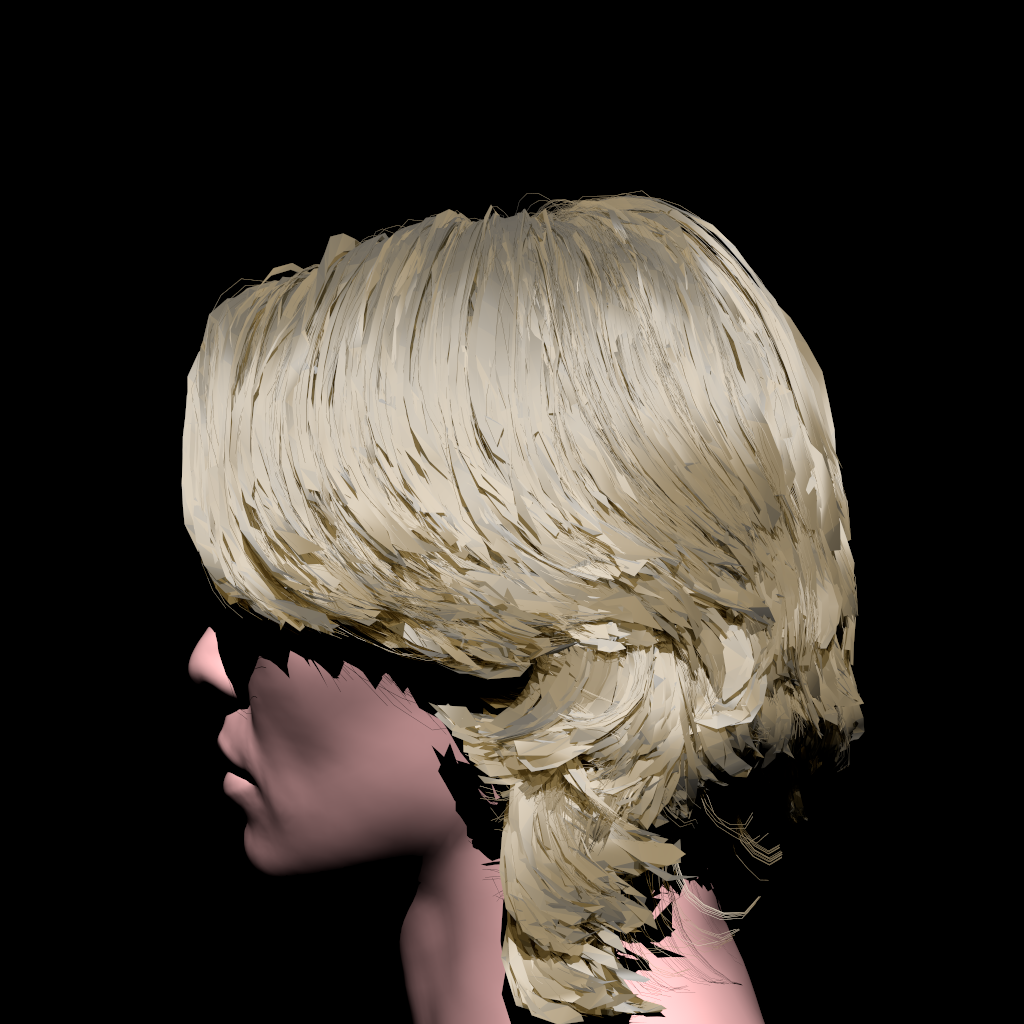}\\
\figcap{\small Full geo. w/ $f^\text{single}$ }\hfill%
\figcap{\small LoD w/ $f^\text{single}$ (Over bright)}\vspace{0.em}\\\vspace{0.5em}
\includegraphics[trim=75 0 75 150,clip, width=0.499\linewidth]{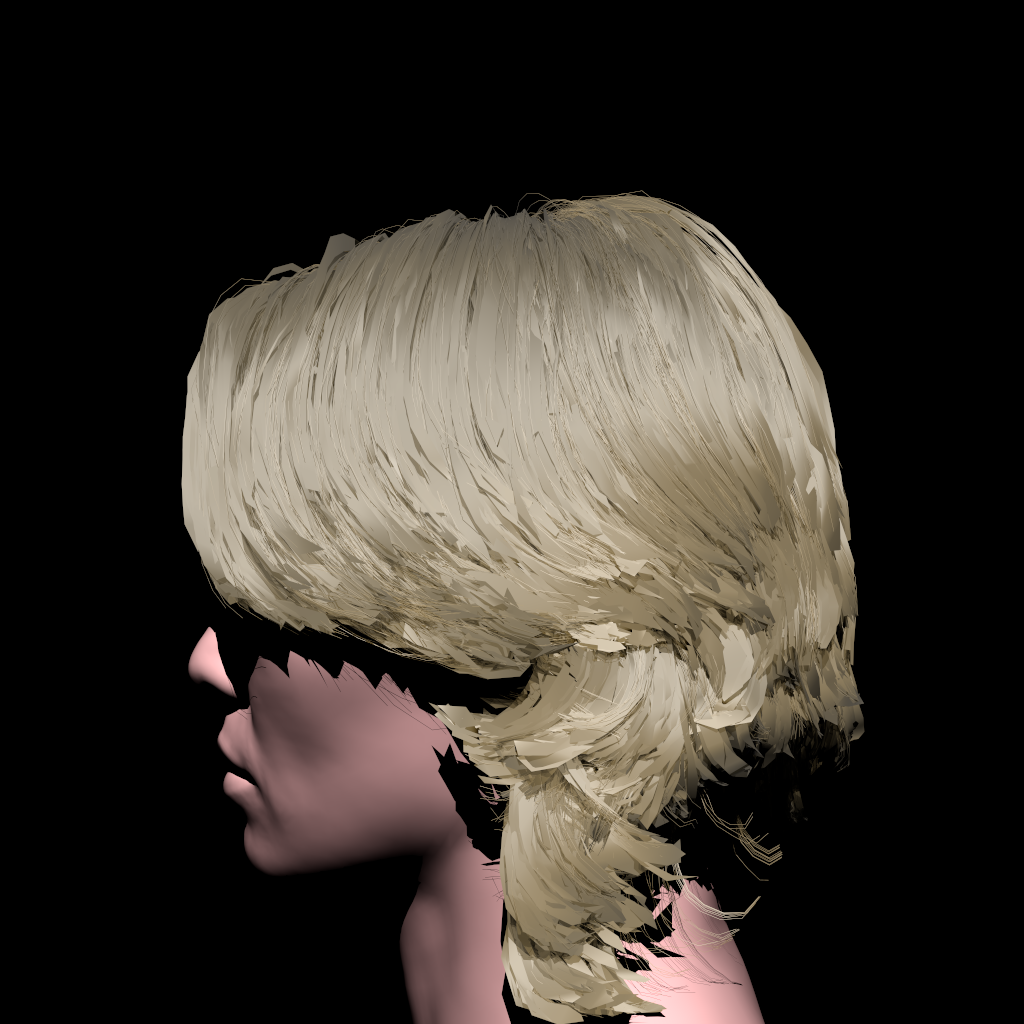}\hfill
\includegraphics[trim=75 0 75 150,clip, width=0.499\linewidth]{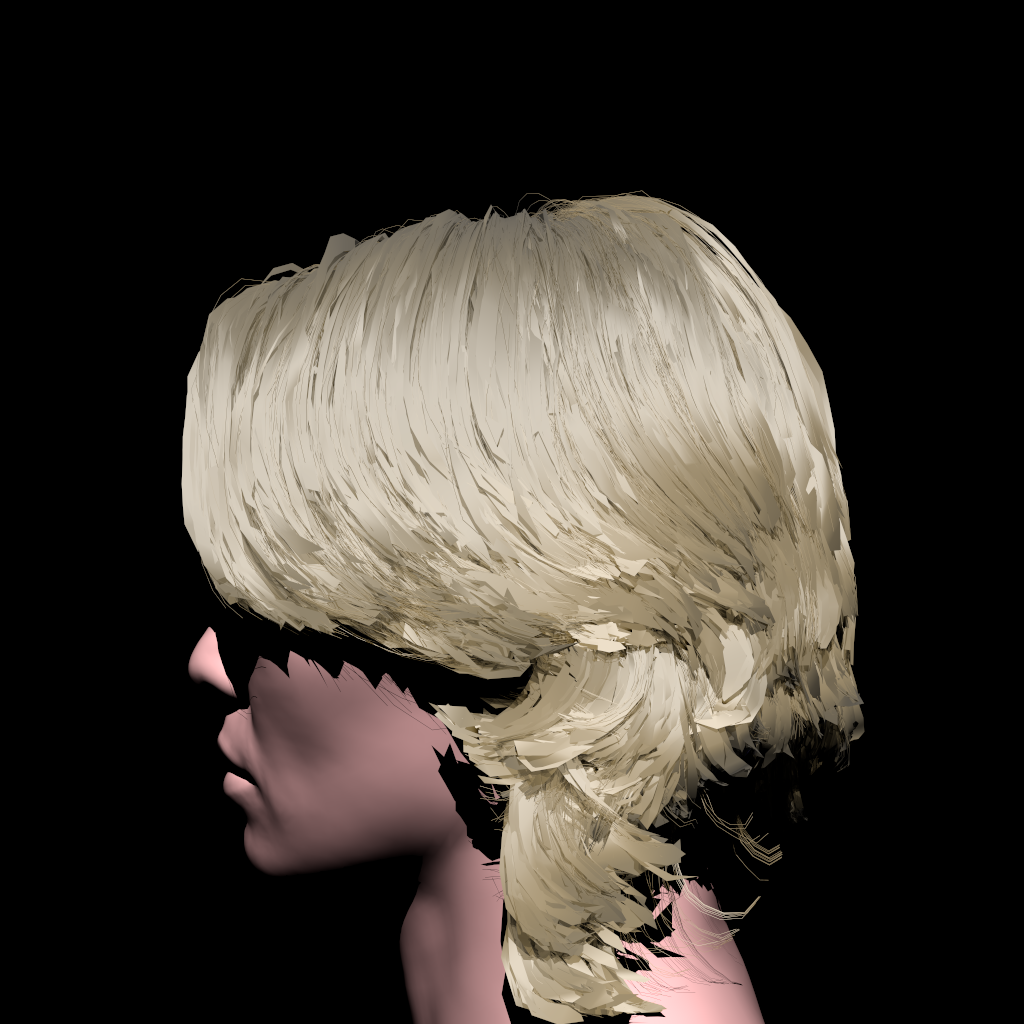}\\
\figcap{\small \textbf{LoD w/ $f^\text{our}$} }\hfill%
\figcap{\small LoD w/o $S(\phi, \theta_d)$ (Over bright)}\\
\caption{Compared to rendering full hair geometry with single BCSDF $f^\text{single}$, using thick hair to reduce the hair size need our aggregated BCSDF $f^\text{our}$ combined with shadow masking term $S(\phi, \theta_d)$.}
\label{fig:sterm}
\Description{}
\end{figure}

\paragraph{The shadowing-masking term and aggregated BCSDF}
To evaluate the impact of the shadowing-masking term in our aggregated BCSDF, we render a full hair geometry with single hair BCSDF $f^\text{single}$ as a reference and render a simplified hair geometry with our aggregated BCSDF $f^\text{our}$ with and without the shadowing-masking term $S(\phi, \theta_d)$. In the simplified hair geometry, each thick hair contains around 256 single hairs. As shown in \autoref{fig:sterm}, omitting the shadowing-masking term results in a brighter output compared to the reference, as the occlusion inside the thick hair is ignored. Further, applying the single hair BCSDF $f^\text{single}$ on the simplified hair geometry results in significant appearance bias, as expected. This bias occurs because using a single BCSDF for a cluster of hairs tends to produce brighter results, failing to account for the actual number of hairs traversed by a ray.

\begin{figure*}[ht]
\centering
\newcommand{\figcap}[1]{\begin{minipage}{0.16\linewidth}\centering#1\end{minipage}}
\centering
\hspace{0.03\linewidth}\hfill%
\figcap{\small Near view (60\%)}\hfill\figcap{\small Middle view (20\%)}\hfill\figcap{\small Far view (2\%)}\hfill
\figcap{\small Near view (60\%)}\hfill\figcap{\small Middle view (20\%)}\hfill\figcap{\small Far view (2\%)}\\
\hspace{0.03\linewidth}\hfill%
\includegraphics[trim={0 0 0 0},        clip,width=0.16\linewidth]{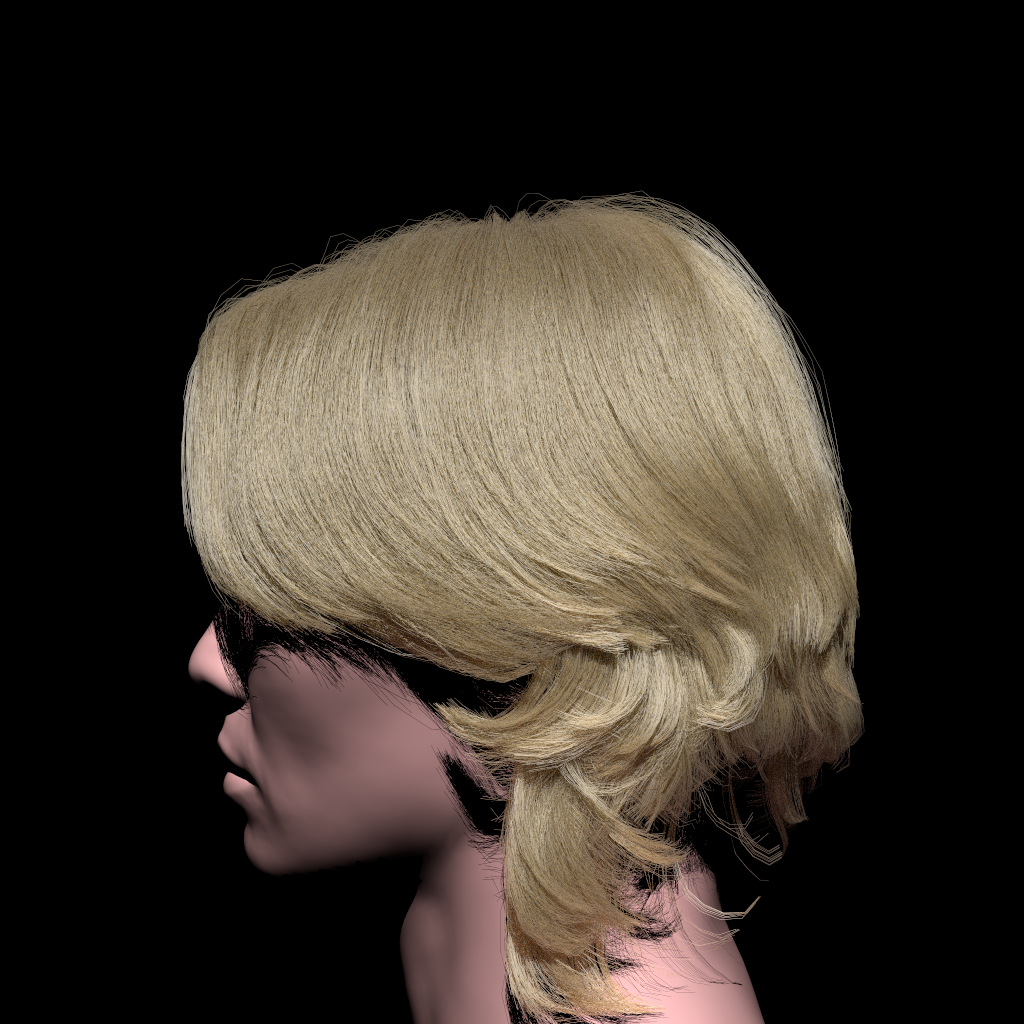}\hfill%
\includegraphics[trim={325 325 325 325},clip,width=0.16\linewidth]{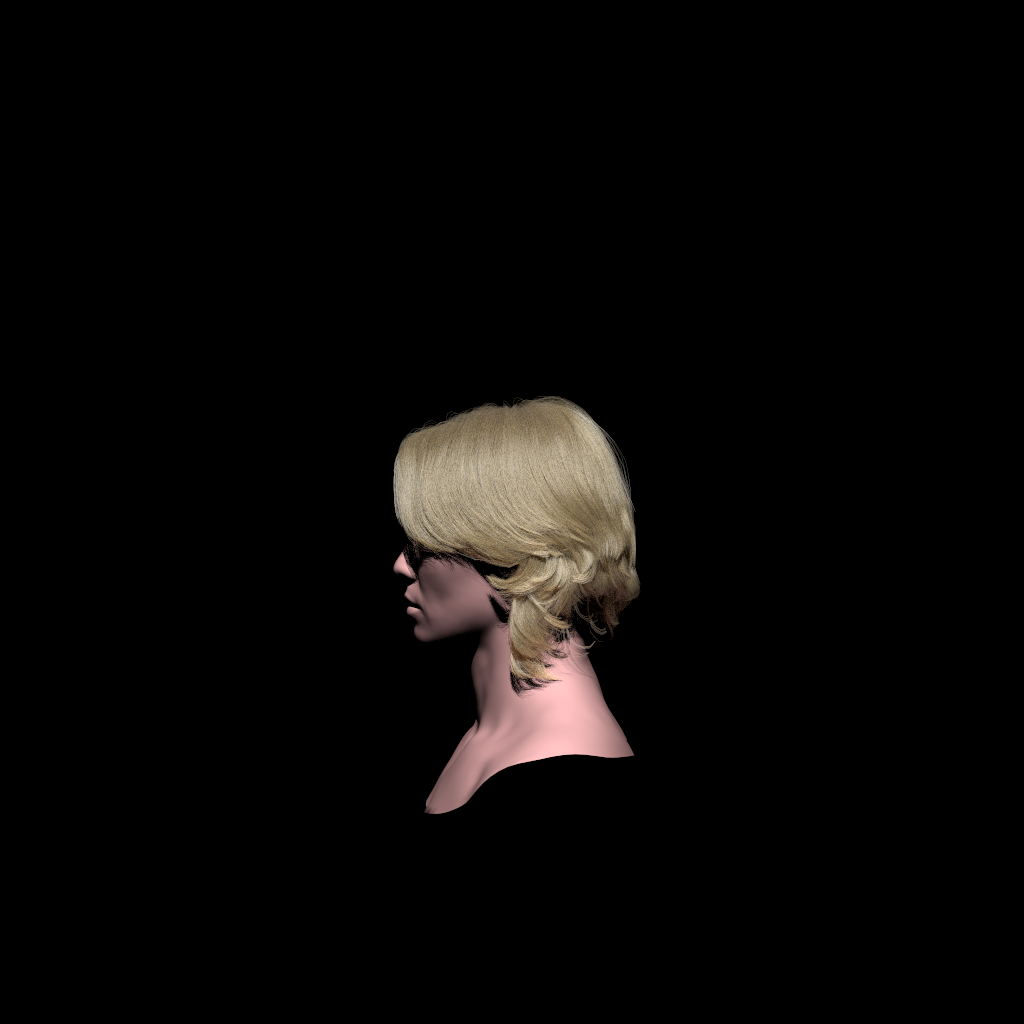}\hfill%
\includegraphics[trim={450 450 450 450},clip,width=0.16\linewidth]{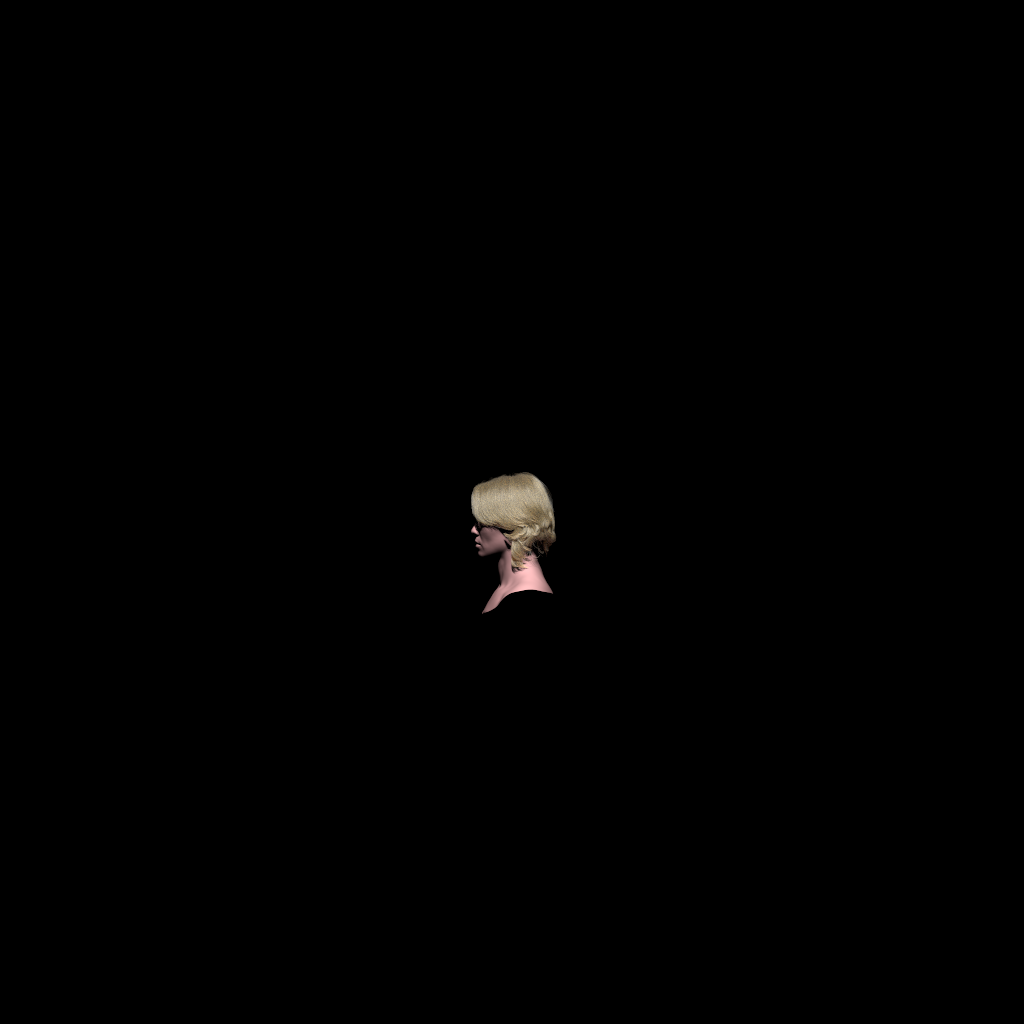}\hfill%
\includegraphics[trim={0 0 0 0},        clip,width=0.16\linewidth]{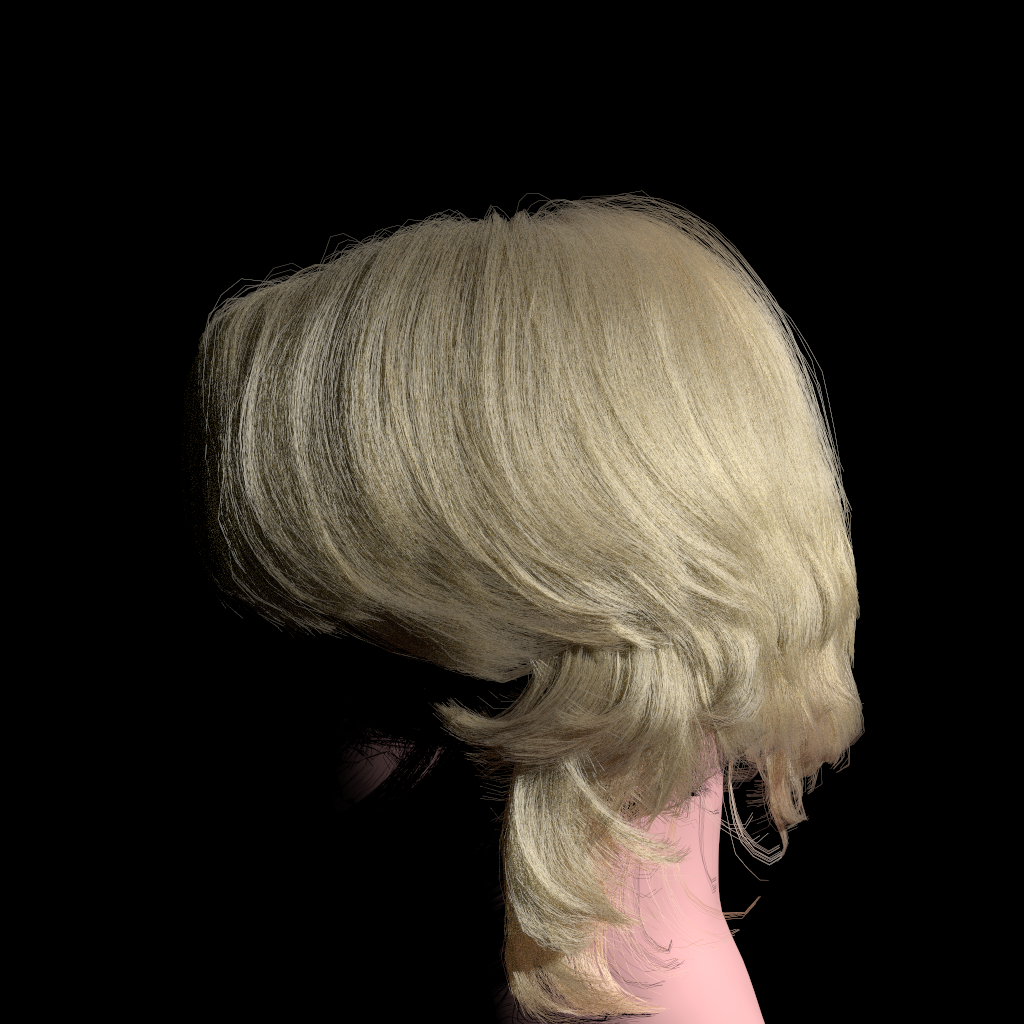}\hfill%
\includegraphics[trim={325 325 325 325},clip,width=0.16\linewidth]{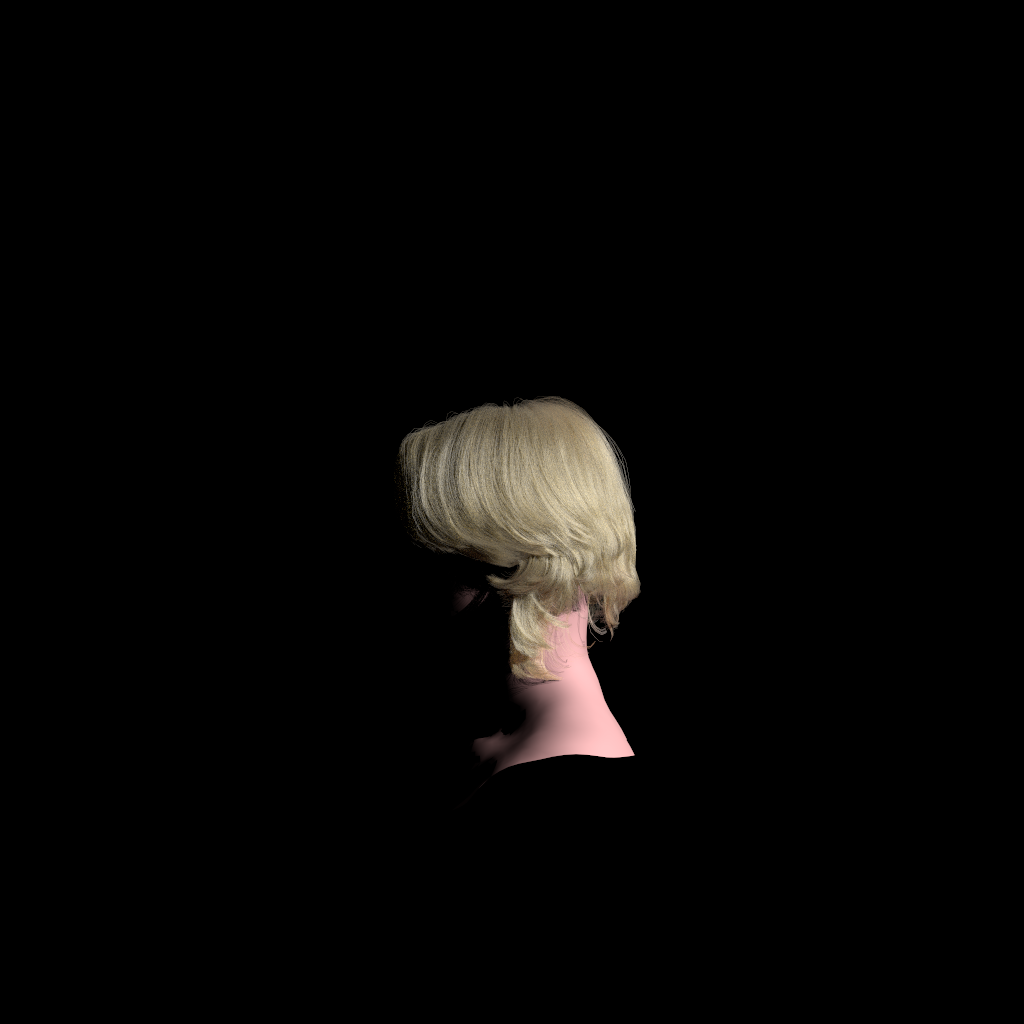}\hfill%
\includegraphics[trim={450 450 450 450},clip,width=0.16\linewidth]{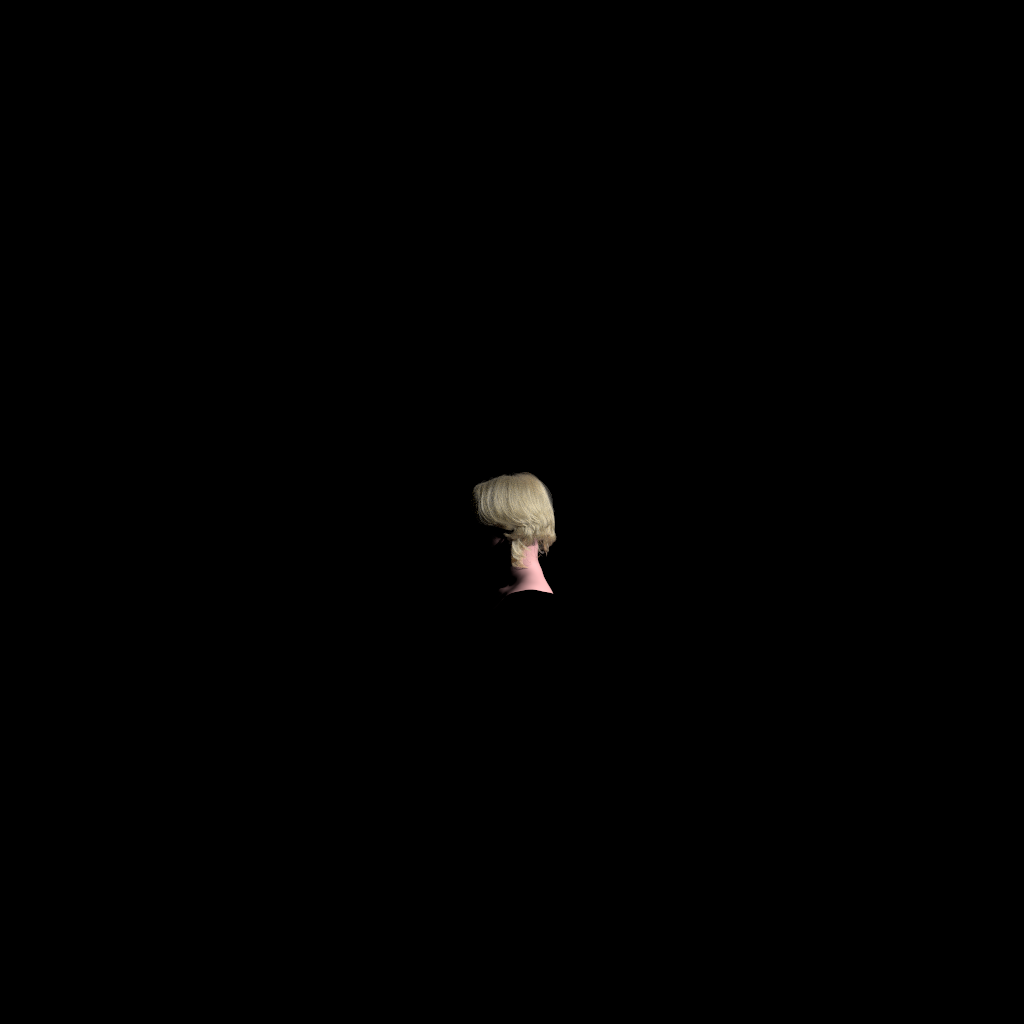}\\
\hspace{0.03\linewidth}\hfill%
\includegraphics[trim={0 0 0 0},        clip,width=0.16\linewidth]{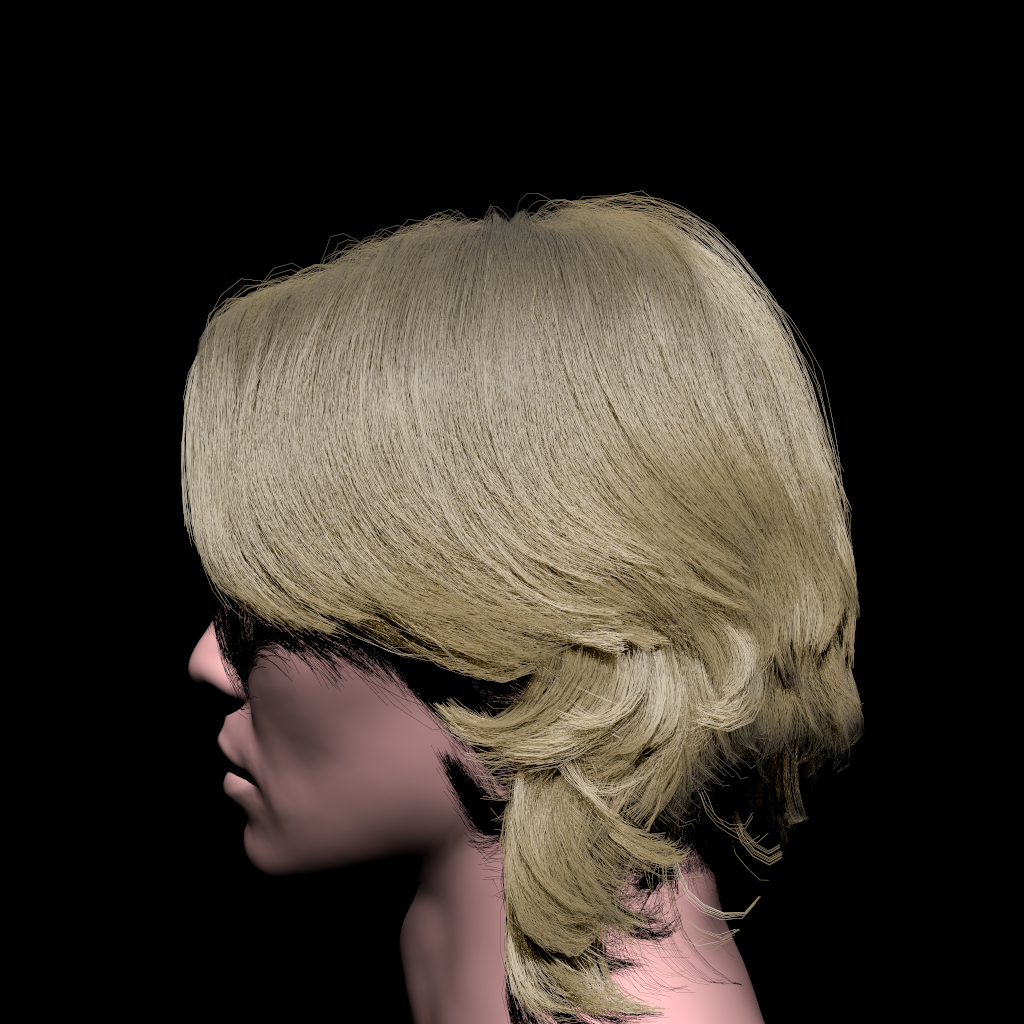}\hfill%
\includegraphics[trim={325 325 325 325},clip,width=0.16\linewidth]{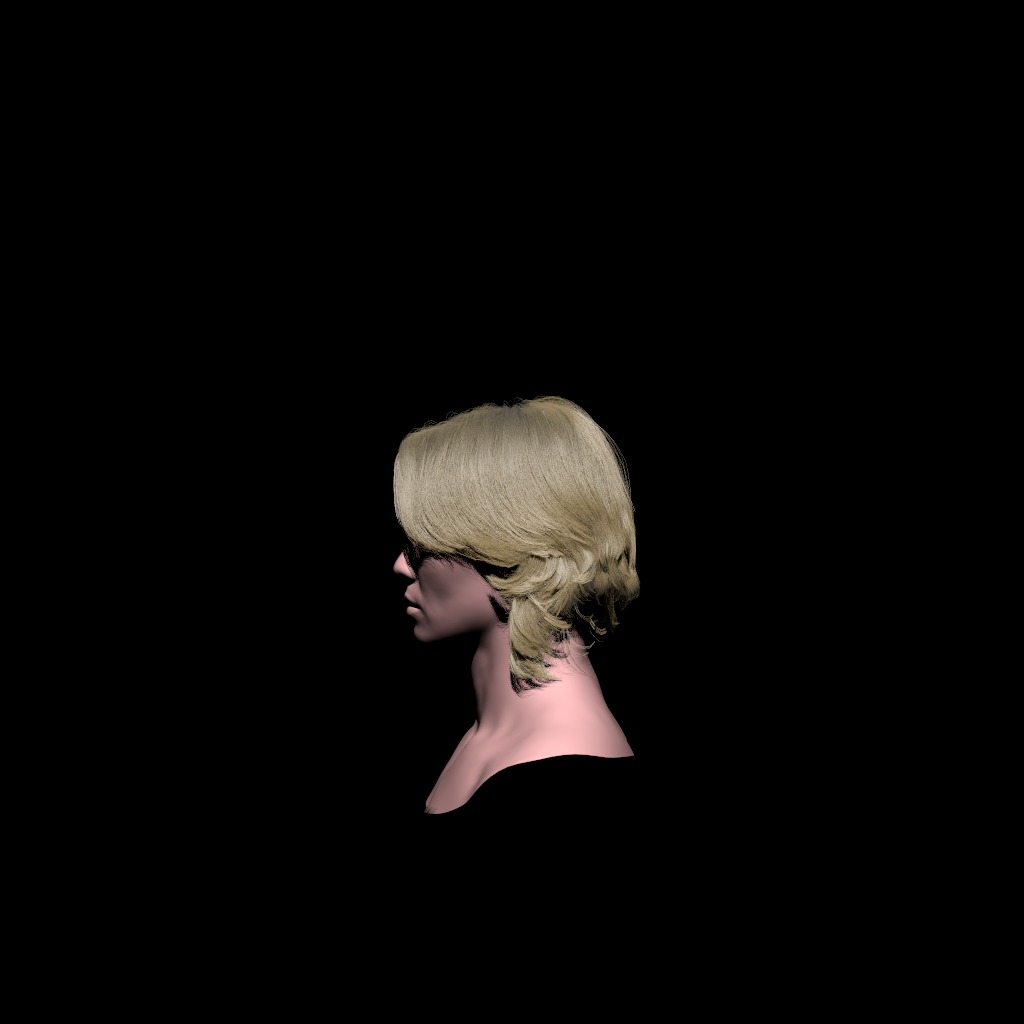}\hfill%
\includegraphics[trim={450 450 450 450},clip,width=0.16\linewidth]{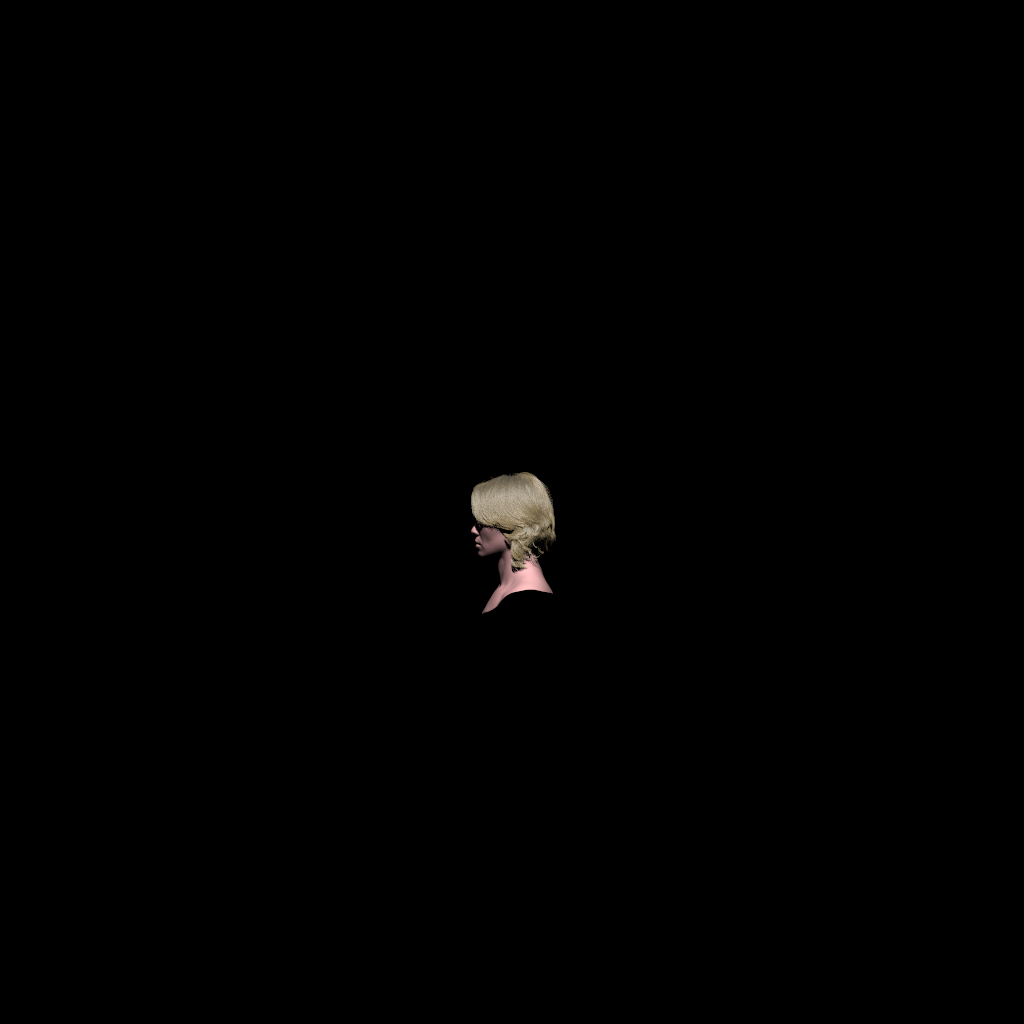}\hfill%
\includegraphics[trim={0 0 0 0},        clip,width=0.16\linewidth]{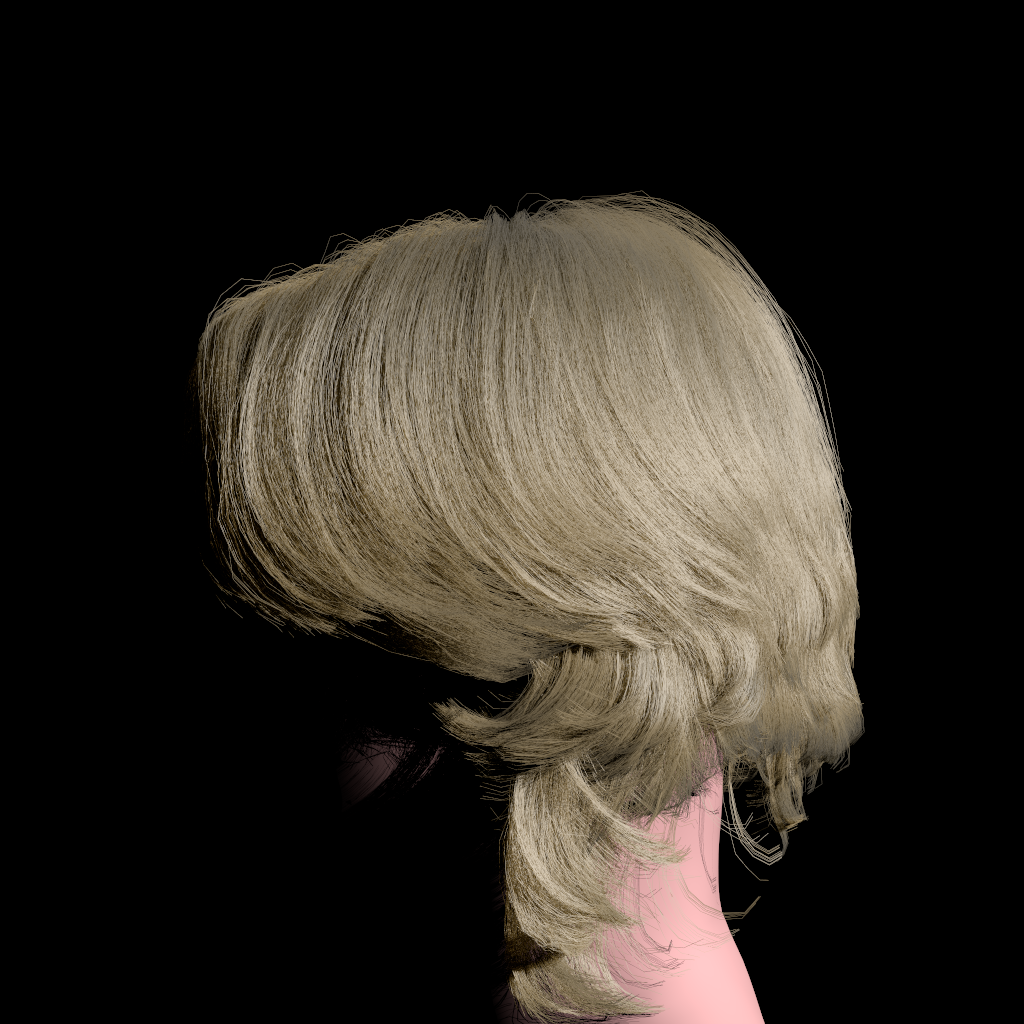}\hfill%
\includegraphics[trim={325 325 325 325},clip,width=0.16\linewidth]{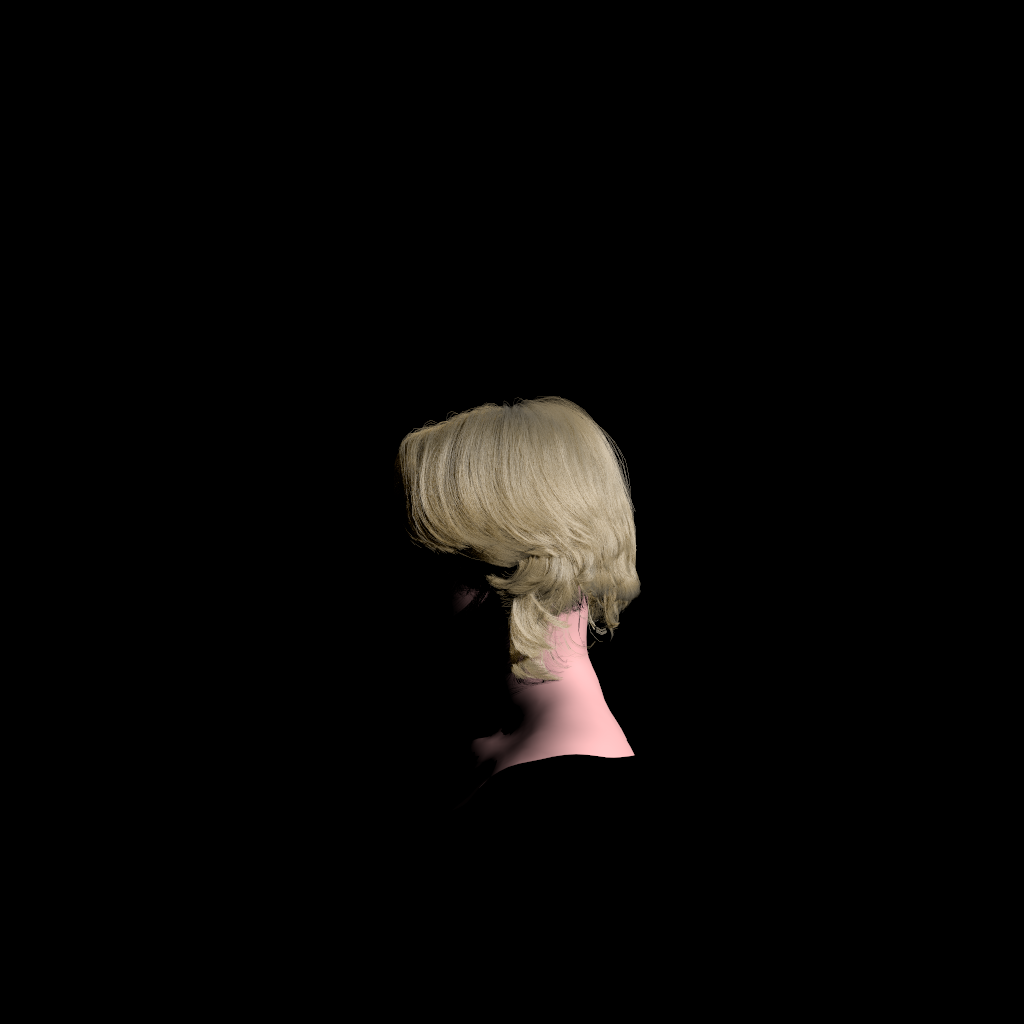}\hfill%
\includegraphics[trim={450 450 450 450},clip,width=0.16\linewidth]{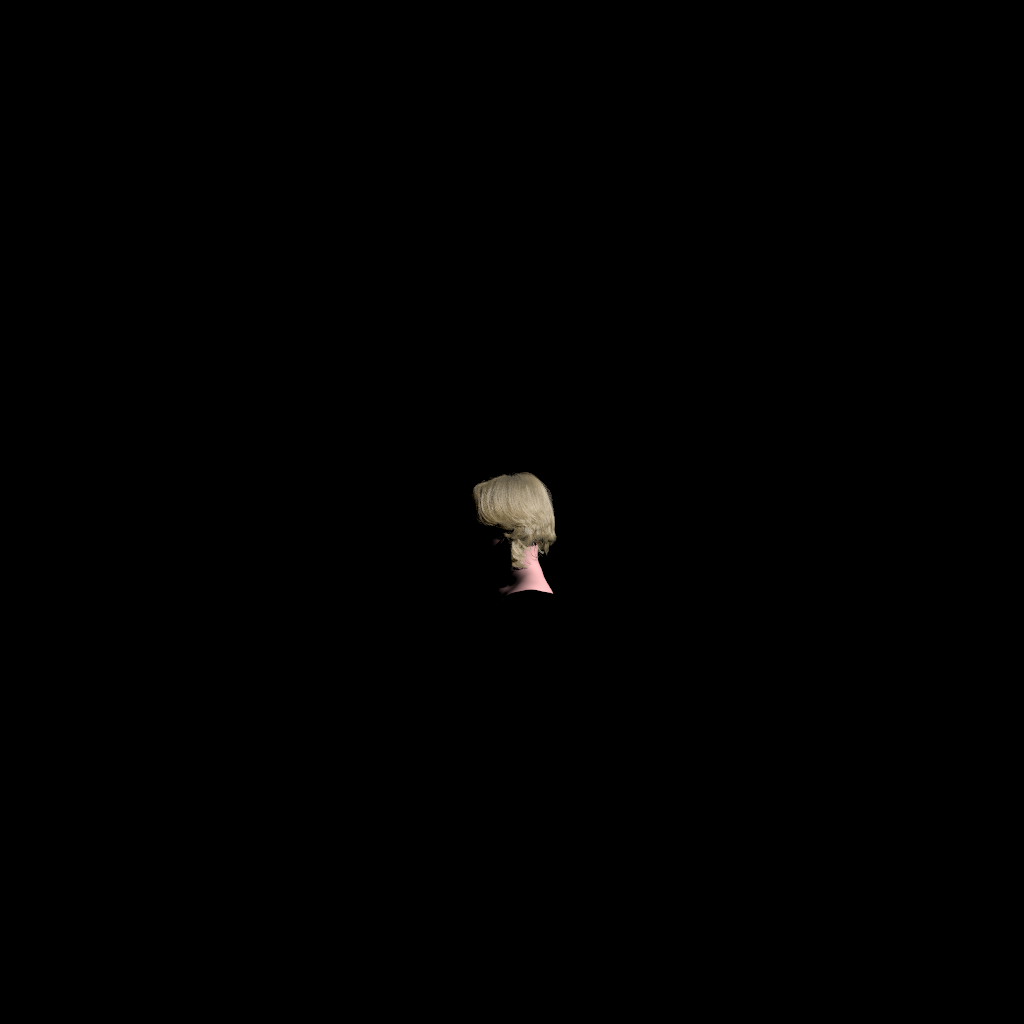}\\
\hspace{0.03\linewidth}\hfill%
\includegraphics[trim={0 0 0 0},        clip,width=0.16\linewidth]{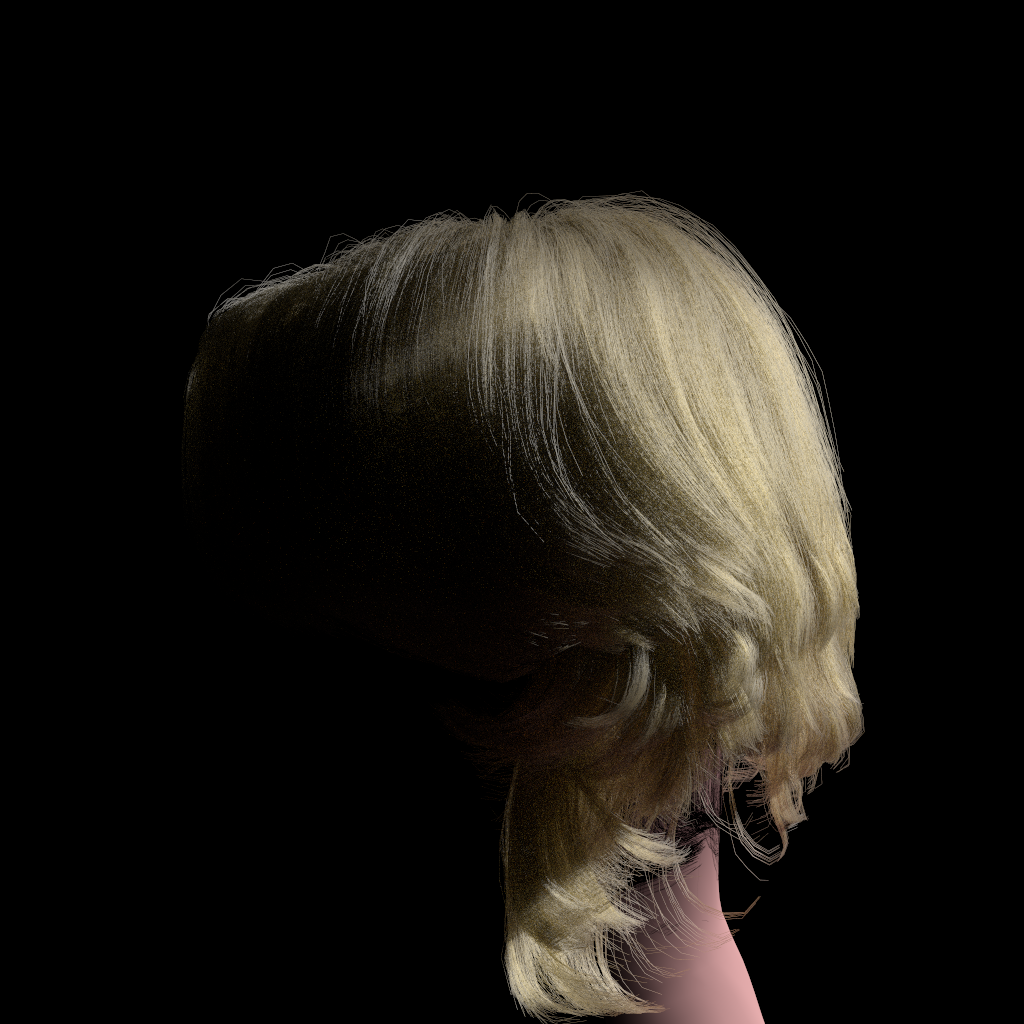}\hfill%
\includegraphics[trim={325 325 325 325},clip,width=0.16\linewidth]{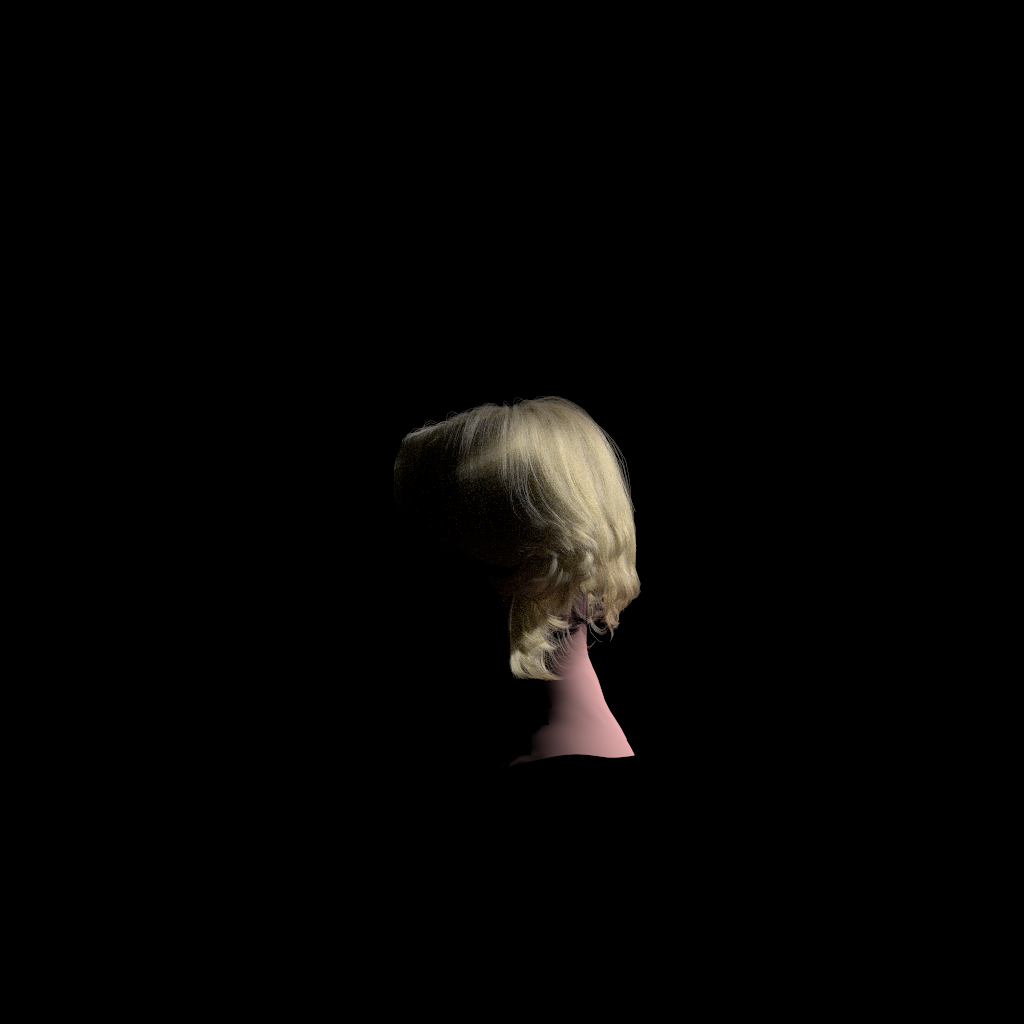}\hfill%
\includegraphics[trim={450 450 450 450},clip,width=0.16\linewidth]{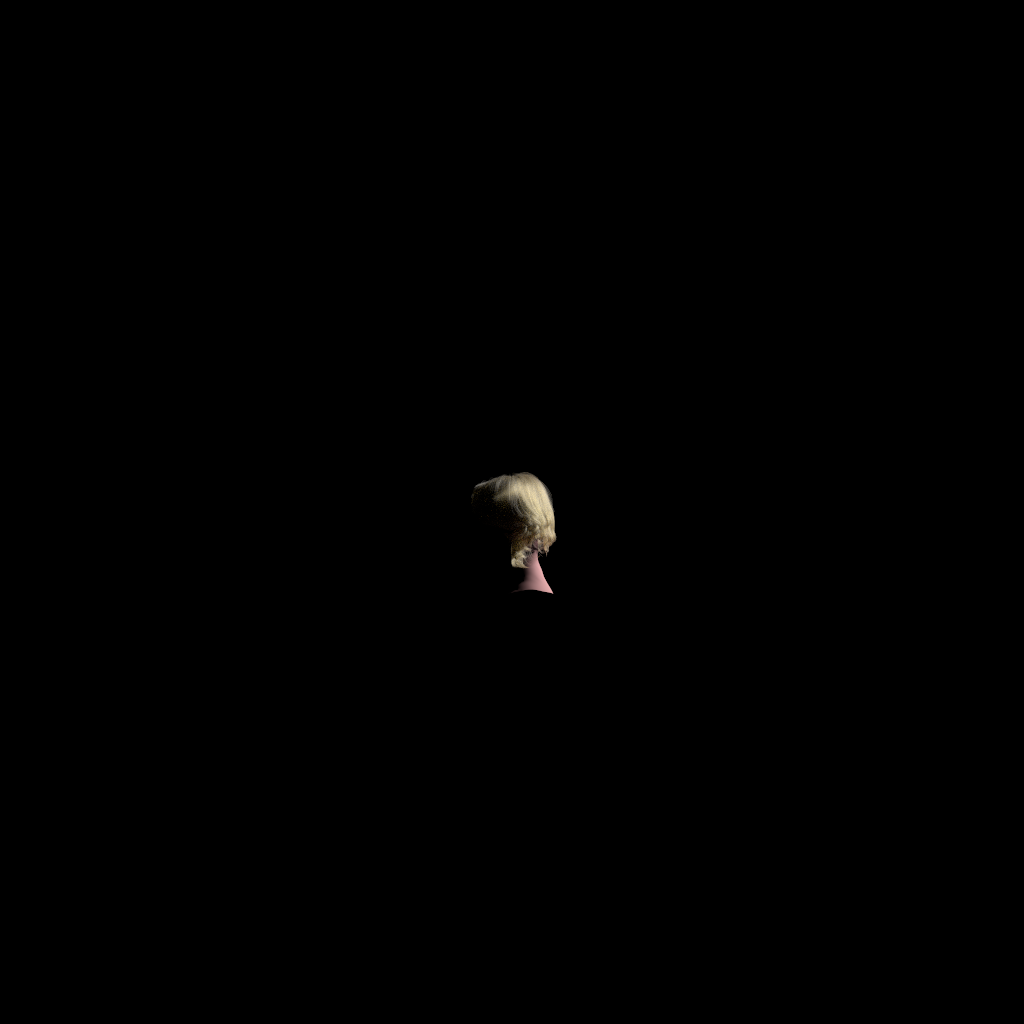}\hfill%
\includegraphics[trim={0 0 0 0},        clip,width=0.16\linewidth]{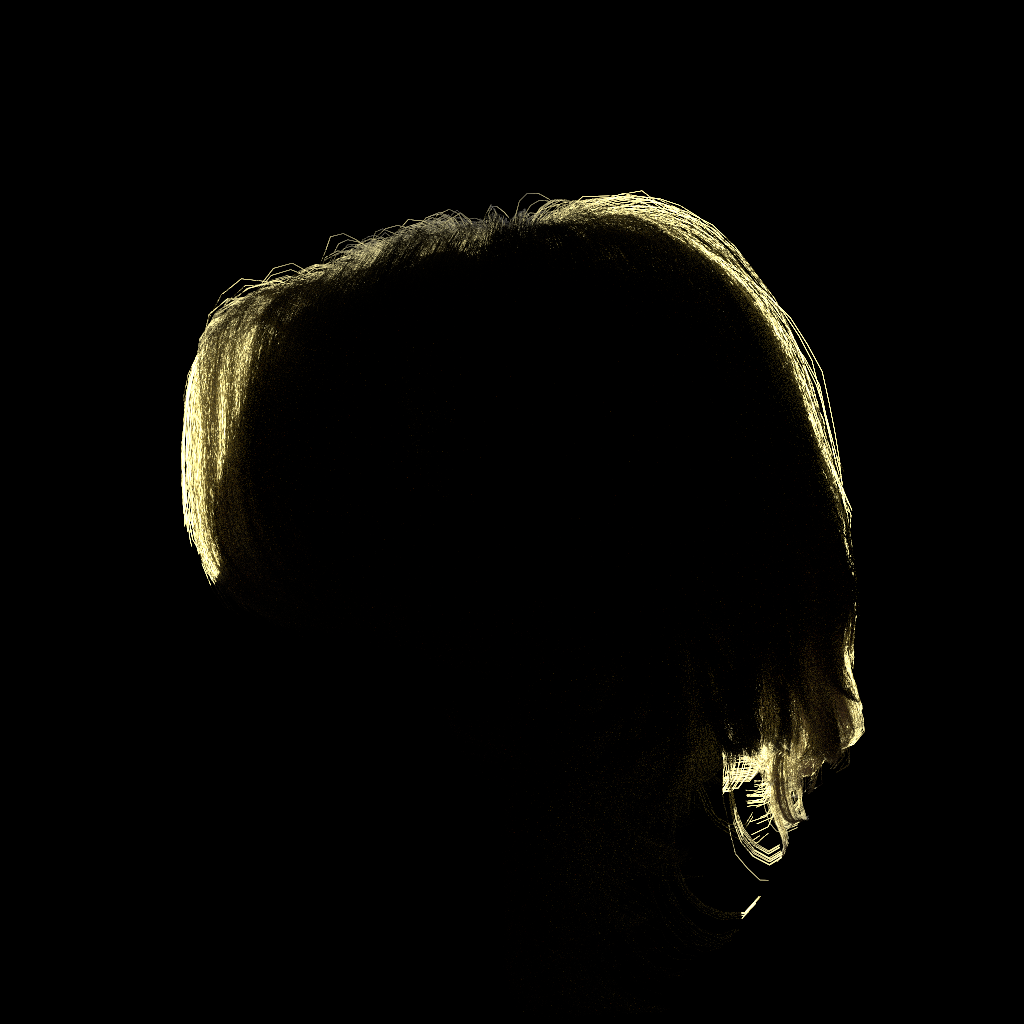}\hfill%
\includegraphics[trim={325 325 325 325},clip,width=0.16\linewidth]{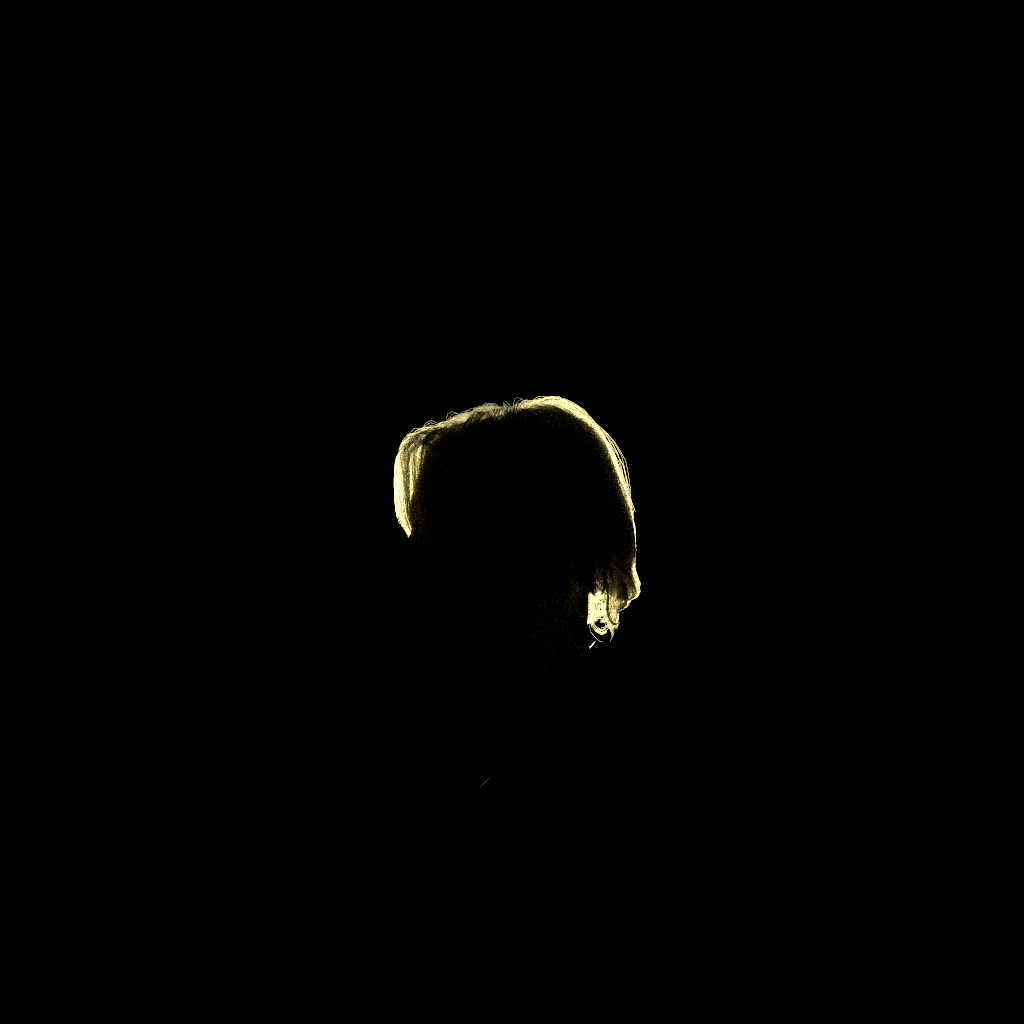}\hfill%
\includegraphics[trim={450 450 450 450},clip,width=0.16\linewidth]{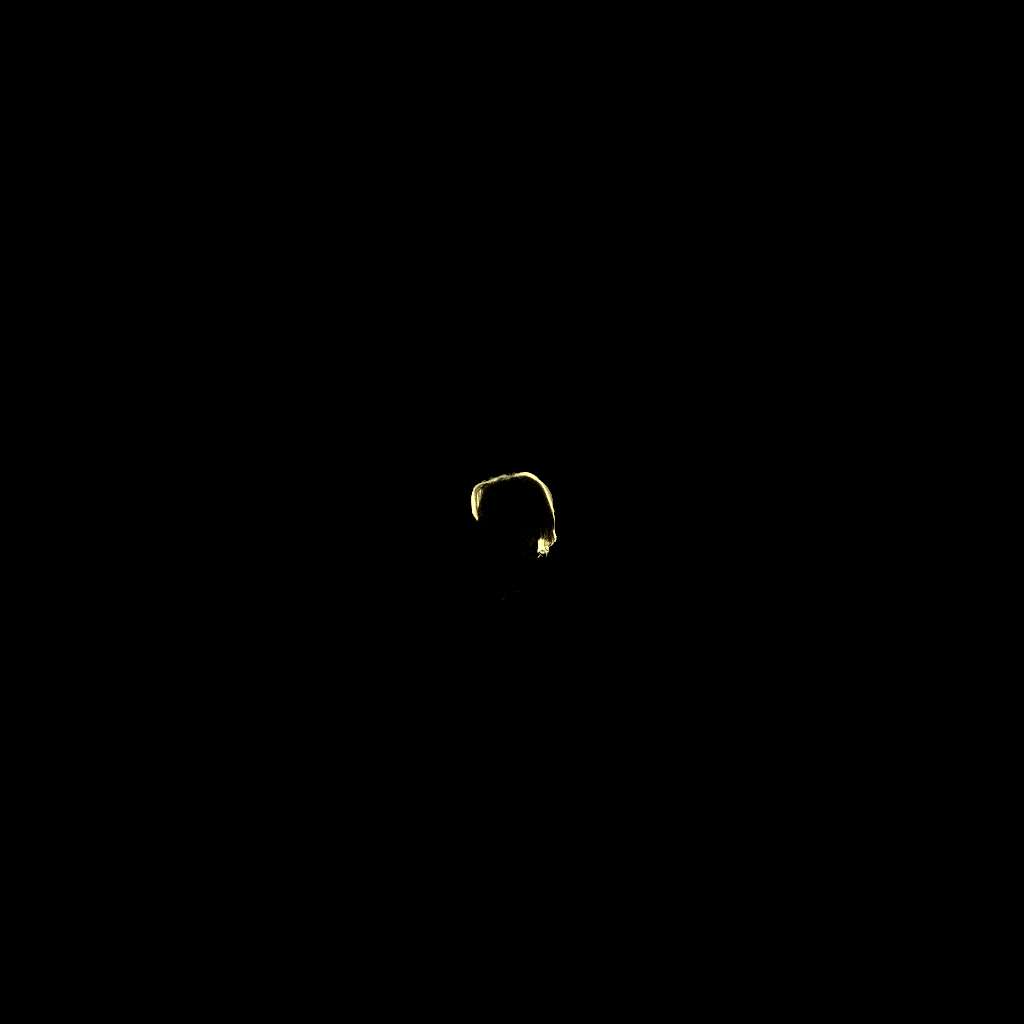}\\
\hspace{0.03\linewidth}\hfill%
\includegraphics[trim={0 0 0 0},        clip,width=0.16\linewidth]{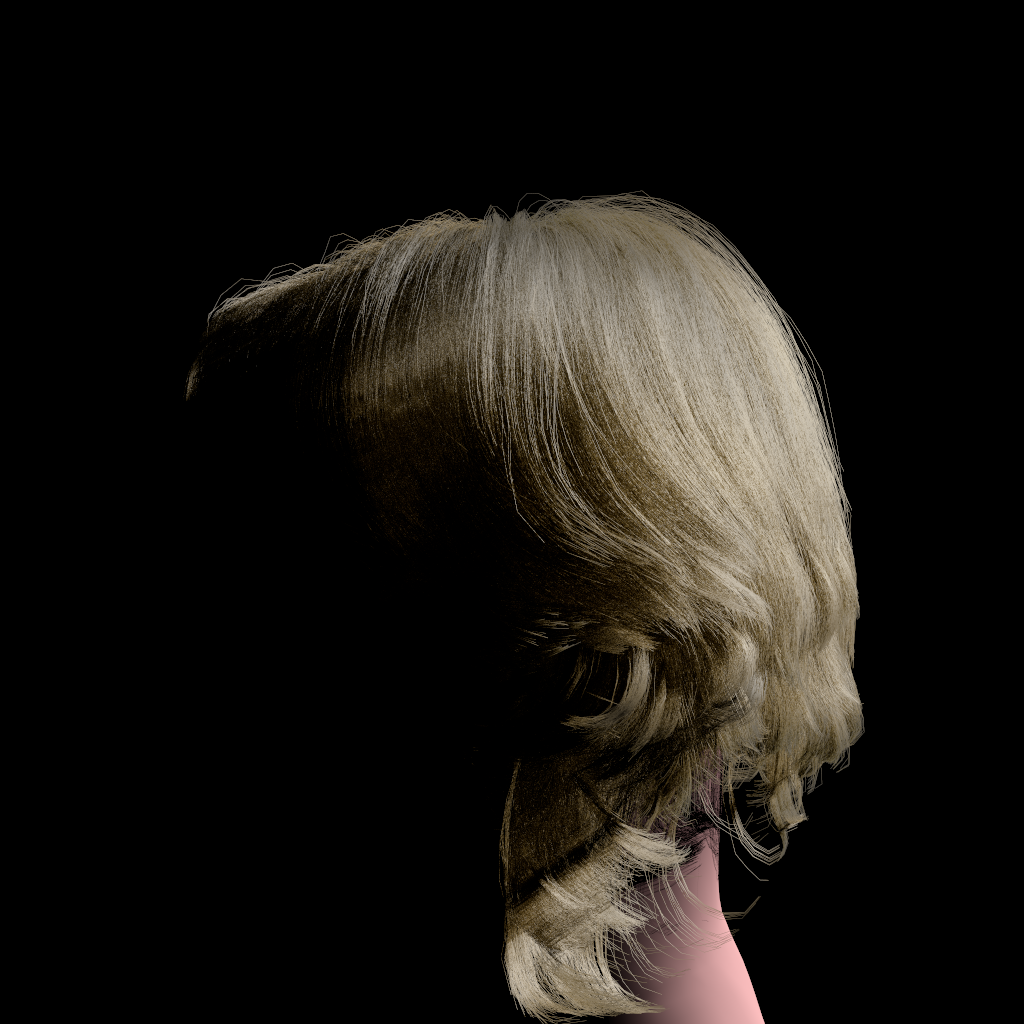}\hfill%
\includegraphics[trim={325 325 325 325},clip,width=0.16\linewidth]{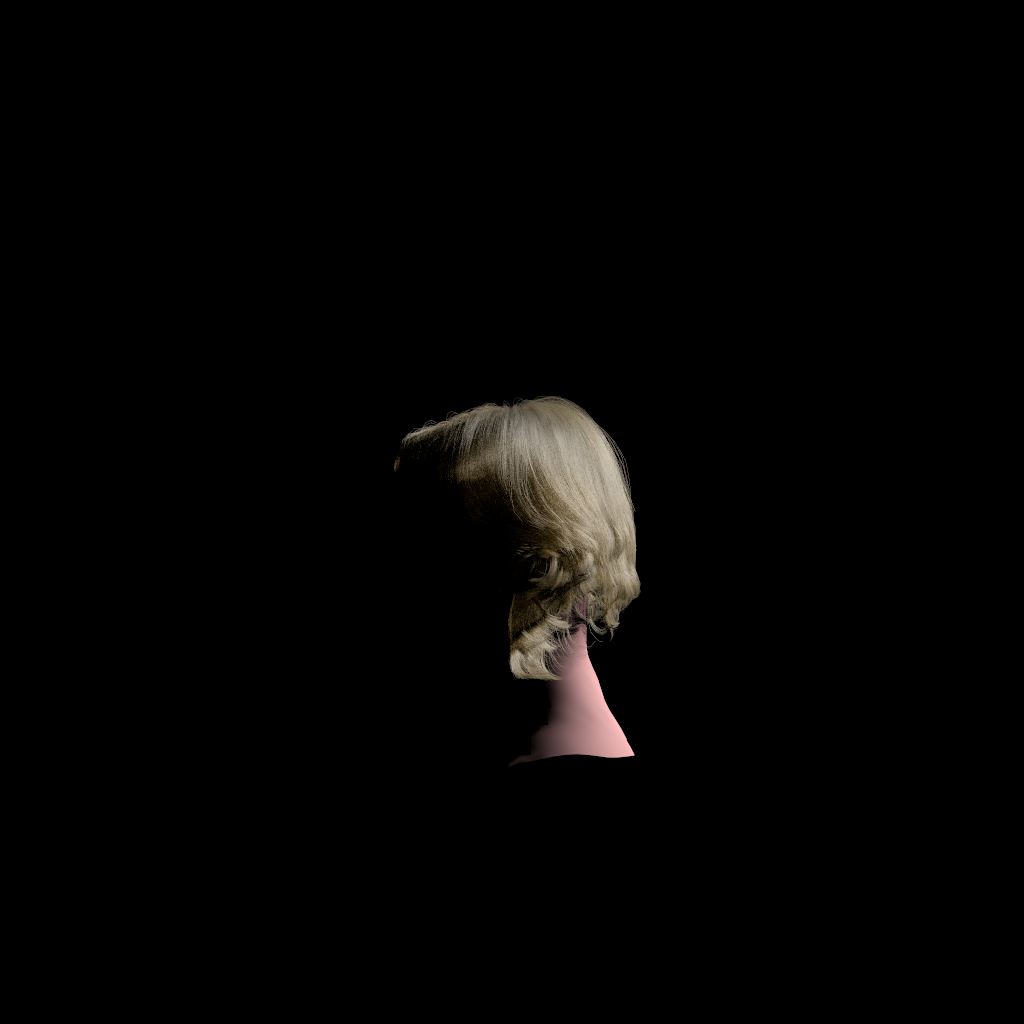}\hfill%
\includegraphics[trim={450 450 450 450},clip,width=0.16\linewidth]{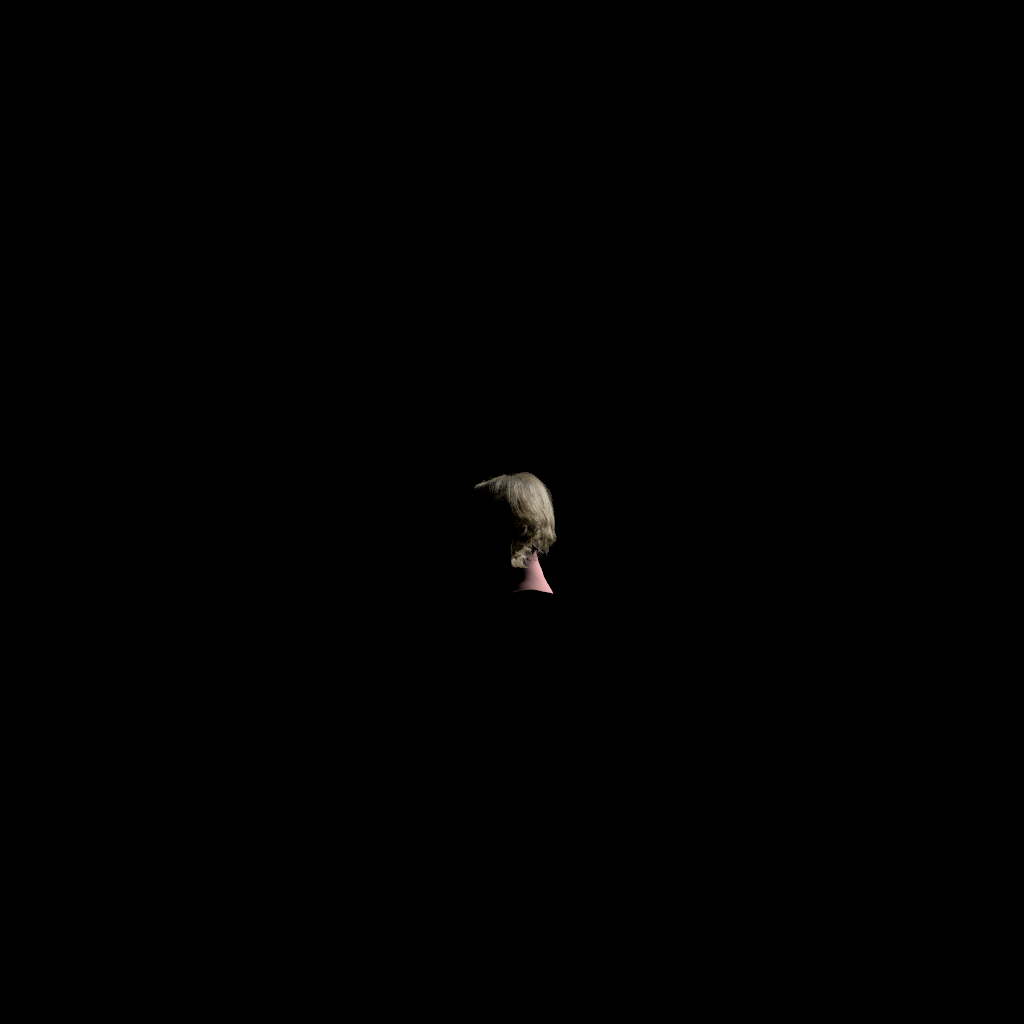}\hfill%
\includegraphics[trim={0 0 0 0},        clip,width=0.16\linewidth]{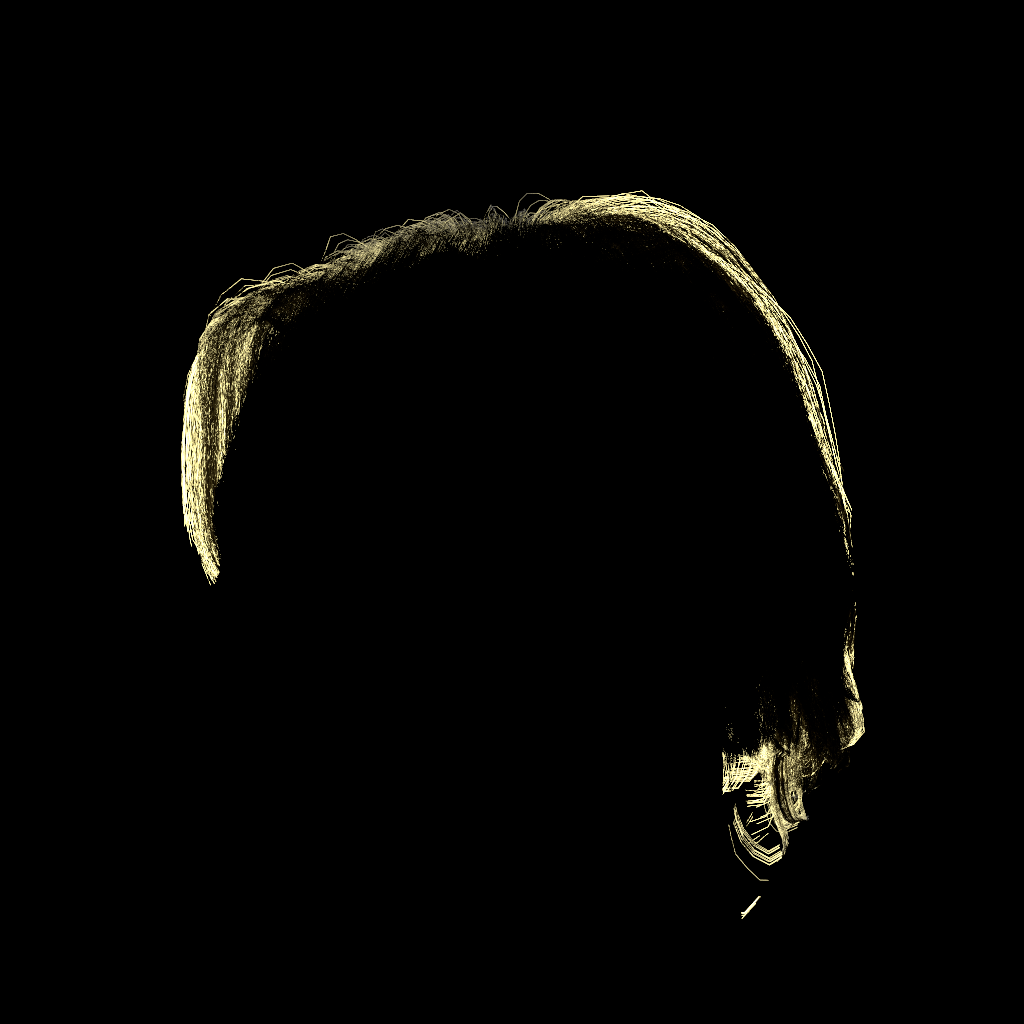}\hfill%
\includegraphics[trim={325 325 325 325},clip,width=0.16\linewidth]{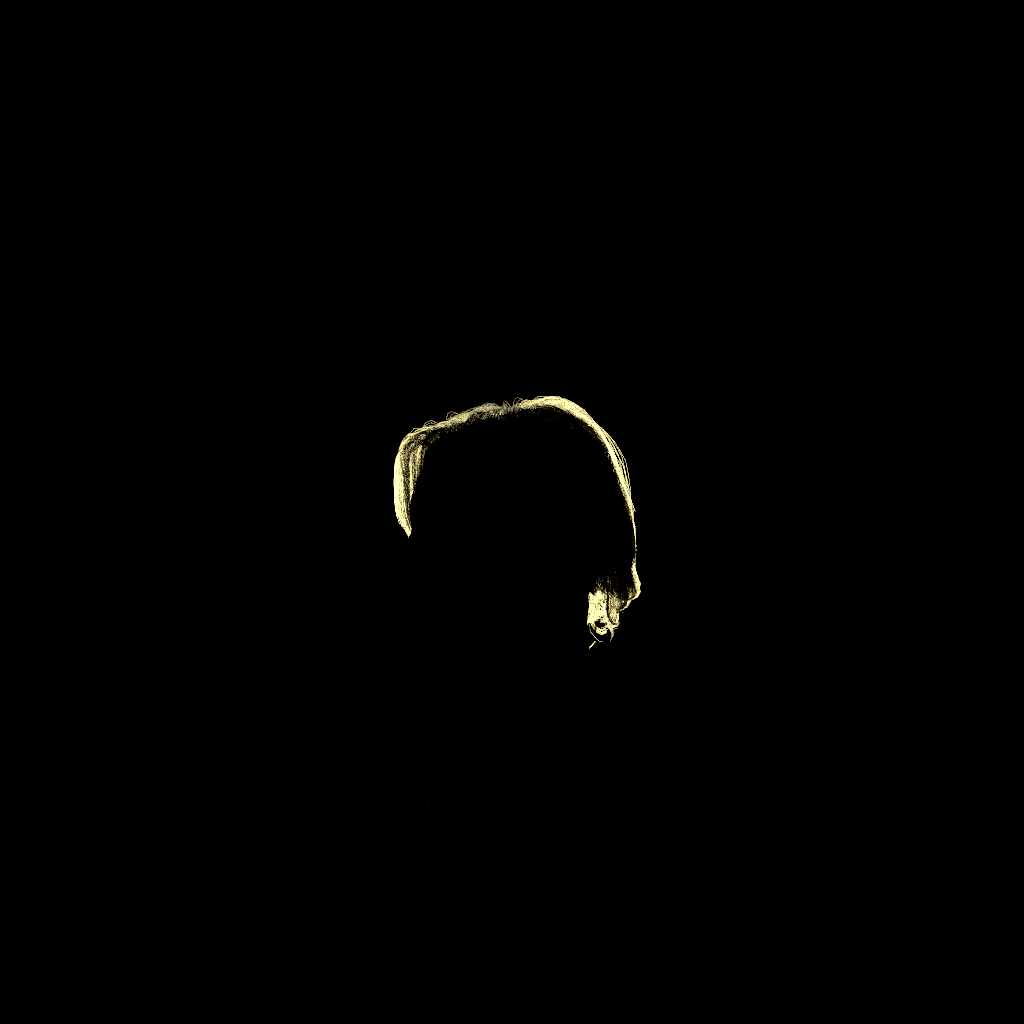}\hfill%
\includegraphics[trim={450 450 450 450},clip,width=0.16\linewidth]{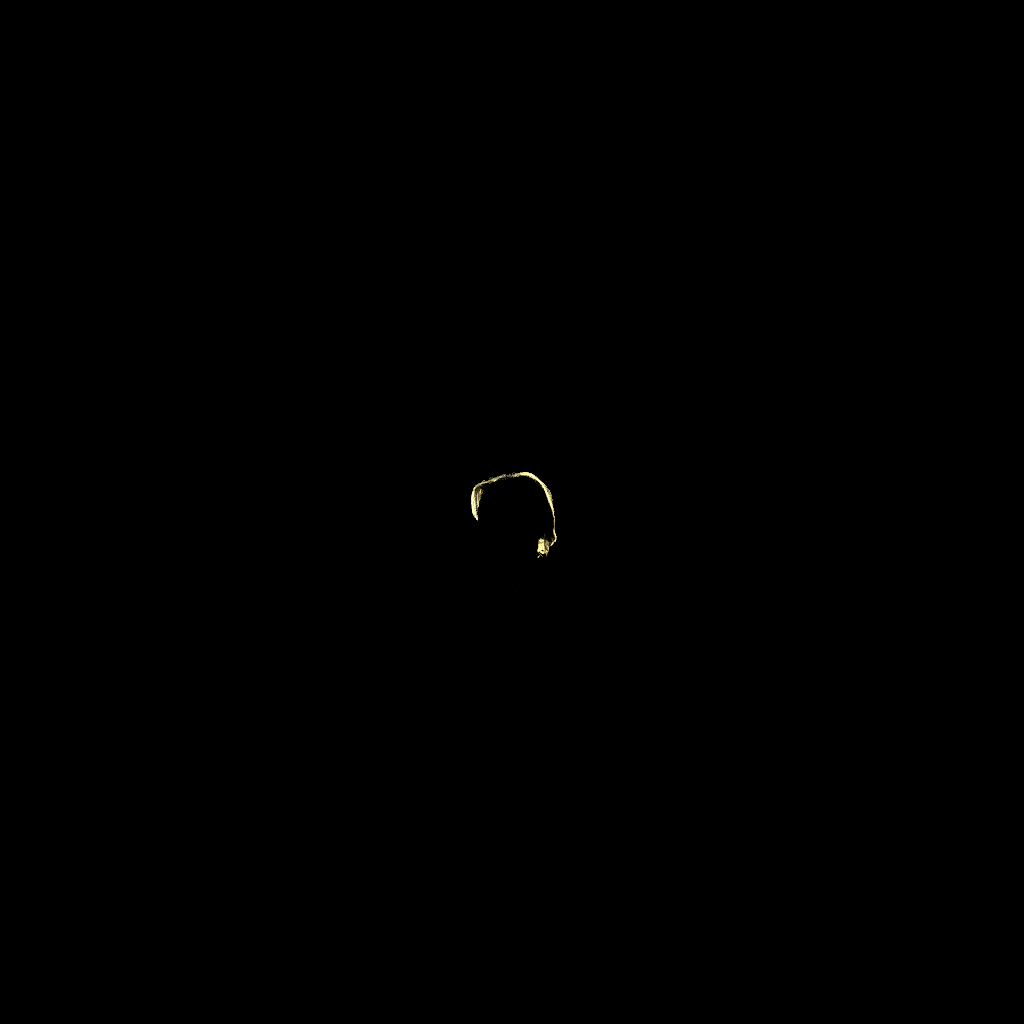}\\
\vspace{-36.5em}
\begin{flushleft}{
\rotatebox{90}{\small \hspace{0.25em} \textbf{Ours (ray shooting)} \hspace{1.5em} Offline path tracing \hspace{1.5em} \textbf{Ours (ray shooting)} \hspace{1.5em} Offline path tracing  }\hfill%
}\end{flushleft}
\caption{Our approach can yield results closely resembling those obtained from offline path tracing when the angles between the view direction and light direction are 0\textdegree, 45\textdegree, 90\textdegree, and 180\textdegree. It's worth noting that offline path tracing employs full hair geometry with 1050K segments, while ours utilizes 1010K, 857K, and 418K segments for near, middle, and far views, respectively. }
\label{fig:lightdir}
\Description{}
\end{figure*}

\paragraph{Abalation study on various light directions}
\autoref{fig:lightdir} demonstrates that our method can produce results closely resembling those obtained from offline path tracing across various light directions, including 0\textdegree~(frontlit), 45\textdegree, 90\textdegree, and 180\textdegree~(backlit), and from different viewing distances.

\begin{figure*}[ht]
\centering
\newcommand{\Typecap}[1]{\begin{minipage}{0.46\linewidth}\centering#1\end{minipage}}
\newcommand{\figcap}[1]{\begin{minipage}{0.23\linewidth}\centering#1\end{minipage}}

\hspace{0.02\linewidth}\hfill%
\figcap{Full w/ PT}\hfill%
\figcap{\textbf{LoD w/ ours}}\hfill%
\figcap{Full w/ PT}\hfill%
\figcap{\textbf{LoD w/ ours}}\vspace{0.25em}
\\
\hspace{0.02\linewidth}\hfill%
\includegraphics[trim={0   0   0   0  },clip,width=0.243\linewidth]{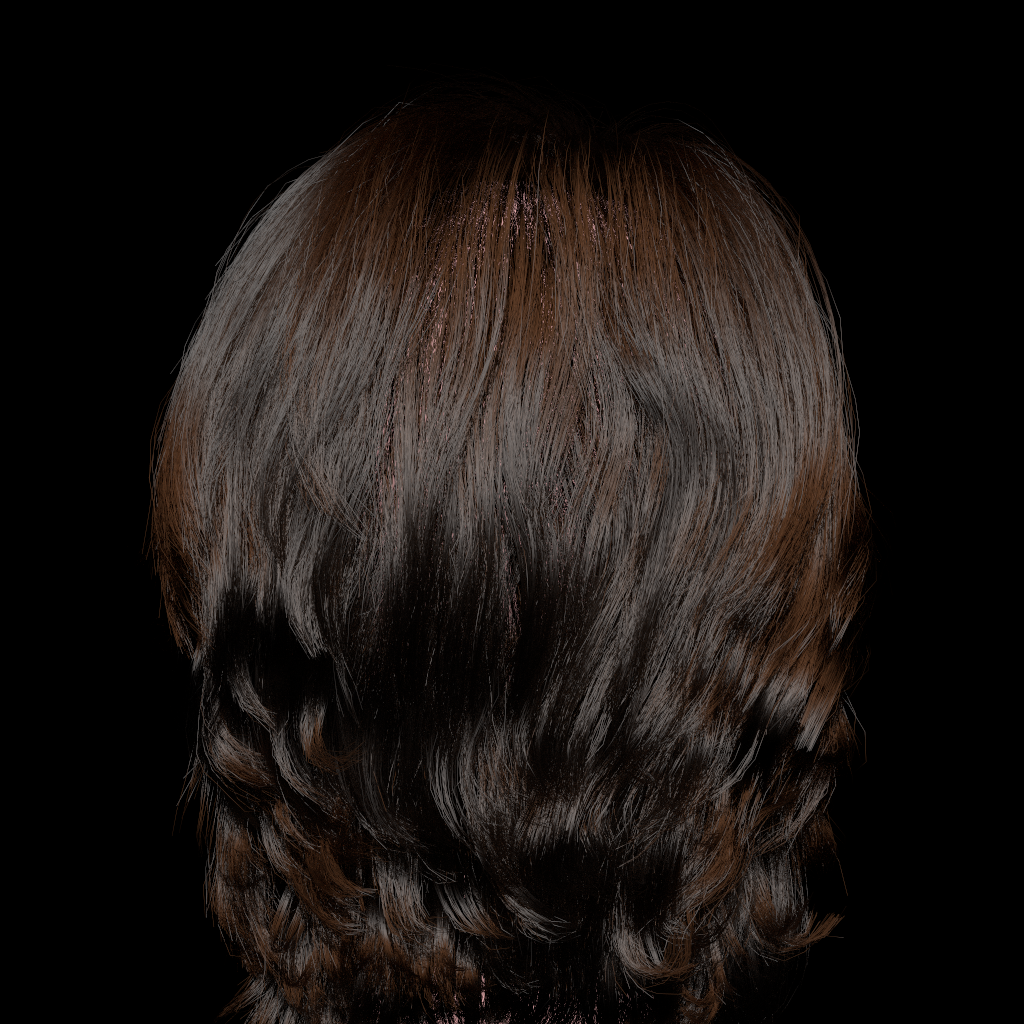}\hfill%
\includegraphics[trim={0   0   0   0  },clip,width=0.243\linewidth]{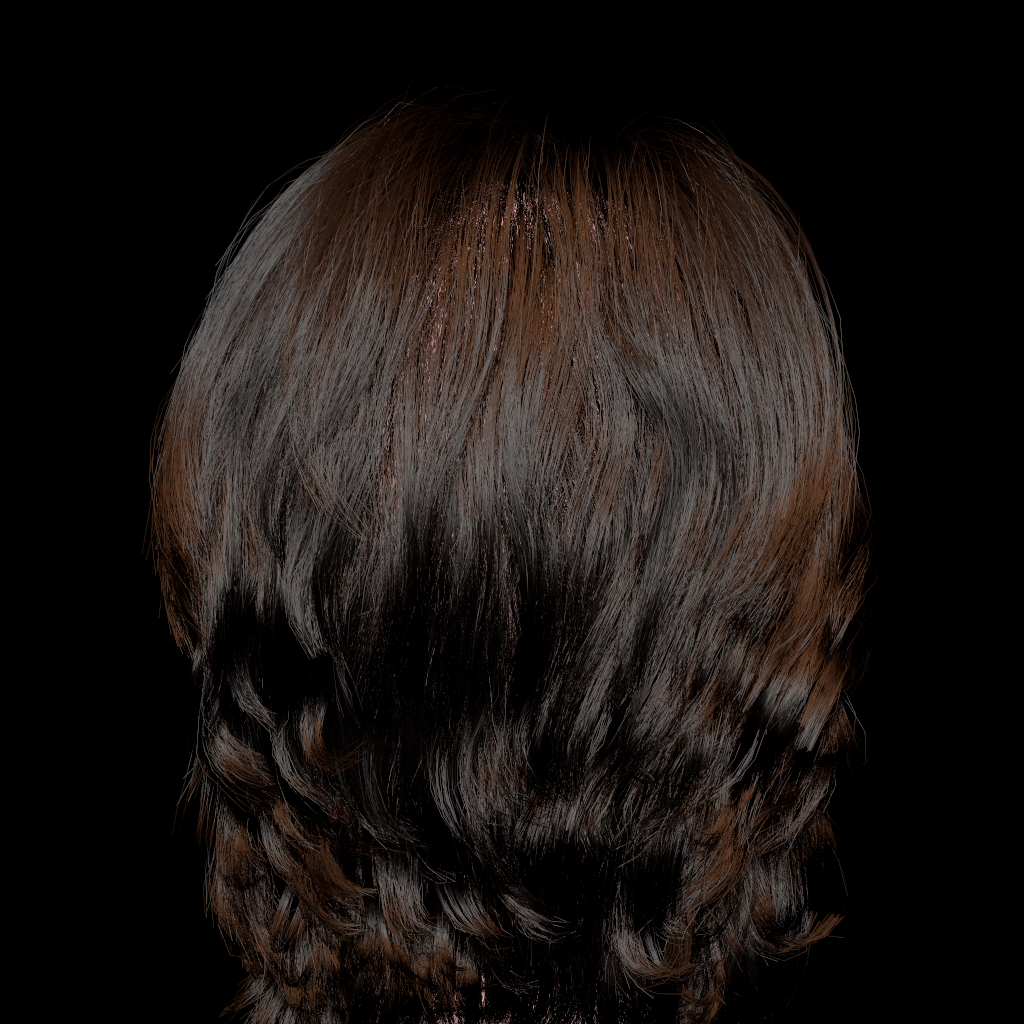}\hfill%
\includegraphics[trim={0   0   0   0  },clip,width=0.243\linewidth]{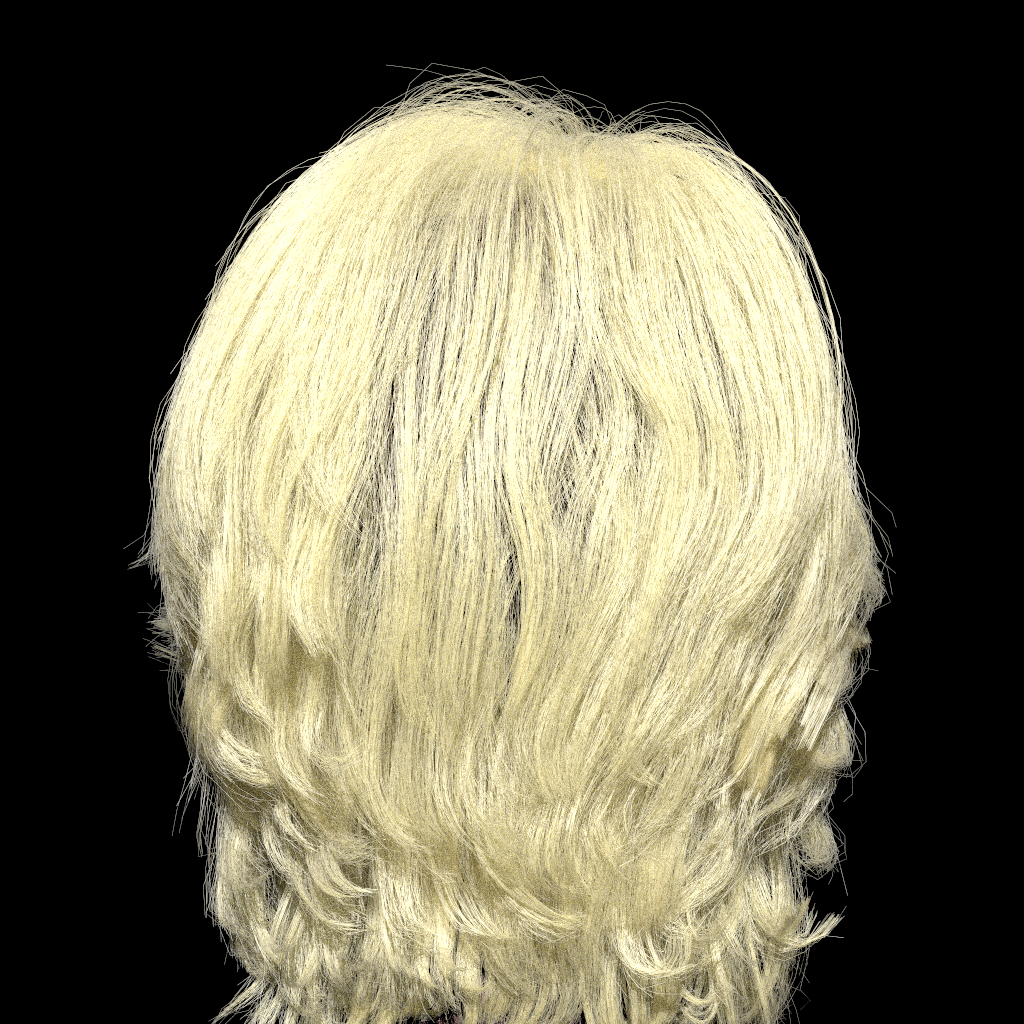}\hfill%
\includegraphics[trim={0   0   0   0  },clip,width=0.243\linewidth]{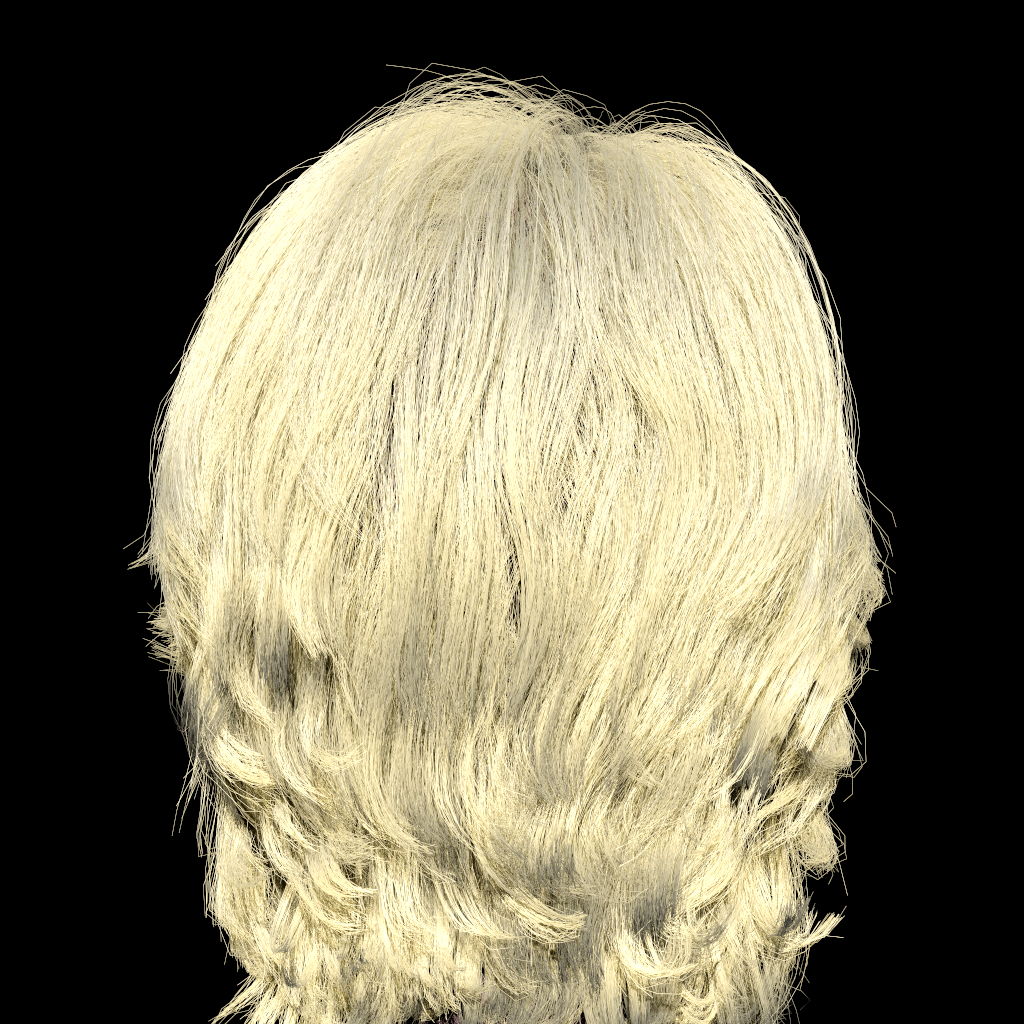}\hfill\\
\hspace{0.02\linewidth}\hfill%
\includegraphics[trim={365 365 365 365},clip,width=0.243\linewidth]{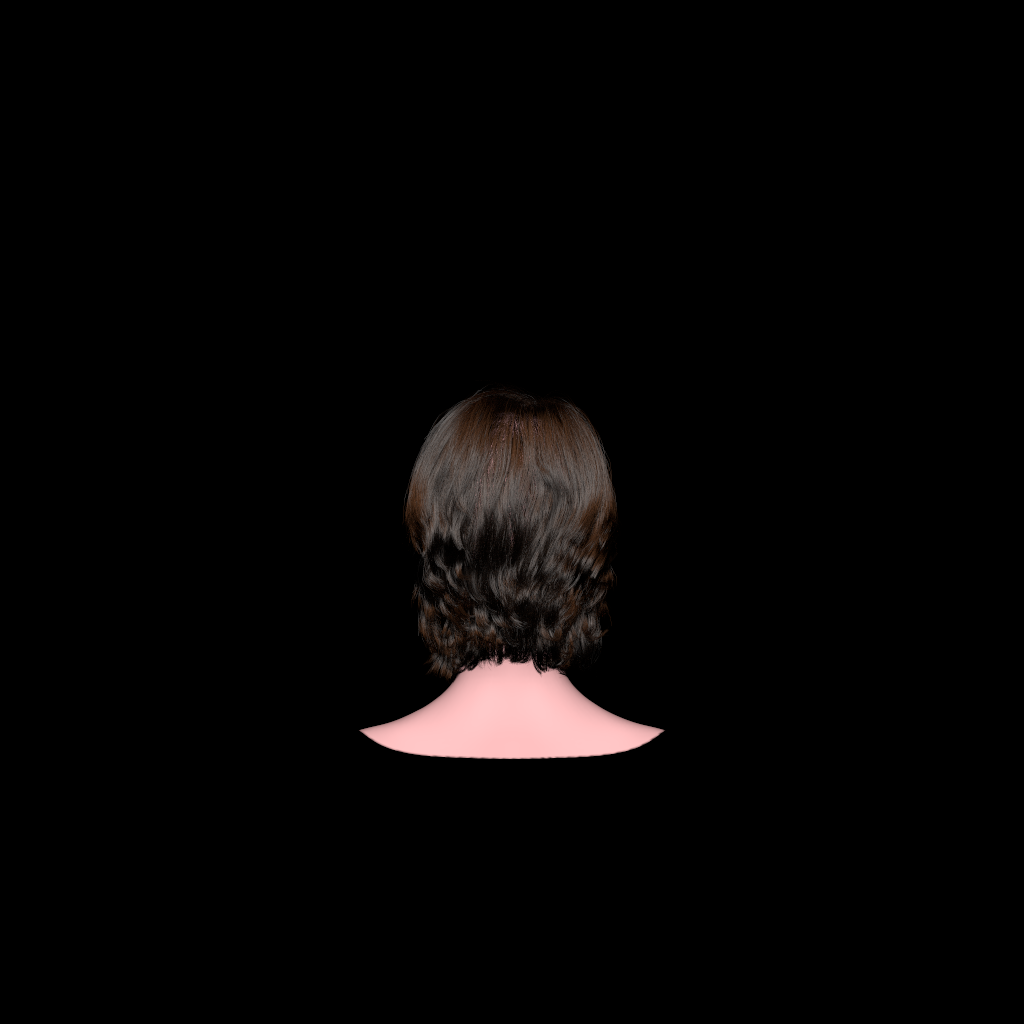}\hfill%
\includegraphics[trim={365 365 365 365},clip,width=0.243\linewidth]{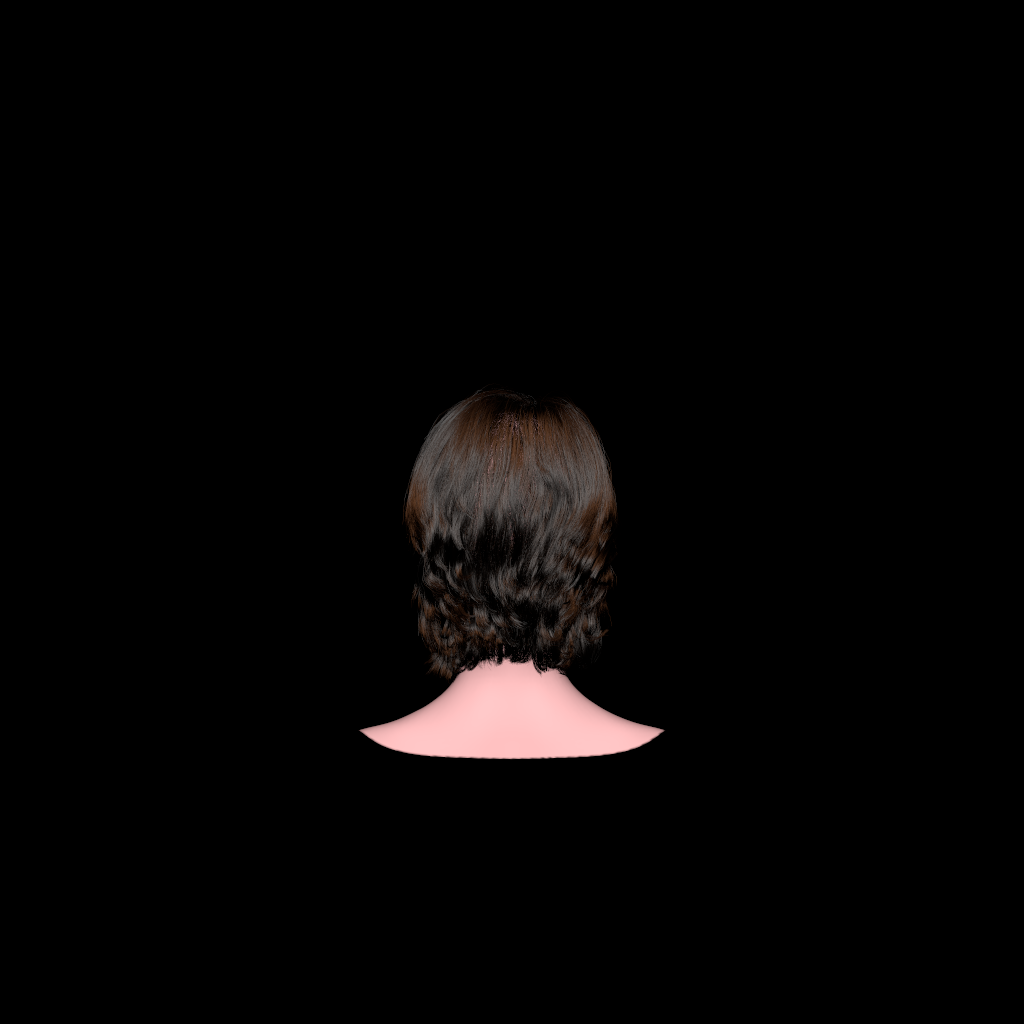}\hfill%
\includegraphics[trim={365 365 365 365},clip,width=0.243\linewidth]{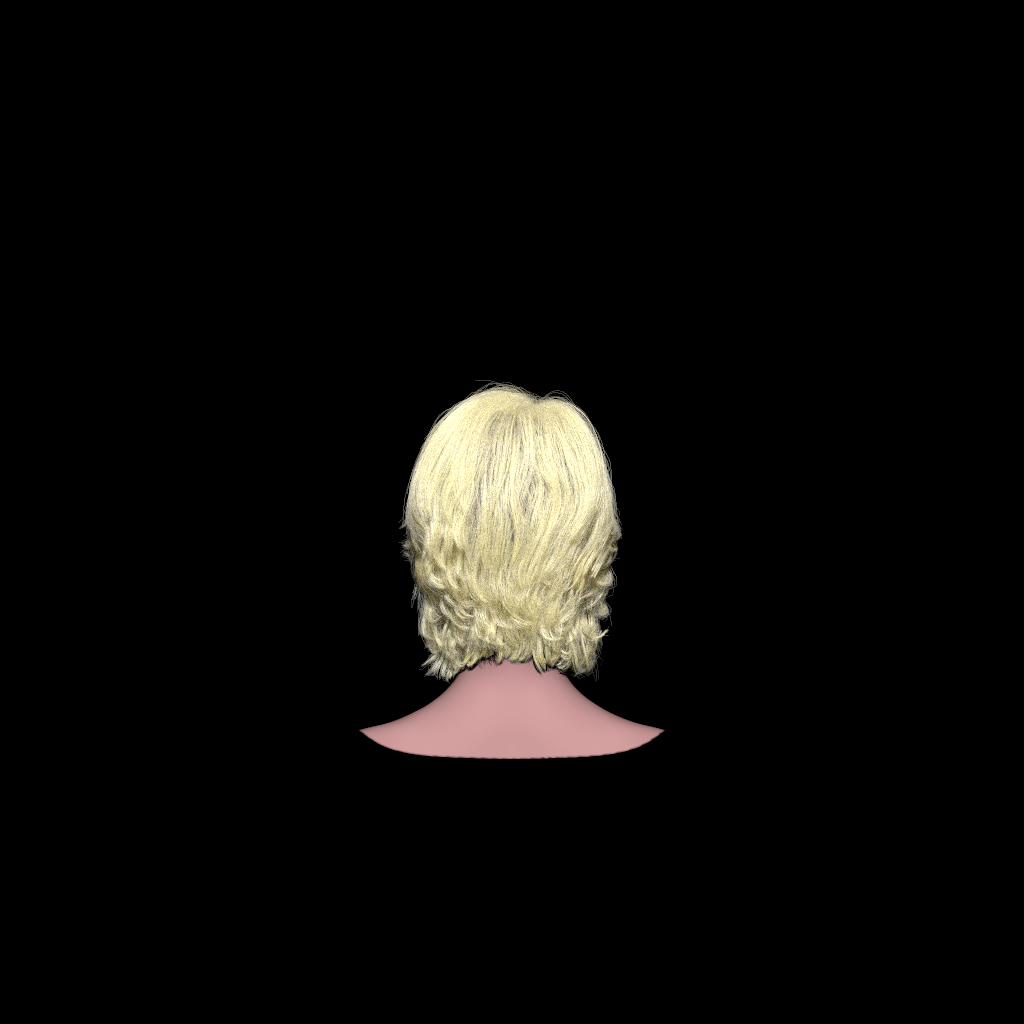}\hfill%
\includegraphics[trim={365 365 365 365},clip,width=0.243\linewidth]{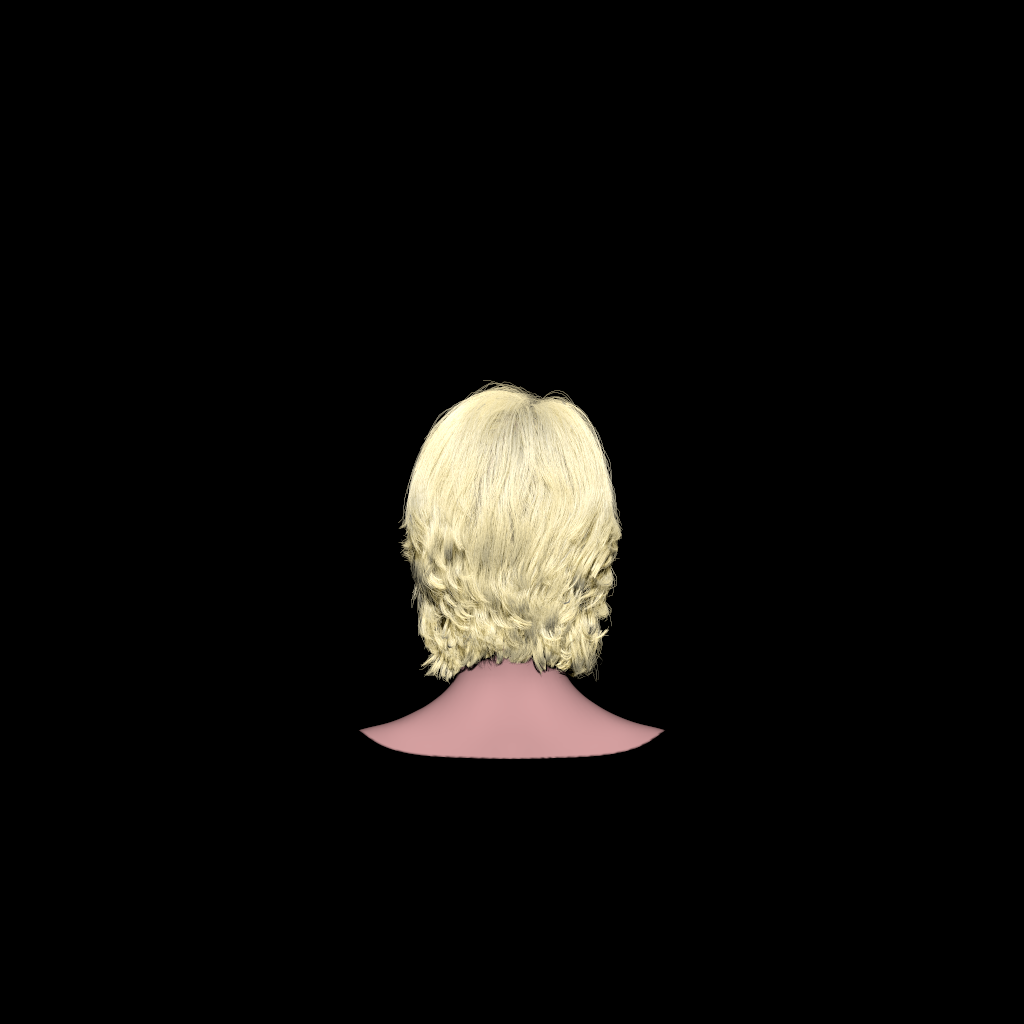}\\
\hspace{0.02\linewidth}\hfill%
\includegraphics[trim={463 463 463 463},clip,width=0.243\linewidth]{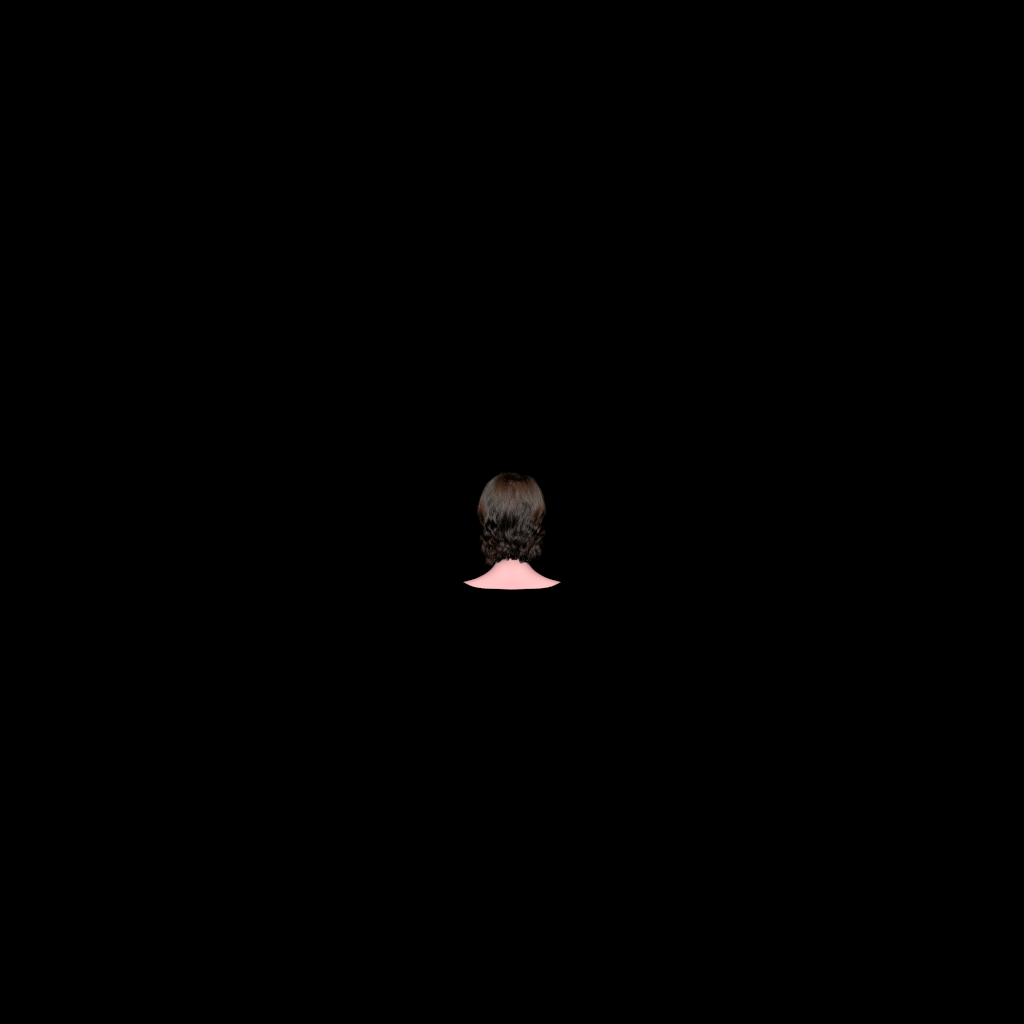}\hfill%
\includegraphics[trim={463 463 463 463},clip,width=0.243\linewidth]{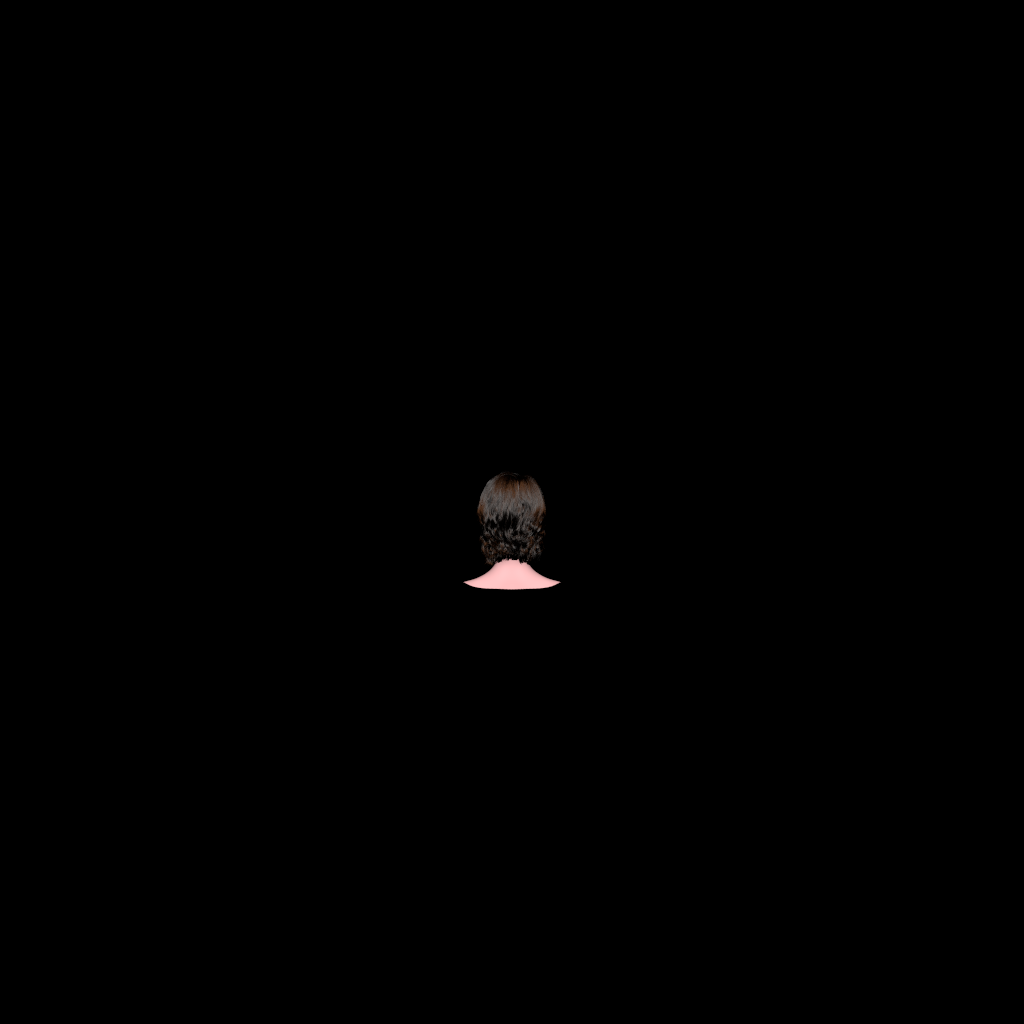}\hfill%
\includegraphics[trim={463 463 463 463},clip,width=0.243\linewidth]{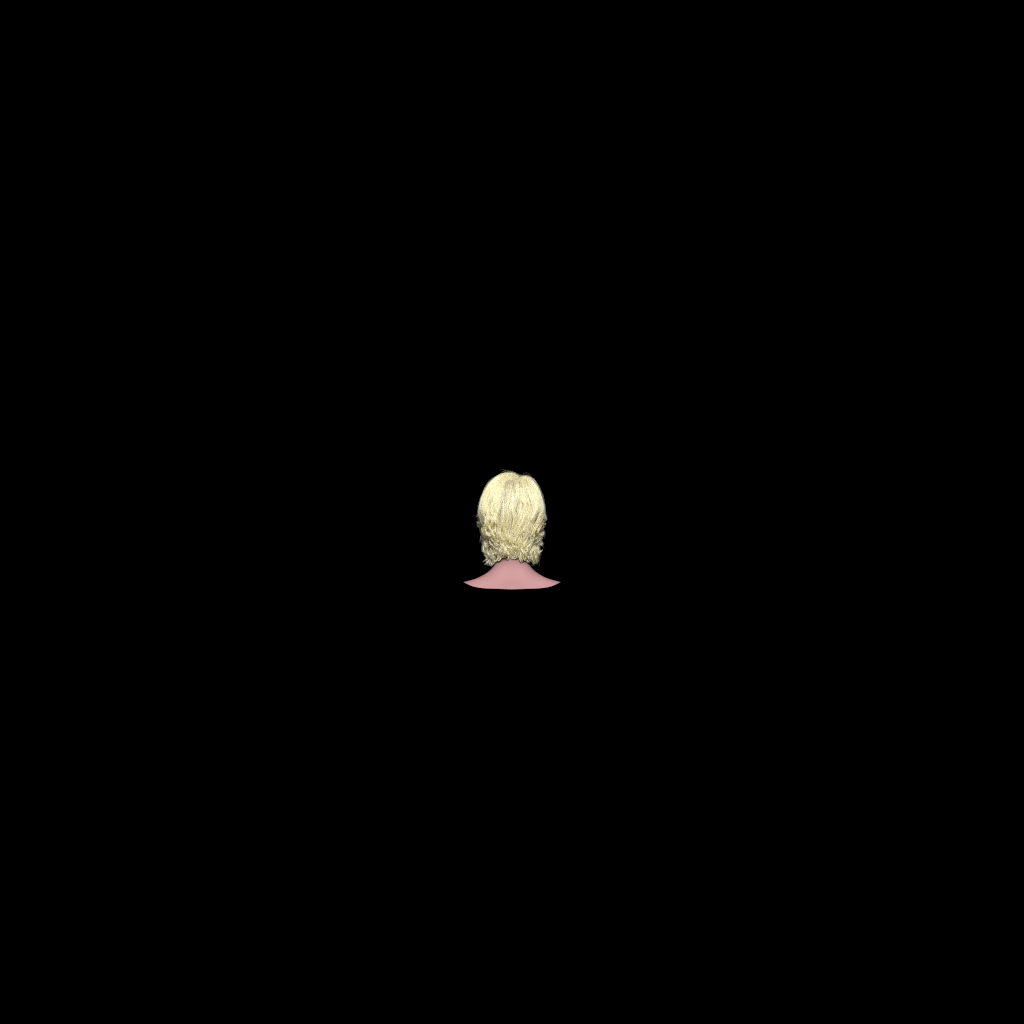}\hfill%
\includegraphics[trim={463 463 463 463},clip,width=0.243\linewidth]{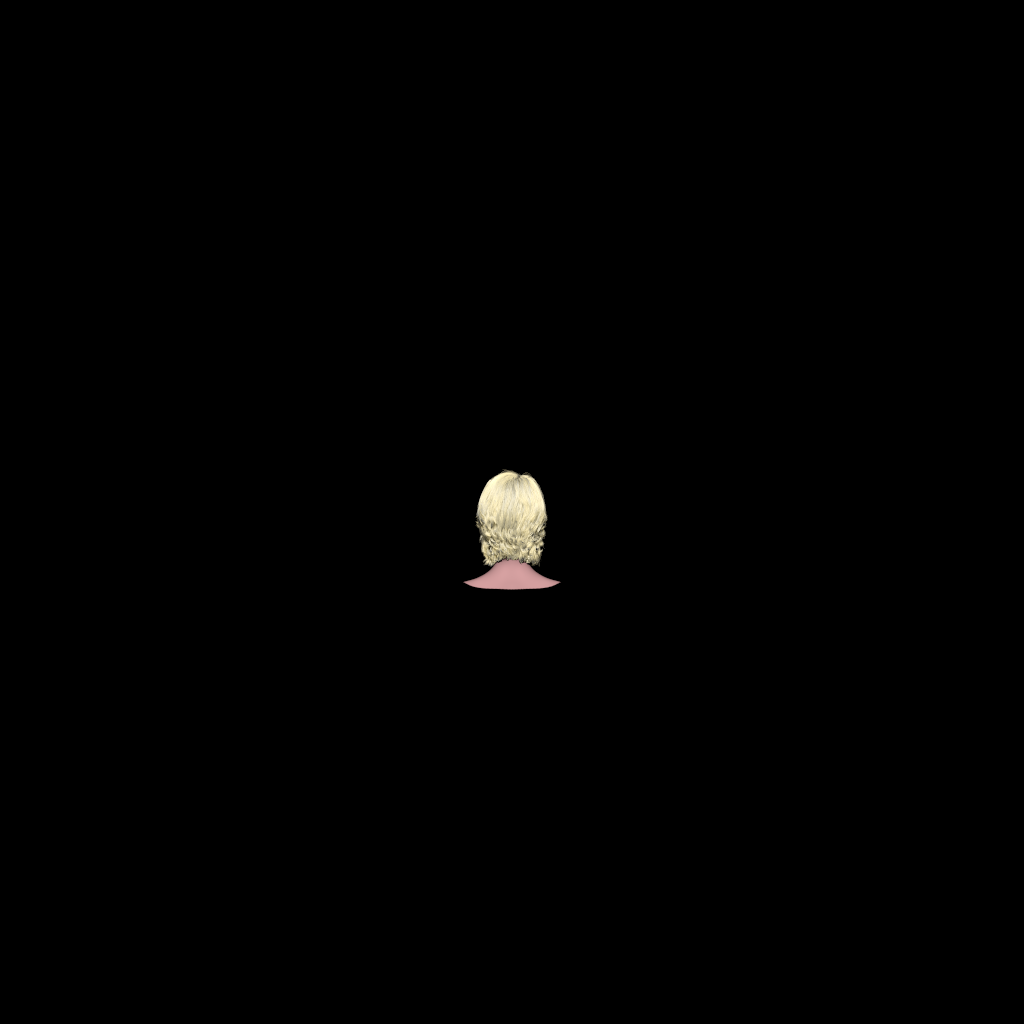}\\
\vspace{-41.5em}
\hspace{0.02\linewidth}\hfill%
\figcap{\color{white}{$1020$K triangles}}\hfill%
\figcap{\color{white}{$990$K triangles}}\hfill%
\figcap{\color{white}{$1020$K triangles}}\hfill%
\figcap{\color{white}{$990$K triangles}}\\
\vspace{12.6em}
\hspace{0.02\linewidth}\hfill%
\figcap{\color{white}{$1020$K triangles}}\hfill%
\figcap{\color{white}{$713$K triangles}}\hfill%
\figcap{\color{white}{$1020$K triangles}}\hfill%
\figcap{\color{white}{$713$K triangles}}\\
\vspace{12.6em}
\hspace{0.02\linewidth}\hfill%
\figcap{\color{white}{$1020$K triangles}}\hfill%
\figcap{\color{white}{$351$K triangles}}\hfill%
\figcap{\color{white}{$1020$K triangles}}\hfill%
\figcap{\color{white}{$351$K triangles}}\\
\vspace{-26.5em}
\begin{flushleft}
{
\rotatebox{90}{Far view~(2\%) \hspace{7em} Middle view~(20\%) \hspace{7em}  Close view~(60\%)   }\hfill%
}
\end{flushleft}
\vspace{3.5em}
\hspace{0.02\linewidth}\hfill%
\Typecap{(a) BCSDF~\cite{MarschnerJCWH03} with $R$, $TT$, and $TRT$}\hfill%
\Typecap{(b) BCSDF~\cite{ZhuZJYA23} with $R$, $TT$, and $D$}\hfill%
\caption{Additional examples: (a) Single hair BCSDF with TRT term \cite{MarschnerJCWH03} for dark-colored hair and (b) BCSDF~\cite{ZhuZJYA23} with $R$, $TT$, and $D$ for light-colored hair. For both shading models, our strand-based LoD model (\textbf{LoD w/ ours}) effectively reduces the number of fibers and preserves appearance compared to path tracing with full hair geometry (full w/ PT). And we mark the number of segments for hair strips in the image. }
\label{fig:extended_validation}
\Description{}
\end{figure*}

\paragraph{Extending to Marschner model with TRT term}
While we are using the BCSDF model~\cite{ZhuZJYA23} with $R$, $TT$, and $D$ (instead of $TRT$) lobes, our framework is generalizable to support the Marschner hair BCSDF~\cite{MarschnerJCWH03} with $R$, $TT$, and $TRT$ lobes. \autoref{fig:extended_validation}a shows an evaluation of our method using \citet{MarschnerJCWH03} with the full geometry in offline path tracing. Our method consistently and effectively reduces the number of hairs while maintaining a high level of appearance accuracy. Additionally, we demonstrate that the BCSDF model~\cite{ZhuZJYA23} without $TRT$ is still capable of supporting light-colored hair, as shown in~\autoref{fig:extended_validation}b.

\paragraph{Memory cost and initialization time}
In addition to the original full-size hair strand data, we store the positions of the centerline control points and the indices of four guiding points for thickness calculation. This results in extra memory consumption of approximately three-quarters of the original model size (78\% for the four models used in the paper). Our initialization time ranges from 6 to 12 minutes for hair models with strands ranging from 60,000 to 120,000. This process is performed only once, and the results can be stored as configuration.

\begin{figure}[ht!]
\newcommand{\figcap}[1]{\begin{minipage}{0.188\linewidth}\centering#1\end{minipage}}
\centering
\hspace{0.03\linewidth}\hfill%
\figcap{\footnotesize Ref. }\hfill%
\figcap{\footnotesize \textbf{Our BCSDF}}\hfill%
\figcap{\footnotesize Ours w/o $S(\phi)$ }\hfill%
\figcap{\footnotesize Fiber BCSDF}\hfill%
\figcap{\footnotesize \cite{ZhuZJYA23}}\\
\hspace{0.03\linewidth}\hfill%
\includegraphics[trim={300 500 600 400},clip,width=0.091\linewidth]{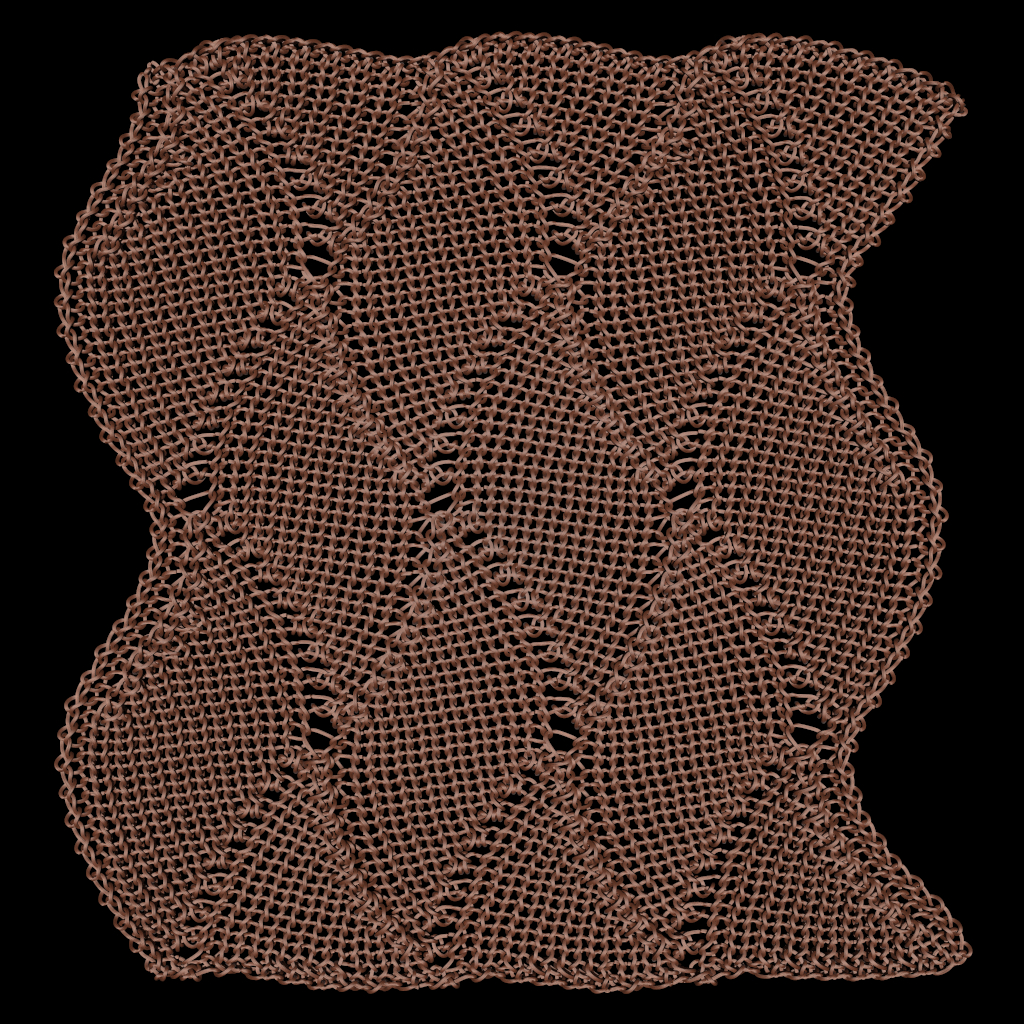}\hfill%
\hspace{0.091\linewidth}\hfill%
\includegraphics[trim={300 500 600 400},clip,width=0.091\linewidth]{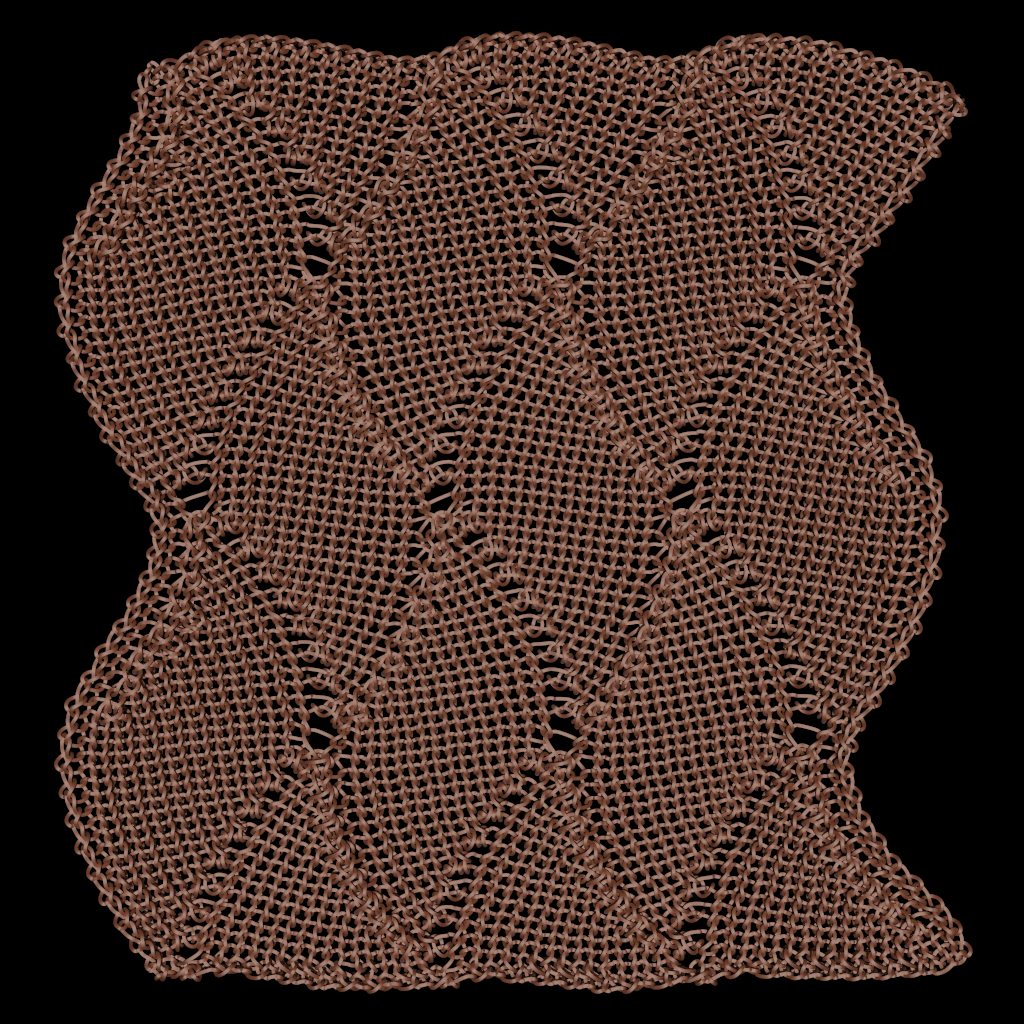}\hfill%
\includegraphics[trim={216 360 432 288},clip,width=0.091\linewidth]{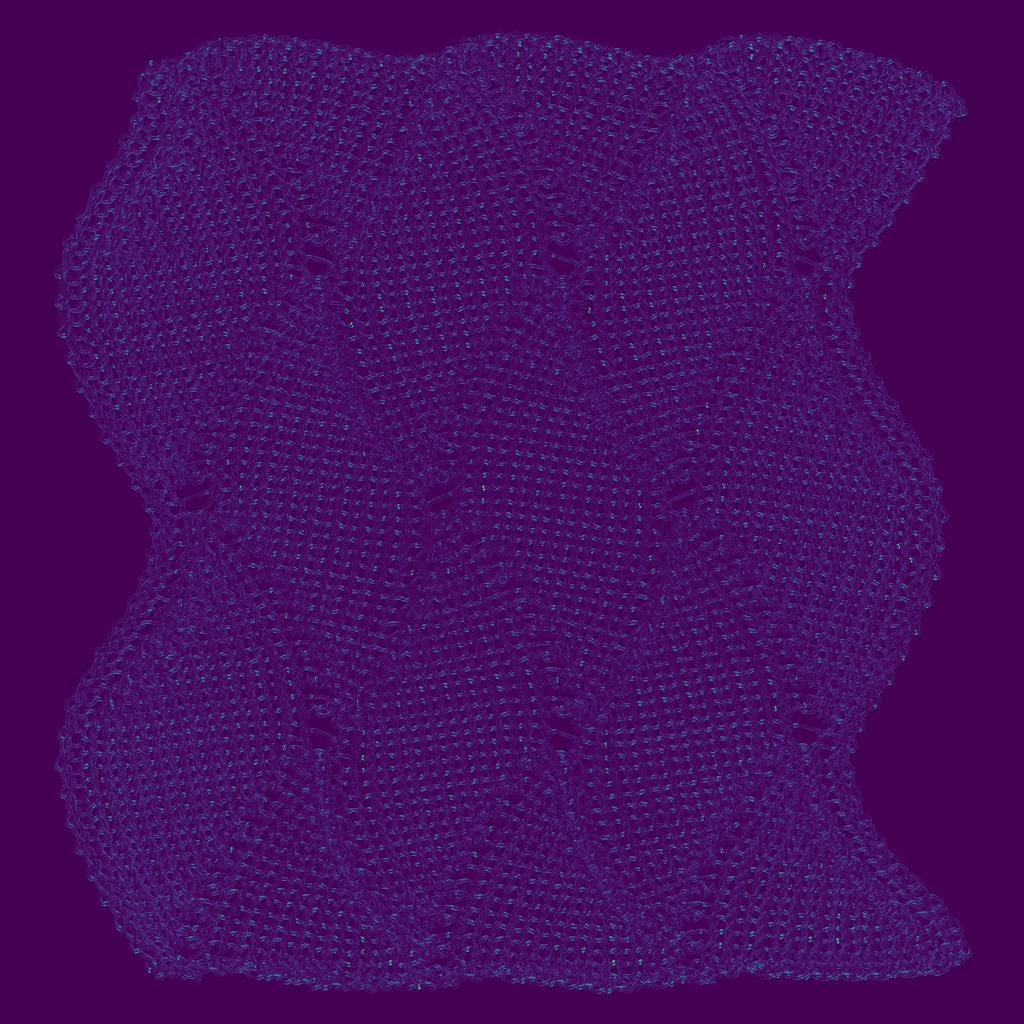}\hfill%
\includegraphics[trim={300 500 600 400},clip,width=0.091\linewidth]{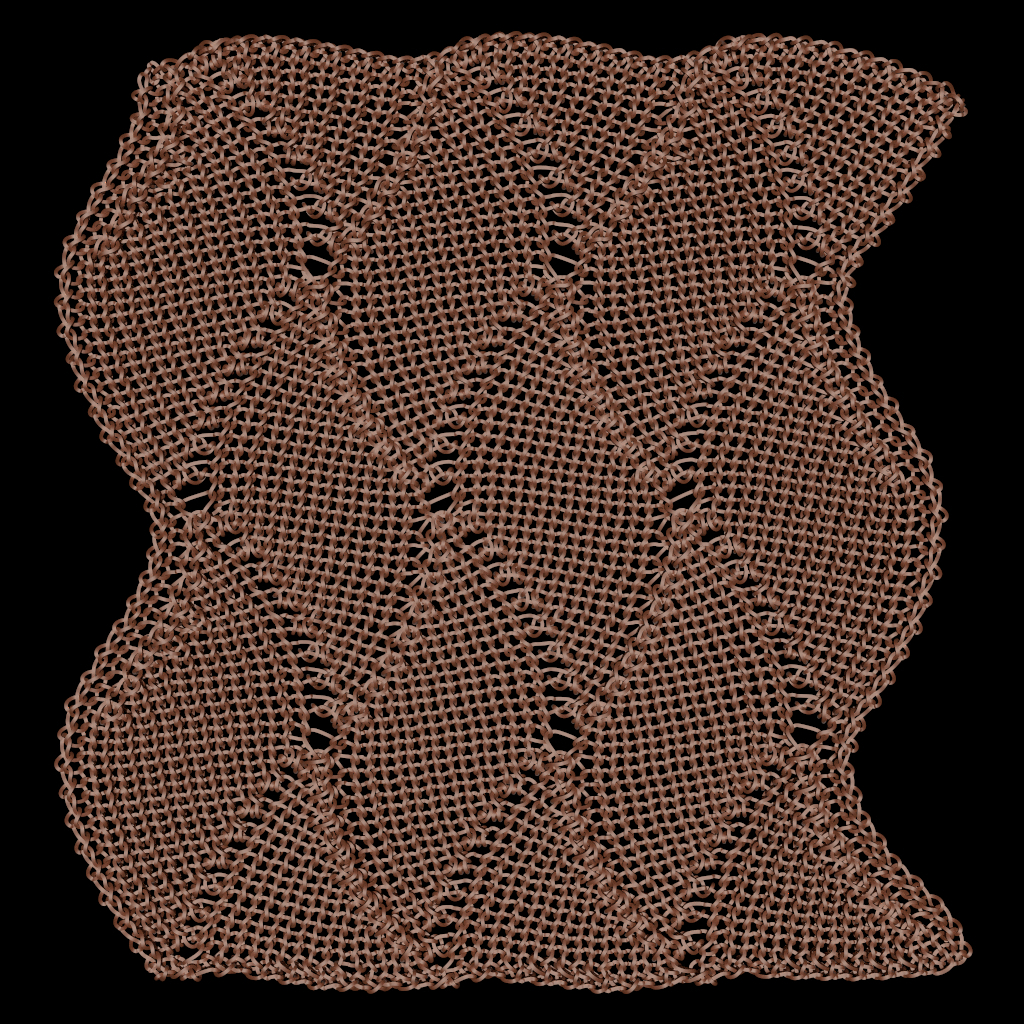}\hfill%
\includegraphics[trim={216 360 432 288},clip,width=0.091\linewidth]{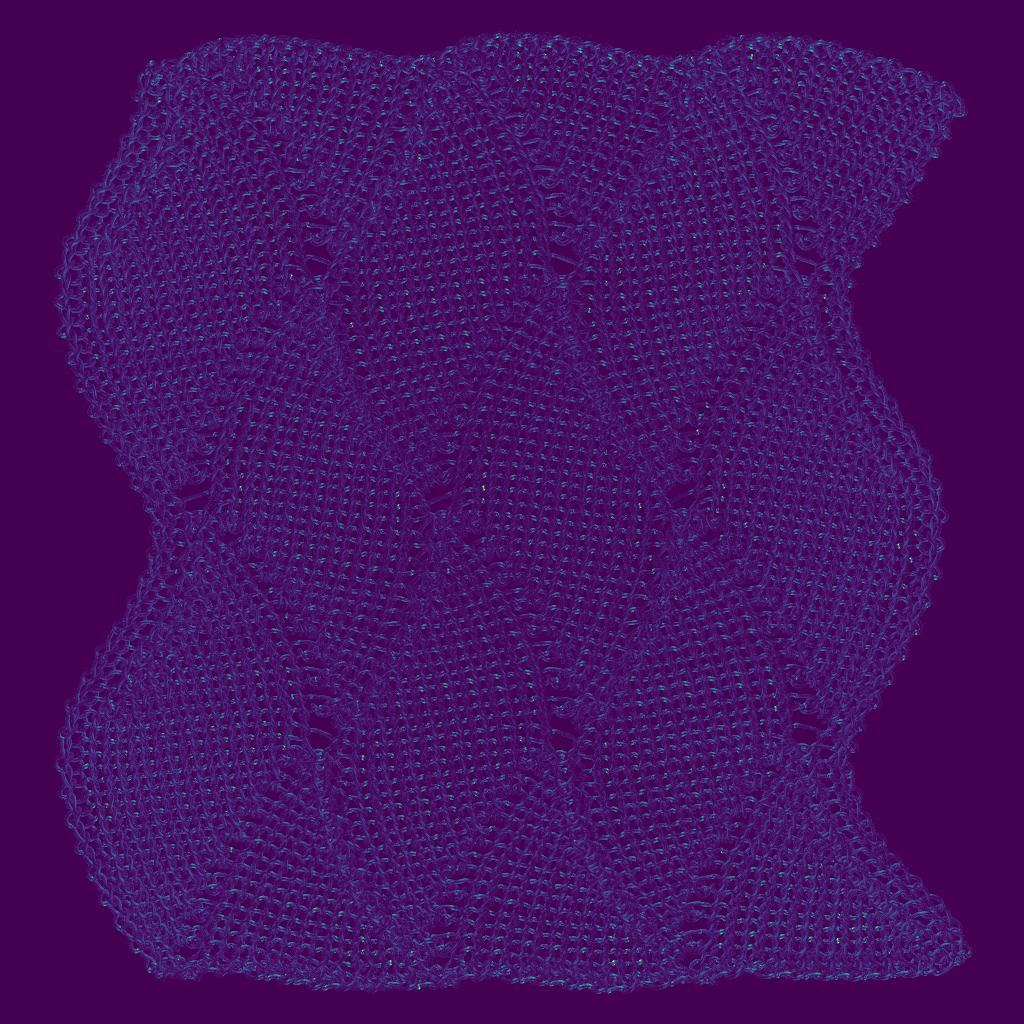}\hfill%
\includegraphics[trim={300 500 600 400},clip,width=0.091\linewidth]{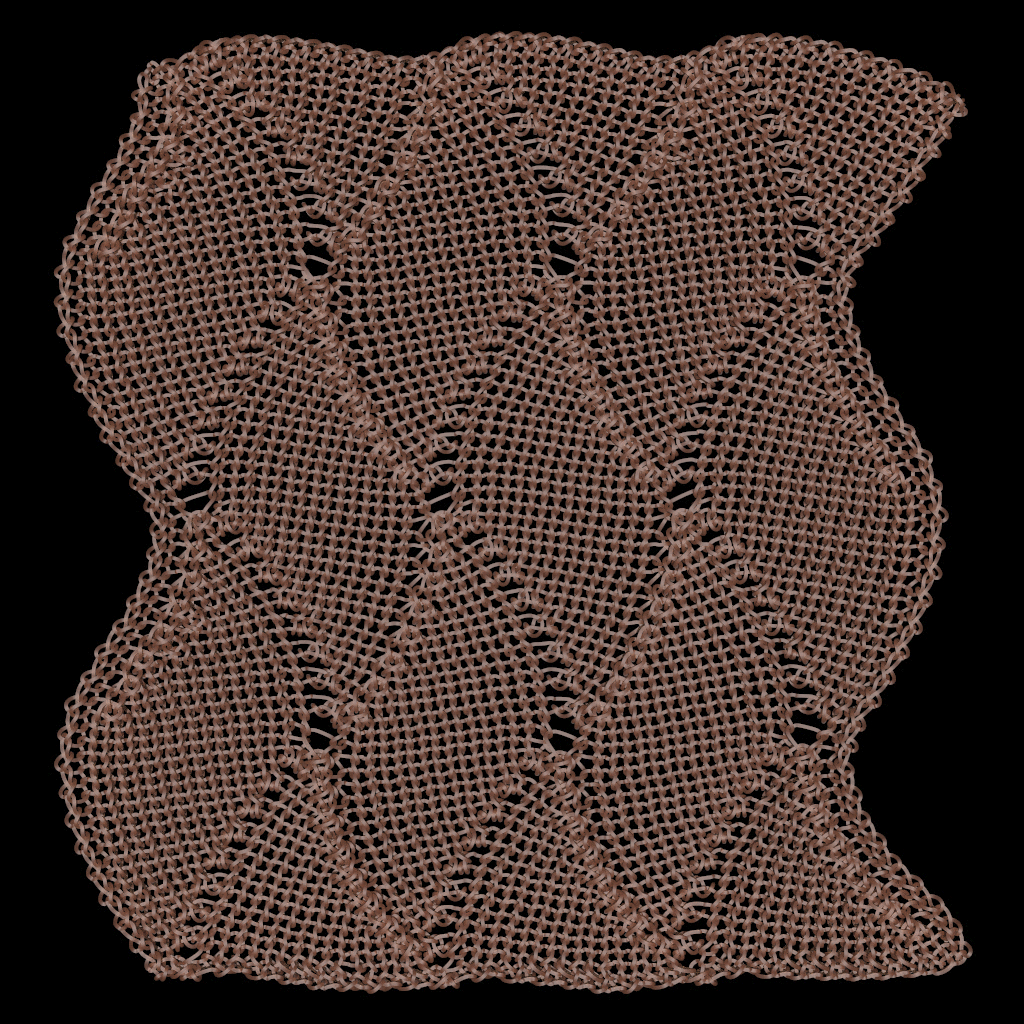}\hfill%
\includegraphics[trim={216 360 432 288},clip,width=0.091\linewidth]{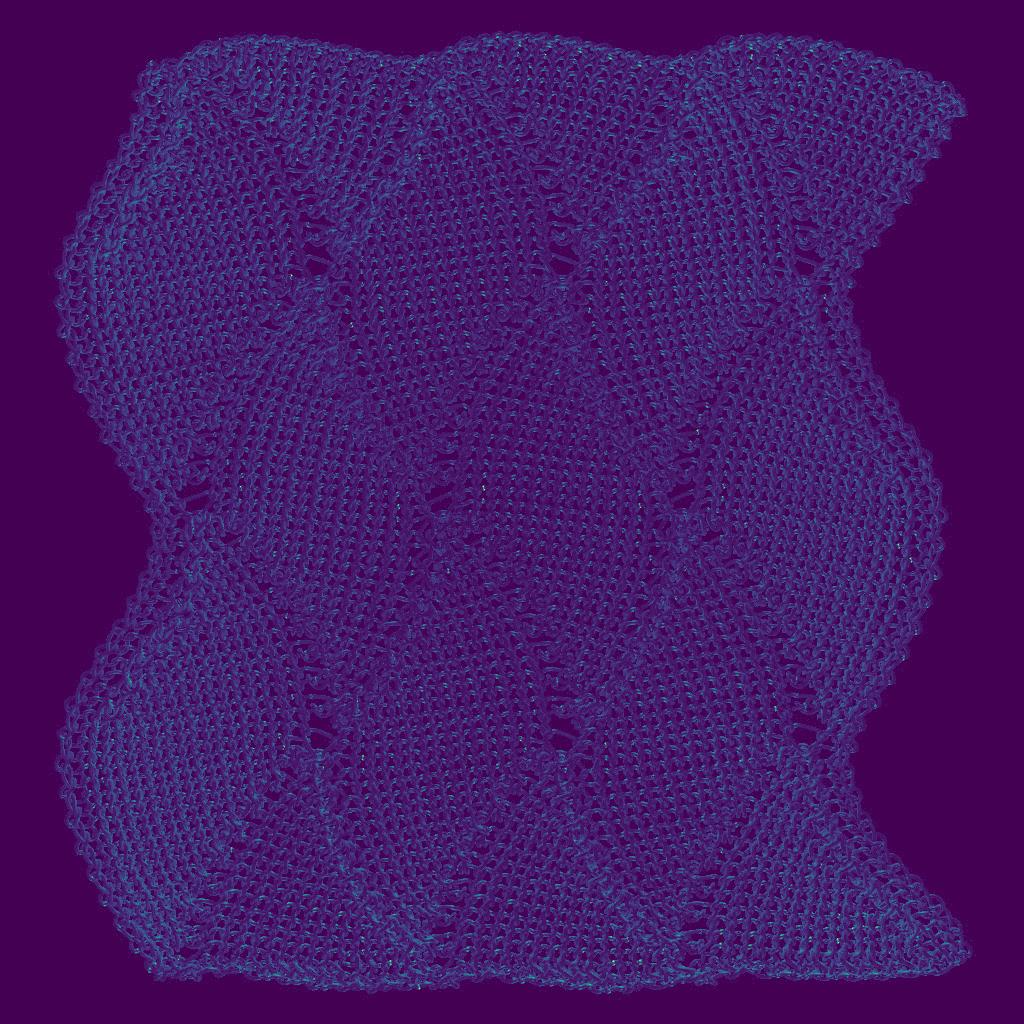}\hfill%
\includegraphics[trim={300 500 600 400},clip,width=0.091\linewidth]{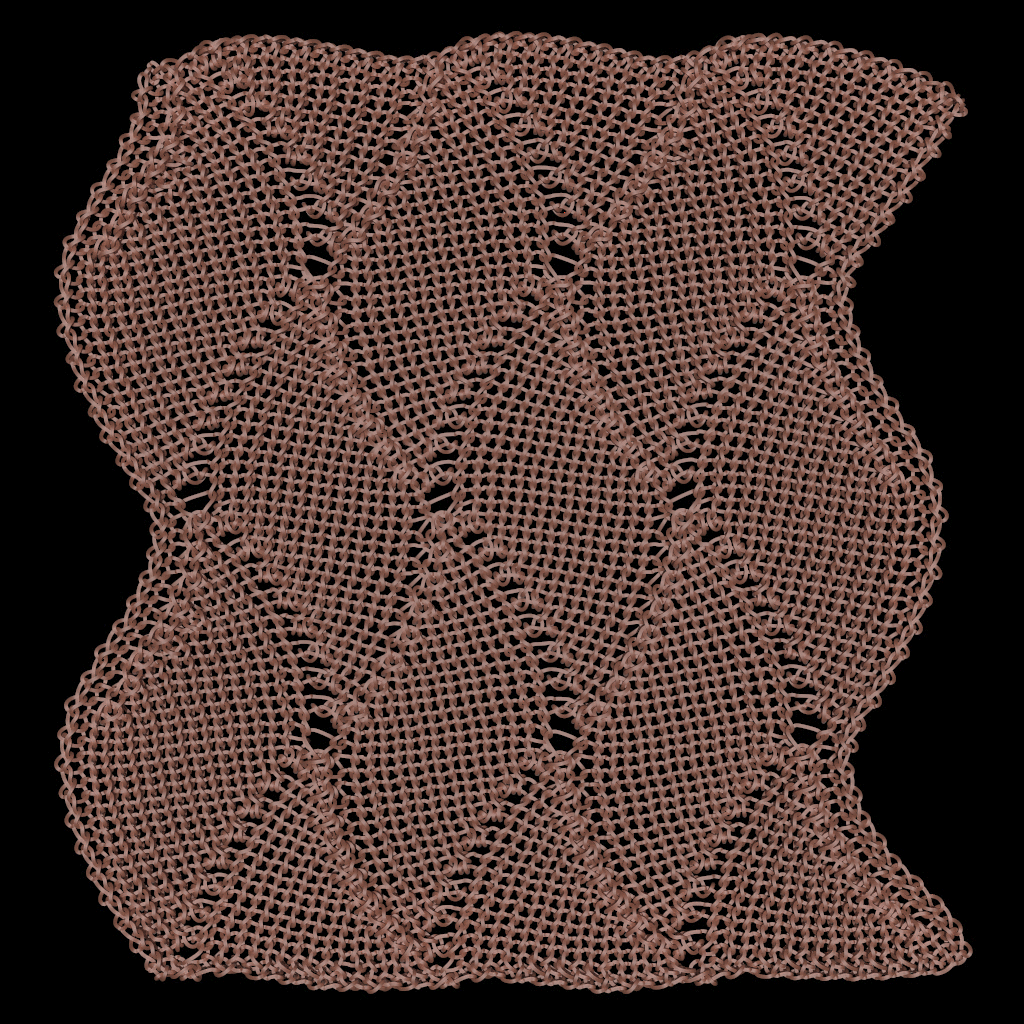}\hfill%
\includegraphics[trim={216 360 432 288},clip,width=0.091\linewidth]{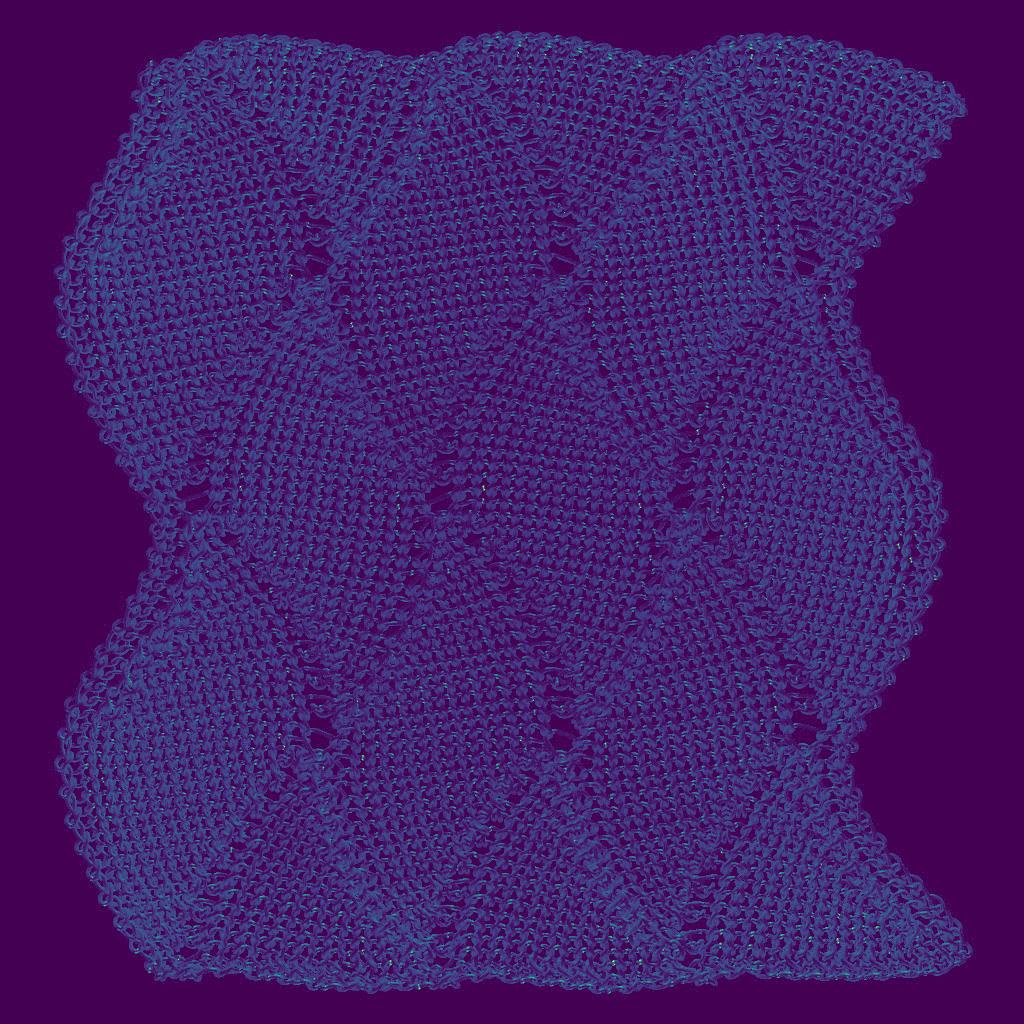}\\
\hspace{0.03\linewidth}\hfill%
\flamewithbox{0.188\linewidth}{50,150,200}{FIG/knit_notwist/Ours.jpg}\hfill\hspace{-0.5em}
\includegraphics[trim={200 200 200 200},clip,width=0.188\linewidth]{FIG/knit_notwist/Ours.jpg}\hfill%
\includegraphics[trim={200 200 200 200},clip,width=0.188\linewidth]{FIG/knit_notwist/no_G.jpg}\hfill%
\includegraphics[trim={200 200 200 200},clip,width=0.188\linewidth]{FIG/knit_notwist/no_aggregated.jpg}\hfill%
\includegraphics[trim={200 200 200 200},clip,width=0.188\linewidth]{FIG/knit_notwist/EG.jpg}\\
\vspace{-0.5em}\hspace{0.03\linewidth}\hfill%
\figcap{\small $\text{PSNR:}$ }\hfill%
\figcap{\small \textbf{62.55}}\hfill%
\figcap{\small 61.61 }\hfill%
\figcap{\small 60.66}\hfill%
\figcap{\small 59.93}\vspace{0.5em}\\
\hspace{0.03\linewidth}\hfill%
\includegraphics[trim={300 500 600 400},clip,width=0.091\linewidth]{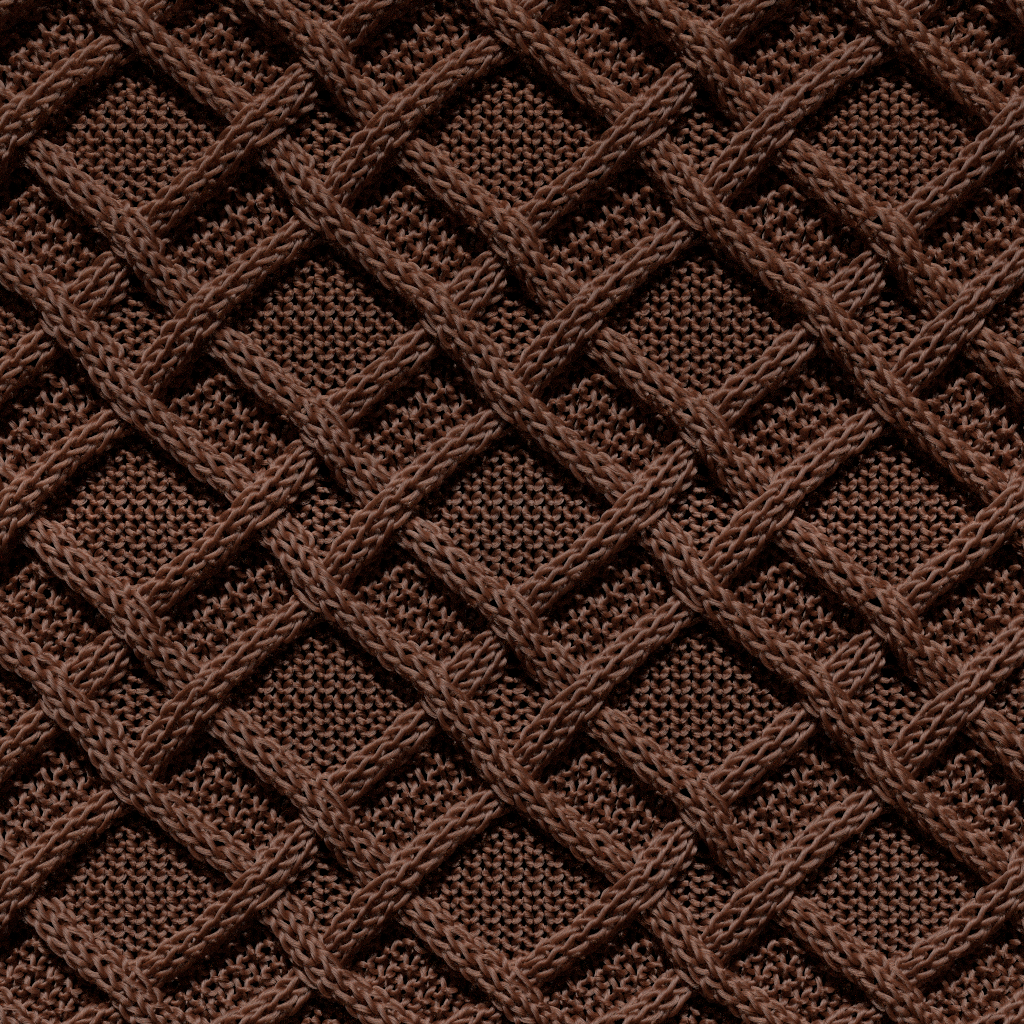}\hfill%
\hspace{0.091\linewidth}\hfill%
\includegraphics[trim={300 500 600 400},clip,width=0.091\linewidth]{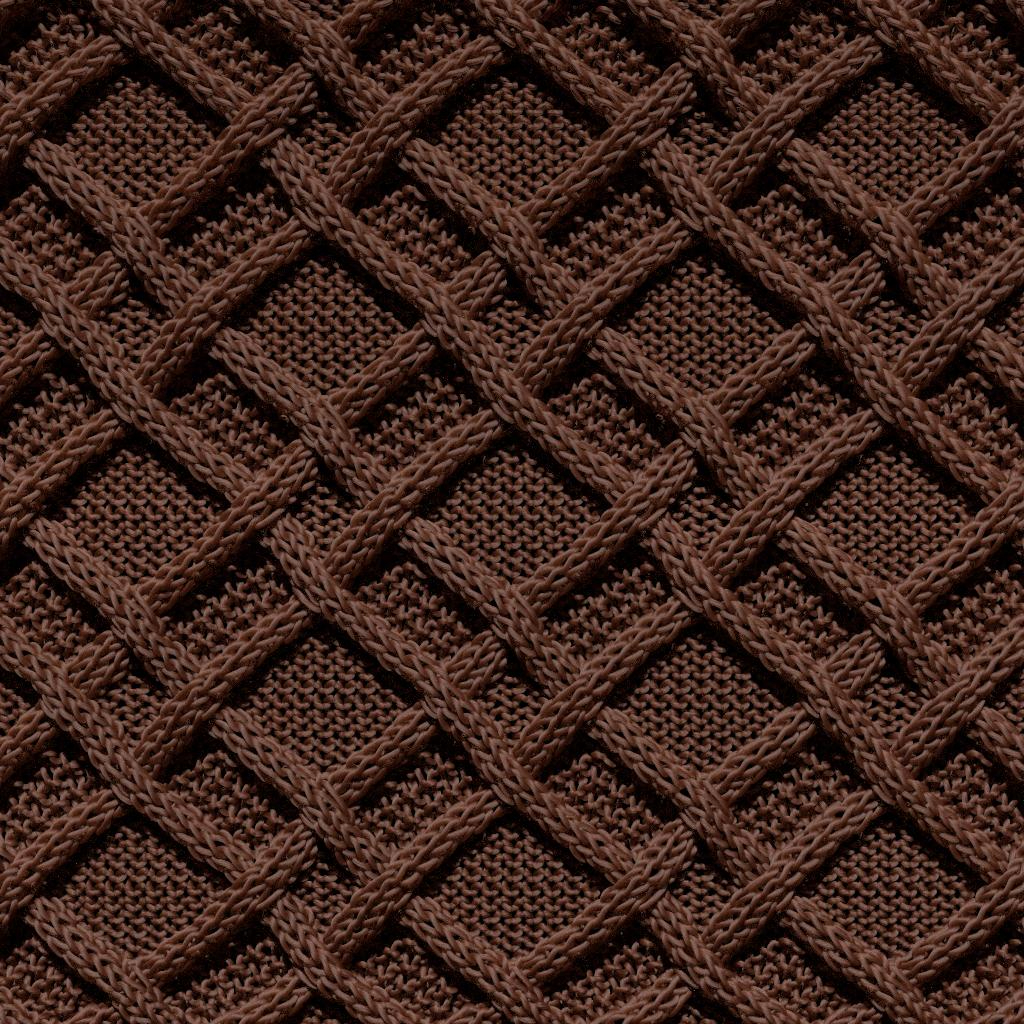}\hfill%
\includegraphics[trim={216 360 432 288},clip,width=0.091\linewidth]{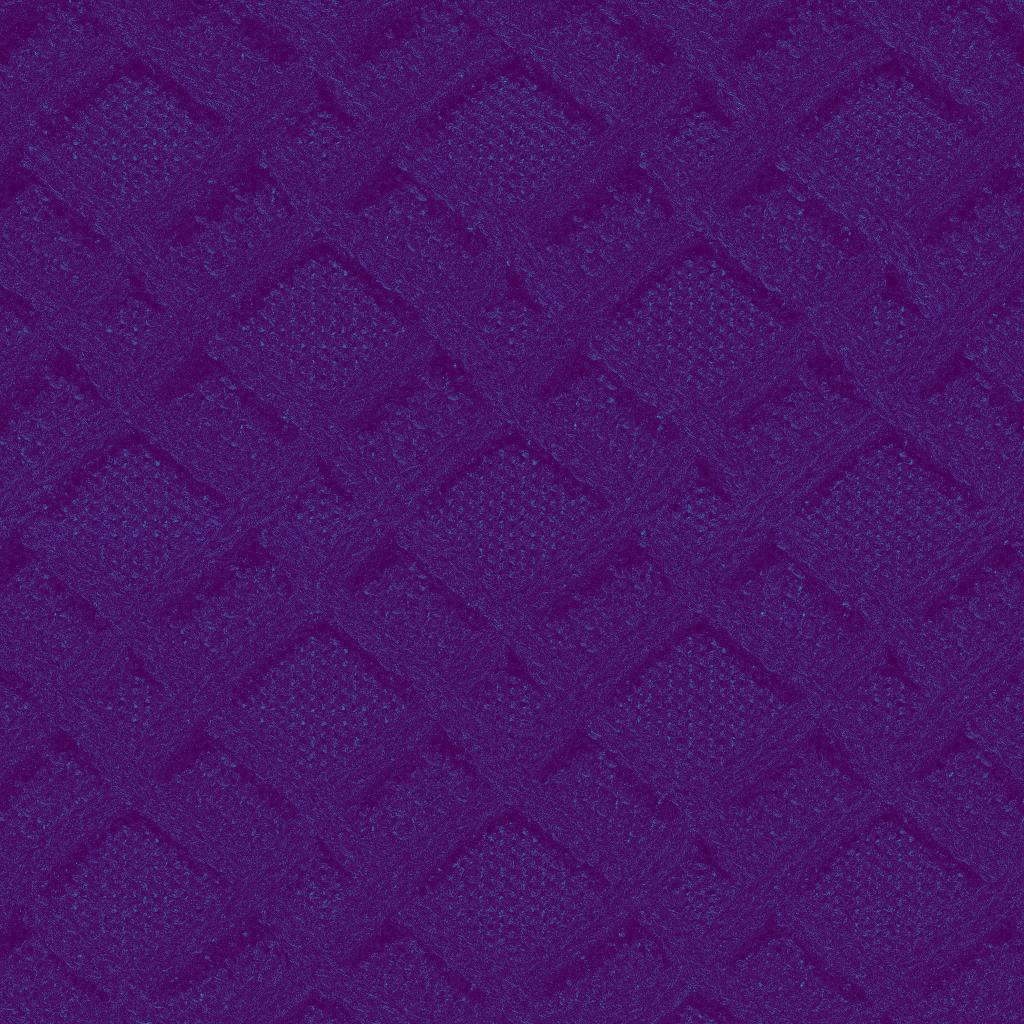}\hfill%
\includegraphics[trim={300 500 600 400},clip,width=0.091\linewidth]{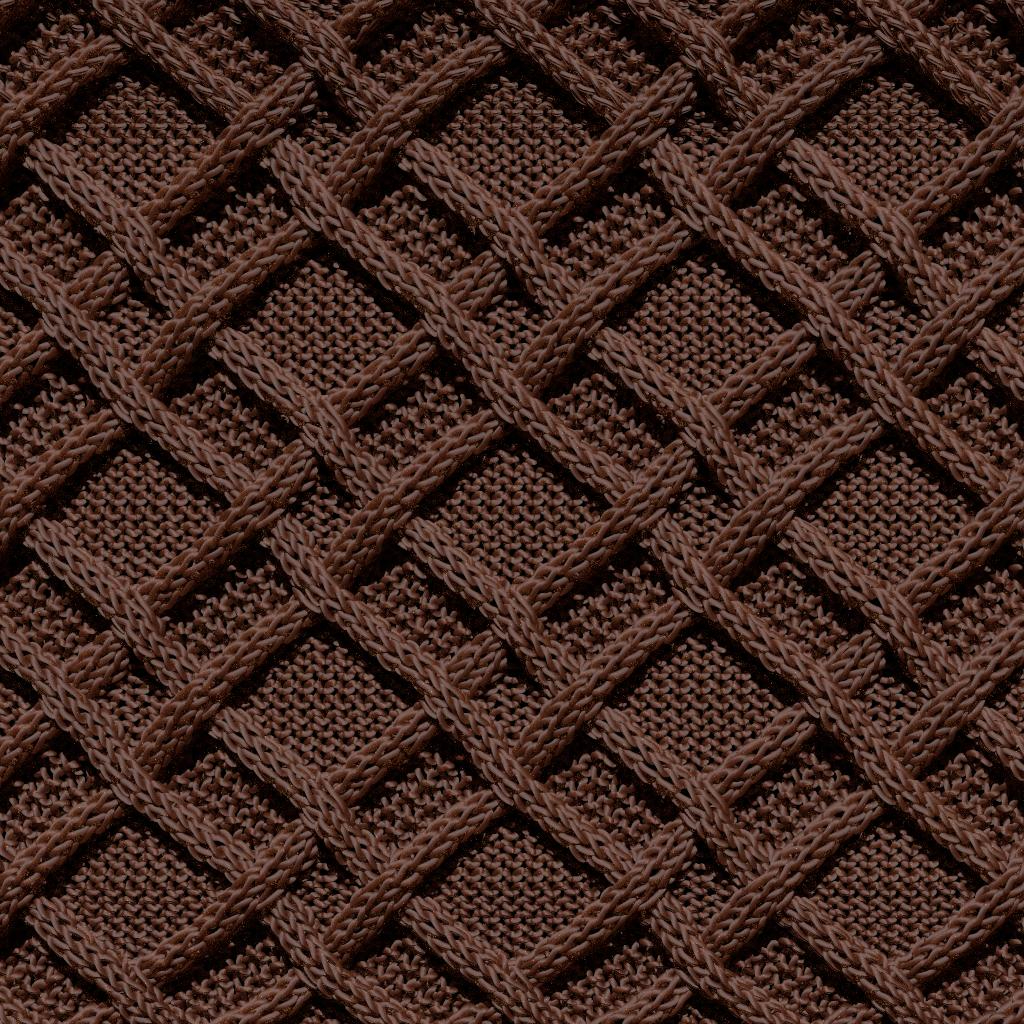}\hfill%
\includegraphics[trim={216 360 432 288},clip,width=0.091\linewidth]{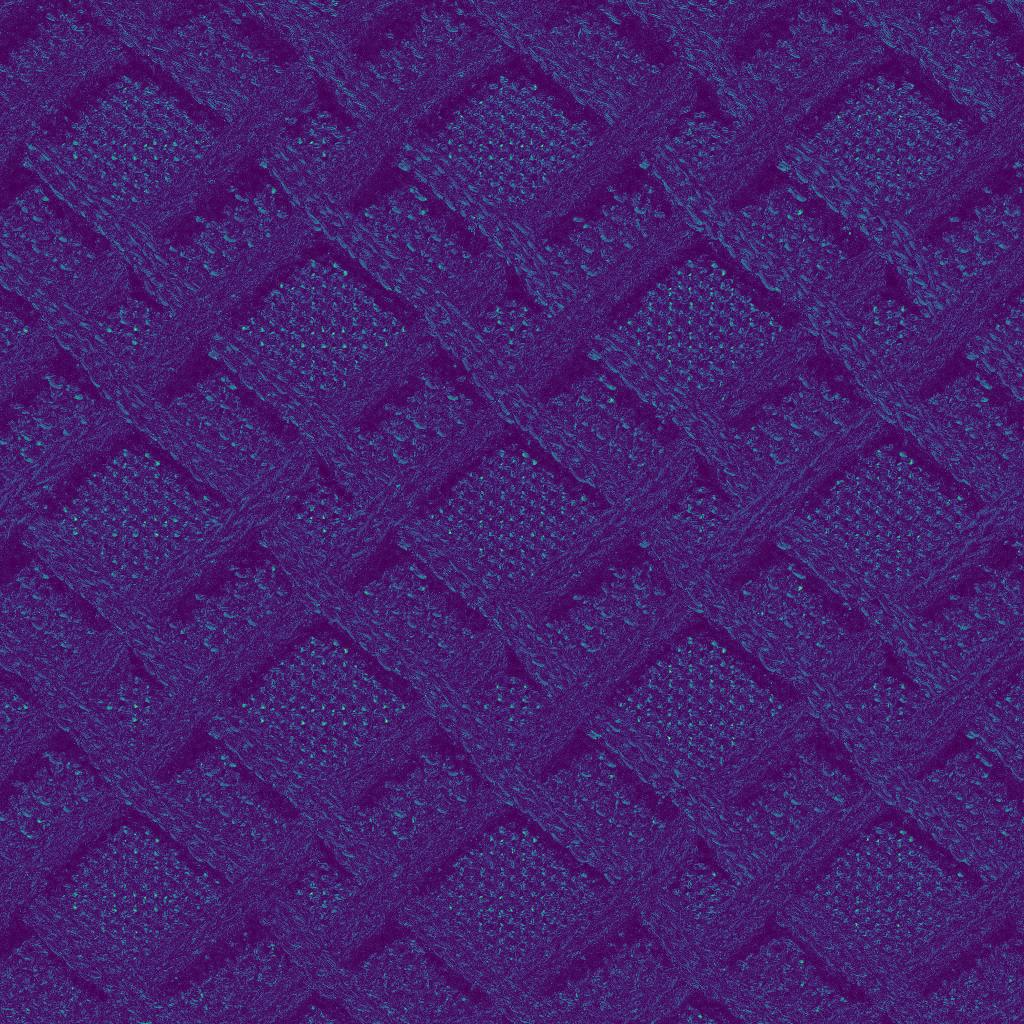}\hfill%
\includegraphics[trim={300 500 600 400},clip,width=0.091\linewidth]{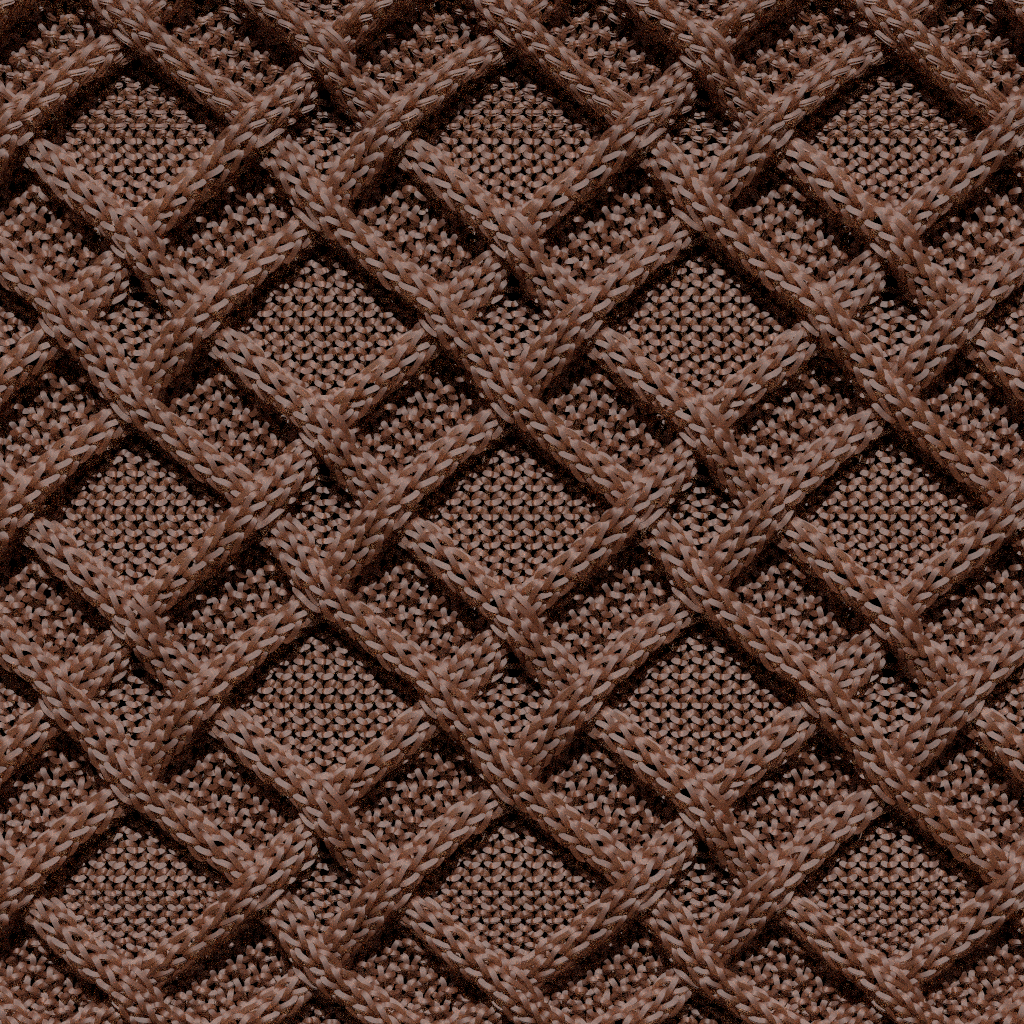}\hfill%
\includegraphics[trim={216 360 432 288},clip,width=0.091\linewidth]{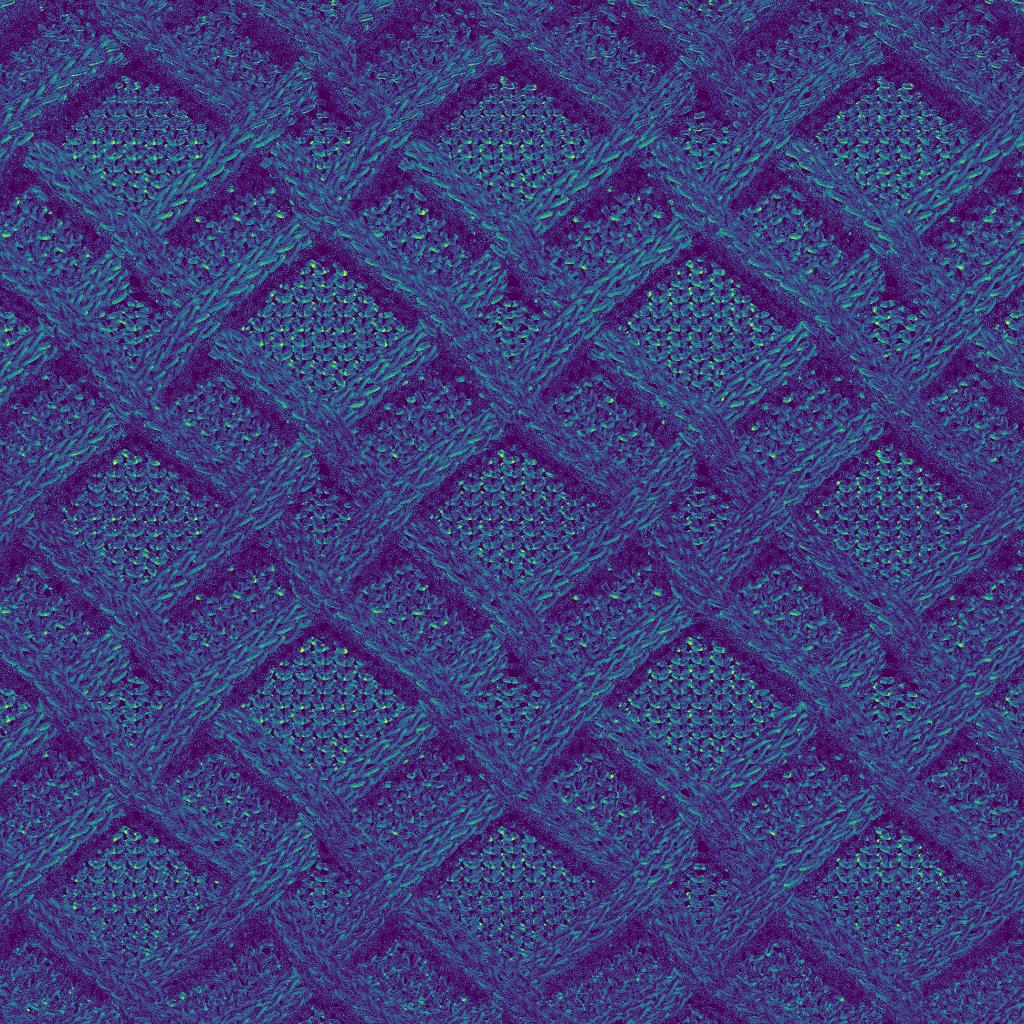}\hfill%
\includegraphics[trim={300 500 600 400},clip,width=0.091\linewidth]{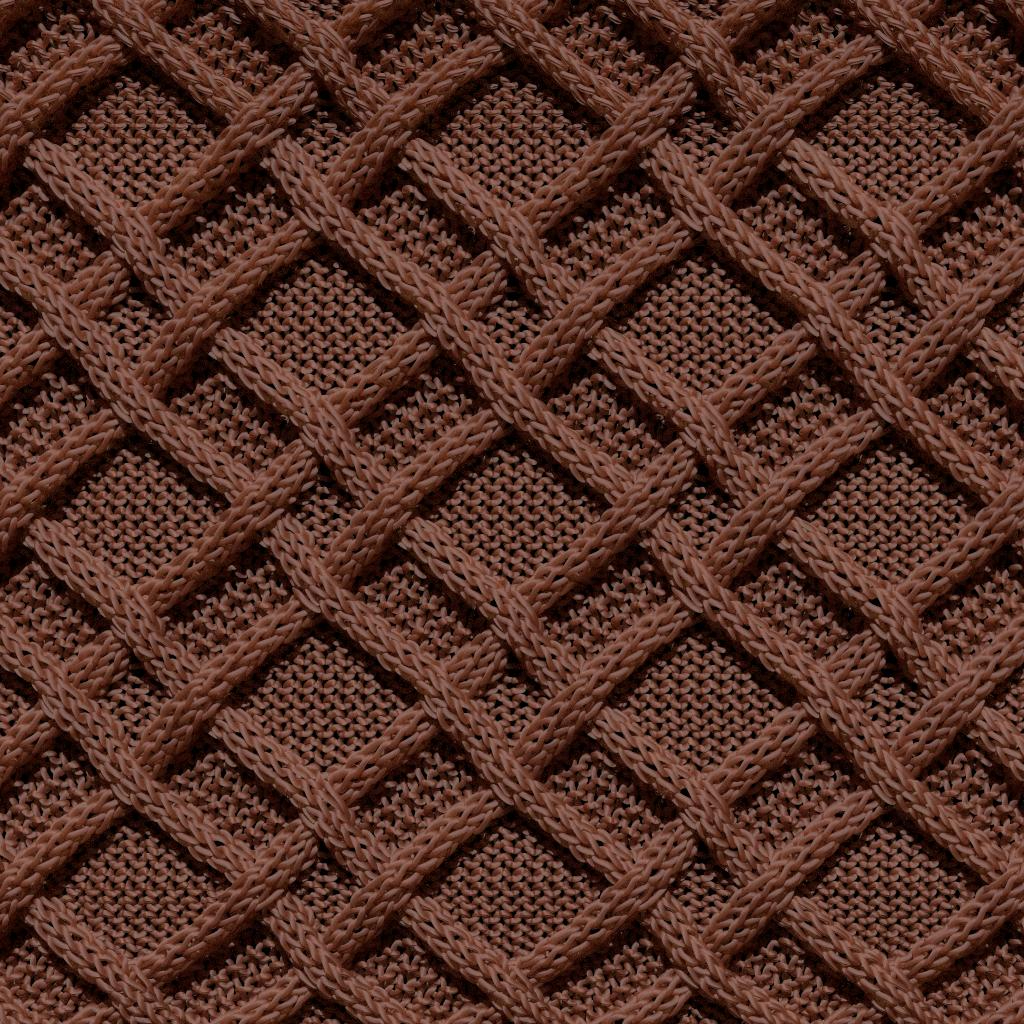}\hfill%
\includegraphics[trim={216 360 432 288},clip,width=0.091\linewidth]{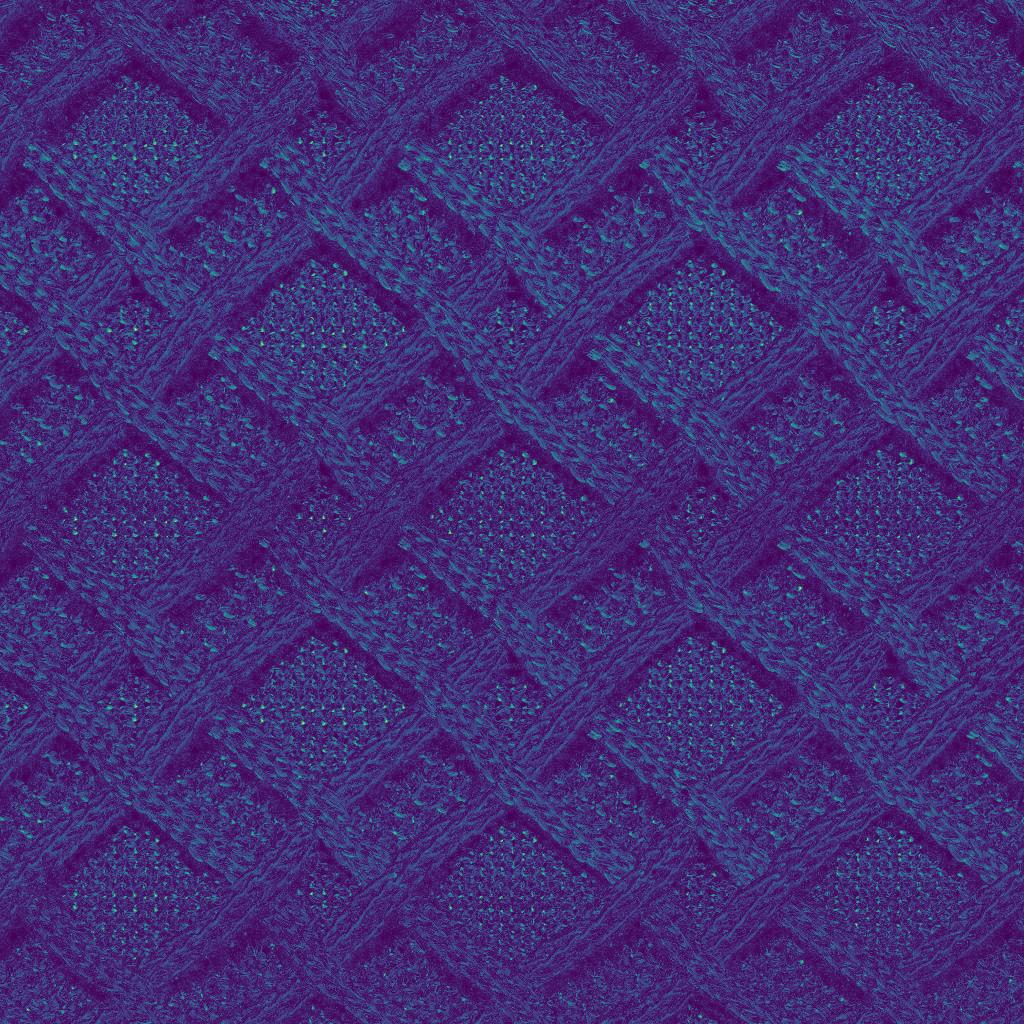}\\
\hspace{0.03\linewidth}\hfill%
\flamewithbox{0.188\linewidth}{50,150,200}{FIG/knit_twist/GT_twist.jpg}\hfill\hspace{-0.5em}
\includegraphics[trim={200 200 200 200},clip,width=0.188\linewidth]{FIG/knit_twist/Ours.jpg}\hfill%
\includegraphics[trim={200 200 200 200},clip,width=0.188\linewidth]{FIG/knit_twist/no_G.jpg}\hfill%
\includegraphics[trim={200 200 200 200},clip,width=0.188\linewidth]{FIG/knit_twist/no_aggregated.jpg}\hfill%
\includegraphics[trim={200 200 200 200},clip,width=0.188\linewidth]{FIG/knit_twist/EG.jpg}\\
\vspace{-0.5em}\hspace{0.03\linewidth}\hfill%
\figcap{\small $\text{PSNR:}$ }\hfill%
\figcap{\small \textbf{61.88}}\hfill%
\figcap{\small 59.86}\hfill%
\figcap{\small 57.12}\hfill%
\figcap{\small 58.62}\\
\vspace{-16.5em}
\begin{flushleft}{
\rotatebox{90}{\small Ply w/ twist \hspace{5em} Ply w/o twist }\hfill%
}\end{flushleft}
\vspace{1.5em}
\caption{{Knit patches:} Ref. is obtained using one fiber-level geometry instance with path tracing and single BCSDF, while others render a collection of fibers as a ply with single BCSDF and the aggregated BCSDF~\cite{ZhuZJYA23}. All results are generated using CPU path tracing. Our approach produces the closest appearance to GT (highest PSNR). Notably, the absence of the shadowing-masking term leads to a noticeable brightness mismatch. Magnified insets (highlighted in blue) and corresponding error maps are provided.}
\label{fig:knits}
\Description{}
\end{figure}

\paragraph{Knit patches} 
Our aggregated BCSDF can also be used for fabric rendering. Our method handles twisted fibers by rotating the local frame at each shading point. We evaluate the efficacy of our aggregated BCSDF using two knit patches. \autoref{fig:knits} showcases the comparison. The reference images, denoted as ref., are generated using fiber-level geometries (9.7M and 8.9M segments for each patch, respectively) with single fiber BCSDF, while other methods render ply-level geometries with significantly fewer segments (130K and 49K segments for each patch, respectively). Our aggregate BCSDF closely resembles the reference appearance. In comparison, the absence of the shadowing-masking term contributes to the observable increase in error. Using single fiber BCSDF results in significantly brighter results. The prior aggregated BCSDF~\cite{ZhuZJYA23} also results in increased brightness by double-counting $R$ and $D$ lobes.

\section{CONCLUSION}

We have presented a Level-of-detail framework for real-time strand-based hair rendering. Our approach incorporates a comprehensive BCSDF that models both single and multiple scattering interactions among individual hairs within a hair cluster. Additionally, we have outlined a method for constructing a hierarchical representation of hair and a strategy for selecting appropriate levels of detail and determining the thick hair width on the fly, minimizing geometry data for efficient real-time rendering without perceptible loss in appearance. Furthermore, we have demonstrated that our proposed system can be implemented efficiently on the GPU to render various hairstyles in real time.

\begin{figure}[ht]
\newcommand{\figcap}[1]{\begin{minipage}{0.495\linewidth}\centering#1\end{minipage}}
\includegraphics[trim=100 0 100 200, clip, width=0.499\linewidth]{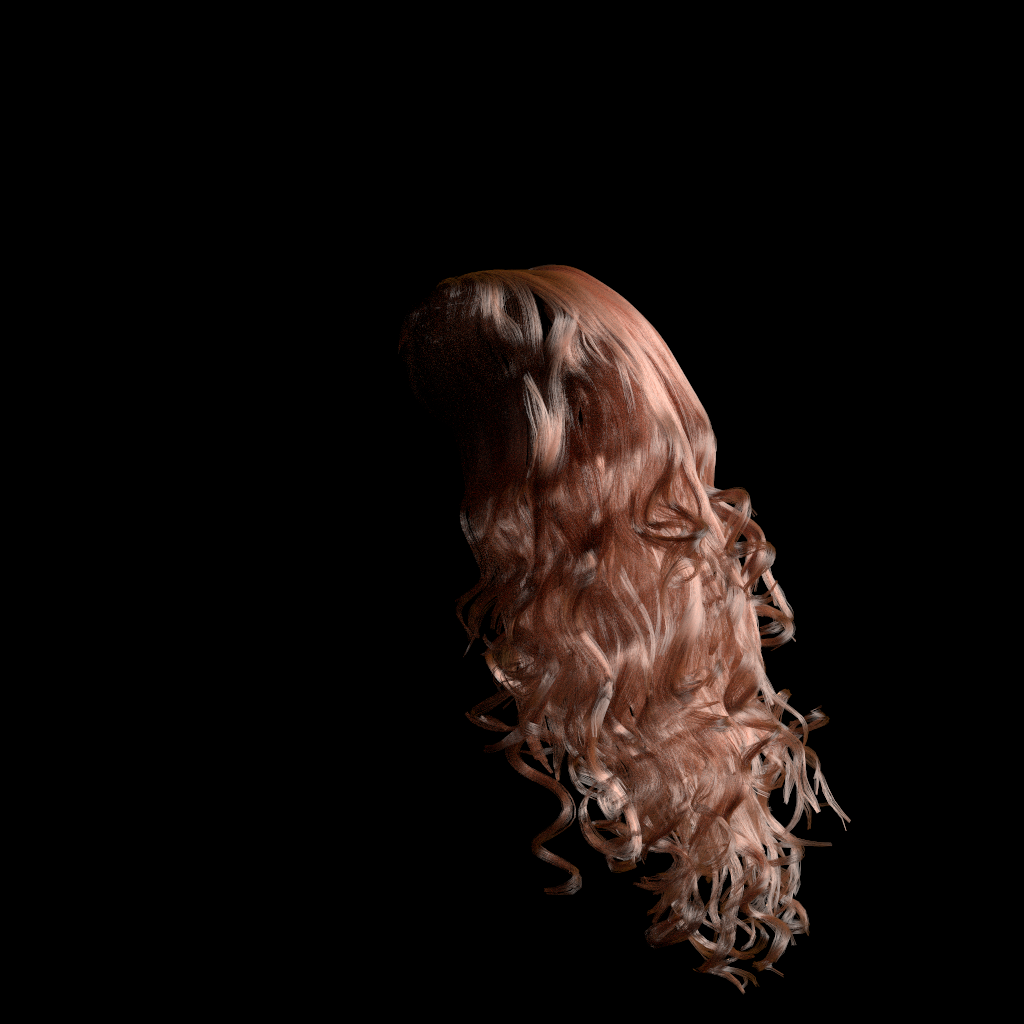}\hfill
\includegraphics[trim=100 0 100 200 , clip, width=0.499\linewidth]{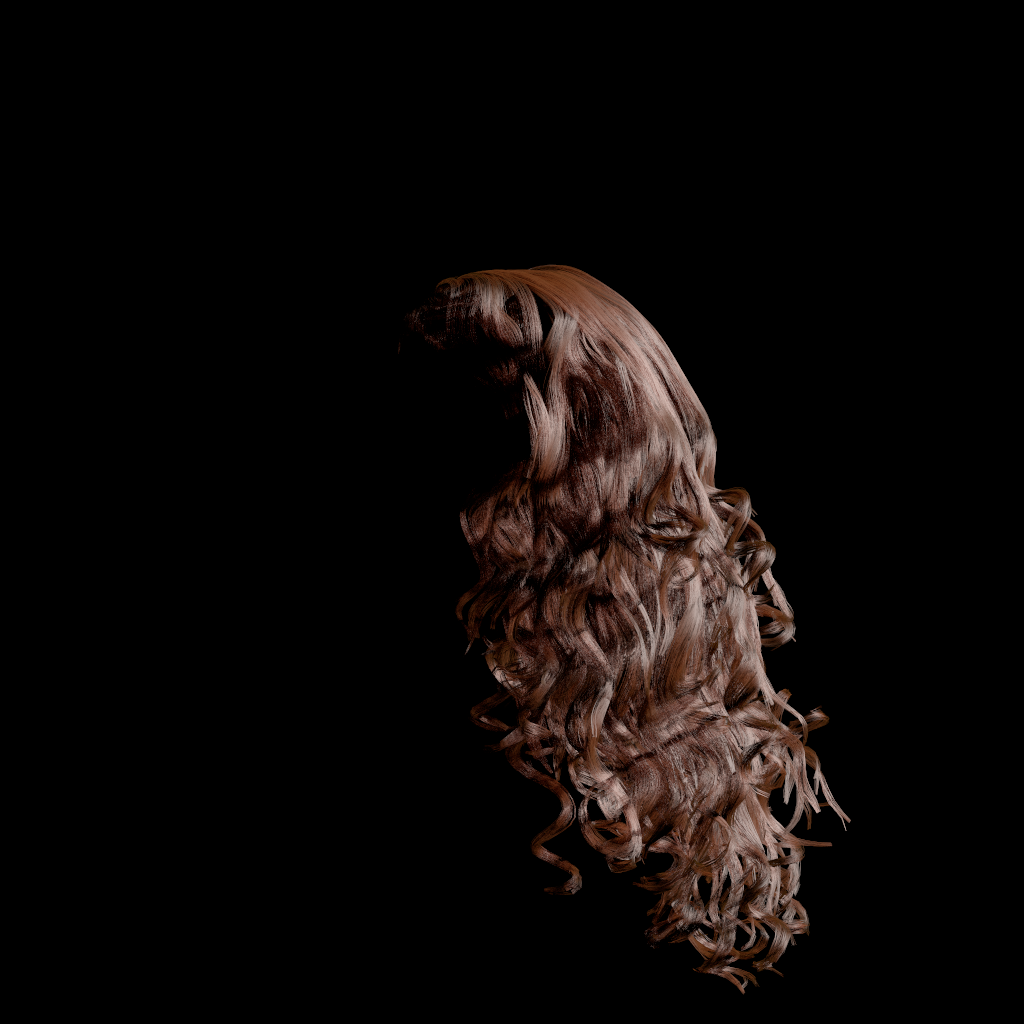}\\\vspace{-1.5em}
\figcap{\small \color{white}{Offline path tracing }}\hfill%
\figcap{\small \color{white}{\textbf{Our ray shooting}}}\\
\caption{Our method inherits limitations from dual scattering, particularly in accurately approximating the appearance of curly, light-colored hairs when illuminated from a side light direction.}
\label{fig:limit}
\Description{}
\end{figure}

\begin{figure}[th!]
\newcommand{\figcap}[1]{\begin{minipage}{0.33\linewidth}\centering#1\end{minipage}}
\includegraphics[trim=315 355 315 285, clip, width=0.33\linewidth]{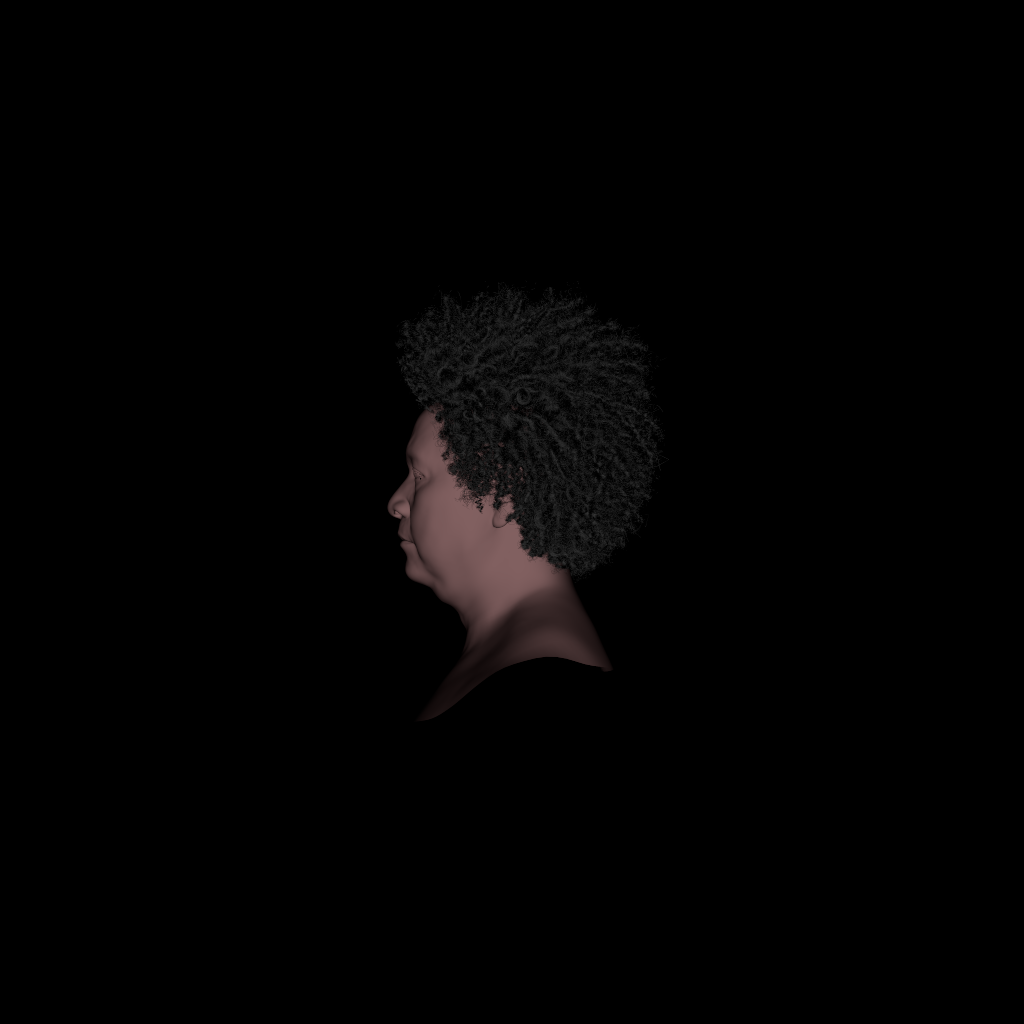}\hfill
\includegraphics[trim=315 355 315 285 , clip, width=0.33\linewidth]{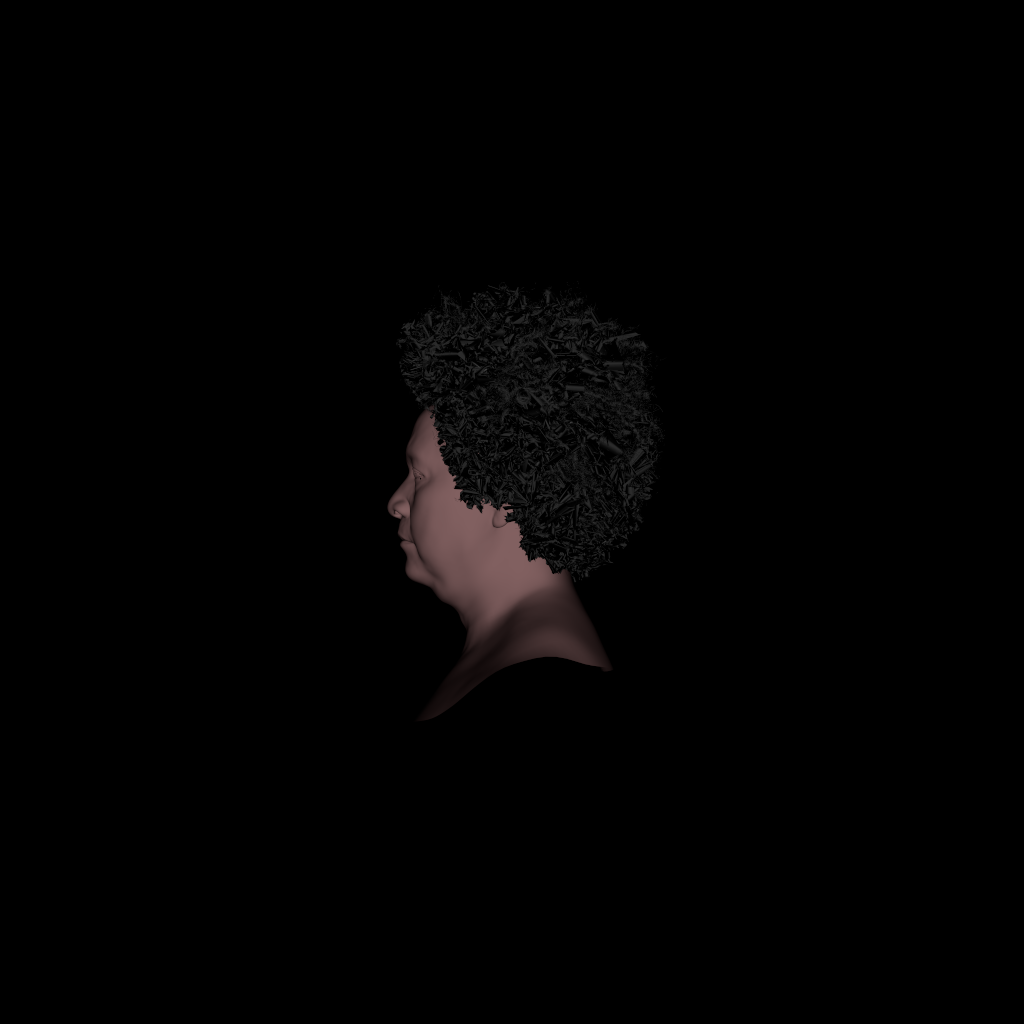}\hfill
\includegraphics[trim=315 355 315 285 , clip, width=0.33\linewidth]{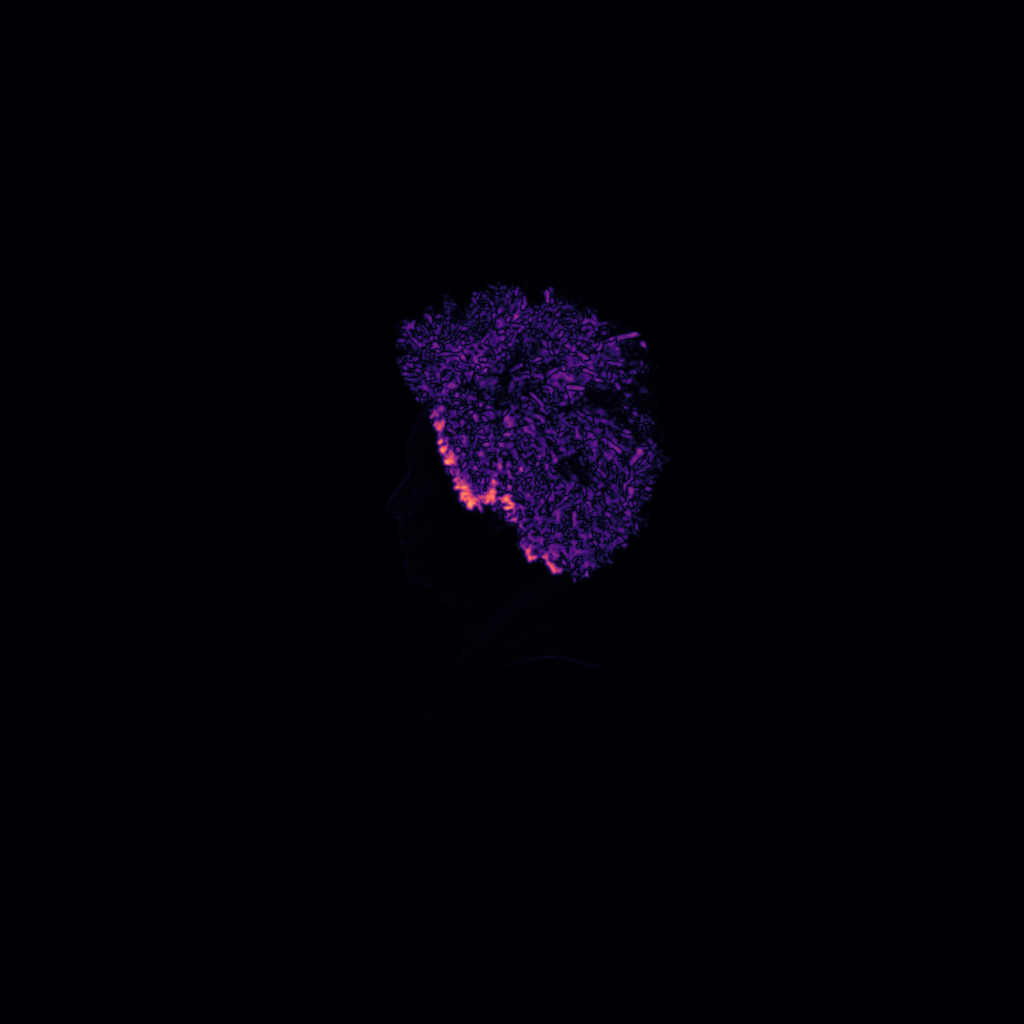}\\\vspace{-1.5em}
\figcap{\small \color{pink}{3300K}}\hfill%
\figcap{\small \color{pink}{1540K}}\hfill%
\figcap{\small \color{cyan}{0.071}}\\\vspace{+.em}
\figcap{\small {Offline path tracing }}\hfill%
\figcap{\small {\textbf{Our ray shooting}}}\hfill%
\figcap{\small {Difference}}\\
\caption{For hairstyles with highly complex geometric structures, when it comes to far view and many strands of hair are replaced by a few thick fibers, a single thick fiber is insufficient to capture all the geometric properties of these hairs, leading to some inconsistency in the results.}
\label{fig:afro}
\Description{}
\end{figure}

\paragraph{Limitations} Our method, aligning with the concept of dual scattering, shares its limitations. In particular, the local scattering in dual scattering theory assumes that the geometric information and reflection model of the fibers surrounding the shading point is essentially consistent, which is not well-suited for hairs with various textured albedo and certain hairstyles, especially curly styles with light hair color, which would introduce biases in the results, as shown in~\autoref{fig:limit}. For hairstyles with highly complex geometric structures, such as the afro shown in \autoref{fig:afro}, it is challenging to accurately maintain all the geometry details of a hair cluster using a single thick fiber. Also, our pipeline cannot preserve glints perfectly, as our LoD selection criteria are based on the assumption that the hair width is around 2 pixels on the screen to reduce hair geometries, while glints that appear usually need a sub-pixel (<0.5 pixels) details. Additionally, our current dynamic LoD selection ensures that rendered thick hairs remain within a certain width threshold (2 pixels), thereby minimizing the likelihood of one "thick hair" being seen through another. However, if the width threshold is increased, order-independent transparency (OIT) would be necessary to achieve correct rendering results, albeit at the cost of reduced performance.
%
Last, we recognize the disparity between our research prototype and established commercial engines like Unreal Engine~\cite{unrealengine}. While our method aligns with existing techniques utilizing spherical radial basis functions for environment lighting~\cite{ren2010}, it would be interesting to implement our LoD solution with aggregated BCSDF in real production and support environment lighting. 



\bibliographystyle{ACM-Reference-Format}
\bibliography{bibliography.bib}

\end{document}